\documentclass[12pt]{article}

\usepackage[dvips]{graphicx}	
\usepackage{amsbsy}		
\usepackage{amsfonts}		
\usepackage{amsmath}		
\usepackage{amssymb}		
\usepackage{upgreek}		
\usepackage{wasysym}
\usepackage{mathrsfs}
\usepackage{multirow}
\usepackage{mciteplus}
\usepackage{psfrag}
\usepackage{url}
\usepackage{float}
\usepackage{rotate}
\usepackage{cite}
\usepackage{soul}
\usepackage{dsfont}
\usepackage{afterpage}
\usepackage{color}		
\usepackage{tabularx}
\usepackage{booktabs}
\usepackage[normalem]{ulem}
\usepackage{threeparttable}
\usepackage{morefloats}
\usepackage[utf8]{inputenc}
\usepackage[font={small}]{caption}

\include{paperdef}
\graphicspath{{figs1/}{figs2/}}

\let\Re\relax
\let\Im\relax

\DeclareMathOperator{\Re}{Re}
\DeclareMathOperator{\Im}{Im}

\newcommand*{\figref}[1]{Fig.~\ref{#1}}

\newcommand{\HB}{{\tt HiggsBounds}}
\newcommand{\HS}{{\tt HiggsSignals}}
\newcommand{\HSv}[1]{{\tt HiggsSignals-#1}}
\newcommand{\HBv}[1]{{\tt HiggsBounds-#1}}
\newcommand{\CM}{{\tt CheckMATE}}
\newcommand{\FH}{{\tt FeynHiggs}}
\newcommand{\SI}{{\tt SuperIso}}
\newcommand{\CMv}[1]{{\tt CheckMATE-#1}}

\newcommand{\pMSSM}{pMSSM\,8}
\newcommand{\CL}[1]{$#1~\mathrm{C.L.}$}
\newcommand{\simMH}{125}
\newcommand{\MHexp}{125}
\newcommand{\btn}{\ensuremath{\br(B_u \to \tau \nu_\tau)}}
\newcommand{\bmm}{\ensuremath{\br(B_s \to \mu^+\mu^-)}}
\newcommand{\bsg}{\ensuremath{\br(B \to X_s \ga)}}
\newcommand{\gmt}{\ensuremath{(g-2)_\mu}}
\newcommand{\ETmiss}{E_T^\text{miss}}
\newcommand{\MS}{M_S}

\newcommand{\lowMH}{low-$\MH$}
\newcommand{\lowMHup}{low-$\MH^{\rm alt}$}
\newcommand{\lowMHuplow}{low-$\MH^{\rm alt-}$}
\newcommand{\lowMHuphigh}{low-$\MH^{\rm alt+}$}
\newcommand{\lowMHupvar}{low-$\MH^{\rm alt\,v}$}

\newenvironment{Eqnarray}%
         {\arraycolsep 0.14em\begin{eqnarray}}{\end{eqnarray}}
\def\beqa{\begin{Eqnarray}}
\def\eeqa{\end{Eqnarray}}
\def\beq{\begin{equation}}
\def\eeq{\end{equation}}
\def\eq#1{Eq.~(\ref{#1})}
\def\eqst#1#2{Eqs.~(\ref{#1})--(\ref{#2})}
\def\eqs#1#2{Eqs.~(\ref{#1}) and (\ref{#2})}

\def\eqs#1#2{Eqs.~(\ref{#1}) and (\ref{#2})}

\def\ls#1{\ifmath{_{\lower1.5pt\hbox{$\scriptstyle #1$}}}}
\def\sbma{s_{\beta-\alpha}}

\def\ifmath#1{\relax\ifmmode #1\else $#1$\fi}

\begin{document}
\mbox{}\hfill{\tt BONN-TH-2016-06}, 
{\tt DESY 16-151}, 
{\tt IFT-UAM/CSIC-16-068}, 
{\tt NIKHEF 2016-033}, 
%
{\tt SCIPP 16/10}

\def\thefootnote{\fnsymbol{footnote}}

\begin{center}
\Large\bf\boldmath
The Light and Heavy Higgs  Interpretation of the MSSM
\unboldmath
\end{center}
\vspace{-0.5cm}
\begin{center}
Philip Bechtle$^{1}$
, Howard E.~Haber$^{2}$
, Sven Heinemeyer$^{3,4,5}$
, Oscar St{\aa}l$^{6}$
,\\[.5em] 
Tim Stefaniak$^{2}$
, Georg Weiglein$^{7}$
~and Lisa Zeune$^{8,}$\footnote{Electronic addresses: 
bechtle@physik.uni-bonn.de,
haber@scipp.ucsc.edu,
Sven.Heinemeyer@cern.ch,\\
\mbox{}\hspace{11em}
tistefan@ucsc.edu, 
Georg.Weiglein@desy.de, 
lisa.zeune@nikhef.nl} \\[0.4cm] 
{\small
{\sl$^1$Physikalisches Institut der Universit\"at Bonn, 
 Nu{\ss}allee 12, D-53115 Bonn, Germany}\\[0.1cm]
 {\sl $^2$ Santa Cruz Institute for Particle Physics (SCIPP) and Department of Physics\\
 University of California, Santa Cruz, 1156 High Street, Santa Cruz, CA 95060, USA} \\[0.1cm]
{\sl$^3$Campus of International Excellence UAM+CSIC, Cantoblanco, E--28049 Madrid, Spain}\\[0.1cm]  
{\sl$^4$Instituto de F\'isica Te\'orica, (UAM/CSIC), Universidad
  Aut\'onoma de Madrid,\\ Cantoblanco, E-28049 Madrid, Spain
}\\[0.1cm]
{\sl$^5$Instituto de F\'isica de Cantabria (CSIC-UC), E-39005 Santander,
  Spain}\\[0.1cm]
{\sl $^6$ The Oskar Klein Centre, Department of Physics, 
Stockholm University,\\ SE-106 91 Stockholm, Sweden (former address)}\\[0.1cm]
{\sl $^7$ Deutsches Elektronen-Synchrotron DESY, 
 Notkestra{\ss}e 85, D-22607 Hamburg, Germany}\\[0.1cm]
{\sl $^8$Nikhef, Theory Group, 
Science Park 105, 1098 XG, Amsterdam, The Netherlands}
}
\end{center}
\vspace{0.2cm}

\renewcommand{\thefootnote}{\arabic{footnote}}
\setcounter{footnote}{0}

\begin{abstract}
  We perform a parameter scan of the phenomenological
    Minimal Supersymmetric Standard Model (pMSSM) with eight
    parameters taking into account the experimental Higgs boson 
    results from Run I of the LHC and 
    further low-energy observables. We
    investigate various MSSM interpretations of the Higgs signal at
    $125\gev$. First, we consider the case where the light $\cp$-even Higgs 
    boson of the MSSM is identified with the discovered Higgs boson. 
    In this case it can impersonate the SM
    Higgs-like signal either in the \emph{decoupling limit}, or in the limit of 
    \emph{alignment without decoupling}. In the latter case, the
    other states in the Higgs sector can also be light, offering good 
    prospects for upcoming LHC searches and for searches at future
    colliders. Second, we demonstrate that the heavy $\cp$-even
    Higgs boson is still a viable candidate to explain the Higgs
    signal --- albeit only in a highly constrained parameter region,
    that will be probed by LHC searches for the $\cp$-odd
    Higgs boson and the charged Higgs boson in the near future. As a
    guidance for such searches we provide new benchmark scenarios that
    can be employed to maximize the sensitivity of the experimental
    analysis to this interpretation.
\end{abstract}
\newpage


\section{Introduction}

The discovery of a Higgs-like scalar boson in Run~I of the Large Hadron
Collider (LHC)~\cite{ATLASdiscovery,CMSdiscovery} marks a milestone in
the exploration of electroweak symmetry breaking (EWSB).  Within
experimental and theoretical uncertainties, the properties of the new
particle are compatible with the Higgs boson of the Standard Model
(SM)~\cite{Aad:2015zhl}.  However, a variety of other interpretations
of the Higgs signal are possible, corresponding to very different
underlying physics.  Here, a prime candidate for the observed scalar boson
is a $\cp$-even Higgs boson of the Minimal
Supersymmetric Standard Model
(MSSM)~\cite{Nilles:1983ge,Haber:1984rc,Barbieri:1987xf}, as it possesses
SM Higgs-like properties over a significant part of the model parameter space with only small deviations
from the SM in the Higgs production and decay rates~\cite{Heinemeyer:2011aa}.

One of the main tasks of the LHC Run~II will be to determine whether
the observed scalar boson forms part of the Higgs sector of an extended
model.
In contrast to the SM, two Higgs doublets are needed in the MSSM to give
mass to up- and down-type fermions.  The extended Higgs sector entails
the existence of five scalar bosons, namely a light and heavy
$\cp$-even Higgs bosons, $h$ and $H$, a $\cp$-odd Higgs boson, $A$,
and a pair of charged Higgs bosons, $H^\pm$.  Mixing between the
neutral $\cp$-even and $\cp$-odd states are possible in the $\cp$-violating
case~\cite{Pilaftsis:1998pe,Pilaftsis:1998dd,Demir:1999hj,Pilaftsis:1999qt,Heinemeyer:2001qd,Schael:2006cr}, which we will not considered here.  At lowest order, the
Higgs sector of the MSSM can be fully specified in terms of the
$W$~and $Z$~boson masses, $\MW$ and $\MZ$, the $\cp$-odd Higgs boson
mass, $\MA$, and $\tb \equiv v_U/v_D$, the ratio of the two neutral Higgs
vacuum expectation values.  However, higher-order corrections are crucial for a
precise prediction of the MSSM Higgs boson properties and introduce dependences on other model parameters, see e.g.\
\citeres{Djouadi:2005gj,Heinemeyer:2004ms,Heinemeyer:2004gx} for
reviews.

Many fits for the Higgs rates in various models and within the
effective field theory approach have been performed over the last
years, see e.g.~\citere{Bechtle:2014ewa,Englert:2014uua}.  Focusing on the
MSSM, recent fits have shown that the interpretation of the observed
scalar as the light $\cp$-even MSSM Higgs boson (``light Higgs case'')
is a viable possibility, providing a very good description of all
data~\cite{Bechtle:2012jw,Scopel:2013bba,Djouadi:2013lra,Cheung:2015uia,Bechtle:2015pma,Bhattacherjee:2015sga};
see also~\citeres{deVries:2015hva,Henrot-Versille:2013yma,Bertone:2015tza}
for global fits including also astrophysical data.
In the light-Higgs case, decoupling of the heavy Higgs bosons
($\MA \gg \MZ$)~\cite{Haber:1989xc,Haber:1995be,Dobado:2000pw,Gunion:2002zf}
naturally explains the SM-like couplings of the light MSSM Higgs boson
$h$~\cite{Heinemeyer:2011aa}.
Another interesting possibility to explain the SM-like behavior of
$h$ without decoupling in the MSSM---the so-called limit of \emph{alignment without decoupling}---has been outlined
in~\citeres{Carena:2013ooa,Carena:2014nza}, relying on an (accidental)
cancellation of tree-level and loop contributions in the $\cp$-even
Higgs boson mass matrix.  This led to the definition
of a specific benchmark scenario~\cite{Carena:2014nza}, which
has since been ruled out in the interesting low $\MA$ region
via $pp \to H/A \to \tau^+\tau^-$ searches~\cite{Bechtle:2015pma}.

Alternatively, it was demonstrated that the heavy $\cp$-even
Higgs boson can also be identified with the observed
signal~\cite{Heinemeyer:2011aa,Hagiwara:2012mga,Benbrik:2012rm,Drees:2012fb,Han:2013mga,Bechtle:2012jw}
(``heavy Higgs case'').%
\footnote{Such a situation is more common in extensions of the MSSM. In
particular, in the NMSSM it occurs generically if the singlet-like
$\cp$-even state is lighter than the doublet-like Higgs bosons, see e.g.\ 
\citeres{Cao:2012fz,King:2014xwa,Domingo:2015eea,Ellwanger:2015uaz,Drechsel:2016jdg}.}
In this scenario all five MSSM Higgs bosons
are relatively light, and in particular the lightest $\cp$-even
Higgs boson has a mass (substantially) smaller than $\simMH\gev$
with suppressed couplings to gauge bosons.  This led to the
development of the \emph{low-$\MH$} benchmark
scenario~\cite{Carena:2013ytb}. This particular scenario has meanwhile
been ruled out by ATLAS and CMS searches for a light charged
Higgs boson~\cite{Aad:2014kga,Khachatryan:2015qxa}. However the heavy
Higgs interpretation in the MSSM
remains viable, as we will discuss in this paper.

The questions arise, whether, and if so by how much, the MSSM can
improve the theoretical description of the experimental data compared
to the SM, and which parts of the MSSM parameter space are favored.  In a
previous analysis~\cite{Bechtle:2012jw} we analyzed these questions
within the MSSM.  We performed a scan over the seven most
relevant parameters for MSSM Higgs boson phenomenology, taking into
account the data up to July 2012, which showed in particular some
enhancement in the measured rate for $h \to \ga\ga$.
We found that both the light and the heavy Higgs case provided a good
fit to the data.  In particular, the MSSM light Higgs case gave a
better fit than the SM when the data in the $\ga\ga$ channel and
low-energy data was included.

The situation has changed in several respects with the release of additional Higgs data by the ATLAS and CMS Collaborations~\cite{Khachatryan:2016vau}.  In
particular, the final data obtained in the LHC Run~I does not show a
significant enhancement over the SM prediction in the $\ga\ga$ channel anymore, and
the heavy Higgs case is much more restricted due to light charged
Higgs boson searches.  The main aim of the present paper is to study the
MSSM Higgs sector in full detail taking into account 
the current experimental data and in particular the final LHC Run I results, and to propose paths towards a complete exploration of the
heavy Higgs case at the LHC in the ongoing Run~II.  We
incorporate the available 
measurements of the Higgs boson mass and signal strengths,
as well as measurements of the relevant low-energy
observables. Furthermore we take into account all relevant
constraints from
direct Higgs and supersymmetric (SUSY) particle searches.  We investigate whether
the MSSM can still provide a good theoretical description of the
current experimental data, and 
which parts of the parameter space of the MSSM are favored.
Within the light Higgs case we analyze the situation with very large
$\MA$ (decoupling), as well as for small/moderate $\MA$ (alignment without decoupling). We
also investigate the feasibility of the heavy Higgs case and define
new benchmark scenarios in which this possibility is realized, in
agreement with all current Higgs constraints. 

The paper is organized as follows.
We employ the phenomenological MSSM with 8 parameters (\pMSSM),
which is introduced in detail in \refse{sect:theory}. In 
this section, we also expand upon the theoretical background of the two possible limits that lead to alignment in the $\cp$ even Higgs sector, i.e.~when one of the $\cp$-even neutral MSSM Higgs 
bosons behaves like the SM Higgs boson. In particular, we outline how leading two-loop effects on the conditions for alignment can be assessed and present a brief quantitative discussion of these effects.\footnote{More details will be presented in a separate publication~\cite{preparation}.} The parameter scan with ${\cal O}(10^7)$ sampling points, the techniques to achieve good coverage, as well as the considered experimental observables and
constraints are described in \refse{sec:ParObsConstraints}. In
\refse{Sect:Results} we present our results for the best-fit points
and the preferred parameter regions for the light Higgs and the
  heavy Higgs interpretation. The effects of the Higgs mass and Higgs
  rates measurements, precision observables, and direct Higgs and SUSY
  searches are discussed, and the phenomenology of the other MSSM Higgs
  states is outlined. In particular, in \refse{Section:HHbenchmark}
  we propose new benchmark scenarios for the study of the heavy Higgs
  case, which can be probed at the LHC Run~II. 
  We conclude in  \refse{Sect:Summary}.  In Appendix A, we discuss the extent of the tuning associated with the 
 regions of the MSSM parameter space that exhibit approximate Higgs alignment without decoupling.  
 Finally in Appendix B, we provide tables listing the
 signal strength measurements from ATLAS, CMS and the Tevatron (D\O\ and CDF) that are 
included in our analysis.

%

\section{Theoretical Background}
\label{sect:theory}
\subsection{The MSSM Higgs sector}\label{sect:theoryparam}
In this section we briefly review the most important features of the
MSSM Higgs sector and motivate the choice of the eight free pMSSM parameters in our scan.
We provide a detailed description of the relevant MSSM parameter sectors and our notations, which remain unchanged compared to~\cite{Bechtle:2012jw}.

In the supersymmetric extension of the SM, an even number of Higgs multiplets consisting of pairs of Higgs doublets with opposite hypercharge is required to avoid anomalies due to the supersymmetric Higgsino partners.
Consequently the MSSM employs two Higgs doublets, denoted by $H_D$ and $H_U$, with hypercharges $-1$ and $+1$, respectively.   After minimizing the scalar potential, the neutral components of 
$H_D$ and $H_U$ acquire vacuum expectation values (vevs), $v_D$ and $v_U$.   Without loss of generality, we assume that the vevs are real and non-negative (this can be achieved by appropriately rephasing the Higgs doublet fields).   The vevs are normalized such that
\beq
v^2\equiv v_D^2+v_U^2\simeq(246~{\rm GeV})^2\,.
\eeq
In addition, we define 
\beq \label{tanbeta}
\tb\equiv v_U/v_D\,.
\eeq
Without loss of generality, we may assume that $0\leq\beta\leq\tfrac12\pi$ (i.e., $\tan\beta$ is non-negative).  This can always be achieved by a rephasing of one of the two Higgs doublet fields.

The two-doublet Higgs sector gives rise to five physical Higgs states.
The mass eigenstates correspond to the neutral Higgs bosons $h$, $H$ (with $M_h<M_H$) and $A$, and the charged Higgs pair $H^\pm$. 
Neglecting possible $\cp$-violating contributions of the 
soft-supersymmetry-breaking terms (which can modify the neutral Higgs properties at the loop level),
$h$ and $H$ are the light and heavy $\cp$-even Higgs bosons,
and $A$ is $\cp$-odd.

At lowest order,  the MSSM Higgs sector is fully described by 
$\MZ$ and two MSSM parameters, 
often chosen as the $\cp$-odd Higgs boson mass, $\MA$, and $\tb$.
In the MSSM at the tree-level the mass of the light $\cp$-even Higgs
boson does not exceed $M_Z$. However, 
higher order corrections to the Higgs masses are known to be sizable
and must be included, in order to be consistent with the observed Higgs signal at
$\MHexp \gev$~\cite{Aad:2015zhl}.
Particularly important are the one- and two-loop contributions from top quarks and their scalar top (``stop'') partners. 
In order to shift the mass of $h$ up to $\MHexp \gev$,
large radiative corrections are necessary, which
require a large splitting in the stop sector and/or heavy stops.
For large values of $\tb$, the sbottom contributions to the radiative corrections also become sizable. 
The stop (sbottom) sector is governed by the soft SUSY-breaking mass
parameter $\MstL$ and $\MstR$ ($\MsbL$ and $\MsbR$), where SU(2) gauge
invariance requires $\MstL=\MsbL$, 
the trilinear coupling $A_t$ ($A_b$) and the Higgsino mass parameter $\mu$.

To achieve a good sampling of the full MSSM parameter space with \order{10^7} points, we restrict ourselves to the eight
MSSM parameters  
\begin{align}
\tb, \quad M_A, \quad \msqd, \quad \Af, \quad \mu, \quad \msld, \quad \mslez, \quad M_2
\label{Eq:fitparameters}
\end{align}
most relevant for phenomenology of the Higgs sector (the scan ranges will be given in \refse{sec:sampling}),
under the assumption that the third generation squark and slepton parameters are universal.  That is, we take
$\msqd := \MstL  (= \MsbL) = \MstR = \MsbR$,  
$\msld := \MstauL = \MstauR = M_{\tilde \nu_\tau}$
and $\Af := \At = \Ab = \Atau$.  Note that the  soft SUSY-breaking mass parameter in the stau sector, $\msld$,  
can significantly impact the Higgs decays as light staus can modify the loop-induced diphoton decay. $\msld$ is therefore taken as an independent parameter in our scans.  
Even though the other slepton and gaugino parameters are generally of less importance for the Higgs phenomenology, we scan over the SU(2) gaugino mass parameter $M_2$
as well as over the mass parameter of the first two generation
sleptons, $\mslez$, (assumed to be equal) as these parameters are important for the low-energy
observables included in our analysis.
The remaining MSSM parameters are fixed,
\begin{align}
\MsqL = \MsqR~(q = c, s, u, d) \; &= \; 1500 \gev, \\
M_3 = \mgl &= 1500 \gev\,. 
\end{align}
We choose relatively high values for the squark and gluino mass parameters, which have a minor impact on the Higgs sector, in order to be in agreement with the limits from direct SUSY searches.
Finally, the U(1)$_{\rm Y}$ gaugino mass parameter, $M_1$, is fixed via the GUT relation 
\BE
M_1 = \frac{5}{3} \frac{\sw^2}{\cw^2} M_2 \approx \frac{1}{2}  \MTwo~,
\label{def:Mone}
\end{equation}
with $\sw = \sqrt{1 - \cw^2}$ and $\cw = \MW/\MZ$.
For more details on the definition of the MSSM parameters, we refer to~\cite{Bechtle:2012jw}.


\subsection{The Higgs alignment limit}
\label{Sec:alignment}

In light of the Higgs data, which indicates that the properties of the
observed Higgs boson are SM-like, we seek to 
explore the region of the MSSM parameter space that yields a SM-like Higgs boson.   In general, a SM-like Higgs boson arises if one of the neutral Higgs mass eigenstates is approximately aligned with the direction of the Higgs vev in field space.
Thus, the limit of a SM Higgs boson is called the \emph{alignment limit}. 

To analyze the alignment limit, it is convenient to define
\beq
(\Phi_1)^i=\epsilon_{ij}(H_D^*)^j\,,\qquad\quad (\Phi_2)^i=(H_U)^i\,,
\eeq
where $\epsilon_{12}=-\epsilon_{21}=1$ and $\epsilon_{11}=\epsilon_{22}=0$,
and there is an implicit sum over the repeated SU(2) index $j=1,2$.  For consistency of the notation, we denote the corresponding neutral Higgs vevs by $v_1\equiv v_D$ and $v_2\equiv v_U$.
We now define the following linear combinations of Higgs doublet fields,
\begin{align}
\cHe = \begin{pmatrix} H_1^+\\H_1^0 \end{pmatrix} \equiv \frac{v_1 \Phi_1 + v_2\Phi_2}{v}, \qquad \cHz = \begin{pmatrix} H_2^+\\H_2^0 \end{pmatrix} \equiv \frac{-v_2 \Phi_1 + v_1\Phi_2}{v}
\end{align}
such that $\langle H_1^0 \rangle = v/\wz$ and $\langle H_2^0\rangle = 0$, which defines the so-called \textit{Higgs basis}~\cite{Georgi:1978ri,Branco:1999fs,Davidson:2005cw}.\footnote{Since the tree-level MSSM Higgs sector is \cp-conserving,
the Higgs basis is defined up to an overall sign ambiguity, where $\mathcal{H}_2\to -\mathcal{H}_2$.  However, since we have adopted the convention in which $\tan\beta$ is non-negative [cf.~the comment below \eq{tanbeta}], the overall sign of the Higgs basis field $\mathcal{H}_2$ is now fixed.}   
It is straightforward to express the scalar Higgs potential in terms of the Higgs basis fields $\mathcal{H}_1$ and~$\mathcal{H}_2$, 
\begin{align}
{\cal V} = \ldots + \edz Z_1 (\cHe^\dagger\cHe)^2 + \ldots +
\KKL 
\tfrac12  Z_5 (\cHe^\dagger\cHz)^2 +
Z_6 (\cHe^\dagger\cHe)(\cHe^\dagger\cHz) + {\rm h.c.} \KKR
+ \ldots\,,
\end{align}
where the most important terms of the scalar potential are highlighted above.  The quartic couplings $Z_1$, $Z_5$ and $Z_6$ are linear combinations of the quartic couplings that appear in the MSSM Higgs potential expressed in terms of $H_D$ and $H_U$.  In particular, at tree-level,
\beq
Z_1=\tfrac{1}{4}(g^2+g^{\prime\,2}) c_{2\beta}^2\,,\qquad\quad
Z_5=\tfrac{1}{4}(g^2+g^{\prime\,2}) s_{2\beta}^2\,,\qquad\quad
Z_6=-\tfrac{1}{4}(g^2+g^{\prime\,2}) s_{2\beta}c_{2\beta}\,,
\eeq
where $g$ and $g'$ are the SU(2) and U(1)$_{\rm Y}$ gauge couplings,
respectively, $c_{2\beta}\equiv\cos 2\beta$ and $s_{2\beta}\equiv \sin 2\beta$.  Hence, the $Z_i$ are $\mathcal{O}(1)$ parameters.

One can then evaluate the squared-mass matrix of the neutral $\cp$-even Higgs bosons, with respect to the neutral Higgs states, $\{\sqrt{2}\,{\rm Re}~H^0_1-v$\,,\,$\sqrt{2}\,{\rm Re}~H^0_2\}$ 
\begin{align}
  \cM^2 = \ML Z_1 v^2 & Z_6 v^2 \\
        Z_6 v^2 & \MA^2 + Z_5 v^2 \MR\,.
\label{HiBa-massmatrix}
\end{align}
If $\sqrt{2}\,{\rm Re}~H^0_1-v$ were a Higgs mass eigenstate, then its tree-level couplings to SM particles would be precisely those of the SM Higgs boson.  This would correspond to the exact alignment limit.  To achieve a SM-like neutral Higgs state,
it is sufficient for one of the neutral Higgs mass eigenstates to be approximately given by $\sqrt{2}\,{\rm Re}~H^0_1-v$.
In light of the form of the squared-mass matrix given in \refeq{HiBa-massmatrix}, we see that a SM-like neutral Higgs boson can arise in two different ways: 
\begin{enumerate}
\item
$\MA^2\gg (Z_1-Z_5)v^2$.  This is the so-called \textit{decoupling limit}, where $h$ is SM-like and $\MA\sim\MH\sim M_{H^\pm}\gg\Mh$. 
\item
$|Z_6|\ll 1$.  In this case $h$ is SM-like if $\MA^2+(Z_5-Z_1)v^2>0$ and
  $H$ is SM-like if $\MA^2+(Z_5-Z_1)v^2<0$. 
\end{enumerate}
In particular, the $\cp$-even mass eigenstates are:
\beq \label{mixing}
\begin{pmatrix} H\\ h\end{pmatrix}=\begin{pmatrix} \cba & \,\,\, -\sba \\
\sba & \,\,\,\phantom{-}\cba\end{pmatrix}\,\begin{pmatrix} \sqrt{2}\,\,{\rm Re}~H_1^0-v \\ 
\sqrt{2}\,{\rm Re}~H_2^0
\end{pmatrix}\,,
\eeq
where $\cba\equiv\cos(\be - \al)$ and $\sba\equiv\sin(\be - \al)$ are defined in terms of the mixing angle $\alpha$ that diagonalizes the $\cp$-even Higgs squared-mass matrix when expressed in the original basis of scalar fields, $\{\sqrt{2}\,{\rm Re}~\Phi_1^0-v_1\,,\,\sqrt{2}\,{\rm Re}~\Phi_2^0-v_2\}$.
Since the SM-like Higgs must be approximately $\sqrt{2}\,{\rm Re}~H_1^0-v$, it follows that
$h$ is SM-like if $|\cba|\ll 1$~\cite{Bernon:2015qea} and
$H$ is SM-like if $|\sba|\ll 1$~\cite{Bernon:2015wef}.
The case of a SM-like $H$ necessarily corresponds to alignment without decoupling.

In the case of exact alignment without decoupling, $Z_6=0$,
the tree-level couplings of the SM-like Higgs boson are precisely
those of the Higgs boson of the Standard Model.  Nevertheless,
deviations from SM Higgs boson properties can arise due to two possible
effects.  First, there might exist new particles that enter in loops and
modify the loop-induced Higgs couplings to $gg$, $\gamma\gamma$ and
$Z\gamma$.  For example, if $H$ is the SM-like Higgs boson, then the
charged Higgs boson mass is not significantly larger than the observed
Higgs mass, in which case the charged Higgs loop can shift the one-loop
induced couplings of the observed Higgs boson to $\gamma\gamma$ and
$Z\gamma$~\cite{Bernon:2015wef}.
Similarly, SUSY particles can give a contribution at the loop-level
to other, at the tree-level SM-like, couplings.
Second, there might exist new
particles with mass less that half the Higgs mass, allowing for new
decay modes of the SM-like Higgs boson.  An example of this possibility
arises if $H$ is the SM-like Higgs boson and $\Mh < \MH/2$, in
which case the decay mode $H\to hh$ is allowed.  Indeed, in the exact
alignment limit where $\sbma=0$, the tree-level $Hhh$ coupling in the
MSSM is given by~\cite{Gunion:1989we}  
\beq
 g\ls{Hhh} =\frac{g\MZ}{2\cw}(1-3\sin^2 2\beta)\,.
\eeq

The possibility of alignment without decoupling has been analyzed in detail in
Refs.~\cite{Gunion:2002zf,Craig:2013hca,Carena:2013ooa,Haber:2013mia,Carena:2014nza,Dev:2014yca,Bernon:2015qea,Bernon:2015wef} (see also the
``$\tau$-phobic'' benchmark scenario in \citere{Carena:2013qia}). It was
pointed out that exact alignment via $Z_6 = 0$ can only happen through an
accidental cancellation of the tree-level terms with contributions arising
at the one-loop level (or higher). In this case the Higgs alignment is
independent of $M_A^2$, $Z_1$ and $Z_5$. This has two phenomenological
consequences.  First, the remaining Higgs states can be light,
which would imply good prospects for LHC searches.
Second, either the light or the heavy neutral Higgs mass eigenstate can be
aligned with the SM Higgs vev and thus be interpreted as the SM-like Higgs
boson observed at $125\gev$.

The leading one-loop contributions to $Z_1$, $Z_5$ and $Z_6$ proportional to $h_t^2 m_t^2$, where
\beq \label{ht}
h_t=\frac{\sqrt{2}m_t}{v s_\beta}\,
\eeq
is the top quark Yukawa coupling, have been obtained in Ref.~\cite{Carena:2014nza} in the limit $M_Z, M_A \ll M_S$ (using results from Ref.~\cite{Haber:1993an}):
\begin{align}
Z_1v^2 &= M_Z^2 c_{2\beta}^2 + \frac{3m_t^4}{2\pi^2v^2} \left[\ln\left(\frac{M_S^2}{m_t^2}\right) +\frac{X_t^2}{M_S^2} \left(1 - \frac{X_t^2}{12M_S^2}\right) \right],\label{zone}\\
Z_5v^2 &=s_{2\beta}^2 \left\{ M_Z^2  + \frac{3 m_t^4}{8\pi^2v^2 s_\beta^4} \left[\ln\left(\frac{M_S^2}{m_t^2}\right) +\frac{X_tY_t}{M_S^2} \left(1 - \frac{X_tY_t}{12M_S^2}\right) \right]\right\},\label{Eq:Z5v2}\\
Z_6v^2 &= -s_{2\beta} \left\{M_Z^2c_{2\beta}  - \frac{3 m_t^4 }{4\pi^2v^2s_\beta^2} \left[\ln\left(\frac{M_S^2}{m_t^2}\right) +\frac{X_t(X_t+Y_t)}{2 M_S^2} - \frac{X_t^3Y_t}{12M_S^4} \right]\right\},
\label{Eq:Z6v2}
\end{align}
where $s_\beta\equiv \sin\beta$, $M_S\equiv \sqrt{m_{\tilde{t}_1}m_{\tilde{t}_2}}$ denotes the SUSY mass scale, given by the geometric mean of the light and heavy stop masses, and
\begin{align}  
X_t \equiv A_t - \mu/\tan\beta,\qquad\qquad Y_t \equiv A_t + \mu\tan\beta. \label{XY}
\end{align}
In \eqst{zone}{XY}, we have assumed for simplicity that $\mu$ and $A_t$ (as well as the gaugino mass parameters that contribute subdominantly at one-loop to the $Z_i$) are real parameters.  That is, we are neglecting CP-violating effects that can enter the MSSM Higgs sector via radiative corrections.

\begin{figure}[t!]
\centering
\includegraphics[width=0.48\textwidth]{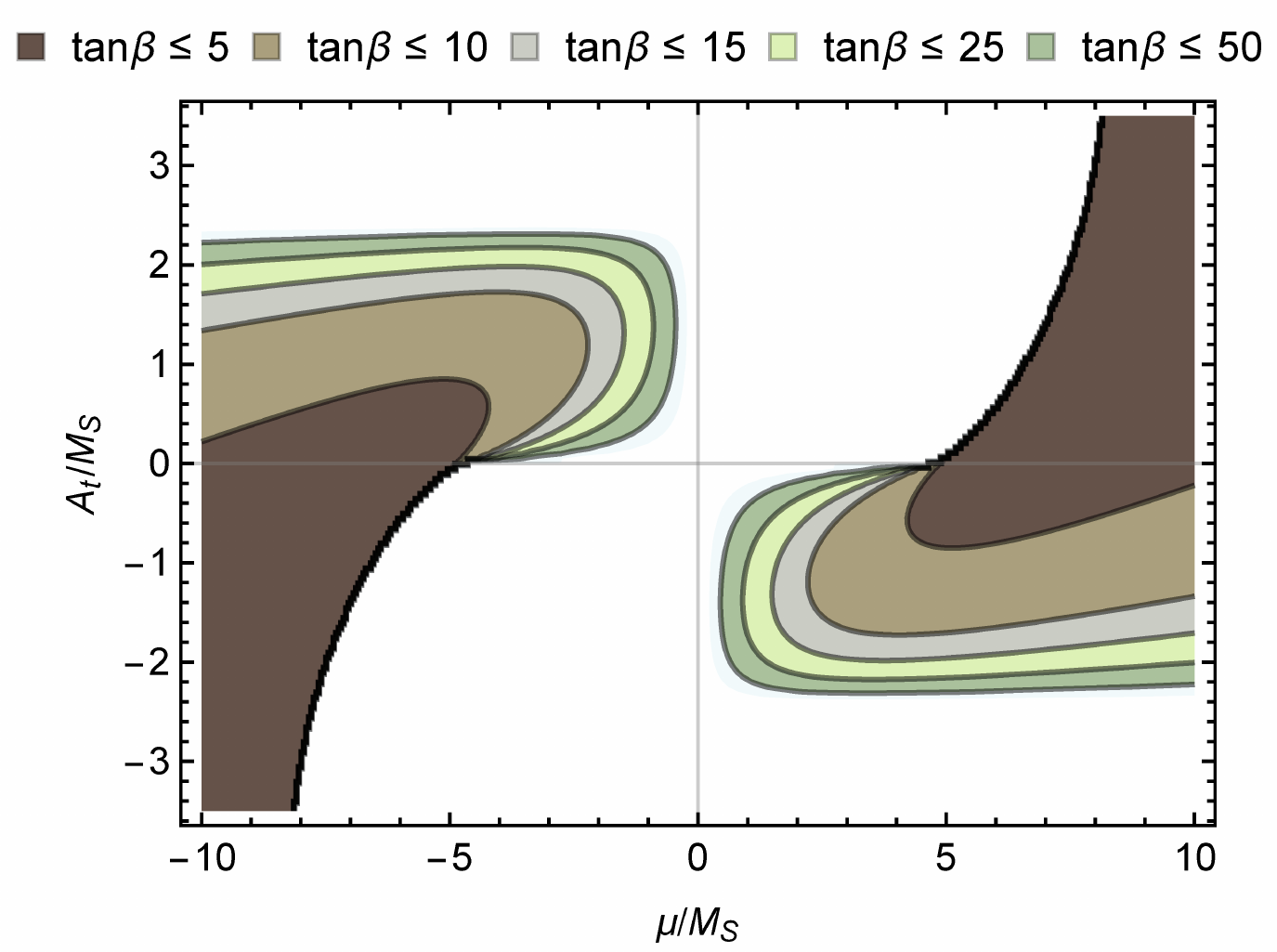}\hfill
\includegraphics[width=0.48\textwidth]{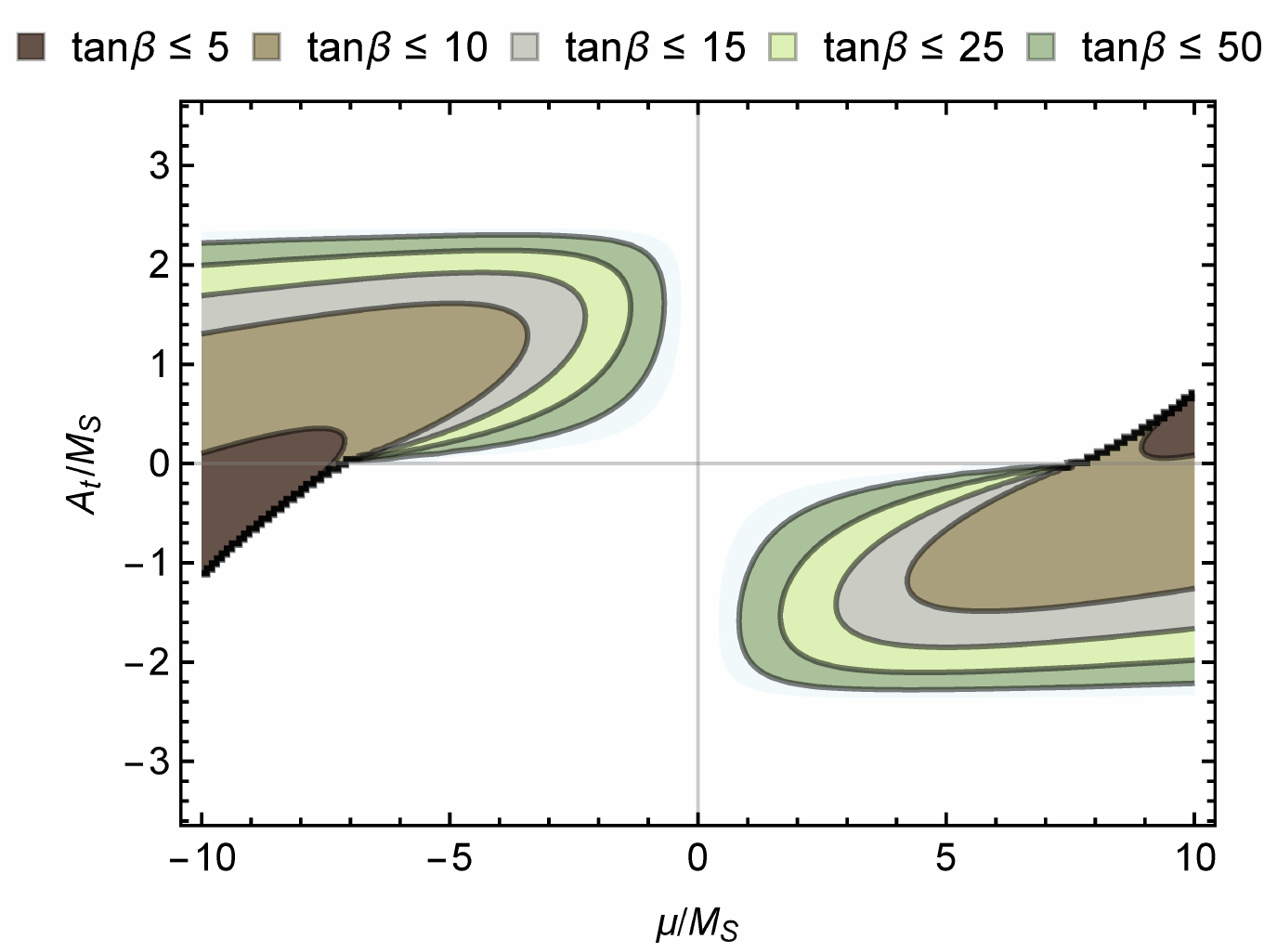}
\includegraphics[width=0.48\textwidth]{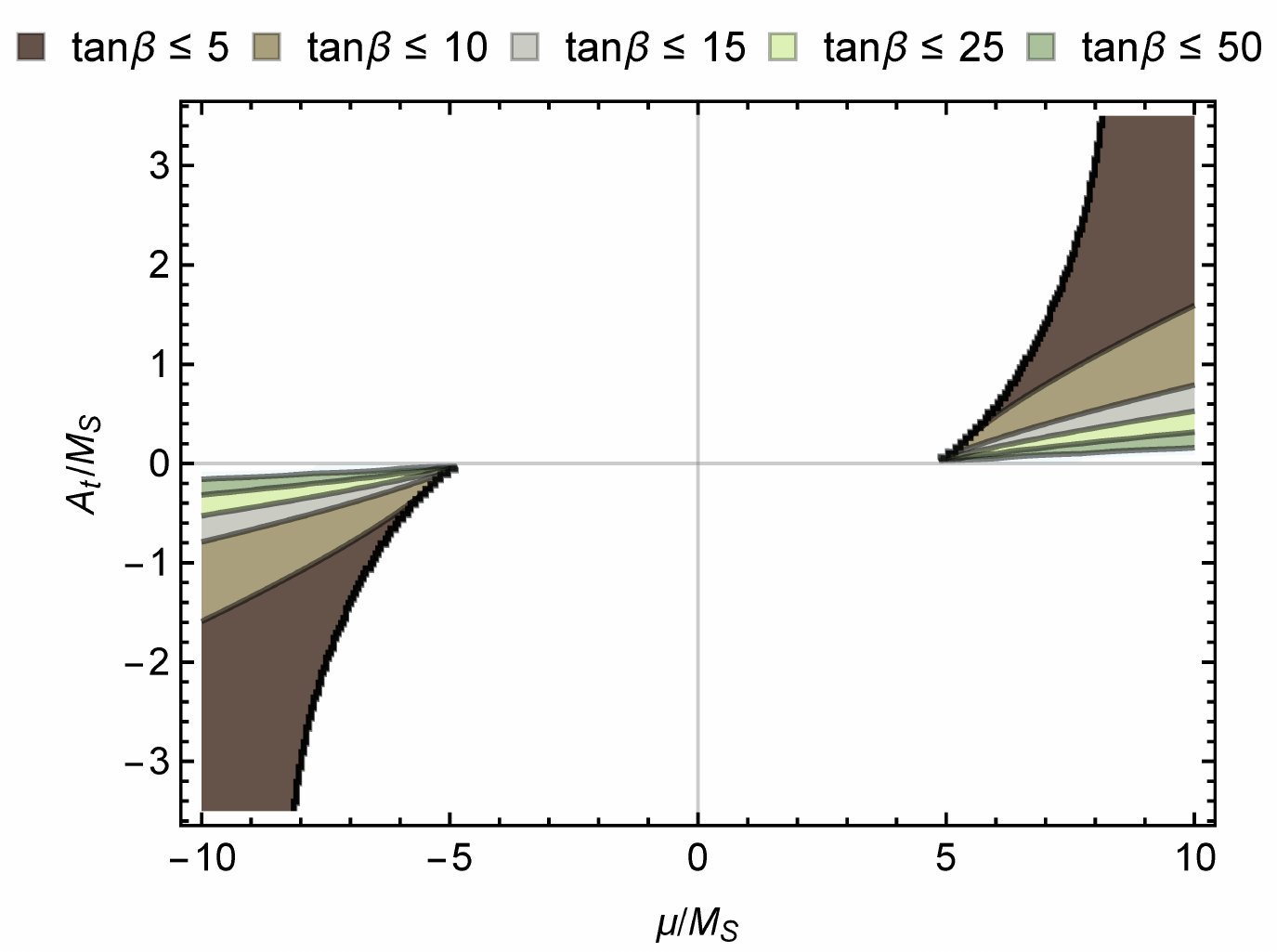}\hfill
\includegraphics[width=0.48\textwidth]{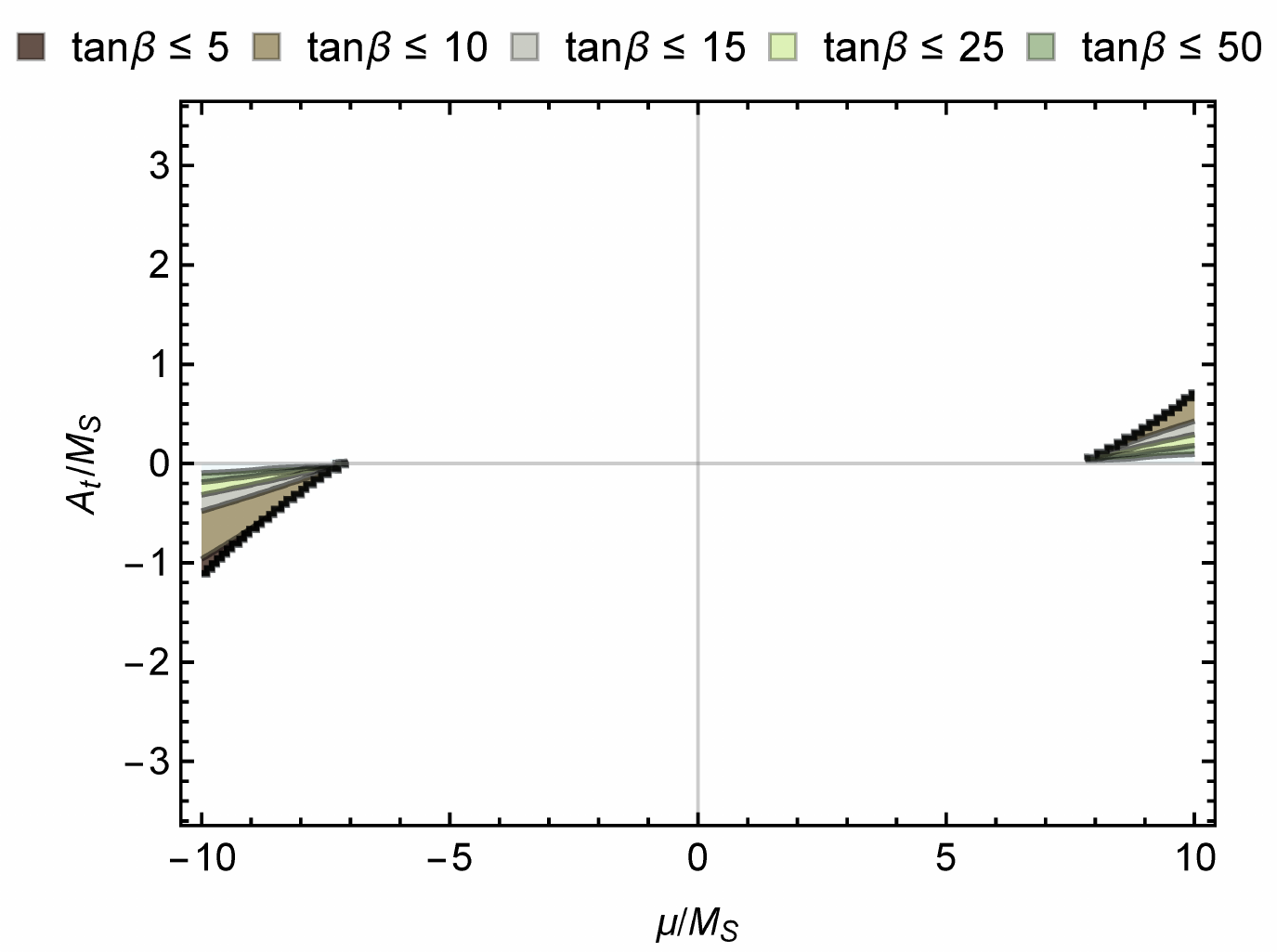}
\includegraphics[width=0.48\textwidth]{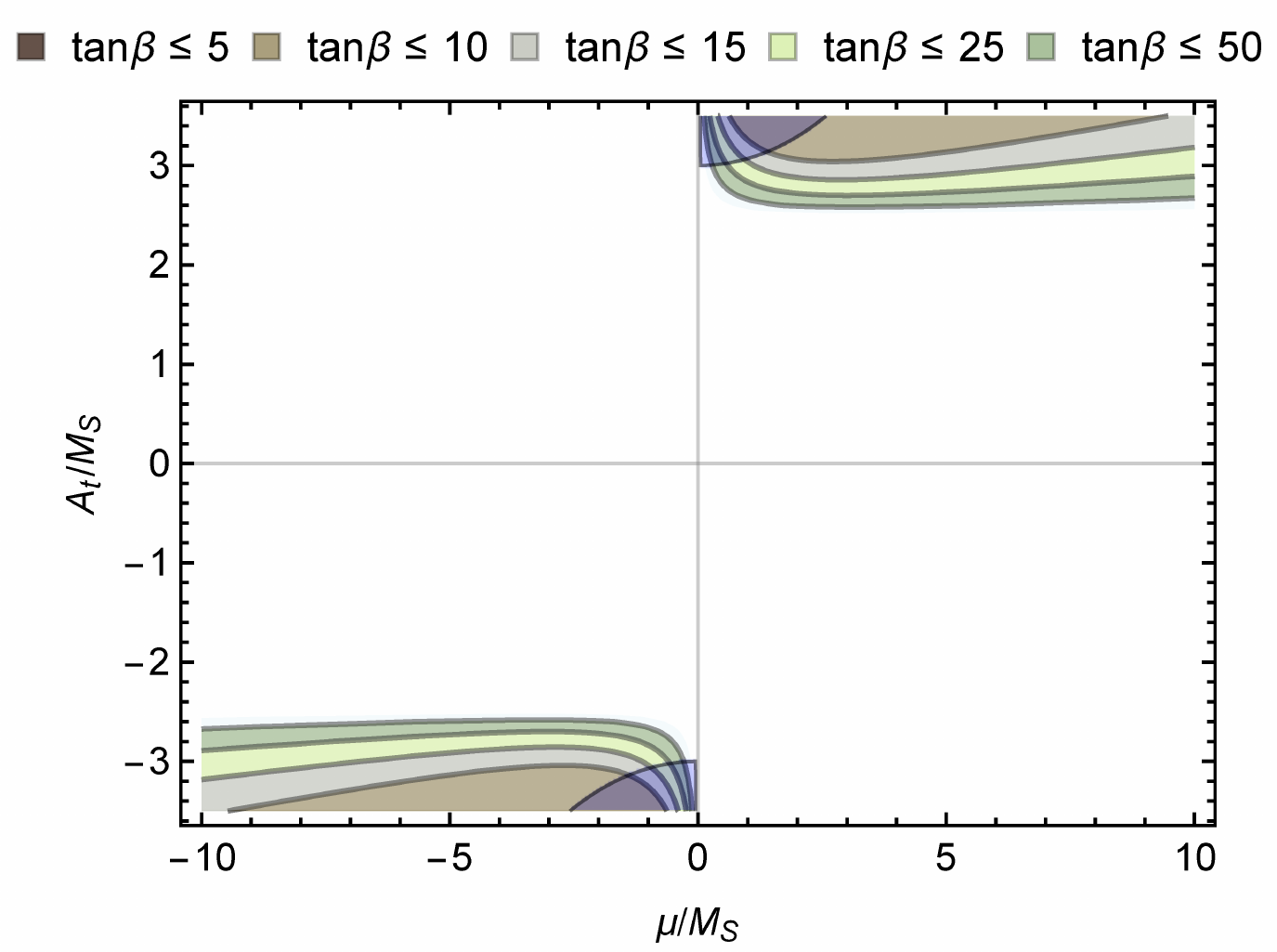}\hfill
\includegraphics[width=0.48\textwidth]{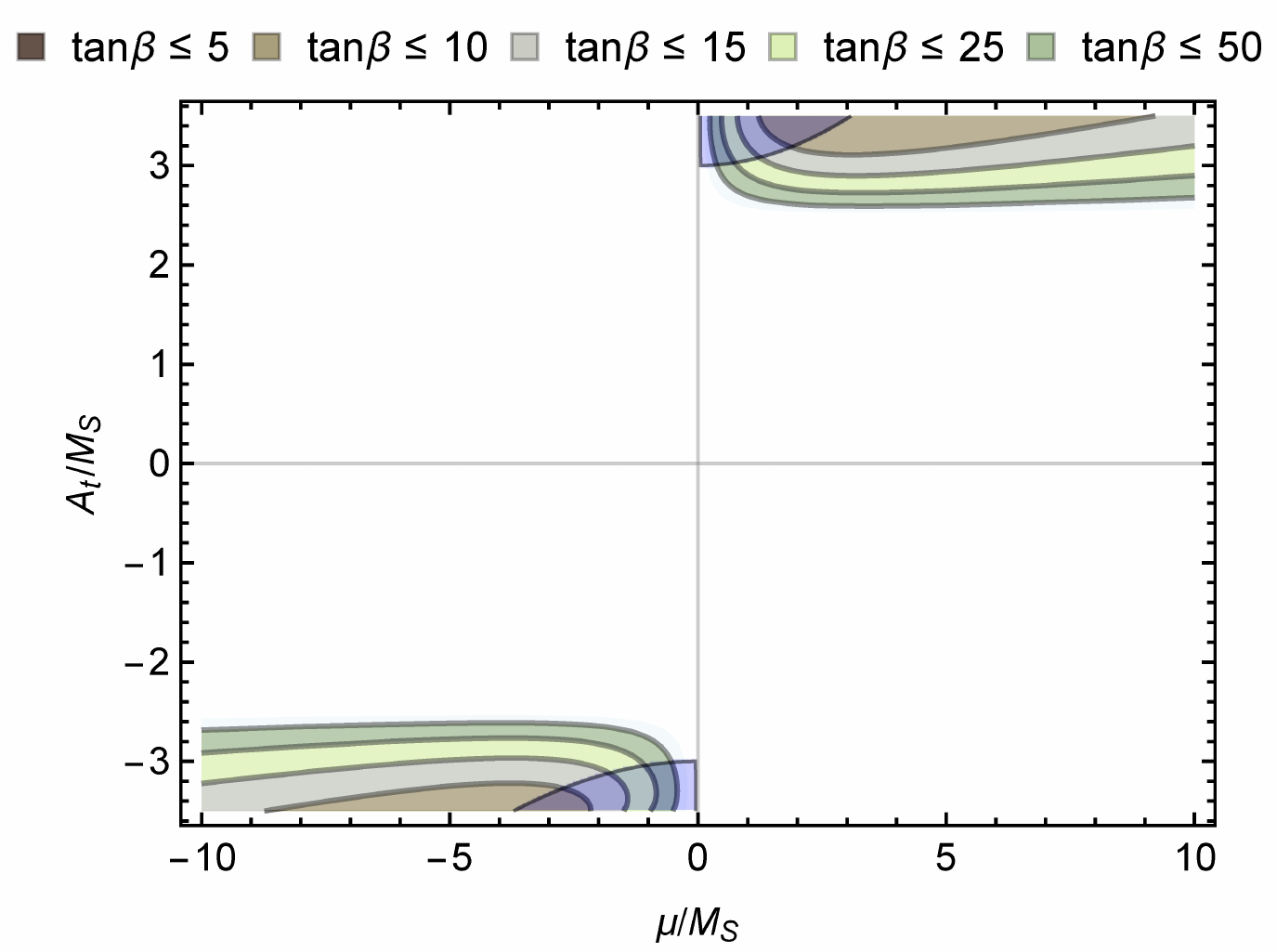}
\caption{Contours of $\tan\beta$ corresponding to exact alignment, $Z_6= 0$,
in the $(\mu/M_S, A_t/M_S)$ plane.
$Z_1$ is adjusted to give the correct Higgs mass.
The three left panels exhibit the approximate one-loop results;
the three right panels exhibit the corresponding two-loop improved results.
Taking the three panels on each side together, one can immediately
discern the regions of zero, one, two and three values of $\tan\beta$ in
which exact alignment is realized. In the overlaid blue regions we have
(unstable) values of $|X_t/\MS| \ge 3$. 
}
\label{Fig:alignment_numerical_TB}
\end{figure}

The approximate expression for $Z_6 v^2$ given in \refeq{Eq:Z6v2} depends only on the unknown parameters $\mu$, $A_t$, $\tan\beta$ and $M_S$.    Exact alignment arises when $Z_6=0$.
Note that $Z_6=0$ trivially occurs when $\beta=0$ or $\tfrac12 \pi$ (corresponding to the vanishing of either $v_1$ or $v_2$).  But, this choice of parameters is not relevant for phenomenology as it leads to a massless $b$ quark or $t$ quark, respectively, at tree-level.   Henceforth, we assume that $\tan\beta$ is non-zero and finite.  In our convention, $\tan\beta$ is positive with $0<\beta<\tfrac12 \pi$.

We can simplify the analysis of the condition $Z_6 = 0$ by solving \eq{zone} for $\ln(M_S^2/m_t^2)$ and inserting this result back into \eq{Eq:Z6v2}.
The resulting expression for $Z_6$ now depends on $Z_1$, $\tan\beta$, and the ratios,
\beq
\widehat{A}_t\equiv \frac{A_t}{M_S}\,,\qquad \quad \widehat{\mu}\equiv\frac{\mu}{M_S}\,.
\eeq
Using \eq{XY} to rewrite the final expression in terms of $\widehat{A}_t$ and $\widehat{\mu}$, we obtain,
\beq \label{z6zero}
Z_6 v^2=-\cot\beta\biggl\{\MZ^2 c_{2\beta}-Z_1 v^2 +\frac{3m_t^4\widehat{\mu}(\widehat{A}_t\tan\beta-\widehat{\mu})}{4\pi^2 v^2 s_\beta^2}
\bigl[\tfrac{1}{6}(\widehat{A}_t-\widehat{\mu}\cot\beta)^2-1\bigr]\biggr\}\,.
\eeq
Setting $Z_6 = 0$, we can identify $Z_1 v^2$ with the mass of the observed (SM-like) Higgs boson (which may be either $h$ or $H$ depending on whether $\sba$ is close to 1 or 0, respectively).   
We can then numerically solve for $\tan\beta$ for given values of  $\widehat{A}_t$ and $\widehat{\mu}$. 
The values of the real positive $\tb$ solutions of $Z_6=0$ obtained by using the one-loop approximate formula given in \eq{z6zero} are illustrated by the contour plots shown in the three left panels of \figref{Fig:alignment_numerical_TB}, where each panel corresponds to a different solution of $Z_6=0$.    Note that at every point in the $(\widehat{\mu}\,,\,\widehat{A}_t)$ plane, the value of $M_S$ has been adjusted according to \eq{zone} such that the squared-mass of the SM-like Higgs boson in the alignment limit is given by $Z_1 v^2\simeq(125~{\rm GeV})^2$.
Taking the three left panels together, one can immediately discern the regions of zero, one, two and three positive $\tan\beta$ solutions of \eq{z6zero}, and their corresponding values.  A more detailed discussion of these solutions will be presented in a separate paper~\cite{preparation}.

It is instructive to obtain an approximate analytic expression for the value of the largest real positive $\tb$ solution. Assuming $|\widehat{\mu}\widehat{A}_t|\tan\beta\gg1$ the following approximate alignment condition, first written in Ref.~\cite{Carena:2014nza}, is obtained,
\begin{align}
\tb &\simeq  \frac{M_{h/H}^2 + \MZ^2 +\displaystyle\frac{3 \mt^4 \widehat{\mu}^2}{8\pi^2 v^2}(\widehat{A}_t^{\,2}-2)}
            {\displaystyle\frac{m_t^4 \widehat{\mu}\widehat{A}_t}{8\pi^2 v^2}(\widehat{A}_t^{\,2}-6)}
            \simeq \frac{127+3\widehat{\mu}^2(\widehat{A}_t^{\,2}-2)}
            {\widehat{\mu}\widehat{A}_t(\widehat{A}_t^{\,2}-6)}\,,
\label{Eq:alignmentcondition}
\end{align}
where $M^2_{h/H}\simeq Z_1 v^2$ denotes the (one-loop) mass of the SM-like Higgs boson obtained from \eq{zone}, which could be either the light or heavy $\cp$-even Higgs boson.  It is clear from Eq.~\eqref{Eq:alignmentcondition} that a positive $\tan\beta$ solution exists if either $\widehat{\mu} \widehat{A}_t (\widehat{A}_t^2 - 6) > 0$ and $\widehat{A}_t^2>2$, or if 
$\widehat{\mu} \widehat{A}_t (\widehat{A}_t^2 - 6) <0$, $\widehat{A}_t^2<2$ and $|\widehat\mu|$ is sufficiently large such that the numerator of 
Eq.~\eqref{Eq:alignmentcondition} is negative.  Keeping in mind that Eq.~\eqref{Eq:alignmentcondition} was derived under the assumption that $\widehat{\mu}\widehat{A}_t\tan\beta\gg1$, one easily verifies that the largest of the three roots of \eq{z6zero} shown in \figref{Fig:alignment_numerical_TB} always satisfies the stated conditions above.
Another consequence of Eq.~\eqref{Eq:alignmentcondition} is that  by increasing the value of $|\widehat{\mu}\widehat{ A}_t|$ (in the region where $2<\widehat{A}^2_t<6$), it is
possible to lower the $\tb$ value at which alignment occurs.

If $|\widehat{A}_t|\ll 1$, then \eq{Eq:alignmentcondition} is no longer a good approximation.   Returning to \eq{z6zero}, we set $\widehat{A}_t=0$ and again assume that $\tan\beta\gg 1$. We can then solve approximately for $\tan\beta$,
\beq
\tan^2\beta\simeq \frac{M_Z^2-M^2_{h/H}+\displaystyle\frac{3m_t^4\widehat{\mu}^2}{4\pi^2 v^2}\bigl(\tfrac16 \widehat{\mu}^2-2\bigr)}{M_Z^2+M^2_{h/H}+\displaystyle\frac{3m_t^4\widehat{\mu}^2}{4\pi^2 v^2}}\,.
\eeq
For example, in the parameter regime where $\widehat{A}_t\simeq 0$ and $|\widehat{\mu}|\gg 1$, we obtain
$\tan\beta\simeq |\widehat{\mu}|/\sqrt{6}$.  

The question of whether the light or the heavy $\cp$-even Higgs boson possesses SM-like Higgs couplings in the alignment without decoupling regime depends on the relative size of $Z_1v^2$ and $Z_5v^2 + M_A^2$.   Combining \eqs{Eq:Z5v2}{Eq:Z6v2}, it follows that in the limit of exact alignment where $Z_6=0$, we identify $Z_1 v^2$ as the squared mass of the observed SM-like Higgs boson and
\beq \label{zfive}
Z_5 v^2=\MZ^2(1+ c_{2\beta}) +\frac{3m_t^4 \widehat{\mu}(\widehat{A}_t-\widehat{\mu}\cot\beta)}{8\pi^2 v^2 s_\beta^4}
\biggl\{s_{2\beta}-\tfrac16\bigl[(\widehat{A}_t^{\,2}-\widehat{\mu}^2)s_{2\beta}-2\widehat{A}_t\widehat{\mu} c_{2\beta}\bigr]\biggr\}\,.
\eeq
We define a critical value of $M^2_A$,
\beq \label{macrit}
M_{A,c}^2\equiv{\rm max}\bigl\{(Z_1-Z_5)v^2\,,\,0\bigr\}\,,
\eeq
where $Z_1 v^2=(125~{\rm GeV})^2$ and $Z_5 v^2$ is given by \eq{zfive}.  Furthermore, since the squared-mass of the non-SM-like $\cp$-even Higgs boson in the exact alignment limit, $M_A^2+Z_5 v^2$, must be positive, it then follows that the minimum value possible for the squared-mass of the $\cp$-odd Higgs boson is
\beq \label{mamin}
M_{A,m}^2\equiv{\rm max}\bigl\{-Z_5 v^2\,,\,0\bigr\}\,.
\eeq
That is, if $Z_5$ is sufficiently large and negative, then the minimal allowed value of $M_A^2$ is non-zero and positive. 

\begin{figure}[t!]
\centering
\vspace{-1.0cm}
\includegraphics[width=0.46\textwidth]{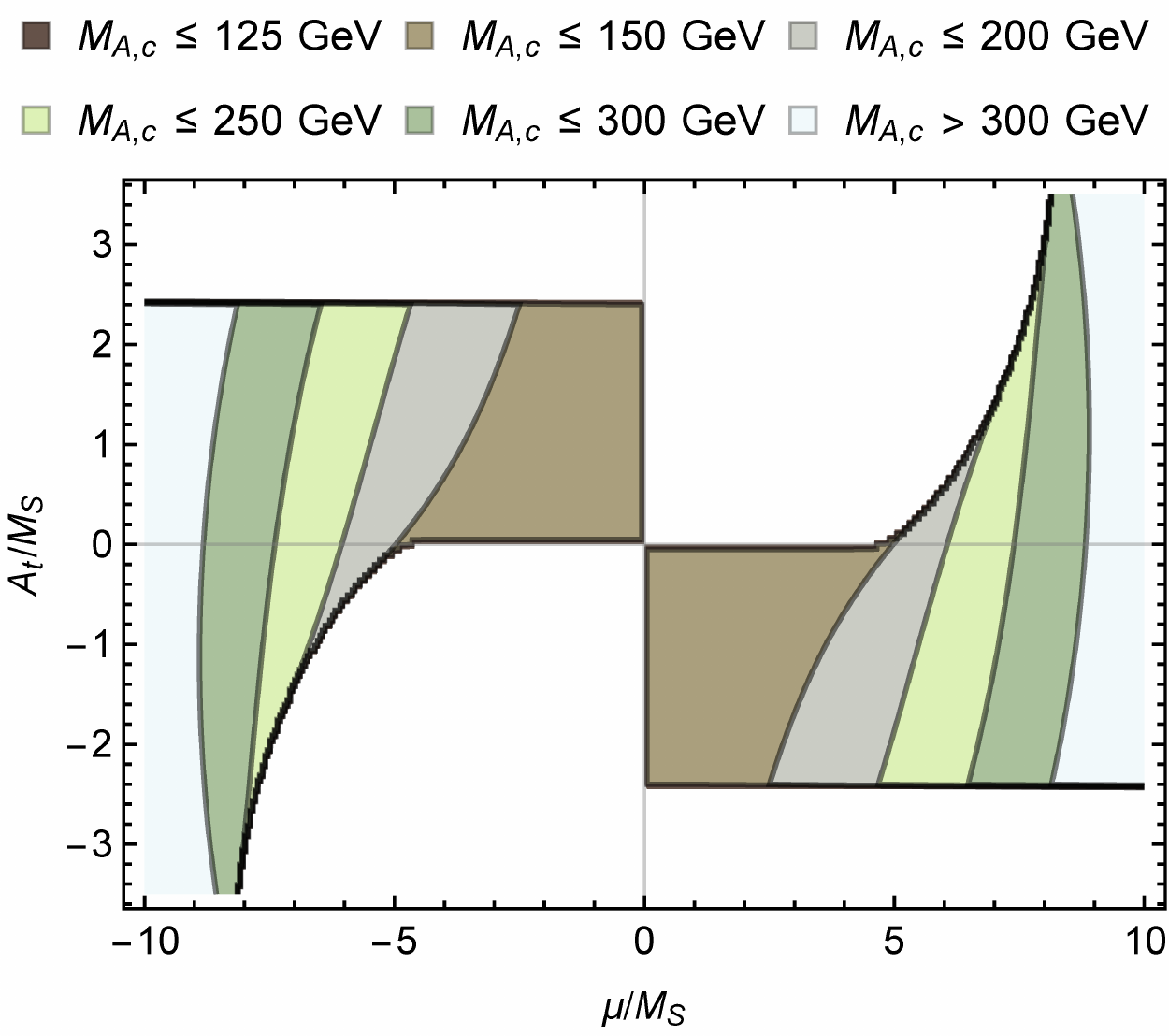}\hfill
\includegraphics[width=0.46\textwidth]{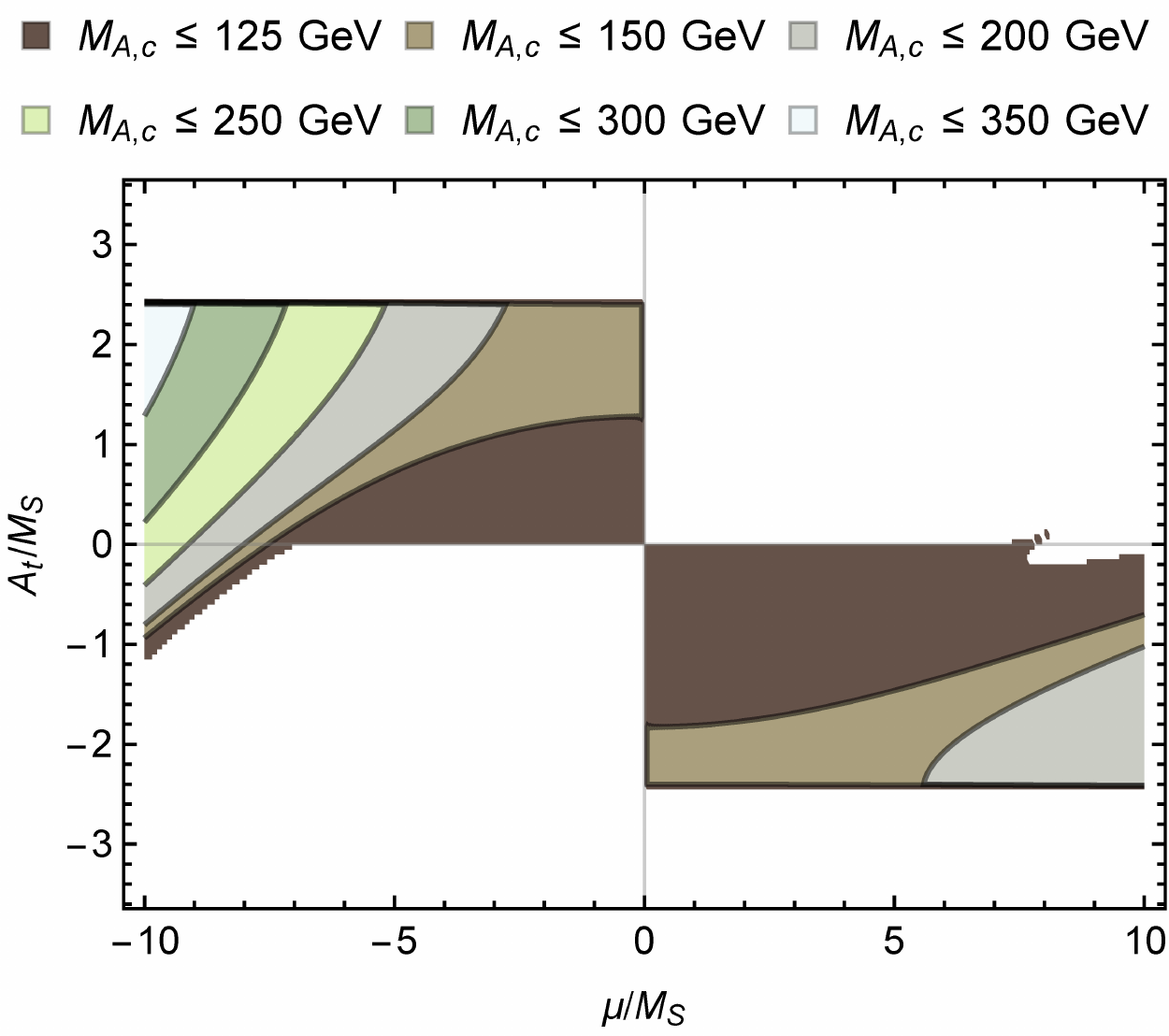}
\includegraphics[width=0.46\textwidth]{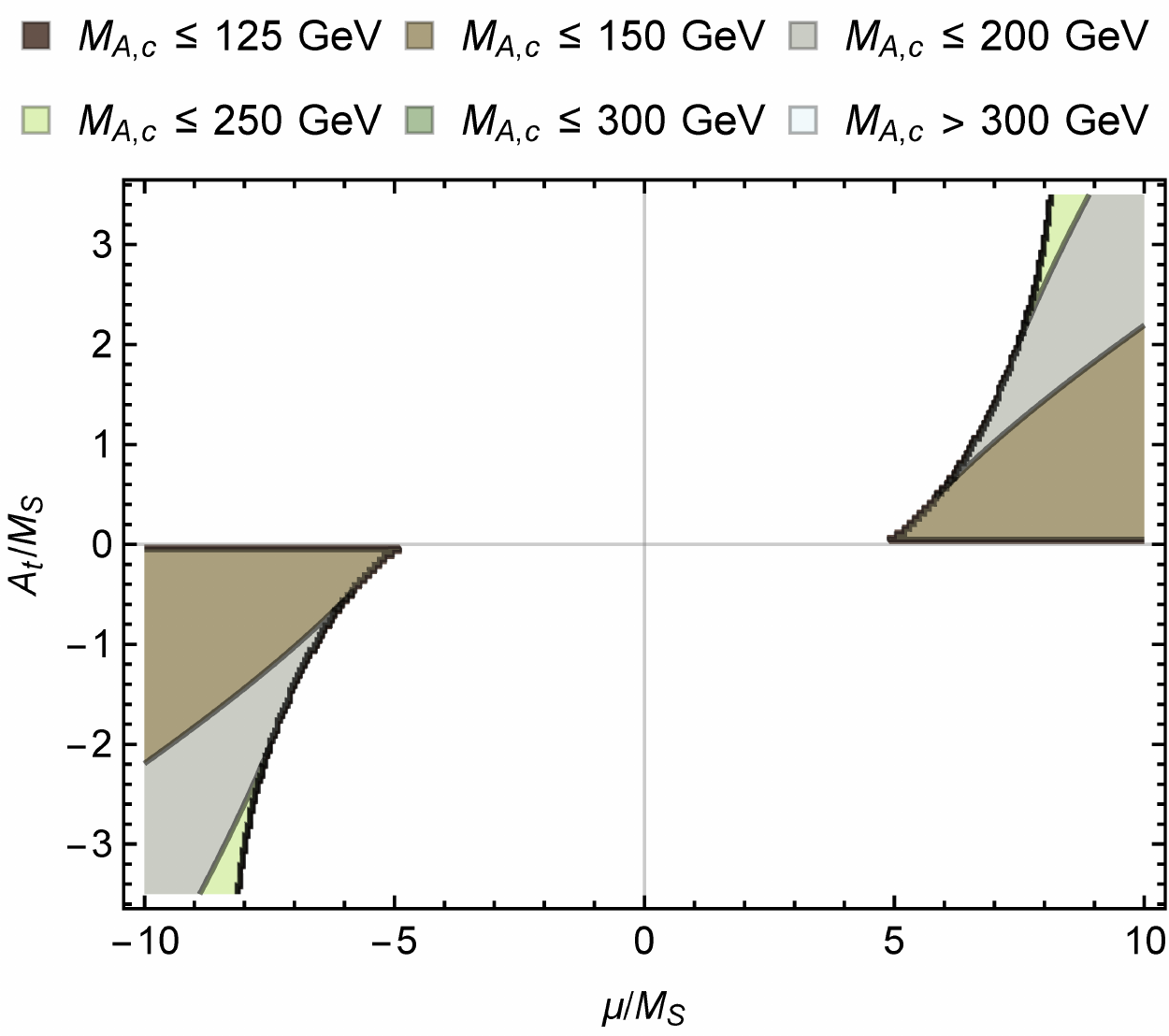}\hfill
\includegraphics[width=0.46\textwidth]{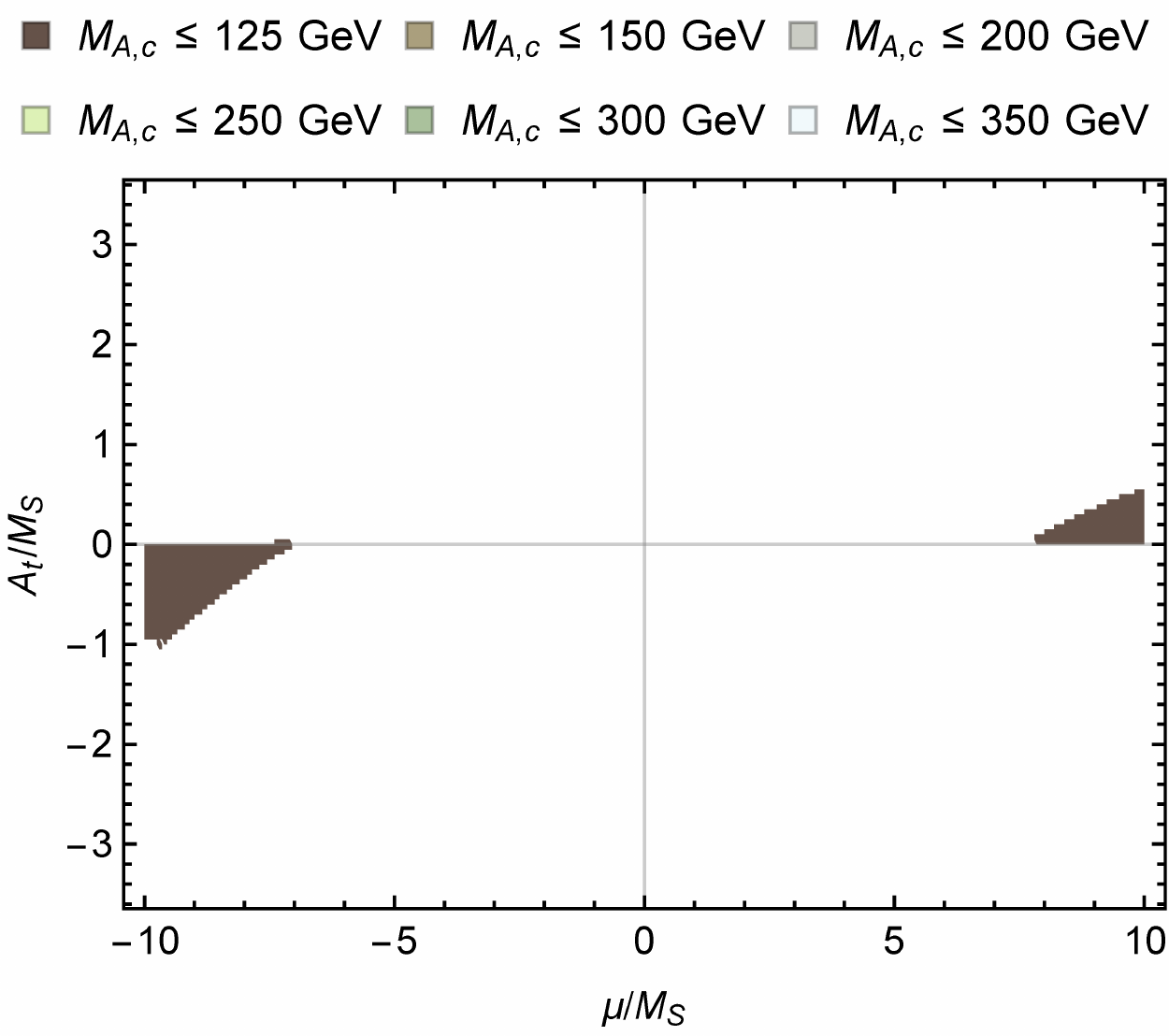}
\includegraphics[width=0.46\textwidth]{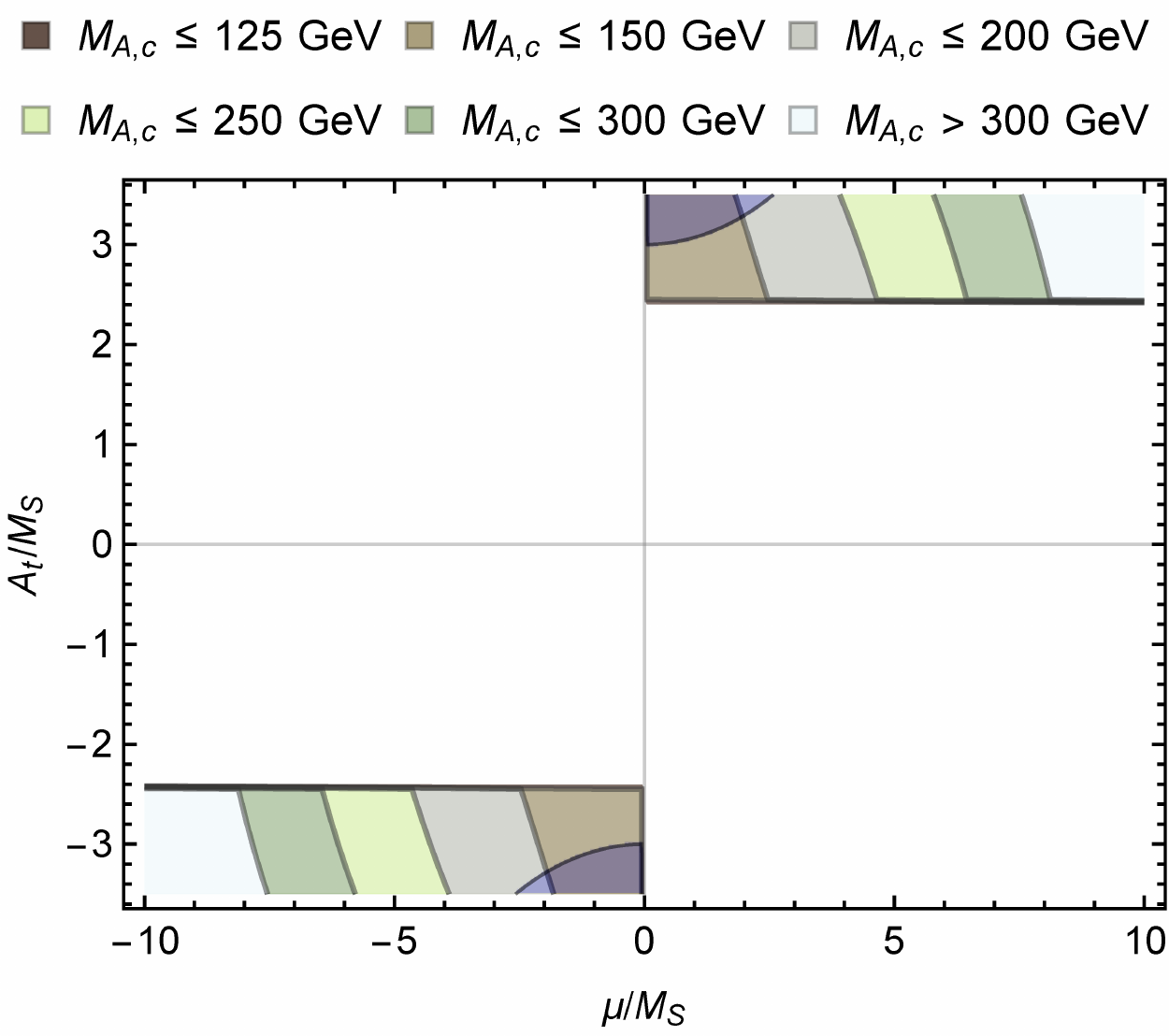}\hfill
\includegraphics[width=0.46\textwidth]{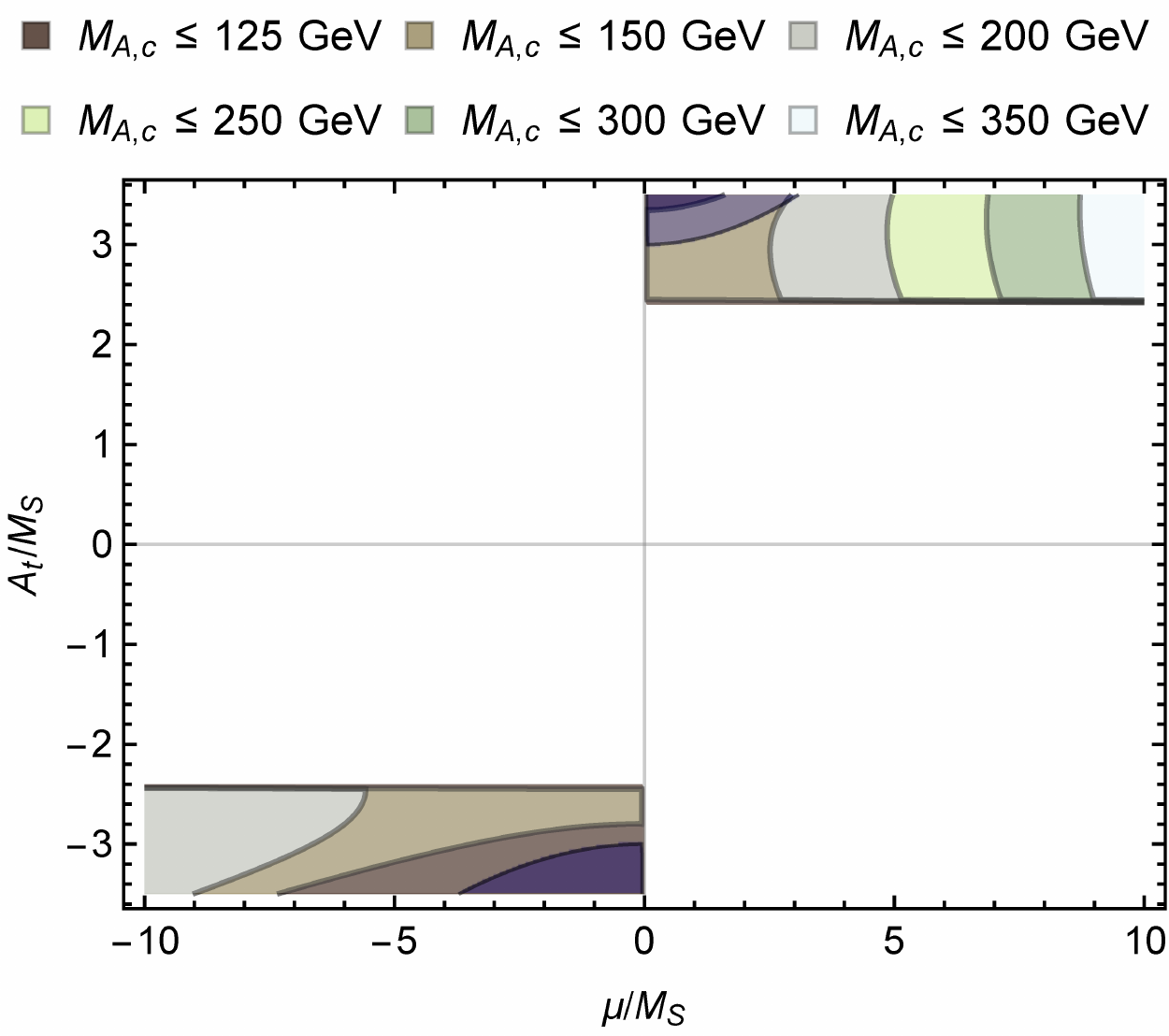}
\caption{Critical $M_A$ value, $M_{A,c}$, in the exact alignment, indicating the maximal $M_A$ value for which the mass hierarchy of the heavy Higgs interpretation is obtained, corresponding to the solutions found in~\figref{Fig:alignment_numerical_TB} in the $(\mu/M_S, A_t/M_S)$ plane.
The three left panels exhibit the approximate one-loop results;
the three right panels exhibit the corresponding two-loop improved results.
In the overlaid blue regions we have (unstable) values of $|X_t/\MS| \ge 3$.
}
\label{Fig:alignment_numerical_MAc}
\end{figure}

We focus again on the parameter region in the $(\widehat{\mu}\,,\widehat{A}_t$) plane,
and compute $Z_5$ from \eq{zfive} using the value of $\tan\beta$ obtained
from setting $Z_6=0$ in \eq{z6zero}.  This allows us to determine the value
of $M_{A,c}^2$ for each point in the
$(\widehat{\mu}\,,\widehat{A}_t$) plane.  The interpretation of
$M_{A,c}^2$ is as follows.  If $M^2_A> M^2_{A,c}$, 
then $h$ can be identified as the SM-like Higgs boson with 
$M_h\simeq 125$~GeV.  
If $M_{A,m}^2<M^2_A<M^2_{A,c}$, then $H$ can be
identified as the SM-like Higgs boson with $M_H\simeq 125$~GeV.     
The corresponding contours of $M_{A,c}$ are exhibited in the three left panels of \figref{Fig:alignment_numerical_MAc}, which are in one-to-one correspondence with the three left panels of \figref{Fig:alignment_numerical_TB}.

\clearpage
As previously noted, the analysis above was based on approximate one-loop formulae given in \eqst{zone}{Eq:Z6v2}, where only the leading terms proportional to $m_t^2 h_t^2$ are included.   In the exact alignment limit, we identify $Z_1 v^2$ given by \eq{zone} as  the squared-mass of the observed SM-like Higgs boson.  However, it is well known that \eq{zone} overestimates the value of the radiatively corrected Higgs mass.  Remarkably, one can obtain a significantly more accurate result simply by including the leading two-loop radiative corrections proportional to $\alpha_s m_t^2 h_t^2$.   

In Ref.~\cite{Carena:2000dp}, it was shown that the dominant part of these two-loop corrections can be obtained from the corresponding one-loop formulae with the following very simple two step prescription.   First, we replace
\beq \label{step1}
m_t^4\ln\left(\frac{M_S^2}{m_t^2}\right)\longrightarrow m_t^4(\lambda)\ln\left(\frac{M_S^2}{m_t^2(\lambda)}\right)\,,\qquad
\text{where $\lambda\equiv\bigl[m_t(m_t)M_S\bigr]^{1/2}$}\,,
\eeq
where $m_t(m_t)\simeq 165.6$~GeV is the $\overline{\rm MS}$ top quark
mass~\cite{Buttazzo:2013uya}, and the running top quark mass in the one-loop approximation is given by
\beq \label{mtmu}
m_t(\lambda)=m_t(m_t)\left[1+\frac{\alpha_s}{\pi}\ln\left(\frac{m_t^2(m_t)}{\lambda^2}\right)\right]\,.
\eeq
In our numerical analysis, we take $\alpha_s = \alpha_s(m_t(m_t)) \simeq 0.10826$.
Second, when $m_t^4$ 
multiplies that threshold corrections due to stop mixing
(i.e., the one-loop terms proportional to $X_t$ and $Y_t$), then
we make the replacement,
\beq
m_t^4\longrightarrow m_t^4(M_S)\,,
\eeq
where
\beq
\label{mtms}
m_t(M_S)=m_t(m_t)\left[1+\frac{\alpha_s}{\pi}\ln\left(\frac{m_t^2(m_t)}{M_S^2}\right)+\frac{\alpha_s}{3\pi}\,\frac{X_t}{M_S}\right]\,.
\eeq
Note that the running top-quark mass evaluated at $M_S$ includes a threshold correction proportional to $X_t$ that enters at the scale of supersymmetry breaking.
Here, we only keep the leading contribution to the threshold correction under the assumption that $m_t\ll M_S$ (a more precise formula can be found in Appendix B of Ref.~\cite{Carena:2000dp}).
The above two step prescription can now be applied to \eqst{zone}{Eq:Z6v2}, which yields a more accurate expression for the radiatively corrected Higgs mass and the condition for exact alignment without decoupling.
Details of this analysis will be presented in a forthcoming work~\cite{preparation}.

The end results are summarized below.  We have derived
analogous expressions to Eqs.~\eqref{z6zero} and~\eqref{zfive} that incorporate the leading two-loop effects at $\mathcal{O}(\alpha_s h_t^2)$.    It is convenient to introduce the following notation,
\beq
C\equiv \frac{3m_t^4}{2\pi^2 v^2}\,,\qquad \overline{\alpha}_s\equiv\frac{\alpha_s}{\pi}\,,\qquad
x_t\equiv X_t/M_S\,,
\eeq
where $m_t\equiv m_t(m_t)$ is the $\overline{\rm MS}$ top quark mass, and
\beq
X_1\equiv x_t^2\bigl(1-\tfrac{1}{12}x_t^2\bigr)\,,\qquad X_5\equiv x_t y_t\bigl(1-\tfrac{1}{12}x_t y_t\bigr)\,,
\qquad
X_6\equiv \tfrac12 x_t(x_t+y_t)-\tfrac{1}{12}x_t^3 y_t\,.
\eeq
Then, the two-loop corrected condition for the exact alignment limit corresponding to $Z_6=0$
is given by,
\beq \label{z6zerocorr}
2M_Z^2 s_\beta^2 c_{2\beta}-(Z_1 v^2-M_Z^2 c^2_{2\beta})\bigl[1+4\overline{\alpha}_s(X_1-X_6)\bigr]
+C(X_1-X_6)\bigl[1+\overline{\alpha}_s(4X_1+\tfrac43 x_t)\bigr]=0\,,
\eeq
which supersedes \eq{z6zero}, and the $\mathcal{O}(\alpha_s)$ correction to \eq{zfive} is given by,
\beq
Z_5 v^2=M_Z^2(1+c_{2\beta})+\frac{C(X_5-X_6)}{\tan^2\beta}\biggl\{1+4\overline{\alpha}_s\bigl(X_6+\tfrac13 x_t-2s_\beta^2 c_{2\beta}C^{-1}M_Z^2\bigr)\biggr\}\,.
\eeq
One can now define two-loop improved versions of $M^2_{A,c}$ and $M^2_{A,m}$ [cf.~\eqs{macrit}{mamin}]. 

In the right panels of Figs.~\ref{Fig:alignment_numerical_TB} and \ref{Fig:alignment_numerical_MAc}, we plot the two-loop improved versions of the corresponding one-loop results shown in the left panels.   There are a few notable changes, which we now discuss.  First, 
in our scan of the $(\widehat\mu\,,\,\widehat{A}_t)$ plane, we have observed numerically that there is a new solution to the alignment condition [cf.~\eq{z6zerocorr}] that is unrelated to the solutions found in the one-loop analysis.
However, this solution always corresponds to a value of $|{X}_t|>3M_S$, which lies outside our region of interest.  Henceforth, we simply discard this possibility.  What remains are solutions that can be identified as the two-loop corrected versions of the one-loop results obtained above.
The right panels of \figref{Fig:alignment_numerical_TB} exhibit the remaining real positive $\tan\beta$ solutions of \eq{z6zerocorr}.  

We can now see the effects of including the leading $\mathcal{O}(\alpha_s h_t^2)$ corrections.  The regions where positive solutions to \eq{z6zerocorr} exist shown in the right panels of \figref{Fig:alignment_numerical_TB} have shrunk somewhat as compared to the corresponding positive solutions to \eq{z6zero} shown in the left panels of \figref{Fig:alignment_numerical_TB}.  For example, only one positive solution for $\tan\beta$ exists for large $\widehat{\mu}$ and $\widehat{A}_t$ in the two-loop approximation, whereas three positive solutions exist in the one-loop approximation.   Using the values of $\tan\beta$ found in the right panels of \figref{Fig:alignment_numerical_TB}, one can now produce the corresponding two-loop corrected plots shown in the right panels of \figref{Fig:alignment_numerical_MAc}.  The qualitative features of the one-loop and two-loop results are similar, after taking note of the slightly smaller regions in which positive solutions for $\tan\beta$ exist in the two-loop approximation.

One new feature of the two-loop approximation not yet emphasized is that we must now carefully define the input parameters $\mu$ and $A_t$.  In the above formulae and plots we interpret these parameters as $\overline{\rm MS}$ parameters.  However, it is often more convenient to re-express these parameters in terms of on-shell parameters.  In Ref.~\cite{Carena:2000dp}, the following expression was obtained for the on-shell squark mixing parameter $X_t^{\rm OS}$ in terms of the $\overline{\rm MS}$ squark mixing parameter $X_t$, where only the leading $\mathcal{O}(\alpha_s)$ corrections are kept,
\beq
X_t^{\rm OS}=X_t-\frac{\alpha_s}{3\pi}M_S\left[8+\frac{4X_t}{M_S}-\frac{X_t^2}{M_S^2}-\frac{3X_t}{M_S}\ln\left(\frac{m_t^2}{M_S^2}\right)\right]\,.
\eeq
Since the on-shell and $\overline{\rm MS}$ versions of $\mu$ are equal at this level of approximation, we also have
\beq
A_t^{\rm OS}=X_t^{\rm OS}+\frac{\mu}{\tan\beta}\,.
\eeq
A more detailed examination of the above results, when expressed in terms of the on-shell parameters, will be treated in Ref.~\cite{preparation}.

The approximations employed in the section capture some of the most
important radiative corrections relevant for analyzing the alignment
limit of the MSSM.   However, it is important to appreciate what has
been left out.
First, higher-order corrections beyond
\order{\als h_t^2} are known to be relevant (see, e.g.,
\citere{Degrassi:2002fi}). In particular the \order{h_t^4}
corrections are in magnitude roughly 
20\% of the $\order{\als h_t^2} $ corrections, and enter with a different sign,
thus leading to potentially 
non-negligible corrections to the approximate two-loop results obtained above.
On more general grounds, the
analysis of this section ultimately corresponds to a
renormalization of $\cos(\beta-\alpha)$, which governs the tree-level
couplings of the Higgs boson and its departure from the alignment limit.
However, radiative corrections also contribute other effects that modify
Higgs production cross sections and branching ratios.  It is well-known
that for $M_A\ll M_S$, the effective low-energy theory below the scale
$M_S$ is a general two Higgs doublet model with the most general
Higgs-fermion Yukawa couplings.  These include the so-called wrong-Higgs
couplings of the MSSM~\cite{Haber:2007dj,Crivellin:2010er}, which ultimately are
responsible for the $\Delta_b$ and $\Delta_\tau$ corrections that can
significantly modify the coupling of the Higgs boson to bottom quarks
and tau leptons.\footnote{For a review of these effects and a guide to
the original literature, see Ref.~\cite{Carena:2002es}.}    
The implication of these couplings will be briefly reviewed in \refse{Sec:wrong}.
In addition, integrating out heavy SUSY particles at the scale $M_S$ can generate
higher dimensional operators that can also modify Higgs production
cross sections and branching ratios~\cite{Dine:2007xi}.  None of these
effects are accounted for in the analysis presented in this section. 

\subsection{Implications of the wrong-Higgs couplings }
\label{Sec:wrong}

At tree-level, the Higgs-fermion Yukawa couplings follow the Type-II pattern~\cite{Hall:1981bc,Gunion:1989we} of the 
two-Higgs doublet model (2HDM), in which
the hypercharge $-1$ Higgs doublet field $H_D$ couples exclusively to right-handed down-type fermions and the 
hypercharge $+1$ Higgs doublet field $H_U$ couples exclusively to right-handed up-type fermions.
When radiative corrections are included, the so-called wrong-Higgs Yukawa couplings are induced by supersymmetry-breaking effects, in which $H_D^*$ couples to right-handed up-type fermions and $H_U^*$ couples to right-handed down-type fermions.
We shall denote by $M_{\rm SUSY}$ a generic scale that characterizes the size of supersymmetric mass parameters. 
In the limit where $M_Z$, $M_A\ll M_{\rm SUSY}$, the radiatively-corrected Higgs-quark Yukawa couplings can 
be summarized by an effective
Lagrangian,\footnote{For simplicity, we ignore the couplings to first and second generation fermions.  We also neglect
weak isospin breaking effects that distinguish between the coupling of neutral and charged Higgs scalars.}
\begin{equation} \label{yuklag}
        -\mathscr{L}_{\rm eff} = \epsilon_{ij} \left[
        (h_b + \delta h_b) \bar b_R 
        {{H_D}}^{\!\!\!\! i}\, Q_L^j
        + (h_t + \delta h_t) \bar t_R Q_L^i
        {{H_U}}^{\!\!\!\! j} \right]
        + \Delta h_t \bar t_R Q_L^k {{H_D}^{\!\!\!\! k \ast}}
        + \Delta h_b \bar b_R Q_L^k {{H_U}^{\!\!\!\! k \ast}}
        + {\rm h.c.}\,,
\end{equation}
which yields a modification of the tree-level relations between
$h_t$, $h_b$ and $m_t$, $m_b$ as
follows~\cite{Hempfling:1993kv,Hall:1993gn,Carena:1994bv,Bartl:1995tx,Coarasa:1995yg,Jimenez:1995wf,Pierce:1996zz,Carena:1999py,Carena:2002es}:
\beqa
        m_b &=& \frac{h_b v}{\sqrt{2}} \cos\beta
        \left(1 + \frac{\delta h_b}{h_b}
        + \frac{\Delta h_b \tan\beta}{h_b} \right)
        \equiv\frac{h_b v}{\sqrt{2}} \cos\beta
        (1 + \Delta_b)\,, \label{byukmassrel} \\[5pt]
        m_t &=& \frac{h_t v}{\sqrt{2}} \sin\beta
        \left(1 + \frac{\delta h_t}{h_t} + \frac{\Delta
        h_t\cot\beta}{h_t} \right)
        \equiv\frac{h_t v}{\sqrt{2}} \sin\beta
        (1 + \Delta_t)\,. \label{tyukmassrel}
\eeqa
The dominant contributions to $\Delta_b$ are $\tan\beta$-enhanced,
with $\Delta_b\simeq (\Delta h_b/h_b)\tan\beta$.  Moreover, in light of our assumption that $M_Z$, $M_A\ll M_{\rm SUSY}$, it follows that $\delta h_b\sim\mathcal{O}(M_Z^2/M_{\rm SUSY}^2)$ is suppressed, whereas $\Delta h_b$ does not decouple.  
This non-decoupling can be explained by the fact that $\Delta h_b$ arises from the radiatively-generated wrong-Higgs couplings.  Below the scale $M_{\rm SUSY}$, the effective low energy theory is the 2HDM which contains the most general set of Higgs-fermion Yukawa couplings allowed by gauge invariance, and is no longer restricted to be of Type-II~\cite{Haber:2007dj}.   
Similarly, $\delta h_t\sim\mathcal{O}(M_Z^2/M_{\rm SUSY}^2)$ is suppressed, whereas $\Delta h_t$ does not decouple.   However, $\Delta_t$ is not $\tan\beta$-enhanced and thus yields only small corrections to the Higgs boson couplings to fermions in the parameter regime of interest (i.e., where $\tan\beta\gsim1$).

In the parameter regime where $M_Z$, $M_A\ll M_{\rm SUSY}$~\cite{Hempfling:1993kv,Hall:1993gn,Carena:1994bv,Pierce:1996zz,Haber:2000kq},
\beq \label{eq:Deltab}
        \Delta_b
        =\left[
        \frac{2 \alpha_s}{3 \pi} \mu m_{\tilde g} \,
        I(m_{\tilde b_1}, m_{\tilde b_2}, m_{\tilde g})
        + \frac{h_t^2}{16 \pi^2} \mu A_t \,
        I(m_{\tilde t_1}, m_{\tilde t_2}, \mu)\right]\tan\beta+\mathcal{O}\left(\frac{M_Z^2}{M_{\rm SUSY}^2}\right)\,,
\eeq
where 
$m_{\tilde g}$ is the gluino mass,
$m_{\tilde b_{1,2}}$ and $m_{\tilde t_{1,2}}$ are the
bottom and top squark masses, respectively, and smaller electroweak
corrections have been ignored.
The loop integral $I(a^2,b^2,c^2)$ is given by
\beq \label{Iabc}
I(a,b,c) = \frac{a^2b^2\ln(a^2/b^2)+b^2c^2\ln(b^2/c^2)+c^2a^2\ln(c^2/a^2)}{(a^2-b^2)(b^2-c^2)(a^2-c^2)}\,,
\eeq
Note that
\beq
I(a,a,c)=\frac{a^2-c^2+c^2\ln(c^2/a^2)}{(c^2-a^2)^2}\,,
\eeq
and $I(a,a,a)=1/(2a^2)$.  Thus, in the limit in which all supersymmetric parameters appearing in \eq{eq:Deltab} are all very large, of $\mathcal{O}(M_{\rm SUSY})$, we see that $\Delta_b\simeq (\Delta h_b/h_b)\tan\beta$ does not decouple, as previously advertised.  

From \eq{yuklag} we can obtain the couplings of the physical Higgs
bosons to third generation fermions.  The resulting interaction
Lagrangian is of the form,
\beq \label{linthff}
\mathscr{L}_{\rm int} =  -\sum_{q=t,b,\tau} \left[g_{h q\bar q} q \bar{q}h +
 g_{Hq\bar q} q \bar{q}H-
 i g_{A q\bar q}  \bar{q} \gamma_5 q A\right] + \left[g_{H^- t\bar b}\bar b 
t H^- + {\rm h.c.}\right]\,.
\eeq
Expressions for the Higgs couplings to the third generation quarks can be found in Ref.~\cite{Carena:2002es}.  In particular, the charged Higgs coupling to the third generation quarks is noteworthy.  It is convenient to write the approximate one-loop corrected $H^- t\bar{b}$ coupling in the following form,
\beq  \label{Ychhiggs}
g_{H^-t\bar b} \simeq \left[h_t\cos\beta\left(1+\frac{\delta h_t}{h_t}\right)-\Delta h_t\sin\beta\right]P_R
+\frac{\sqrt{2}m_b}{v}\tan\beta\left[1+
\frac{1}{(1+\Delta_b)\sin^2\beta}\left(\frac{\delta h_b}{h_b}
-\Delta_b\right)\right]P_L\,,
\eeq
where $P_{R,L}\equiv\tfrac12(1\pm\gamma_5)$.   

One of the important constraints on the MSSM Higgs sector is derived from the decay rate for $b\to s\gamma$ due to the presence of one loop diagrams involving a charged Higgs boson.  At large $\tan\beta$, it is important to incorporate SUSY corrections to the charged Higgs couplings to quarks\footnote{By including the radiative corrections via \eq{Ychhiggs}, we are effectively incorporating the leading two-loop contributions to the decay matrix element for $b\to s\gamma$ induced by SUSY vertex corrections.}
 in the computation of  ${\rm BR}(b\to s\gamma)$~\cite{Carena:2000uj}.  Including the radiatively corrected $H^+\bar{t}b$ and $H^- t\bar{s}$ couplings using \eq{Ychhiggs}, suitably generalized to include intergenerational quark mixing, and taking $\tan\beta\gg 1$,  
\beq \label{BRratio}
\frac{{\rm BR}(b\to s\gamma)_{{\rm MSSM}, H^\pm}}{{\rm BR}(b\to s\gamma)_{{\rm 2HDM-II}}}\simeq \frac{1}{1+\Delta_b}\left[1-\frac{\Delta h_t}{h_t}\tan\beta\right]\,,
\eeq
after comparing the result obtained from the contribution of the charged Higgs loop in the MSSM, including the leading SUSY radiative corrections to the charged Higgs-fermion couplings, to the corresponding results of the 2HDM with Type-II Yukawa couplings.  
In \eq{BRratio}, $\Delta_b$ is given by \eq{eq:Deltab} and $\Delta h_t$ is given by~\cite{Carena:2000uj}
\beq \label{eq:Deltahtb} 
 \frac{\Delta h_t}{h_t} \simeq \frac{2\als}{3\pi} \mu \mgl \bigl[\cos^2\theta_{\tilde{t}} ~I(m_{\tilde{s}_L},m_{\tilde{t}_2}, \mgl) + \sin^2\theta_{\tilde{t}}~I(m_{\tilde{s}_L},m_{\tilde{t}_1}, \mgl) \bigr]\,,
\eeq
where 
$m_{\widetilde{s}_L}$ is the mass of the SUSY partner of the left-handed strange quark, $\theta_{\tilde{t}}$ is the $\widetilde{t}_L$--$\widetilde{t}_R$ mixing angle~\cite{Ellis:1983ed}, and $I$ is defined in \eq{Iabc}.  Once again, the non-decoupling behavior of $\Delta h_t$ is evident
in the limit in which all supersymmetric parameters appearing in \eq{eq:Deltahtb} are of $\mathcal{O}(M_{\rm SUSY})$.  
As previously emphasized, the non-decoupling properties $\Delta_b$ and $\Delta h_t$  arise due to the wrong-Higgs Yukawa couplings, and are responsible for the significance of the deviation from Type-II behavior of the two-Higgs doublet sector of the MSSM.

\section{Parameter sampling, Observables and Constraints}
\label{sec:ParObsConstraints}
\subsection{Sampling of the parameter space}
\label{sec:sampling}

We sample the \pMSSM\ parameter space with uniformly distributed random
values in the eight input parameters. Scans are performed separately
for the light Higgs and heavy Higgs interpretation of the observed Higgs
signal (see below for details)
over the parameter ranges given in Table~\ref{tab:param}. Besides the
scan parameters listed in Table~\ref{tab:param}, the remaining MSSM
parameters are chosen as described in \refse{sect:theoryparam}.

\begin{table}[h!]
\centering
\begin{tabular}{|r|cc|cc|}
\toprule
 & \multicolumn{2}{c|}{Light Higgs case} & \multicolumn{2}{c|}{Heavy Higgs case} \\
\midrule
Parameter &  Minimum &  Maximum &  Minimum &  Maximum \\
\midrule
$\MA$ [GeV]    & 90       & 1000  & 90 & 200 \\
$\tb$ \phantom{[GeV]}         & 1        & 60  &1 & 20 \\
$M_{\tilde{q}_3}$ [GeV]  & 200      & 5000 & 200      & 1500 \\
$\msld$ [GeV]  & 200      & 1000  & 200      & 1000 \\
$\mslez$ [GeV] & 200      & 1000  & 200      & 1000 \\
$\mu$ [GeV]    & $-3\,M_{\tilde{q}_3}$      & $3\,M_{\tilde{q}_3}$ & $-5000$ & 5000 \\
$\Af$ [GeV]    & $-3\,M_{\tilde{q}_3}$ & $3\,M_{\tilde{q}_3}$ & $-3\,M_{\tilde{q}_3}$ & $3\,M_{\tilde{q}_3}$\\
$M_2$ [GeV]    & 200      & 500 & 200      & 500 \\
\bottomrule
\end{tabular}
\caption{Ranges used for the free parameters in the \pMSSM\ scan.}
\label{tab:param}
\end{table}

In both cases, we start with \order{10^7} randomly sampled points in the
ranges given in Table~\ref{tab:param} and identify interesting regions where
either $h$ or $H$ has a mass close to the observed signal at $\MHexp\gev$ (i.e.~we
select points with $M_{h/H} \in [120, 130]\gev$) and
  the global $\chi^2$ function is low, see \refse{sec:observables} for details on how the global $\chi^2$ function is evaluated. In a second step we perform dedicated
smaller scans 
over more restricted
parameter ranges in order to obtain high sampling densities in the
interesting regions of the parameter space. 

The choices of the parameter ranges for the light Higgs and heavy Higgs case
differ in particular for $\MA$ and $\tb$, where the ranges in the heavy
Higgs case are quite restricted. This is because $\MH\sim \simMH\gev$ can
only be obtained in a rather small region of the parameter space, and a high
sampling density in this region is desired. Furthermore, while we scan
the third generation squark masses, $\msqd$, up to $5\tev$ in the light
Higgs case, we restrict $\msqd$ to be at most $1.5\tev$ in the heavy Higgs
case. As mentioned before, the SM-like Higgs boson mass can be
lifted to the observed value of $\sim 125\gev$ by radiative
corrections from either a large stop mass scale, $M_S$, or from a
large stop mixing parameter, $X_t$. We consider a larger $\msqd$
range in the light Higgs case in order to allow for solutions
with small to moderate $X_t$, $\mu$  and $A_t$ values. In contrast, in
the heavy Higgs case, the SM-like properties can only be obtained in the
alignment  limit (without decoupling) which already requires large
values of $\mu/M_S$ and/or $A_t/M_S$ (see~\refse{Sec:alignment}), and we
restrict ourselves to $\msqd < 1.5 \tev$ in this case.
Lastly, the choice of the scanning range in the Higgsino mass
parameter, $\mu$, differs in the two cases. In the light Higgs case we
restrict $|\mu| \le 3 \msqd$, thus allowing $\mu \lesssim 15\tev$ for very
large third generation squark masses $\msqd\sim 5\tev$. Parameter points
with more extreme values of $|\mu/\msqd|$ beyond $\sim 3$ often face severe
constraints from vacuum stability 
requirements~\cite{ccb1,ccb2,ccb3,ccb4,ccb5,ccb6,ccb7,Chowdhury:2013dka,Bagnaschi:2015pwa} 
(for a recent
analysis see also \citere{Hollik:2016dcm}). Nevertheless, in the heavy Higgs case
we include
such more extreme values of $|\mu/\msqd|$ and do not impose a
specific upper limit on this ratio. As we discussed in~\refse{Sec:alignment}, $|\mu/M_S|$ greatly influences the $\tb$ value where the alignment limit occurs as
well as the critical $M_A$ value that indicates the crossover of the light and heavy Higgs case mass hierarchies. As we will see, a large ratio
$\mu/\msqd$ will be crucial to obtain an acceptable fit of the heavy Higgs
to the observed Higgs signal. We will comment on the fit outcome in the case
where the requirement $|\mu/\msqd| \le 3$ is imposed.

In our scans we allow both signs of the Higgsino mass parameter $\mu$. The
sign of the SUSY contributions to the anomalous magnetic moment of the muon,
$(g-2)_\mu$, is given by the sign of $\mu$, thus the negative $\mu$ branch
is significantly disfavored in the light of this observable, receiving a $\chi^2$ penalty like the SM
or higher, depending on the mass scale of the relevant SUSY particles. In this work we will
therefore present fit results where $(g-2)_\mu$ is either included or
excluded from the global $\chi^2$ function, see below for details. 
%
In addition to the eight \pMSSM\ scan parameters we sample the top quark pole mass  from a Gaussian distribution with $\mt=173.34\pm 0.76\gev$~\cite{ATLAS:2014wva}, using a cutoff at $\pm 2\sigma$. Effects from other parametric 
uncertainties of SM quantities are estimated to be small and therefore neglected in this analysis. 

We employ \FH\ (version 2.11.2)\footnote{%
Recent updates in the Higgs boson mass calculations~\cite{FHwww} lead to a
downward shift in $\Mh$, in particular for large values of $X_t/M_S$. These
changes range within the estimated uncertainties and should not have a
drastic impact on our analysis.}%
~\cite{Heinemeyer:1998np,*Heinemeyer:1998yj,Degrassi:2002fi,Frank:2006yh,Hahn:2009zz,Hahn:2013ria}
to calculate the SUSY particle spectrum and the MSSM Higgs masses.  The 
remaining theoretical uncertainty (e.g.~from unknown higher-order corrections) in the Higgs mass calculation is estimated 
to be $3\gev$~\cite{Degrassi:2002fi}.
Following \citeres{Benbrik:2012rm, Bechtle:2012jw}, we demand that all points fulfill a $\matr{Z}$-matrix criterion, $\left||Z_{k1}^{\mathrm{2L}}| - |Z_{k1}^{\mathrm{1L}}|\right|/
|Z_{k1}^{\mathrm{1L}}|<0.25$ (with $k=1$ (2) in the light (heavy) Higgs case),
in order to ensure a reliable and stable perturbative behavior in the 
calculation of propagator-type contributions in the MSSM Higgs sector.\footnote{The $\matr{Z}$-matrix is defined in Ref.~\cite{Frank:2006yh}.} In the Feynman-diagrammatic approach of \FH\ all model parameters (except for $\tb$, which is a $\overline{\rm DR}$ parameter defined at the scale $m_t$) are defined in the on-shell (OS) renormalization scheme, which we adopt for the definition of our fit parameters [cf.~\refeq{Eq:fitparameters}].

\subsection{Observables}
\label{sec:observables}
In our scan we take into account the following experimental measurements (we denote all experimental measurements with a hat, while unhatted quantities correspond to the model predictions of the respective quantity):

\begin{itemize}
\item \textbf{Higgs boson mass:}

We use the combined result from the ATLAS and CMS Higgs mass measurements~\cite{Aad:2015zhl},
\begin{align}
\MHhat = (125.09 \oplus 0.21~(\text{stat.}) \oplus 0.11~(\text{syst.}))\gev,
\end{align}
where we linearly combine the uncertainties. In our \pMSSM\ scans, where the measured Higgs mass corresponds to either the light Higgs or the heavy Higgs mass, we linearly add the theoretical Higgs mass uncertainty of $3\gev$. Thus, the total mass uncertainty in the MSSM case is $\sigma_{\MHhat} = 3.32\gev$.

\item \textbf{Higgs signal rates:}

We employ the public code \HSv{1.4.0}~\cite{Bechtle:2013xfa,Stal:2013hwa,Bechtle:2014ewa} to evaluate a $\chi^2$ value,
$\chi^2_\text{HS}$, for the compatibility of the \pMSSM\ predictions
with rate measurements in $85$ different Higgs signal channels from the LHC
experiments ATLAS and CMS, as well as the Tevatron experiments CDF and D\O.
A detailed list of all Higgs rate observables is given in
Appendix~B. \HS\ takes into account the correlations among
major systematic uncertainties, including the uncertainties of the
integrated luminosity and the theoretical uncertainties for the cross
section and branching ratio predictions for a SM Higgs boson. It furthermore
takes into account a potential overlap of signals from nearby Higgs bosons
by simply adding the signal rates if the mass difference of the Higgs bosons
is less than the experimental mass resolution of the search channel.%
\footnote{Interference effects can be incorporated in this context
using the method developed in \citere{Fuchs:2014ola}. Since in our analysis
non-negligible interference contributions only occur between the $\cp$-even 
states $h$ 
and $H$ in the parameter regions where they are nearly mass-degenerate, we
neglect these effects here.}
 This feature is of relevance for the heavy Higgs interpretation where all three neutral Higgs bosons can be within the mass range $\sim (100 - 150)\gev$. For instance, the mass resolution of the $H_\text{SM}\to \tau^+\tau^-$ analyses is typically assumed to be $\sim 25\gev$, thus the $\tau^+\tau^-$ signal rates of Higgs bosons within the above mass range will potentially be added by \HS.

\begin{table}[t!]
\centering
\small
\begin{tabular}{| c| rrr |}
\toprule
Observable & \multicolumn{1}{c}{Experimental value} & \multicolumn{1}{c}{SM value}  & \multicolumn{1}{c|}{MSSM uncertainty} \\
\midrule
$\mathrm{BR}(B\to X_s\gamma)$ & $(3.43\pm 0.21\pm 0.07)\times 10^{-4}$~\cite{Amhis:2014hma}  & 
$(3.40 \pm 0.22)\times 10^{-4}$
& $\pm~0.15 \times 10^{-4}$  \\
$\mathrm{BR}(B_s\to \mu^+\mu^-)$ & $ (2.8 \pm 0.7) \times 10^{-9}$~\cite{CMS:2014xfa} & 
$(3.54 \pm 0.2)\times 10^{-9}$
& -- \\
$\mathrm{BR}(B^+\to \tau^+ \nu_\tau)$ & $(9.1\pm 1.9 \pm 1.1)\times 10^{-5}$~\cite{Lees:2012ju,Kronenbitter:2015kls}  & 
$(8.09 \pm 0.7)\times 10^{-5}$
& -- \\
$\delta a_{\mu}$ & $(30.2\pm 9.0)\times 10^{-10}$~\cite{Bennett:2004pv,Bennett:2006fi,Davier:2010nc}  & -- & --\\
$M_W$ & $(80.385\pm 0.015) \gev$~\cite{Group:2012gb,ALEPH:2005ab}& $(80.358 \pm 0.007) \gev$ & $\pm~0.003 \gev$  \\
\bottomrule
\end{tabular}
\caption{The experimental values and SM theory predictions for the low-energy observables (LEOs) that are used in the \pMSSM\ scan. The last column lists additional uncertainties intrinsic to the MSSM predictions.}
\label{tab:flav}
\vskip -0.1in
\end{table}

The Higgs production cross sections are evaluated,
both in the MSSM and the SM, with the code \FH\ (version 2.11.2)
\cite{Heinemeyer:1998np,*Heinemeyer:1998yj,Degrassi:2002fi,Frank:2006yh}. 
This includes an implementation of the SM cross sections of the LHC Higgs cross section
working group (LHCHXSWG)~\cite{Dittmaier:2011ti,*Dittmaier:2012vm,*LHCHXSWG} (using 
the $gg\to H$ cross section prediction from Ref.~\cite{deFlorian:2009hc,*ggHgrazzini}).
The MSSM Higgs production cross sections are calculated
in the effective coupling approximation~\cite{Hahn:2006my}. 
More details on the calculation of the production cross section in the
various channels can be found in Refs.~\cite{Hahn:2006my,Hahn:2010te}. The
Higgs decay widths are also calculated with \FH, including the full
one-loop corrections for
the Higgs decay to fermions and leading higher-order 
contributions~\cite{Heinemeyer:2000fa,Williams:2011bu}.

\item \textbf{Low energy observables (LEOs):}

We include the rare $B$ meson decays $B\to X_s \gamma$ ($X_s$ represents any hadronic system containing a strange quark), $B_s \to \mu^+\mu^-$ and $B^+ \to \tau^+ \nu_\tau$, 
which have small branching ratios in the SM, being either loop or helicity suppressed.
In SUSY, however, they can be mediated 
through SUSY particles and/or charged Higgs bosons,
which can give sizable contributions.
Thus these observables feature a high sensitivity to physics beyond
the Standard Model (BSM).\footnote{
We do not include the $B \to D^{(*)}\tau^-\bar{\nu}_\tau$ measurements\cite{Lees:2012xj,Abdesselam:2016cgx,Aaij:2015yra}, which show some tension with respect to the SM prediction. 
For an explanation within the MSSM (requiring a mass degeneracy between the
lightest chargino and neutralino), see~\cite{Boubaa:2016mgn}.
}

We evaluate the SM prediction --- using a top mass value of $m_t = 173.34\gev$ and a Higgs mass value of $\MH^{\mathrm{SM}} = 125.09\gev$ as input --- and MSSM predictions for these flavor observables with the public code \SI\ (version 3.5)~\cite{Mahmoudi:2007vz,Mahmoudi:2009zz,Mahmoudi:2008tp}. These predictions are listed besides the latest (combinations of) experimental measurements in Table~\ref{tab:flav}. 
For $\br(B\to X_s\gamma)$ we use the current world average of \citere{Amhis:2014hma}.
Here we assign an additional uncertainty on the MSSM prediction for $\br(B\to X_s\gamma)$ of $\pm 0.15\times 10^{-4}$~\cite{Gambino:2001ew,Degrassi:2000qf}.
The process $B_s \to \mu^+ \mu^-$ was observed for the first time by LHCb and CMS~\cite{CMS:2014xfa,CMS-PAS-BPH-13-007} and recently also by ATLAS~\cite{Aaboud:2016ire}.\footnote{The value used in this work does not include the ATLAS measurement yet.}
For $B^+\to \tau^+\nu_\tau$ we use a Belle combination of measurements using hadronic and semi-leptonic tagging methods with a combined significance of $4.6\sigma$~\cite{Lees:2012ju,Kronenbitter:2015kls}. 
For all observables in Table~\ref{tab:flav}, theoretical and experimental uncertainties are combined linearly.

The anomalous magnetic moment of the muon, $a_\mu = \edz \gmt$, comprises another very sensitive low energy probe of BSM physics. The experimentally observed value exhibits a very persistent deviation from the Standard Model prediction at the level of 3---4$\sigma$~\cite{Blum:2013xva,Dorokhov:2014,Dorokhov:2016ejs}.
We obtain the MSSM contribution to the anomalous magnetic moment of the muon from {\tt SuperIso}, 
which includes the one-loop result \cite{Martin:2001st} as well as leading two-loop contributions \cite{
Degrassi:1998es,Heinemeyer:2003dq,Heinemeyer:2004yq,Stockinger:2006zn}.
We cross-checked the {\tt SuperIso} result with results from \FH\ and found good agreement.

Besides the flavor observables and $a_\mu$, we also include the MSSM prediction of the $W$~boson mass into our fit.
The SM value for $M_W$ shows a $1.8 \sigma$ deviation~\cite{Stal:2015zca} from the latest experimental value~\cite{Group:2012gb,ALEPH:2005ab}.
Our MSSM evaluation of $\MW$ follows~Refs.~\cite{Heinemeyer:2013dia,Stal:2015zca} 
and includes, besides the most advanced SM calculation, the full SUSY one-loop contributions as well as leading SUSY two-loop contributions.
The uncertainties from unknown higher-order corrections have
been estimated to be around $4$~MeV in the SM~\cite{Awramik:2003rn} and 
somewhat larger ($\sim (4-9)$ MeV) in the MSSM~\cite{Haestier:2005ja,Heinemeyer:2006px}, depending on the SUSY mass scale. The main parametric uncertainty on $M_W$ stems from the top quark mass and does not need to be included in our $\chi^2$ evaluation because we vary $m_t$ within its $2 \sigma$ uncertainty in the scan.
The remaining parametric uncertainties from $M_Z$ and $\Delta \alpha_{\rm had}$ are $\sim 3$ MeV. 
Combining these two sources of theoretical uncertainties linearly we estimate $10 \mev$ for the MSSM uncertainty and $7 \mev$ for the SM uncertainty.

\end{itemize}

From these observables and their predictions we evaluate for every parameter point in the scan the global $\chi^2$ function
\begin{align}
\chi^2 &= \frac{(M_{h,H}-\MHhat)^2}{\sigma_{\MHhat}^2}
         + \chi^2_\text{HS}
         +\sum_{i=1}^{n_{\mathrm{LEO}}} \frac{(O_i-\hat{O}_i)^2}{\sigma_i^2} 
         - 2\ln\mathcal{L}_\text{limits}.
\label{eq:totchi2}
\end{align}
As mentioned above, we denote all experimental measurements with a hat. Unhatted
quantities correspond to the model predictions of the respective quantity.
The sum over the low energy observables $O_i$ runs over the five
observables mentioned above. The last term, $-
2\ln\mathcal{L}_\text{limits}$, denotes the contribution from Higgs
search limits at LEP and LHC, for which the likelihood information about
the level of exclusion is available. Details will be given below in
\refse{Sect:Constraints}.

The total number of degrees of freedom, $\nu$, is given by the number of
observables, $n_{\text{obs}}$, minus the number of scan parameters,
$n_\text{para}$. We count every observable and constraint that contributes
to the global $\chi^2$ function, Eq.~\eqref{eq:totchi2}, to $n_\text{obs}$,
thus we have in total $n_\text{obs}=93$ if all observables are included in
the fit. In the SM we have only one free parameter ($n_{\mathrm{para}}=1$),
namely the Higgs mass, whereas in both MSSM cases we have eight fit
parameters ($n_{\mathrm{para}}=8$).


\subsection{Constraints}
\label{Sect:Constraints}

\begin{itemize}

\item \textbf{Exclusion limits from Higgs collider searches:}

For every scan point we test the neutral and charged Higgs bosons against the exclusion limits from Higgs searches at the LEP, Tevatron and LHC experiments by employing the public computer code \HBv{4.2.1}~\cite{Bechtle:2008jh,Bechtle:2011sb,Bechtle:2013gu,Bechtle:2013wla,Bechtle:2015pma}. \HB\ determines for each model parameter point the most sensitive exclusion limit, based on the expected exclusion limit given by the experiments. It then judges whether the parameter point is excluded at the \CL{95\%} by comparing the signal prediction against the observed exclusion limit from the most sensitive analysis. In this way, the quoted C.L.~of the limit is preserved even though many different Higgs analyses are considered at the same time.

Besides the hard cut imposed by testing the parameter points at the
standard \CL{95\%} limit, \HB\ enables us to obtain a likelihood value
for the model exclusion by LEP Higgs searches~\cite{Schael:2006cr}, as
well as by the CMS search for non-standard Higgs bosons decaying into
$\tau$ lepton pairs~\cite{Khachatryan:2014wca,CMS:2015mca} (see
Ref.~\cite{Bechtle:2015pma} for details). While the LEP Higgs searches
are only relevant in the heavy Higgs case, i.e.~where the heavier
Higgs state is the SM-like Higgs boson at $\simMH\gev$, the CMS search
yields important constraints in either case, and in particular at
larger values of $\tan\beta$. Each of these likelihoods, which we
commonly denote as $-2\ln\mathcal{L}_\text{limits}$, approximately
resembles a $\chi^2$ contribution corresponding to one degree of
freedom, and can therefore simply be added to the global $\chi^2$
function, Eq.~\eqref{eq:totchi2}.

\item \textbf{Exclusion limits from SUSY collider searches:}

Lower limits on sfermion and chargino masses from mostly model-independent direct searches at LEP are typically at the level of $\sim 100 \gev$ (summarized in the PDG review~\cite{Olive:2016xmw}) and are applied in our scan. We furthermore require the lightest supersymmetric particle (LSP) to be the lightest neutralino, however, we do not apply any dark matter relic density constraints.

Exclusions from SUSY searches at Run 1 of the LHC are tested by employing the public computer code \CMv{1.2.2}~\cite{Drees:2013wra}, which includes all relevant $8\tev$ SUSY analyses from ATLAS and CMS. However, due to the large computational effort and the large scan samples it is neither feasible nor relevant to test all scan points with \CM. Therefore, in a post-processing step, we select the most interesting parameter points, i.e.~points with a $\chi^2$ difference to the minimal $\chi^2$ of less than $\sim10$, and only test the LHC SUSY search constraints on these points.\footnote{We explicitly check that the point with minimal $\chi^2$ is not excluded by \CM, or, in case it is excluded, we select more points for the \CM\ test in order to retain the maximal $\chi^2$ difference of $\sim 10$ to the minimum.} For each of these points we evaluate the sparticle decay spectrum with \texttt{SUSY-HIT-1.5}~\cite{Djouadi:2006bz} and feed these into \texttt{Herwig++} (version 2.7.1)~\cite{Bahr:2008pv,Bellm:2013hwb} for Monte-Carlo generation of inclusive sparticle pair production and the evaluation of the leading-order production cross section. It is not computationally feasible to evaluate the NLO corrections to the leading-order cross section for each parameter point. Instead, we multiply the leading-order cross section by an estimated global $k$-factor of $1.5$ to approximately account for these corrections. \CM\ processes the MC events through the implemented ATLAS and CMS analyses and follows a similar statistical procedure as \HB: It first determines which analysis is the most sensitive one, based on the expected exclusion limit, and then applies the observed exclusion limit from only this search in order to judge whether the parameter point is excluded at the $95\%~\mathrm{C.L.}$ or not.

\end{itemize}



\section{Results}\label{Sect:Results}

In this section we discuss the results of our numerical analysis.
We first discuss the results for the best fit points in both the light
and heavy Higgs case in order to give an impression on the overall fit
quality. Then we discuss the preferred parameter space for the 
light Higgs case in \refse{sec:LightHiggsResults}. We include a
dedicated discussion of the alignment without decoupling scenario
for which we select only parameter points with $M_A \le 350\gev$.
In \refse{sec:HeavyHiggsResults} we present the results for the heavy
Higgs case. New benchmark scenarios for the heavy Higgs case for LHC
Higgs searches during Run~II are presented in
\refse{Section:HHbenchmark}. 

Recall that
approximate alignment without decoupling relies on an approximate cancellation between tree-level
and loop level contributions to the effective Higgs basis parameter $Z_6$ as discussed in \refse{Sec:alignment}.
The extent of the tuning associated with the 
regions in the \pMSSM\ scan that exhibit approximate Higgs alignment without decoupling
is discussed in Appendix~A.

\subsection{Best-fit points and fit quality}

The minimal $\chi^2$ over the number of
degrees of freedom, $\nu$, indicates the best achievable level of agreement with the
observations within a specific model and defines our best-fit (BF) points. These are summarized in
Table~\ref{tab:totchi2} for the SM and the two MSSM fits that are
separately performed for the light Higgs interpretation ($h$) and heavy
Higgs interpretation ($H$).  The results are given for fits to three
different selections of observables (cf.~\refse{sec:observables}):
\emph{(i)} only Higgs data (i.e.~Higgs mass and rate measurements as
well as the Higgs exclusion likelihoods, \emph{right column}),
\emph{(ii)} all observables except $a_\mu$ \emph{(middle column)},
and \emph{(iii)} all observables \emph{(left column)}. We furthermore provide
the reduced $\chi^2$ value, $\chi^2_\nu \equiv \chi^2/\nu$, as well as
the corresponding $p$-value (assuming an idealized $\chi^2$ probability
distribution) for each scenario in Table~\ref{tab:totchi2}. 

\begin{table}[h!]
\centering
\begin{tabular}{|c|ccc|ccc|ccc|}
\toprule
 & \multicolumn{3}{c|}{full fit} & \multicolumn{3}{c|}{fit without $a_\mu$} & \multicolumn{3}{c|}{fit without all LEOs} \\
 Case & $\chi^2/\nu$ & $\chi^2_{\nu}$ & $p$ & $\chi^2/\nu$ & $\chi^2_{\nu}$ & $p$ & $\chi^2/\nu$ & $\chi^2_{\nu}$ & $p$  \\
\midrule
SM  & $83.7/91$ & $0.92$ & $0.69$ & $72.4/90$ & $0.80$ & $0.91$ & $70.2/86$ & $0.82$ & $0.89$ \\
$h$ & $68.5/84$ & $0.82$ & $0.89$ & $68.2/83$ & $0.82$ & $0.88$ & $67.9/79$ & $0.86$ & $0.81$  \\
$H$ & $73.7/85$& $0.87$ & $0.80$ & $71.9/84$ & $0.86$ & $0.82$ & $70.0/80$ & $0.88$ & $0.78$  \\ 
\bottomrule
\end{tabular}
\caption{Global $\chi^2$ results with $\nu$ degrees of freedom from the
  fits of the SM and the MSSM with either $h$ or $H$ as the LHC signal, 
  the reduced $\chi^2_\nu \equiv \chi^2/\nu$, 
  and the corresponding
  $p$-values. The number of degrees of freedom, $\nu$, are estimated by subtracting the number of free model parameters from the number of observables.   }
\label{tab:totchi2}
\end{table}

In total, we have $92$ ($93$)
observables in the SM and light Higgs MSSM
interpretation (heavy Higgs MSSM interpretation)  contributing to the
global $\chi^2$ value (cf.~Sec.~\ref{sec:observables}): $85$ Higgs
signal rate measurements, one Higgs mass measurement, and one (two)
Higgs exclusion observable(s) for the SM and light Higgs MSSM
interpretation (heavy Higgs MSSM interpretation --- here
also the LEP exclusion bounds apply), as well as five low
energy observables. 

We treat the SM as a one-parameter model, where the free parameter is
the Higgs mass, $M_H$. Its best-fit value is mainly set by the Higgs mass
measurement. 
The $\chi^2$ contribution from the Higgs mass
measurement is therefore negligible in the SM. 
In both MSSM cases we have eight free model parameters. 

Taking into account only the Higgs data, 
the minimal $\chi^2$ values found in all three cases are very similar,
with the lowest value being found in the light Higgs case of the
MSSM. However, accounting for the additional degrees of freedom in the
two MSSM cases, the overall fit quality is slightly better in the
SM. Nevertheless, all three scenarios give very high $p$-values,
indicating excellent agreement with the observations in Higgs searches
in each case.\footnote{It should be noted that the
    Higgs signal rates of certain search
      channels are based on the
    same physical degrees of freedom of the model and thus their
      predictions cannot be varied independently. For example, in the MSSM,
      the predicted rates in $h\to ZZ^*$ and $h\to WW^*$ searches are directly related.
      The effective number of degrees of freedom is thus lower, see
    e.g.~Ref.~\cite{Bechtle:2015nua} for a detailed discussion and
    analysis. However, since the naive $p$-value found in the study
    presented here is of ${\cal O}(>50\,\%)$, no significant change of
    the conclusion on the validity of the model can be expected from a
    pseudo-data based study as performed
    in~Ref.~\cite{Bechtle:2015nua}.}

The picture does not change much when the three flavor observables
\bsg, \bmm\ and \btn, as well as the $W$~boson mass are included in the fit.
In the SM the largest
$\chi^2$ contributions from these additional observables come from the
$W$~boson mass ($\chi^2 \sim 1.5$) and \bmm\ 
($\chi^2 \sim 0.6$). Both MSSM best-fit points yield slightly
better agreement with these observables in comparison to the SM. 

\begin{table}[t!]
\centering
\begin{tabular}{|c|cc|cc|}
\toprule
           &  \multicolumn{2}{c|}{Light Higgs case}  &  \multicolumn{2}{c|}{Heavy Higgs case} \\
Observable &  Prediction & Pull &  Prediction & Pull \\
\midrule
$M_{h/H}~[\mathrm{GeV}]$     & $125.20$ & $+0.034$ & $124.15$ & $-0.29$\\ 
$\bsg$ & $3.55 \times 10^{-4}$ & $+0.185$ & $4.17\times 10^{-4}$ & $+1.138$ \\
$\mathrm{BR}(B_s\to \mu^+\mu^-)$ & $3.03\times 10^{-9}$ & $+0.247$ & $3.48\times 10^{-9}$ & $+0.731$ \\
$\mathrm{BR}(B^+\to \tau^+\nu)$ & $7.53\times 10^{-5}$ & $-0.424$ & $7.38\times 10^{-5}$ & $-0.465$\\
$\delta a_\mu$ & $28.8\times 10^{-10}$ & $-0.151$ & $27.6\times 10^{-10}$ & $-0.289$\\
$M_W~[\mathrm{GeV}]$ & $80.383$ & $-0.080$ & $80.373$ & $-0.480$ \\
\bottomrule
\end{tabular}
\caption{Pull table for the best-fit (BF) points of the two MSSM Higgs interpretations.}
\label{tab:pull}
\end{table}

Taking into account also the anomalous magnetic moment of the muon,
$a_\mu$, as observable in the fit, 
the SM receives a large $\chi^2$ penalty
$(\chi^2\sim 11.3$) and thus becomes disfavored with respect to the two
MSSM interpretations. Both the light and heavy Higgs case of the MSSM
are capable to accommodate the $a_\mu$ measurement, receiving $\chi^2$
contributions of only $\sim 0.3$ and $\sim 1.8$ from this observable at
the best fit point, respectively. The various Higgs mass and LEO predictions and the
respective pull, defined as $(\hat{O}_i - O_i)/\sigma_i$, for the BF points of the two MSSM interpretations are summarized in
Table~\ref{tab:pull}.

The parameters for the best-fit points in the light and heavy Higgs case
are shown in Table~\ref{tab:BFparameters}. Naturally, the $\MA$ values
differ significantly, with a value in the decoupling regime for the
light Higgs case and a low value $\sim 170 \gev$ in the heavy Higgs
case. Large mixing in the stop sector is required in the light Higgs
case to yield $\Mh \sim 125 \gev$. 
As explained above, large higher-order corrections are also required in
the heavy Higgs case. For the best fit value in the latter case the
trilinear coupling $A_t$ is small, while $\Xt$ is still sizable because of
the large contribution from the term $\mu/\tb$.
We find large
and positive values of $\mu$ for the best fit points in both scenarios.
In the light Higgs case 
we find parameter points providing a very good fit in the entire positive $\mu$ range (as we will discuss below),
whereas in the heavy Higgs case large values for $\mu / M_S$ are crucial to achieve the approximate alignment limit.
The scalar
leptons of the first and second generation are relatively light to
accommodate $a_\mu$, 
and also the preferred value for $M_2$ is relatively low.

\begin{table}[t]
\centering
\begin{tabular}{|c|cccccccc|}
\toprule
 & $M_A$ & $\tan\beta$ & $\mu$ & $A_t$ & $M_{\tilde{q}_3}$ & $M_{\tilde{\ell}_3}$ & $M_{\tilde{\ell}_{1,2}}$ & $M_2$ \\
 Case & (GeV) &             & (GeV) & (GeV) & (GeV)             & (GeV)              & (GeV)                    & (GeV) \\
\midrule
$h$ & $929$ & $21.0$ & $7155$ & $4138$ & $2957$ & $698$ & $436$  & $358$ \\
$H$ & $172$ & $6.6$ & $4503$ & $-71$ & $564$ & $953$ & $262$ & $293$ \\
\bottomrule
\end{tabular}
\caption{MSSM parameters for the BF points found for the light Higgs ($h$) and heavy Higgs ($H$) interpretation in the full fit.}
\label{tab:BFparameters}
\end{table}

\begin{figure}[ht!]
\centering
\includegraphics[width=1.10\textwidth]{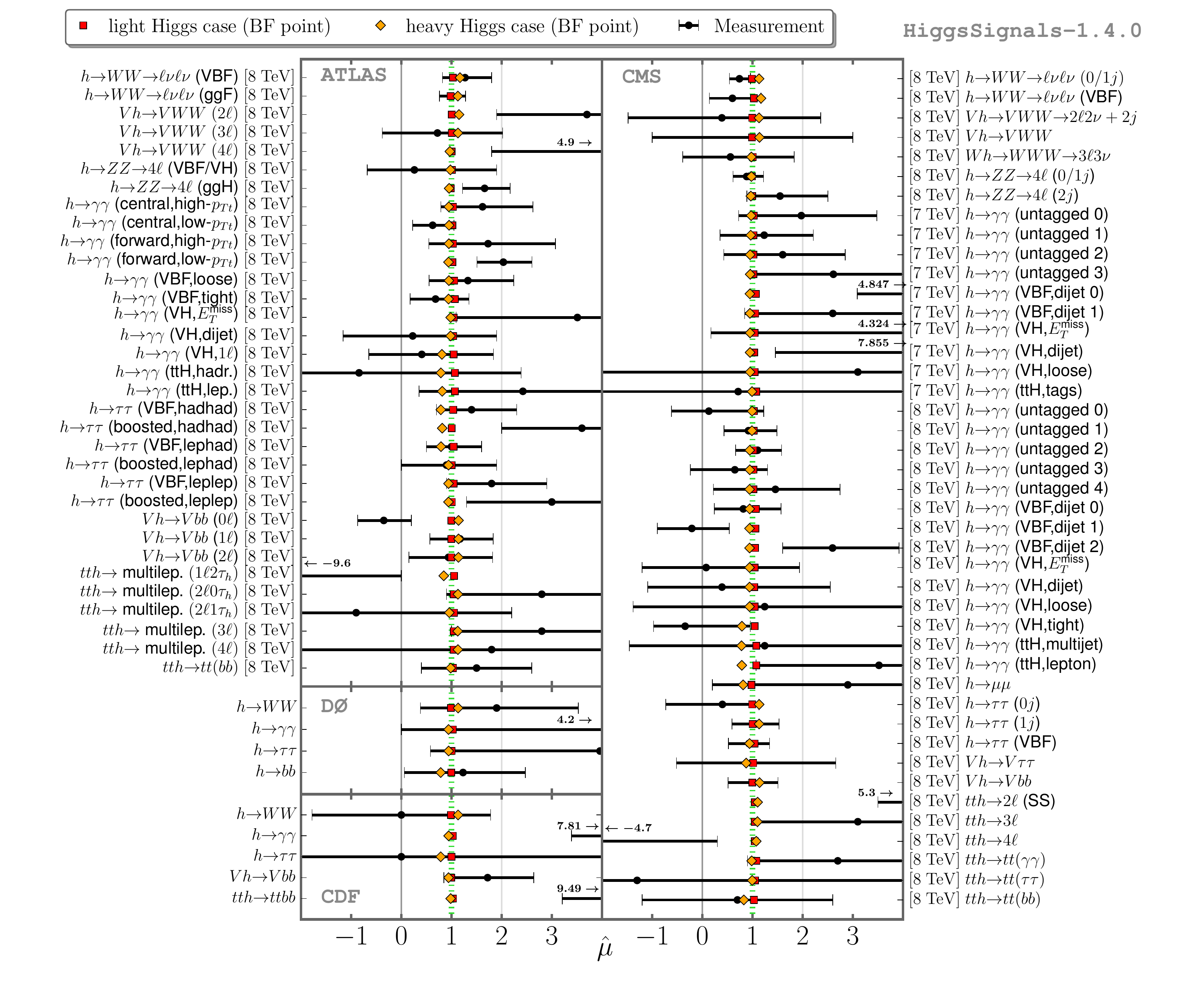}\hfill
\caption{
Comparison of Higgs signal rates in terms of signal strength modifiers $\mu$ between the BF point predictions for the MSSM light Higgs case (\emph{red squares}) and  heavy Higgs case (\emph{orange diamonds}) and the measurements from the LHC and Tevatron experiments ({black dots with error bars}). Displayed are all 85 Higgs rate observables that are provided by \HSv{1.4.0} and included in our fit.
}
\label{Fig:rates}
\end{figure}

A comparison of the Higgs signal rates --- given in terms of signal
strength modifiers $\mu$, which are defined by the measured value of
$\sigma \times {\rm BR}$ in the respective channel normalized to the SM
prediction (see e.g.~Ref.~\cite{Bechtle:2013xfa}) --- between the predictions of the BF points in the light and heavy Higgs case and the Tevatron and LHC measurements is displayed in \reffi{Fig:rates}. The BF point predictions are closely centered around the SM prediction ($\mu =1$), especially in the light Higgs case, where deviations from the SM prediction are $\lesssim\mathcal{O}(2-3\%)$. For the heavy Higgs case BF point the deviations are slightly larger $\lesssim\mathcal{O}(10\%)$, however, these mostly appear in less accurately measured channels.


\subsection{The light Higgs interpretation}
\label{sec:LightHiggsResults}

We start our presentation of the fit results in the light Higgs interpretations of the MSSM with the predicted Higgs signal rates and their correlations for the preferred parameter points. We then give an overview of the preferred MSSM parameter regions. Here we separate the discussion between the full parameter space and a region with low $\cp$-odd Higgs mass, $M_A\le 350\gev$. The latter selects most of the preferred parameter points that feature the limit of alignment without decoupling.
The last two subsections provide dedicated discussions of the impact of the low energy observables and direct LHC SUSY searches on the fit.


\subsubsection{Higgs signal rates}

\begin{figure}
\centering
\includegraphics[width=0.44\columnwidth]{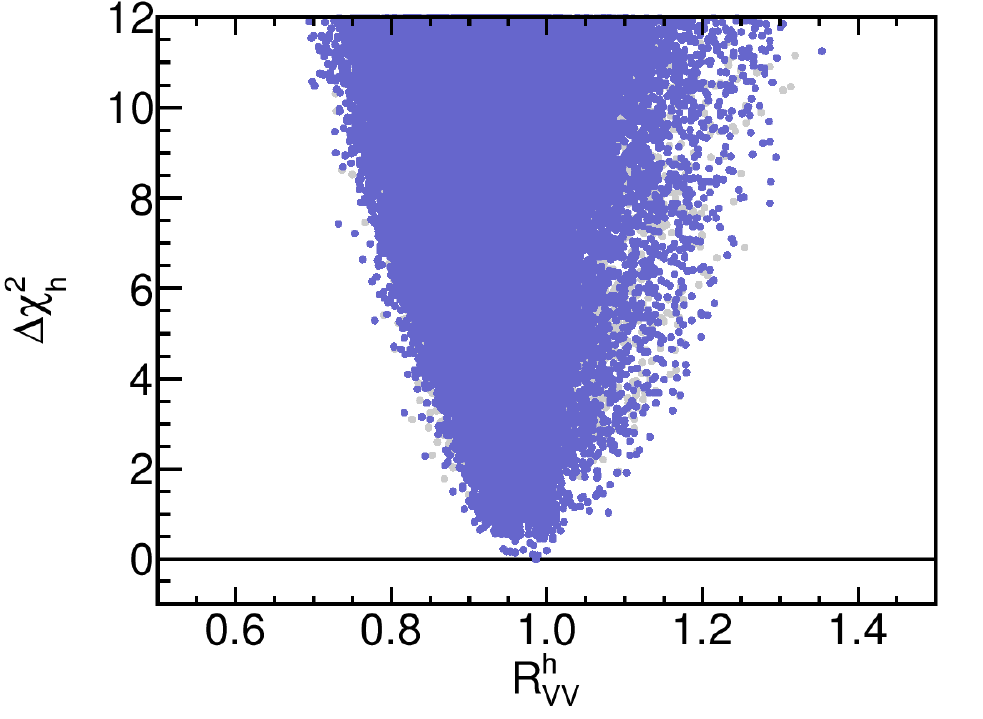}\hspace{0.5cm}
\includegraphics[width=0.44\columnwidth]{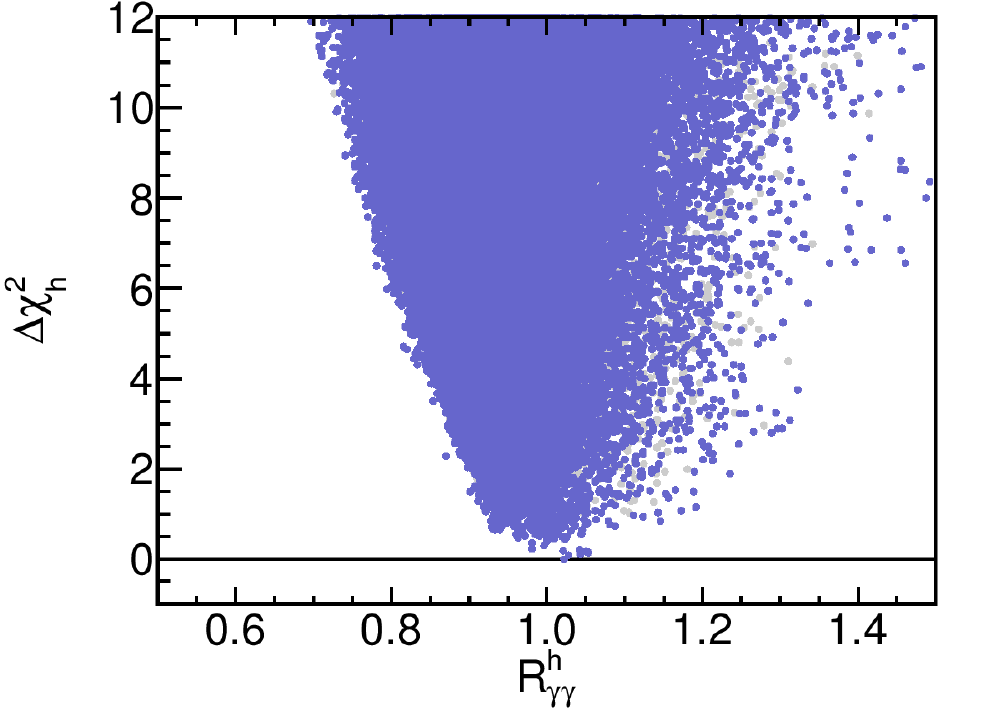}\\
\includegraphics[width=0.44\columnwidth]{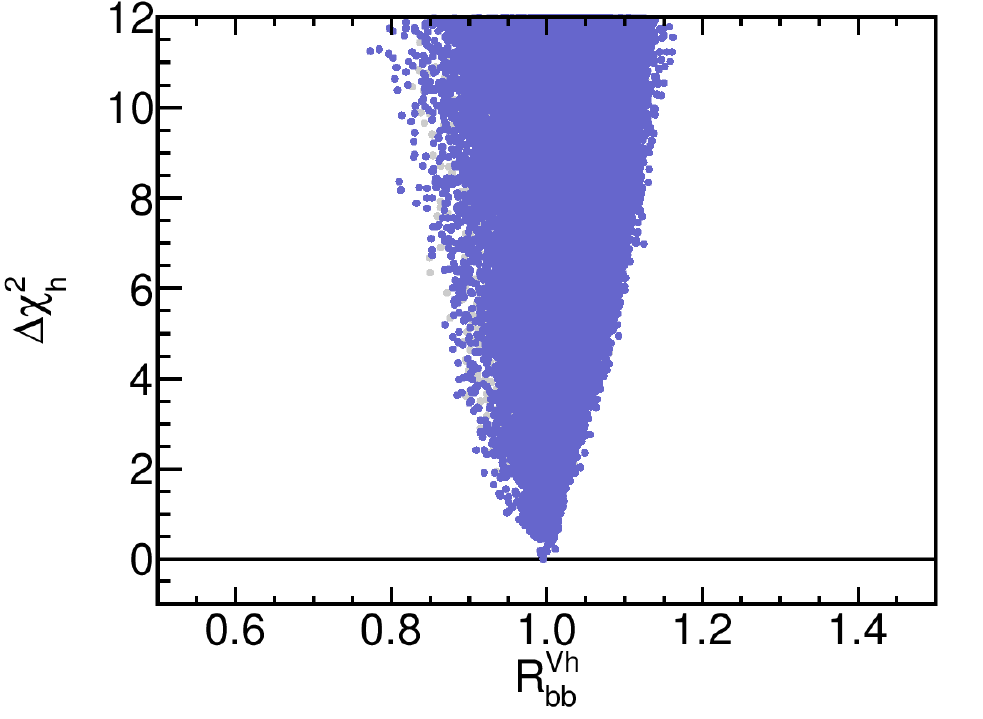}\hspace{0.5cm}
\includegraphics[width=0.44\columnwidth]{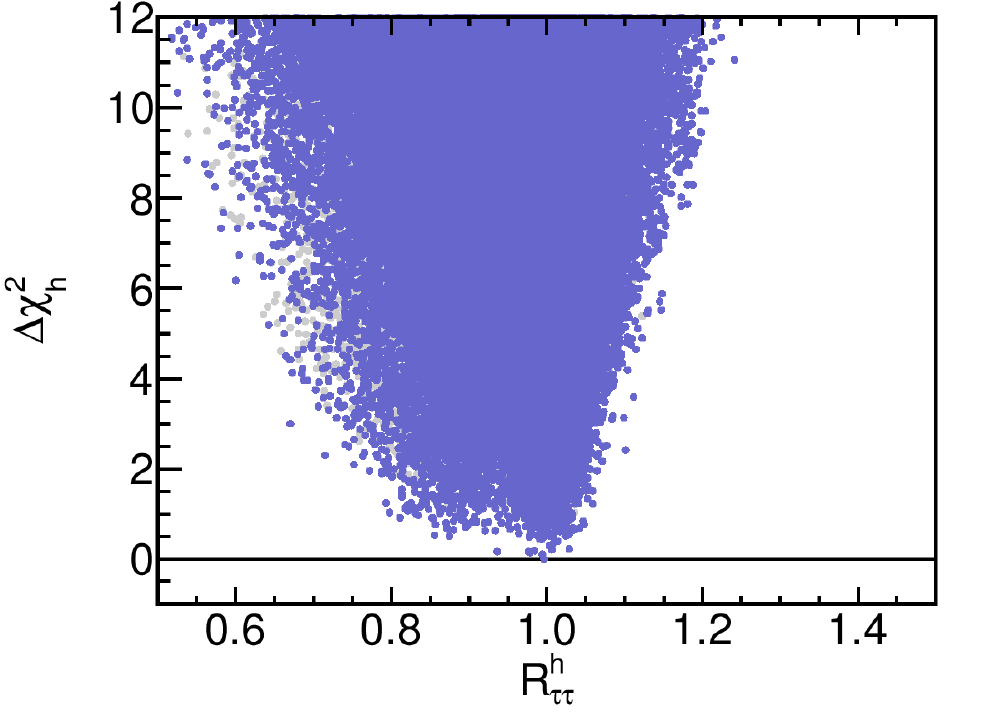}
\caption{$\Delta\chi_h^2 = \chi_h^2-\chi^2_{h,\mathrm{min}}$ distributions in the four dominant Higgs signal
  rates (defined in the text) for the light Higgs interpretation. Points
that pass (do not pass) 
the direct constraints from Higgs searches from \HB\ and
from LHC SUSY particle searches from \CM\ are shown in \emph{blue} (\emph{gray}). }  
\label{fig:totchi2h_rates}
\end{figure}

In Fig.~\ref{fig:totchi2h_rates} we show the 
$\Delta\chi_h^2 = \chi_h^2-\chi^2_{h,\mathrm{min}}$ distributions 
(the subscript '$h$'
refers to the light Higgs ($h$) interpretation), based on all
observables, for four different Higgs signal rates, defined by 
\begin{align}
R_{XX}^{P(h)} = \frac{\sum_{P(h)} \sigma(P(h)) \times \br(h\to XX)}{\sum_{P(h)} \sigma_\mathrm{SM}(P(h)) \times \br_\mathrm{SM}(h\to XX)}.
\label{Eq:Rvalues}
\end{align}
Here $XX= VV, \gamma\gamma, bb, \tau\tau$ (with $V=W^\pm,Z$) denotes the final state from the Higgs decay 
and $P(h)$
denotes the Higgs production mode. For inclusive Higgs production, $P(h)
\equiv h$, the sum in Eq.~\eqref{Eq:Rvalues} runs over the five dominant
Higgs production modes: gluon-gluon fusion (ggf), vector boson fusion
(VBF), associated Higgs production with a $W$ or $Z$ boson, ($VH$, with
$V=W^\pm,Z$) and Higgs production in association with a top quark pair
($t\bar{t}H$). The subscript 'SM' denotes the quantities as predicted in
the SM, whereas no subscript refers to the quantity predicted in the
model. 
Parameter points that pass (do not pass) the constraints from Higgs
exclusion limits, tested via \HB, and direct SUSY LHC searches, tested via \CM, are
given as blue (gray) points.

The preferred Higgs signal rates are
\begin{align}
R_{VV}^h = 0.99 \substack{+0.09 \\ -0.08},\qquad
R_{\gamma\gamma}^h = 1.02 \substack{+0.16 \\ -0.10},\qquad
R_{bb}^{Vh} = 1.00 \substack{+0.02 \\ -0.05},\qquad
R_{\tau\tau}^{h} = 1.00 \substack{+0.06 \\ -0.20},
\end{align}
where the upper and lower values are determined by
$\Delta\chi^2_h \le 1$, which approximately 
corresponds to the one-dimensional
\CL{68\%} interval. All
central values lie very close to the SM prediction ($R=1$). 

The narrowest range is found for the Higgs signal rate $R_{bb}^{Vh}$,
i.e.~for $Vh$ production with the Higgs boson decaying into
$b\bar{b}$. This is mainly because a direct variation of $\br(h\to
b\bar{b})$ through a modification of the partial width $\Gamma(h\to
b\bar{b})$ also leads to a substantial change of the total Higgs decay
width, $\Gamma_{h,\text{tot}}$, and thus affects significantly the
branching ratios of all other Higgs decay modes. This global rescaling
of the branching ratios of all decay modes except $h\to b\bar{b}$ could
only be compensated by an inverse rescaling in the rates of all Higgs
production modes. However, this cannot be accomplished within the MSSM,
and hence the modification of $\br(h\to b\bar{b})$ is severely
constrained. 

In Fig.~\ref{fig:totchi2h_rates} we can furthermore observe a spread of
parameter points deviating substantially from the SM value, $R=1$, within the
$\Delta\chi^2\le 4$ interval (approximately corresponding to the
\CL{95\%} interval), e.g.~$R_{\gamma\gamma}^{h} \lesssim 1.35$ and 
$R_{\tau\tau}^h \gtrsim 0.65$.
Such modifications can easily
appear in the MSSM: An enhancement of the $h\to \gamma\gamma$ partial
decay width can appear through the loop contribution of light charged
SUSY particles such as light scalar tau leptons (staus) or
charginos.\footnote{In the analysis of \citere{Bechtle:2012jw}, based on
the data available at that time, contributions of this kind leading to a substantial
  increase in $R_{\ga\ga}^h$ were favored, whereas with the new data this
  enhancement turns out to be much smaller.} 
  The observed reduction of $R_{\tau\tau}^h$ for some points in \reffi{fig:totchi2h_rates} originates from a simultaneous (but small) suppression of the gluon fusion production cross section, $\sigma(gg\to h)$, and the decay rate $\text{BR}(h\to\tau^+\tau^-)$ with respect to their SM predictions.
  
%

\begin{figure}
\centering
\includegraphics[width=0.44\columnwidth]{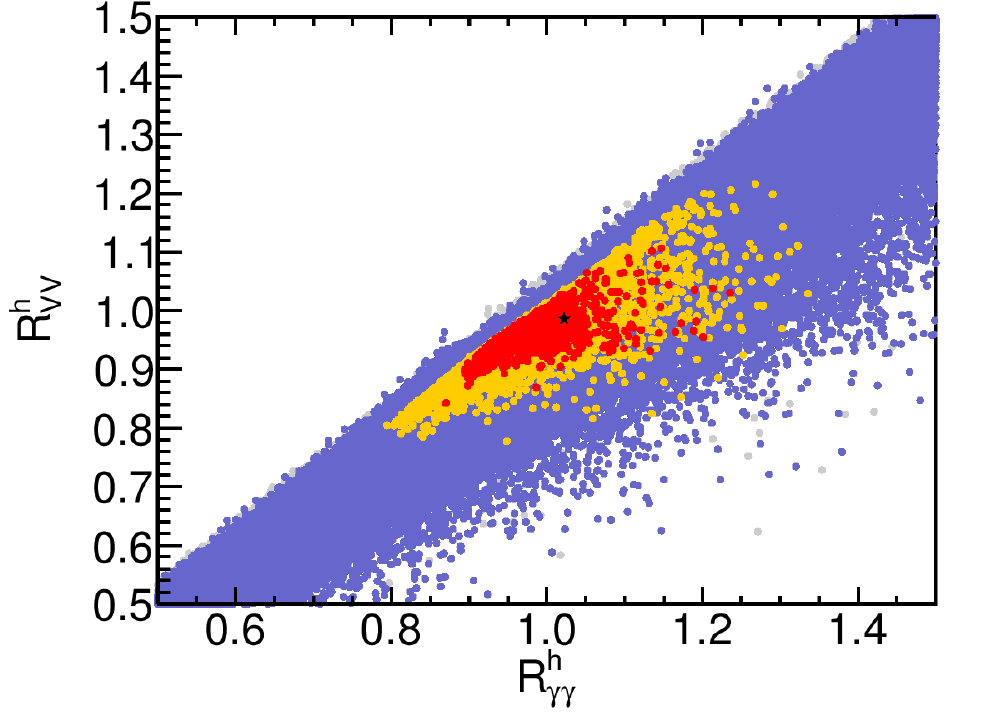} \hspace{0.5cm}
\includegraphics[width=0.44\columnwidth]{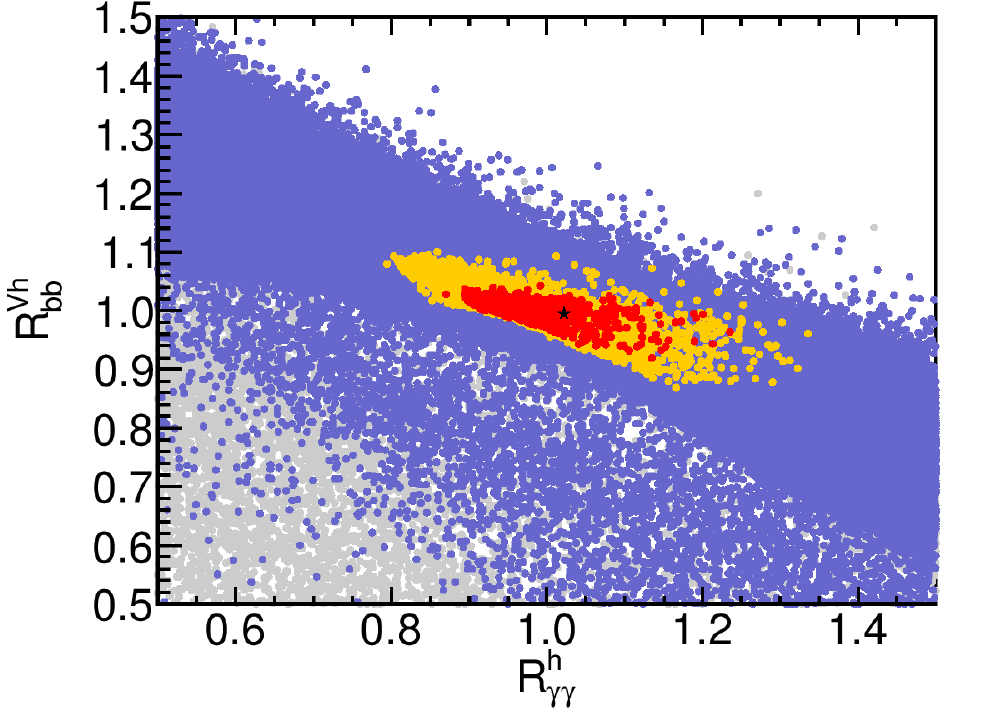}\\
\includegraphics[width=0.44\columnwidth]{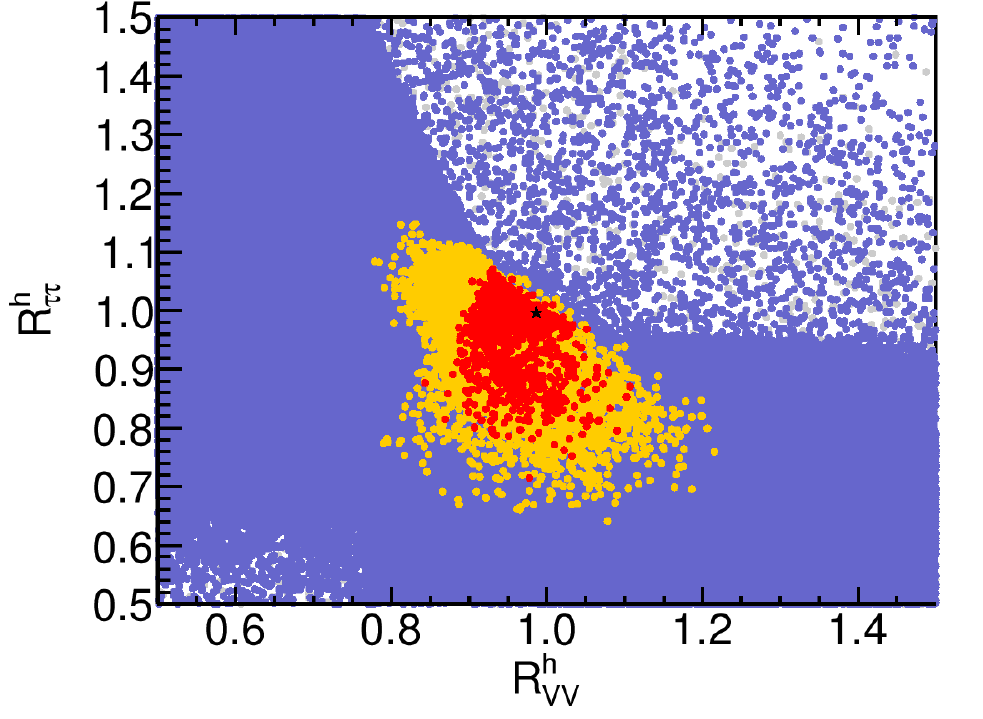} \hspace{0.5cm}
\includegraphics[width=0.44\columnwidth]{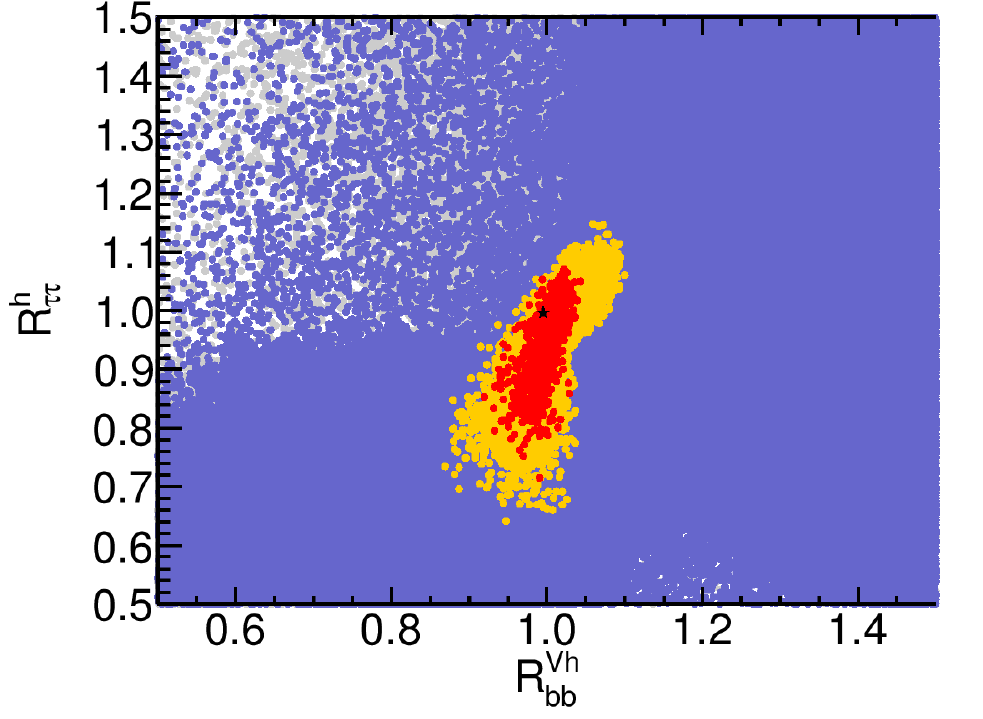}
\caption{Correlations between Higgs signal rates for the
  light Higgs case. The color coding follows that of
  \figref{fig:totchi2h_rates}, with the addition of the favored regions
  with $\Delta \chi_h^2 < 2.3$ (\emph{red}) and $\Delta \chi_h^2 < 5.99$
  (\emph{yellow}). The best fit point is indicated by a \emph{black star}.} 
\label{fig:hrates_corr}
\end{figure}

The two-dimensional correlations among the Higgs signal rates are shown in \figref{fig:hrates_corr}.
Compared to \figref{fig:totchi2h_rates} we introduced two new colors in order to indicate regions close to the minimum 
$\chi^2$. Points with $\Delta\chi^2_{h}<2.30~(5.99)$ are highlighted in
red (yellow), corresponding to points in a two-dimensional ${68}\%$
($95\%$) C.L.~region in the Gaussian limit. We shall denote these
regions simply by \emph{favored/preferred} (yellow) and \emph{most
favored/preferred regions} (red). The best fit point is indicated by a
black star in the figures.

As already noticed in our previous analysis~\cite{Bechtle:2012jw} the
diphoton rate $R_{\gamma\gamma}^h$ exhibits a strong correlation with
$R_{VV}^h$ and a strong anti-correlation with $R_{bb}^{Vh}$. The latter
arises through the strong influence of the $h\to b\bar{b}$ partial width
on the total Higgs decay width (see the discussion above). In
contrast, the 
rate $R_{\tau\tau}^h$ only shows mild correlations with $R_{VV}^{h}$,
$R_{\gamma\gamma}^{h}$ (not shown here) and $R_{bb}^{Vh}$. 
The latter is easily understood from the
fact that the same Higgs doublet couples to down-type quarks and leptons
in the MSSM, thus, at tree-level, the light Higgs coupling to $\tau$
leptons and $b$ quarks is affected in the same way (and enhanced at
large $\tan\beta$).
However,
we also find favoured points that feature a 
significant suppression of $R_{\tau\tau}^h$ while $R_{bb}^h$ shows no
(or only a small) suppression. 
Differences between $R_{\tau\tau}^h$ and $R_{bb}^h$ arise from loop contributions, 
which modify the $h\tau^+\tau^-$ and $hb\bar{b}$ couplings in a different manner, see e.g the discussion of $\Delta_b$ corrections in \citere{Bechtle:2012jw}. 


\subsubsection{Parameter space}

\begin{figure}[b!]
\centering
\includegraphics[width=0.44\columnwidth]{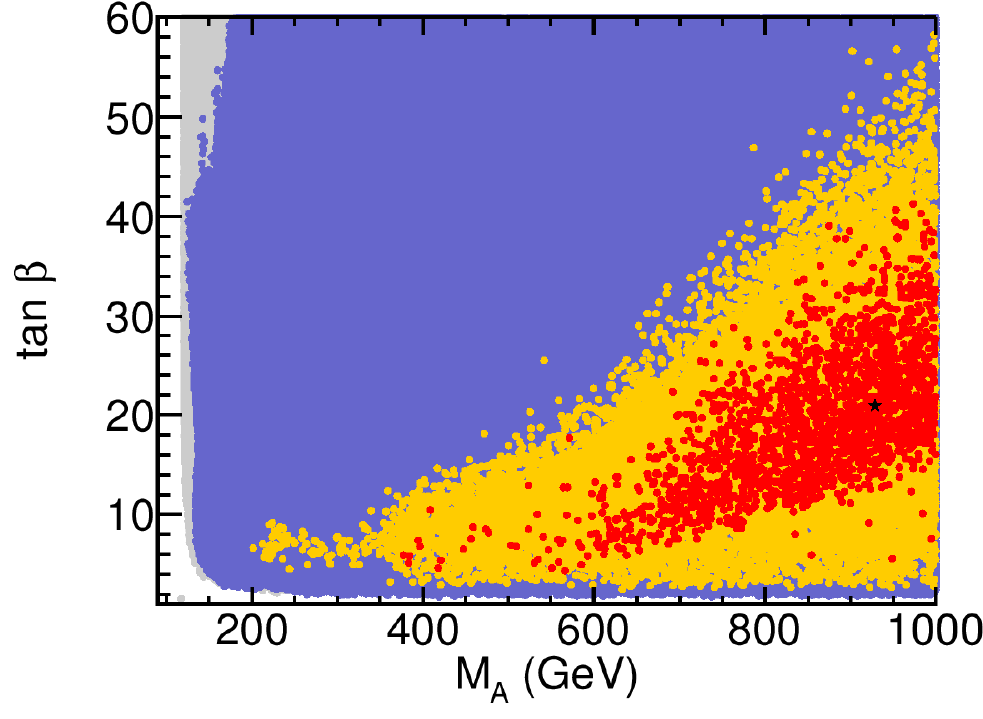}\hspace{0.5cm}
\includegraphics[width=0.44\columnwidth]{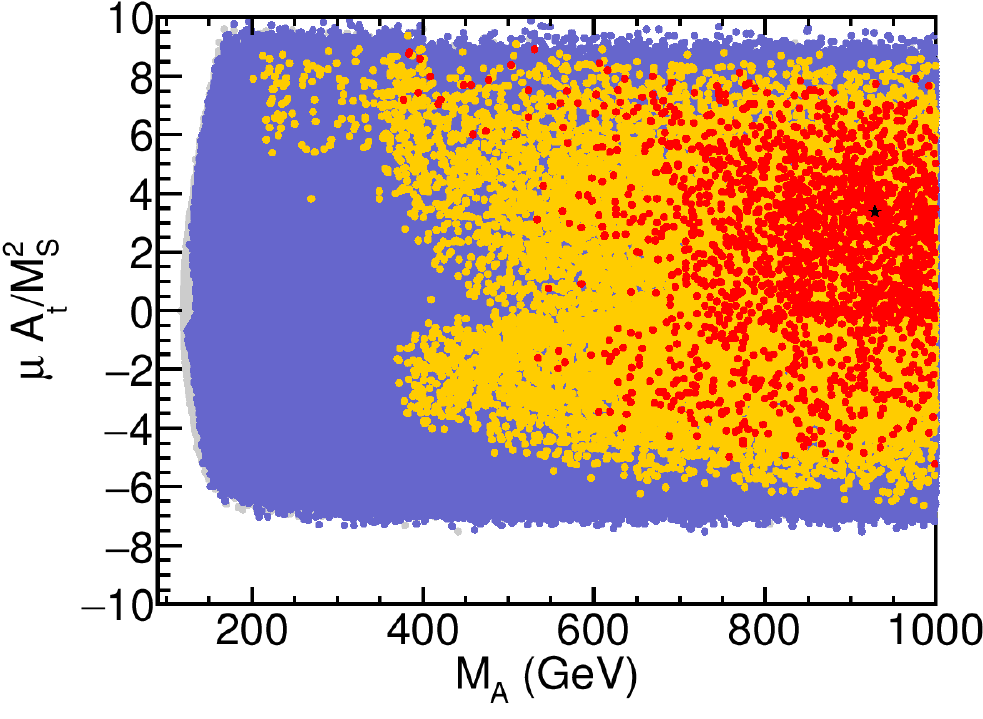}
\caption{Preferred parameter regions in the ($\MA$, $\tb$) plane (\emph{left}) and
the ($\MA$, $\mu A_t/M_S^2$) plane (\emph{right}) for the light
  Higgs case. The color coding is the same as in Fig.~\ref{fig:hrates_corr}. }
\label{fig:h_paramregions}
\end{figure}

The distribution of preferred parameter points in the 
plane of the parameters $M_A$ and $\tan\beta$,
which determine the Higgs sector at lowest order, is shown in
Fig.~\ref{fig:h_paramregions} (left). The bulk of the favored points is
found at large $\cp$-odd Higgs mass values, $\MA \gtrsim 350\gev$,
i.e.~in a region where the decoupling 
limit is already approximately realized. Large $\tan\beta$ values at moderate values of $M_A$ are
disfavored by the non-observation of a signal in LHC $H/A\to
\tau^+\tau^-$ searches, which is incorporated in our study by adding the
exclusion likelihood from CMS (provided by \HB) to the global $\chi^2$
function instead of applying the hard cut at the \CL{95\%} (see \refse{Sect:Constraints})~\cite{Bechtle:2015pma}. 
Thus, points excluded (e.g.~at the \CL{95\%}) by the CMS search alone can still appear as blue
points here, but are unlikely to show up as yellow or red points.

While all most favored (red) points are found for $\MA \gtrsim 350 \gev$,
some preferred parameter points (yellow) are found at low $\cp$-odd Higgs
masses down to $ \MA \gtrsim 170\gev$ in a narrow range of $\tan\beta$
values $\sim 4-10$. 
These points are far away from the decoupling limit,
however, they turn out to feature an (approximate) realization of the limit of \emph{alignment
without decoupling}.  As discussed in \refse{Sec:alignment}
these points must  have either large values of $\mu A_t/M_S^2$ or, in the case where $|A_t/M_S|$ is small, even larger values of $|\mu/M_S|$ (beyond $\sim 3$) in order to
achieve the Higgs alignment limit at reasonably small values of $\tan\beta$
which are unexcluded by LHC $H/A\to \tau^+\tau^-$ searches. This
is indeed the case, as can be seen in
Fig.~\ref{fig:h_paramregions} (right), where we show the preferred
parameter points in the ($\MA$, $\mu A_t/M_S^2$) plane. All preferred
parameter points with $M_A \lesssim 350\gev$ have $\mu A_t/M_S^2$ values
of at least $3$, but typically between $5$ and $9$. It should be
noted that in the light Higgs scan we constrain both scan parameters $\mu$ and $A_t$ to be
$\le 3 M_{\tilde{q}_3}$ (see Table~\ref{tab:param}), limiting the
possible values of $\mu A_t/M_S^2$ to $\lesssim 9$. 
In the following, in
order to study the parameter points close to the limit of \emph{alignment without
  decoupling}, we apply a cut $M_A \le 350\gev$ (denoted as
\emph{low-$\MA$ selection}) whenever relevant to isolate these points from the
other  parameter points.

\begin{figure}
\centering
\includegraphics[width=0.44\columnwidth]{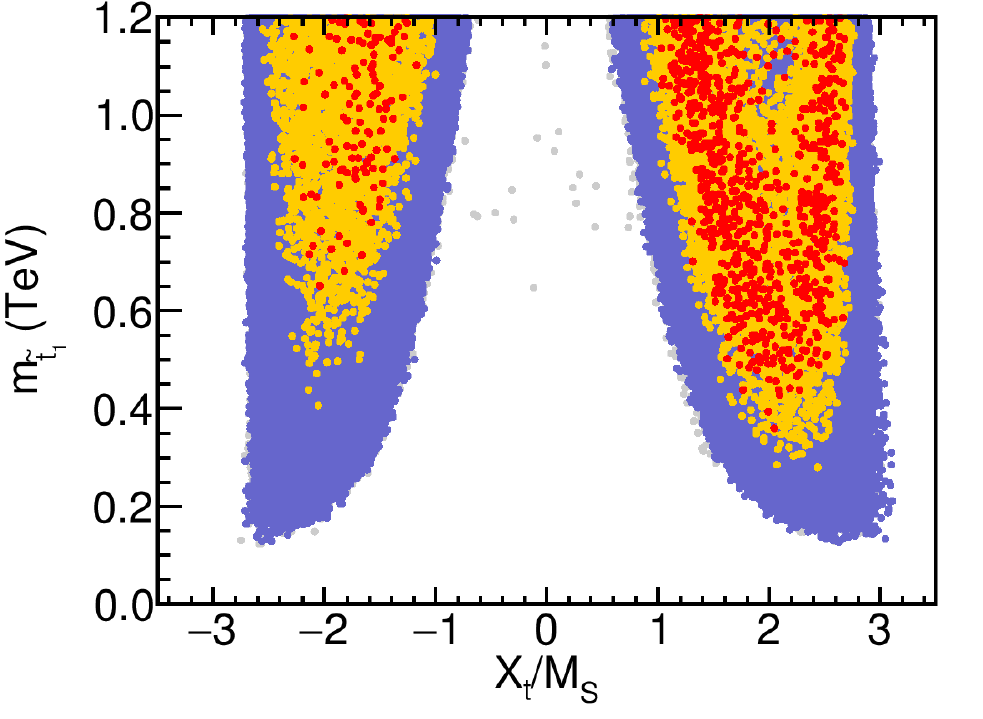}\hspace{0.5cm}
\includegraphics[width=0.44\columnwidth]{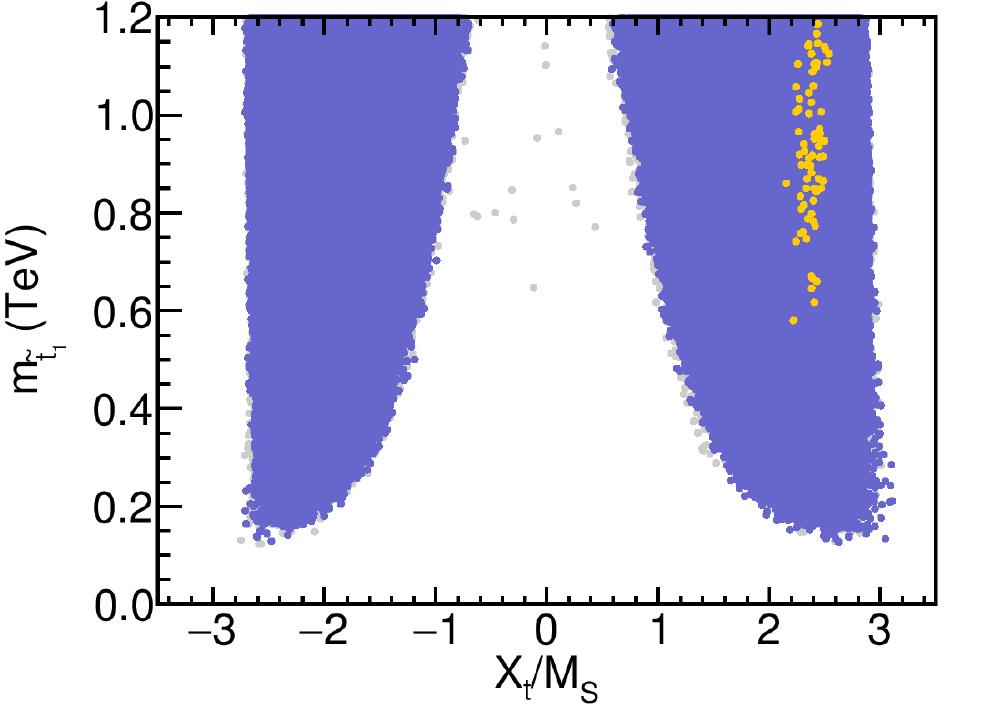}
\caption{Stop mixing parameter, $\Xt/M_S$, versus the light stop mass, $m_{\tilde{t}_1}$, for the light Higgs case for the full scan (\emph{left}) and the low-$M_A$ selection (\emph{right}). The color coding is the same as in Fig.~\ref{fig:hrates_corr}.} 
\label{fig:hl_mstop}
\end{figure}

The MSSM parameters from the stop sector, namely the stop mixing
parameter, $X_t/M_S$, and the light stop mass, $m_{\tilde{t}_1}$, are
shown in Fig.~\ref{fig:hl_mstop} for the full scan (left) and the
low-$\MA$ selection (right). In the full scan we find preferred
parameter points in both the positive and negative $X_t/M_S$ branches
near the value where the contribution to the Higgs mass from stop
mixing is maximized, $|X_t/M_S| \sim 2$.%
\footnote{The highest $\Mh$ values are reached for
$|X_t/M_S| \sim 2$ due to the inclusion of the higher-order corrections
in the on-shell renormalization scheme, see
\citere{Heinemeyer:1999be} for details.} Light stop masses, $m_{\tilde{t}_1}$, down to values $\gtrsim
300~(400)\gev$ are possible in the positive (negative) $X_t/M_S$
branch.\footnote{Note that we assumed universality of the left- and
right-handed soft-breaking stop mass parameter here. Lower light stop
masses even below the top quark mass can be obtained while being
consistent with the Higgs rates in the presence of a large mass
splitting in the stop sector~\cite{Liebler:2015ddv}.} In the low-$M_A$
selection preferred parameter points are found only in the positive
$X_t/M_S$ branch for $X_t/M_S \gtrsim 2$. Here, the lowest light stop mass
value in the preferred parameter region is found at around
$m_{\tilde{t}_1} \sim 580\gev$. 

\begin{figure}
\centering
\includegraphics[width=0.44\columnwidth]{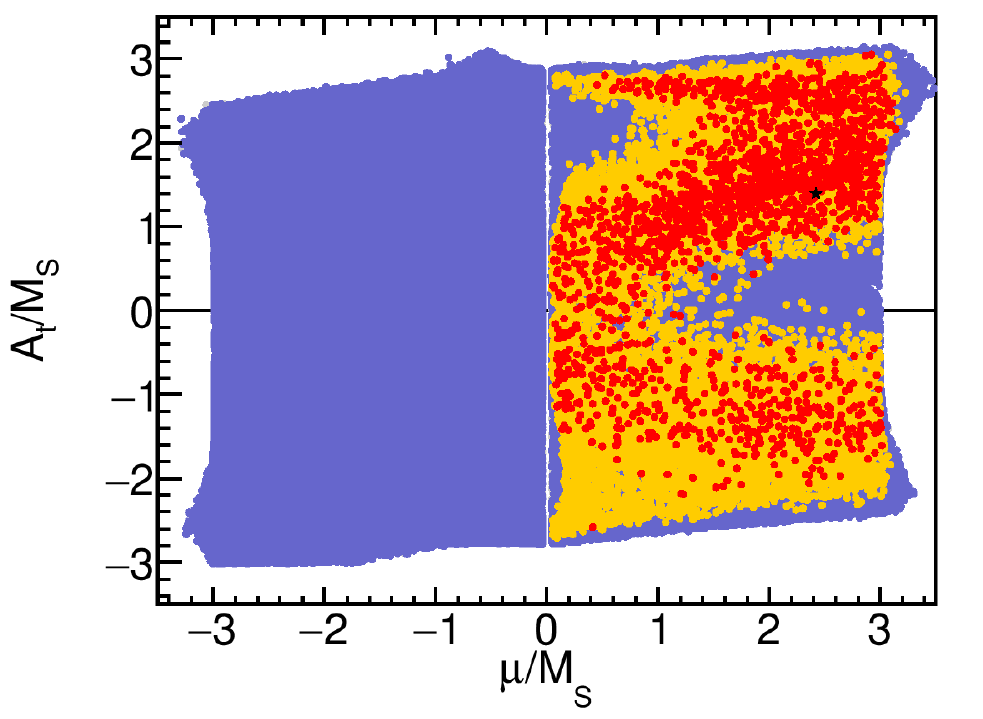}\hspace{0.5cm}
\includegraphics[width=0.44\columnwidth]{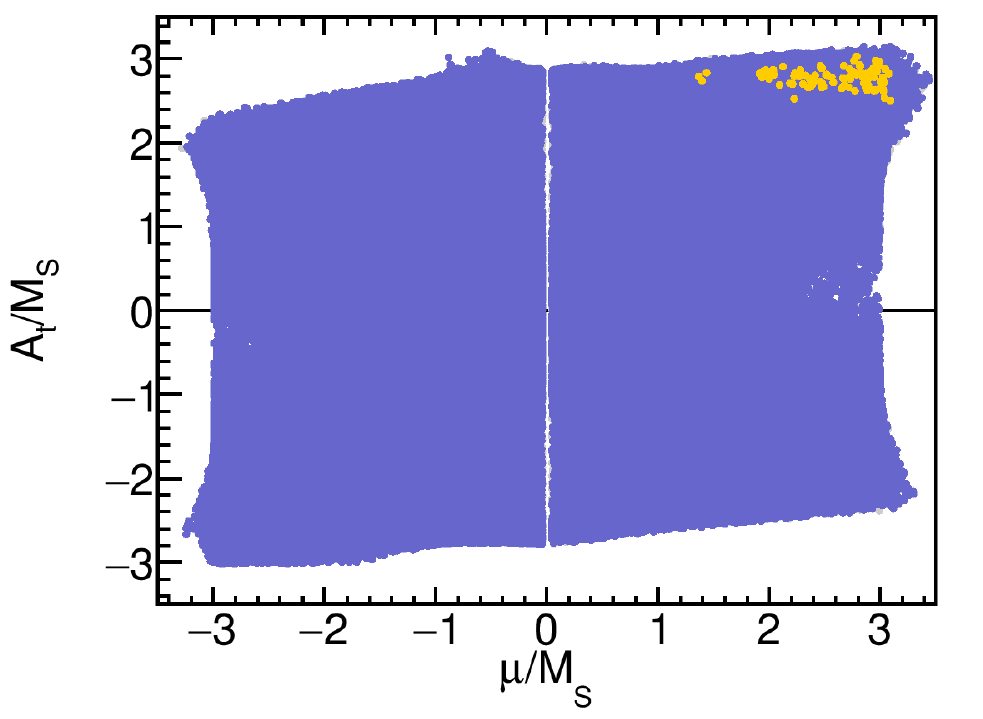}
\caption{Preferred parameter regions in the ($\mu/M_S$, $A_t/M_S$) plane for the full fit (\emph{left}) and the low-$\MA$ selection (\emph{right}). The color coding is the same as in Fig.~\ref{fig:hrates_corr}.}
\label{fig:h_Atmu}
\end{figure}

In order to understand why the favored parameter points near the limit of 
alignment without decoupling are found only at $X_t/M_S =
A_t/M_S - (\mu/M_S) \cot\beta \gtrsim 2$, we show the correlations of
the parameters $A_t/M_S$ and $\mu/M_S$ in Fig.~\ref{fig:h_Atmu} for the
full scan (left) and the low-$\MA$ selection (right). While we find
preferred parameter points at both positive and negative $A_t/M_S$
values for $\mu/M_S > 0$ in the full scan, the low-$M_A$ selection 
features favored points only for very large and positive values of
$A_t/M_S$ and $\mu/M_S$. In particular, we find most of the preferred
parameter points in the low-$\MA$ region in a narrow range $2.4 \lesssim
A_t/M_S \lesssim 3$, while the range in $\mu/M_S$ is larger (roughly
between $1.4$ and $3$). 
In the full scan (left) it can be seen that even values with $A_t/M_S$
and $\mu/M_S$ close to zero can yield a very good fit. Consequently, the quite
large best fit value of $\mu$, see also Table~\ref{tab:BFparameters}, should be
regarded as accidental.

%
%
\begin{figure}[ht!]
\centering
\includegraphics[width=0.48\columnwidth]{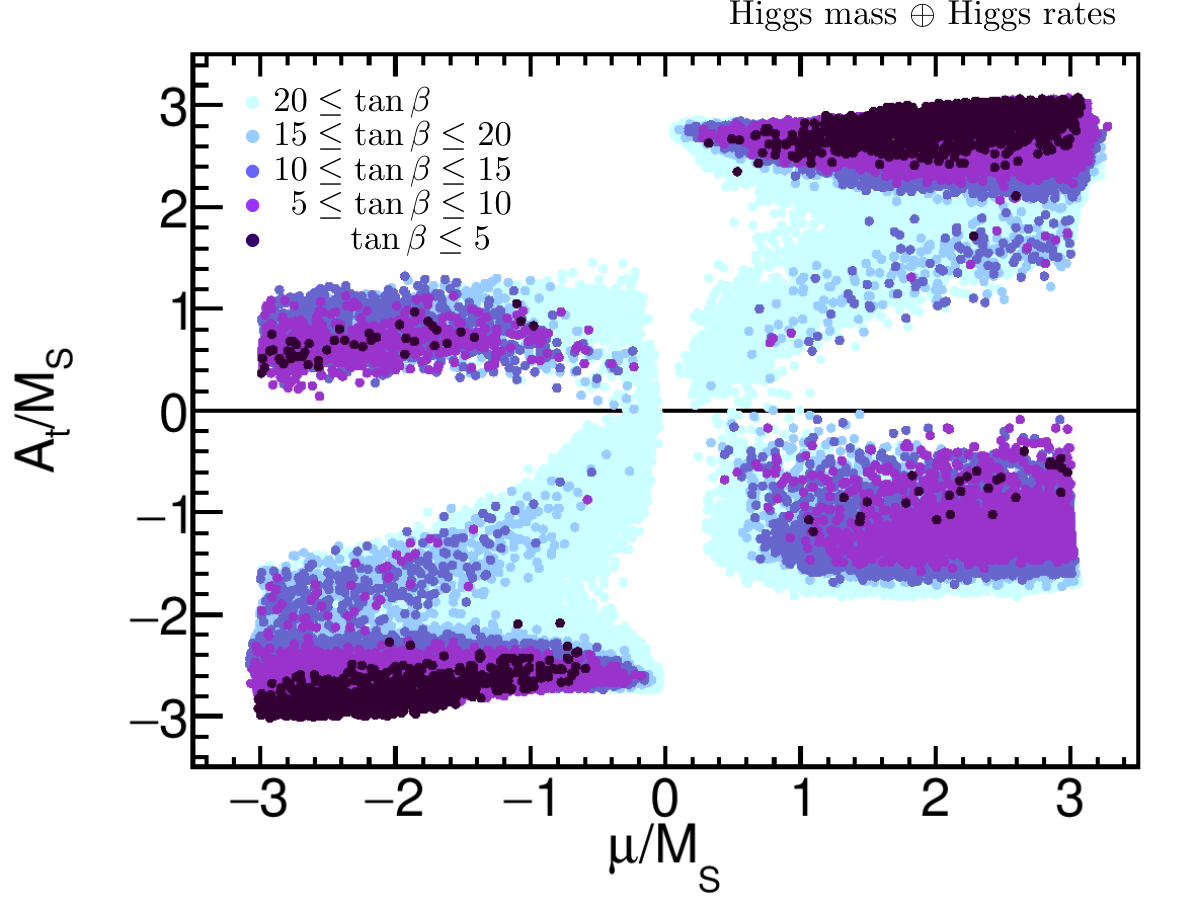}\hfill
\includegraphics[width=0.48\columnwidth]{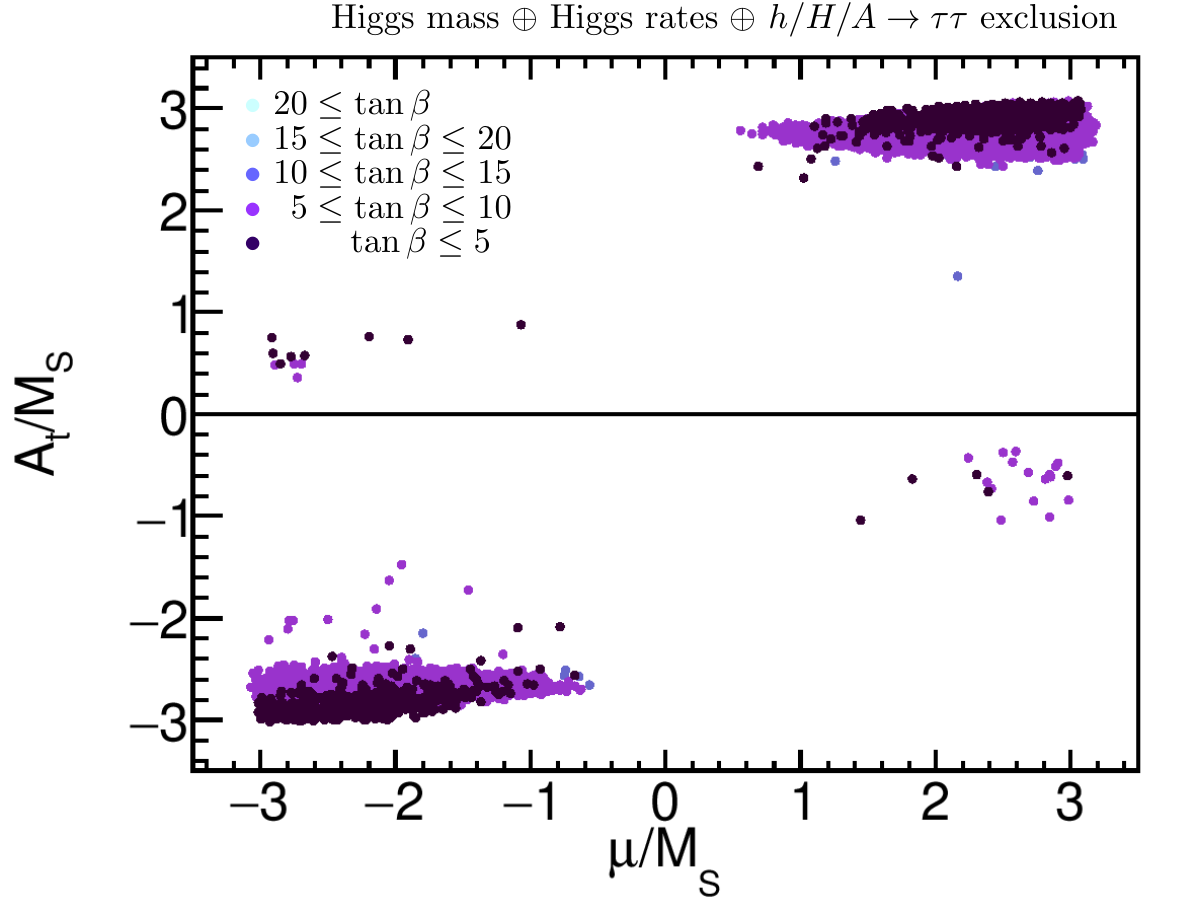}\\
\includegraphics[width=0.48\columnwidth]{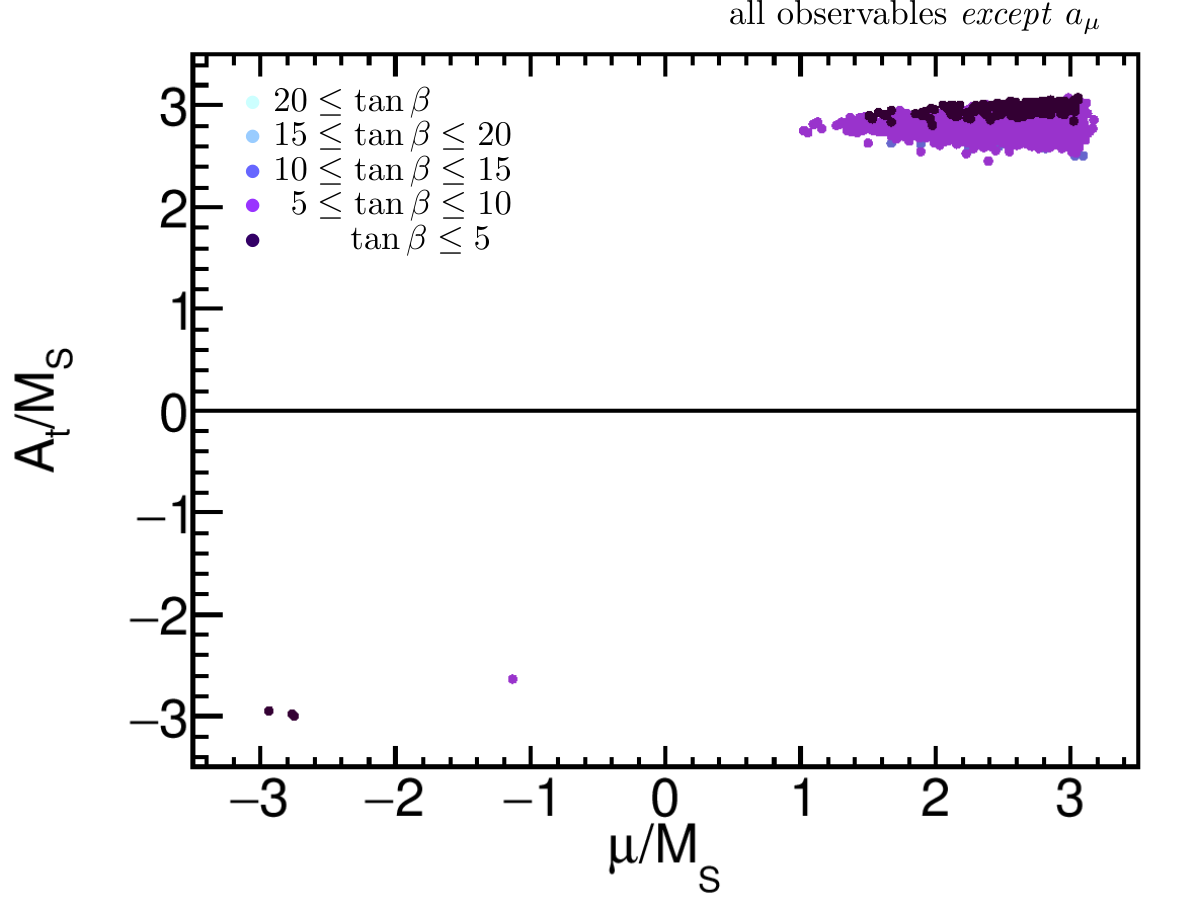}\hfill
\includegraphics[width=0.48\columnwidth]{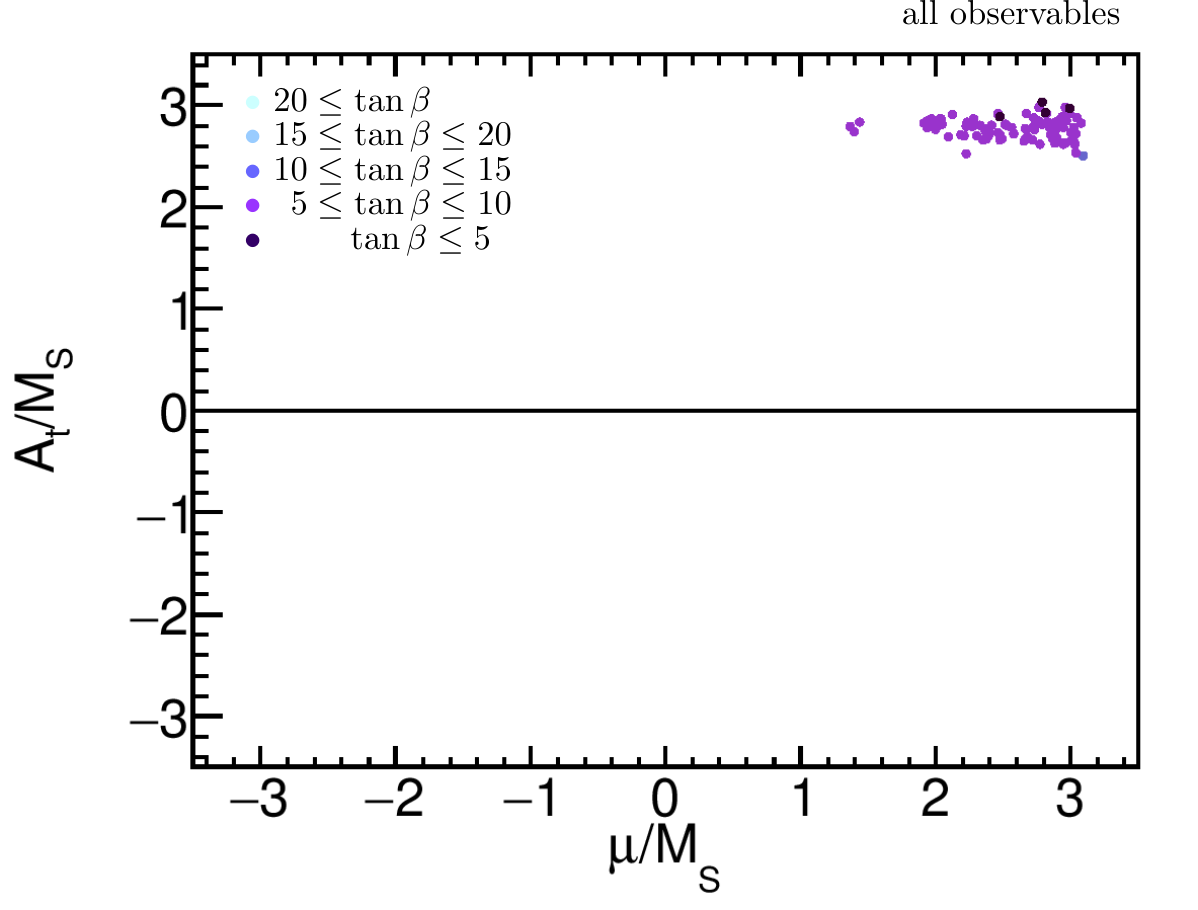}
\caption{$\mu/M_S$ vs. $A_t/M_S$ for the preferred points with low
$\cp$-odd Higgs mass, $M_A \le 350\gev$, for different selection of
observables. The color indicates the $\tan\beta$ value of the parameter
points (see legend). The points are within the (approximate) 
\CL{95\%} region, based on the following selection of observables: only Higgs mass and signal rates (\emph{upper left}), Higgs mass, signal rates and $h/H/A\to \tau^+\tau^-$ exclusion likelihood (\emph{upper right}), all observables except $a_\mu$ (\emph{lower left}), all observables (\emph{lower right}).} 
\label{fig:h_Atmu_tanb}
\end{figure} 

In the following we will analyze in detail which observables and constraints
lead to this particular favored region of parameter space in the alignment limit. We recall the parametric dependence of the
approximate one-loop alignment condition in the limit $|\mu A_t|\tan\beta\gg M_S^2$,~\refeq{Eq:alignmentcondition},
\begin{align}
\tan\beta \sim \left( \frac{\mu A_t}{M_S^2} \left[ \frac{A_t^2}{M_S^2}
  - 6\right] \right)^{-1}.
\label{Eq:alignment_analytic}
\end{align}
In order to find viable solutions of this condition (while
  restricting ourselves to $|A_t|/M_S \le 3$ and $|\mu|/M_S\le 3$) we
need to have%
\footnote{
    Note that higher-order corrections as discussed in \refse{Sec:alignment}
    can modify the numbers given here.
}
\begin{align}
A_t  \quad\left\{ \begin{array}{ll} > \sqrt{6}~M_S & \mbox{if}\quad\mu A_t > 0, \\ < \sqrt{6}~M_S & \mbox{if}\quad\mu A_t < 0.
\end{array} \right.
\label{Eq:alignment_parametric}
\end{align}
We illustrate the impact of the various constraints and observables on
the possible solutions for the limit of alignment without decoupling
in the ($\mu/M_S$, $A_t/M_S$) plane in Fig.~\ref{fig:h_Atmu_tanb}. Here
we plot only the parameter points in the low-$\MA$ selection with
$\Delta\chi^2 \le 5.99$ (approximately
corresponding to the \CL{95\%} region) based on
different sets of observables: only Higgs mass and signal rates (top
left), Higgs mass, signal rates and $h/H/A\to \tau^+\tau^-$ exclusion
likelihood (upper right), all observables except $a_\mu$ (lower left)
and all observables (lower right). In color we indicate the $\tan\beta$
value of the parameter points (see legend).
The upper left plot of Fig.~\ref{fig:h_Atmu_tanb} shows all possible regions where the alignment condition, $Z_6v^2 = 0$, (cf.~\refeq{Eq:Z6v2}), is approximately fulfilled. These are found in each quadrant labeled by the
algebraic signs of ($\mu/M_S$, $A_t/M_S$): $(+,+)$, $(+,-)$, $(-,+)$,
$(-,-)$ (cf.~also the discussion of Fig.~\ref{Fig:alignment_numerical_TB} in \refse{Sec:alignment}). It can clearly be seen that $|A_t/M_S| < \sqrt{6}$ is required in the
$(+,-)$ and $(-,+)$ quadrants, consistent with \refeq{Eq:alignment_parametric}, and that these parameter points tend to have larger $\tan\beta$ values than those in the $(+,+)$ and $(-,-)$
quadrants.
 Once the constraints from LHC $H/A\to \tau^+\tau^-$
searches are taken into account,
as shown in the upper right plot of \reffi{fig:h_Atmu_tanb},
the large $\tan\beta$ points --- and thus the $(+,-)$ and $(-,+)$
quadrants --- become strongly disfavored. 
Small $|A_t|$ values in the $(+,+)$ and $(-,-)$
quadrants also require larger $\tb$ and are thus
equally disfavored. 
Adding also the flavor
observables $\br(B \to X_s \gamma)$, $\br(B_s\to \mu^+\mu^-)$ and
$\br(B^+\to \tau^+\nu_{\tau})$ as well as the $W$ boson mass
observable to the fit, the negative $\mu$ region (as well as the regions with $\mu A_t < 0$)
become mostly
disfavored, as shown in the lower left plot of Fig.~\ref{fig:h_Atmu_tanb}.
Interestingly, this feature emerges already {\em
without} including the observable $a_\mu$ in the fit.
The negative $\mu$ region is
mostly disfavored by $\br(B\to X_s \gamma)$, while the negative $\mu A_t$
region is disfavored by $\br(B_s \to \mu^+\mu^-)$, as we will discuss in more detail in the next section. 
 In addition, the
negative $\mu$ region becomes strongly disfavored after adding
$a_\mu$ to the fit, as 
the sign of the SUSY contribution to $a_\mu$ depends on the sign of
$\mu$. Thus $\mu$ needs to be positive in order to account for the currently observed
discrepancy between measurement and theory prediction.
This is shown in the lower right plot of Fig.~\ref{fig:h_Atmu_tanb} where only points with
  positive $\mu$ and $\At$ at low $\tb$ values remain, reproducing the distribution of favored points in the right plot of \reffi{fig:h_Atmu}.


\subsubsection{Impact of low energy observables}\label{sec:h_leo}

In this section we discuss the interplay of the Higgs
observables and limits from Higgs searches with the low-energy observables, 
in particular the rare $B$ decays, in the global fit.
The light Higgs case features a very good fit to all low-energy observables
as we have already seen in Tabs.~\ref{tab:totchi2} and \ref{tab:pull}.
For the most part we concentrate on the low-$M_A$ selection and study the
low-energy observables for the parameter points close to the limit of \emph{alignment
without decoupling}. This is particularly interesting since the low
$M_A$ value implies that all MSSM Higgs bosons, and in particular the
charged Higgs boson, are relatively light, which can lead to large
contributions to the $B$ decays. In contrast, the MSSM (loop-)contributions
to the anomalous magnetic moment of the muon and the $W$ boson mass are
dominated by light squarks and sleptons, effects from MSSM Higgs bosons are
not very pronounced.

The leading contribution to the FCNC process $b \to s \gamma$ occurs in the SM via a $W^{\pm}$--$t$ loop, allowing new-physics contributions to be of similar size. The branching ratio $\bsg$ can receive sizable positive contributions from an $H^\pm$--$t$ loop if the charged Higgs boson mass is not too large. It was recently pointed out~\cite{Misiak:2015xwa} that $b \to s \gamma$ excludes charged Higgs bosons with $\MHp < 480 \gev$ at the 95$\%$ C.L.~in a 2HDM with Type-II Yukawa couplings~\cite{Donoghue:1978cj,Hall:1981bc,Gunion:1989we}.
However, as discussed in \refse{Sec:wrong}, wrong-Higgs Yukawa couplings are induced radiatively by SUSY-breaking effects in the MSSM. These lead to important modifications of the charged Higgs couplings to up- and down-type quarks [cf.~\eq{Ychhiggs}], which in turn change the $\bsg$ prediction in the MSSM from the corresponding prediction in the  Type-II 2HDM according to \eq{BRratio}. In particular, if $\Delta_b$ [given by \eq{eq:Deltab}] is positive, then the MSSM prediction for $\bsg$ is smaller than in the Type-II 2HDM.
Furthermore, supersymmetric particles in the loop can also  contribute
to the $b \to s \gamma$ amplitude.  For example, chargino-stop ($\tilde{\chi}^\pm$--$\tilde{t}$) loops can
contribute with either sign, depending on sign and magnitude of the parameters $\mu$, $M_2$ and $A_t$, and thus may partially cancel the effects of the $H^\pm$--$t$ loop.
\begin{figure}[t!]
\centering
\includegraphics[width=0.46\columnwidth]{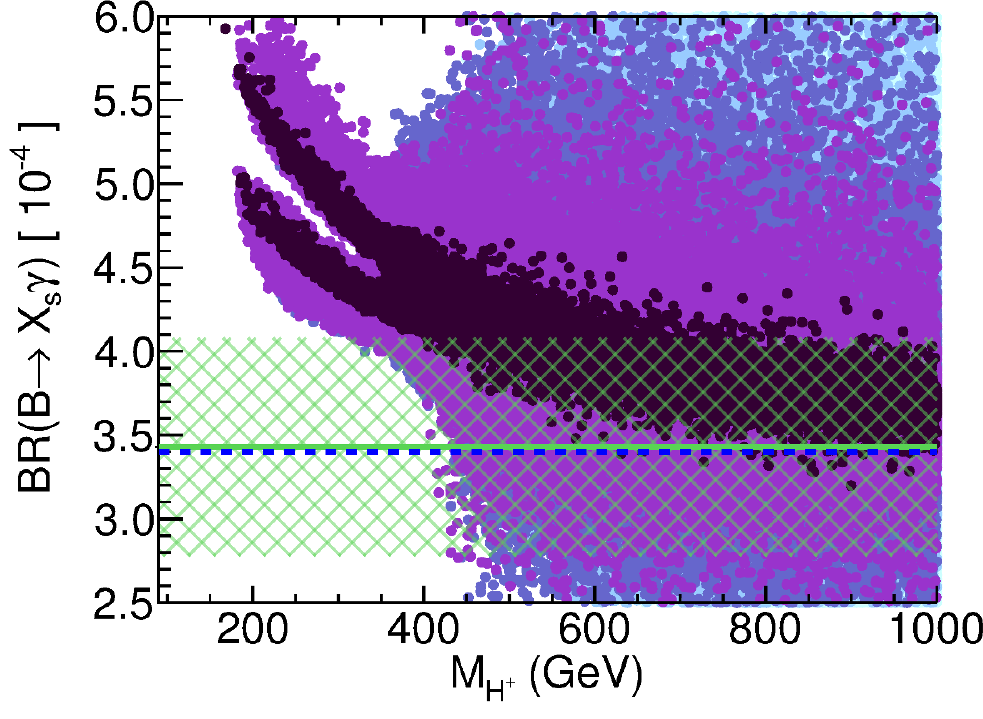}\hfill
\includegraphics[width=0.46\columnwidth]{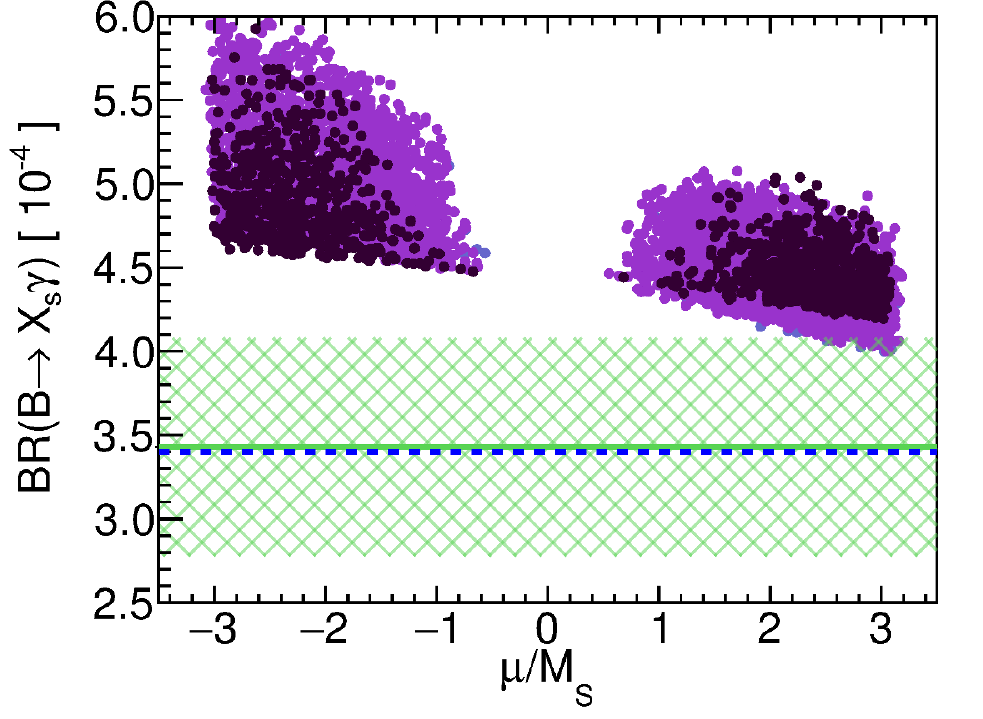}\hfill
\caption{$\bsg$ vs. $\MHp$ (\emph{left}) and
$\mu/M_S$ (\emph{right}) for the favored points in the fit 
without taking
into account the LEOs. In the right plot only points with $M_A < 350 \gev$
are shown. The green line and hatched region indicate the
corresponding experimental measurements and the total $1\sigma$ uncertainty
region, while the SM prediction is indicated by the blue dashed line. 
The color coding of the displayed points
is the same as in \reffi{fig:h_Atmu_tanb}.}
\label{fig:h_LEO_1}
\end{figure} 

The impact of the LEOs on the global fit is illustrated best by
studying the LEO predictions of the parameter points that are preferred
\emph{before} the LEOs are included in the fit. In the following we
therefore focus on the (approximate) \CL{95\%} preferred parameter points in the low-$M_A$ selection after the Higgs signal rates, Higgs mass and $h/H/A\to \tau^+\tau^-$ exclusion likelihood are included in the fit, i.e.~the points in the upper right plot of Fig.~\ref{fig:h_Atmu_tanb}.
The left plot in \reffi{fig:h_LEO_1} shows the charged Higgs mass dependence of $\bsg$.
Generally we observe that for light charged Higgs values corresponding
to the low-$M_A$ selection (slightly) too large predictions for
$\bsg$ are obtained. For $\MHp \lsim 350\gev$
we do not find any preferred points within the $1\sigma$ region of the
experimental measurement, which is indicated by the green band. One can see
that the $\bsg$ prediction tends to increase
when going to smaller charged Higgs masses, $\MHp$.
The two branches in this parameter region visible at low $M_A$ values correspond to $\mu > 0$
(lower branch) and $\mu < 0$ (upper branch), which can also be seen in the
right plot in \reffi{fig:h_LEO_1}, where $\bsg$ is
shown as a function of $\mu/M_S$ (for $\MA < 350 \gev$). 
The lower branch lies mostly within the 2$\sigma$ region of the
experimental measurement whereas most of the upper branch is
inconsistent with the measurement at the 2$\sigma$ level. This confirms
our previous statement that negative $\mu$ is disfavored by $B\to X_s
\gamma$.

\begin{figure}[t!]
\centering
\includegraphics[width=0.46\columnwidth]{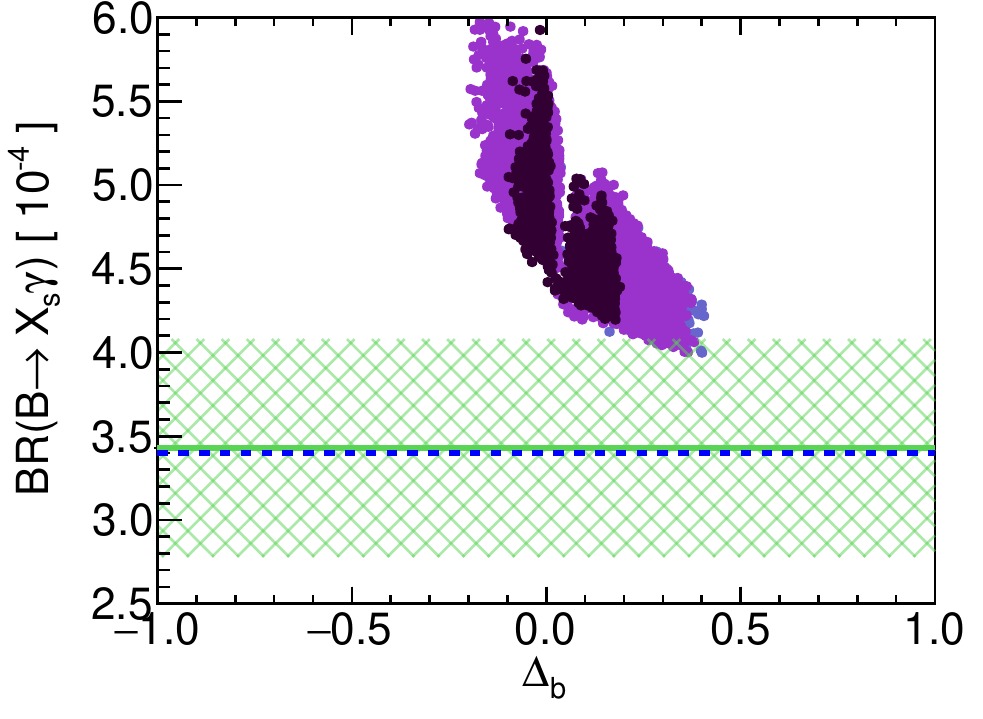}
\caption{$\bsg$ vs.~$\Delta_b$ for the favored points in the fit 
without taking into account the LEOs, for the low $M_A$ selection ($M_A < 350 \gev$). The green and blue line and the hatched region are the same as in Fig.~\ref{fig:h_LEO_1}. The color coding of the displayed points is the same as in \reffi{fig:h_Atmu_tanb}.}
\label{fig:h_bsg_Deltab}
\end{figure}

The $\mu$ dependence of the $\bsg$ prediction is easily understood from the discussion in \refse{Sec:wrong}. In the approximation $M_Z,M_A \ll M_S$, $\Delta_b$  is large and positive [negative] for large positive [negative] $\mu$, thereby leading to a substantial decrease [increase] of $\bsg$ with respect to the Type-II 2HDM prediction [cf.~\eq{BRratio}]. This $\Delta_b$ dependence is also directly shown in Fig.~\ref{fig:h_bsg_Deltab} for the parameter region with $\MA \le 350\gev$. The color coding furthermore illustrates that $\Delta_b \propto \tan\beta$, i.e.~the largest suppression of $\bsg$ is obtained for large, positive $\mu/M_S$ and large $\tan\beta$.
Note that the chargino-stop contributions to $\bsg$ are found to be relatively small in this parameter region and thus play only a minor role in this discussion.

%

\begin{figure}[h!]
\centering
\includegraphics[width=0.46\columnwidth]{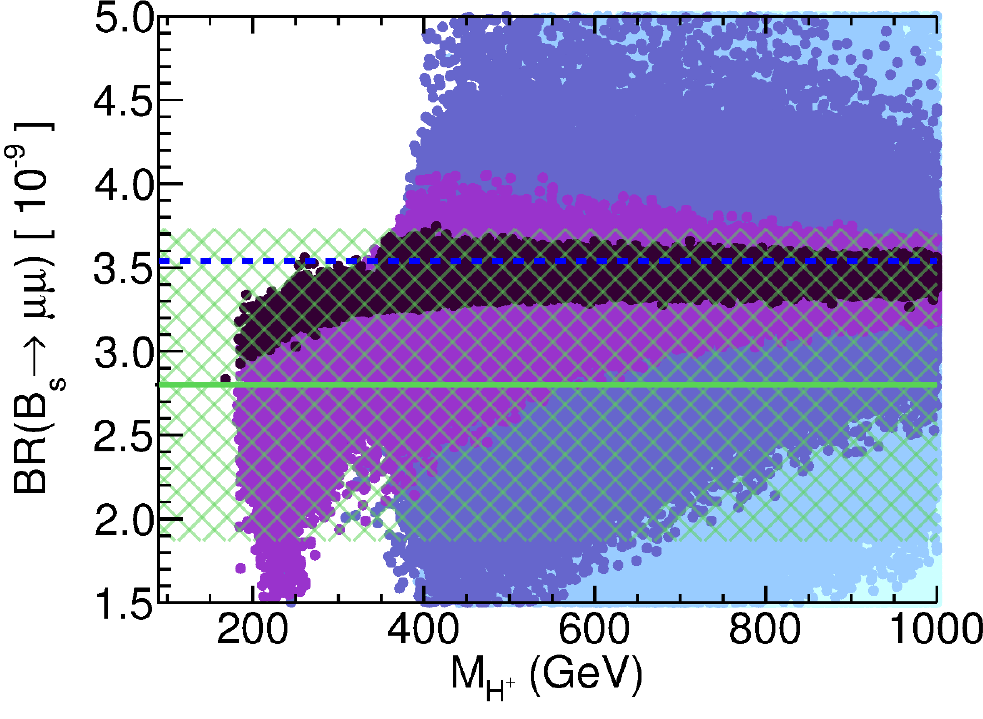}\hfill
\includegraphics[width=0.46\columnwidth]{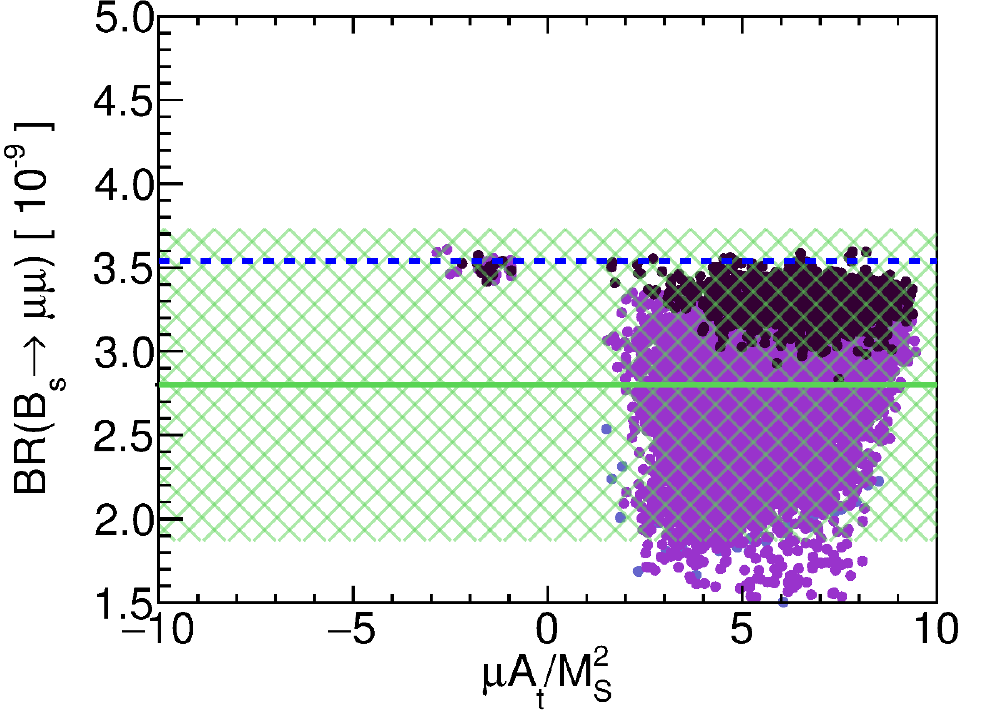}\hfill
\caption{
$\bmm$ in dependence of the charged Higgs mass, $\MHp$,
(\emph{left}) and $\mu A_t/M_S^2$ (\emph{right})
for the favored points in the fit 
without taking into account the LEOs. 
In the right plot only points with $M_A < 350 \gev$
are shown. The green line and hatched region indicate the
corresponding experimental measurements and the total $1\sigma$ uncertainty
region, while the SM prediction is indicated by the blue dashed line.
The color coding of the displayed points
is the same as in \reffi{fig:h_Atmu_tanb}.}
\label{fig:h_LEO_1b}
\end{figure} 


The decay $B_s\to \mu^+\mu^-$ can be mediated in the MSSM by a neutral
Higgs boson and receives loop corrections involving squarks, sleptons and
electroweakinos. 
As can be seen in \reffi{fig:h_LEO_1b},
in the limit of alignment without decoupling we find a very good fit to
$\mathrm{BR}(B_s\to \mu^+\mu^-)$, while for larger values of $\MHp$
larger deviations are possible. In the right plot 
of \reffi{fig:h_LEO_1b}, where 
$\bmm$ is shown as a function of $\mu A_t / M_S^2$
(for $\MA < 350\gev$), one can see
that the few points with negative $\mu A_t$ yield a value of $\bmm$ close
to the SM result, whereas the many points with positive $\mu A_t$ 
predict a smaller $\bmm$ than in the SM, which are in better agreement
with the current experimental central value.
These negative corrections to $\bmm$ become more sizable for larger values of $\tb \gtrsim 5$.
It is interesting to note that predictions for  $\mathrm{BR}(B_s\to \mu^+\mu^-)$ that precisely match the experimental central value are possible over the whole range of $\MHp$ values displayed in the
left plot of \reffi{fig:h_LEO_1b}.

The decay $B^+\to \tau^+\nu$ is helicity suppressed in the SM. The tree-level exchange of a charged Higgs boson constitutes the dominant MSSM contribution.
These contributions can be sizable for large values of $\tb$ and small $\MHp$.
In the low-$M_A$ selection we
have small $\tb$ after including the $h/H/A \to \tau^+ \tau^-$ exclusion likelihood and thus the 
contributions to $\mathrm{BR}(B^+\to \tau^+\nu)$ are small. We have
checked that the MSSM predictions are close to the SM value over the entire
range of $\MHp$ covered in our scan.

%
\begin{figure}
\centering
\includegraphics[width=0.46\columnwidth]{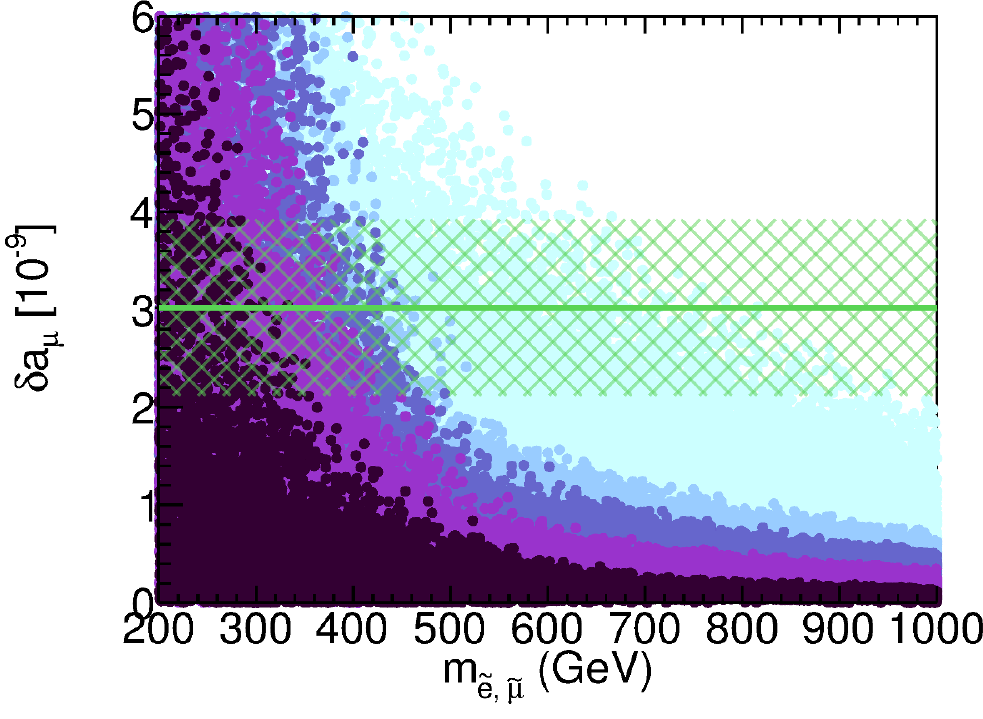}\hfill
\includegraphics[width=0.46\columnwidth]{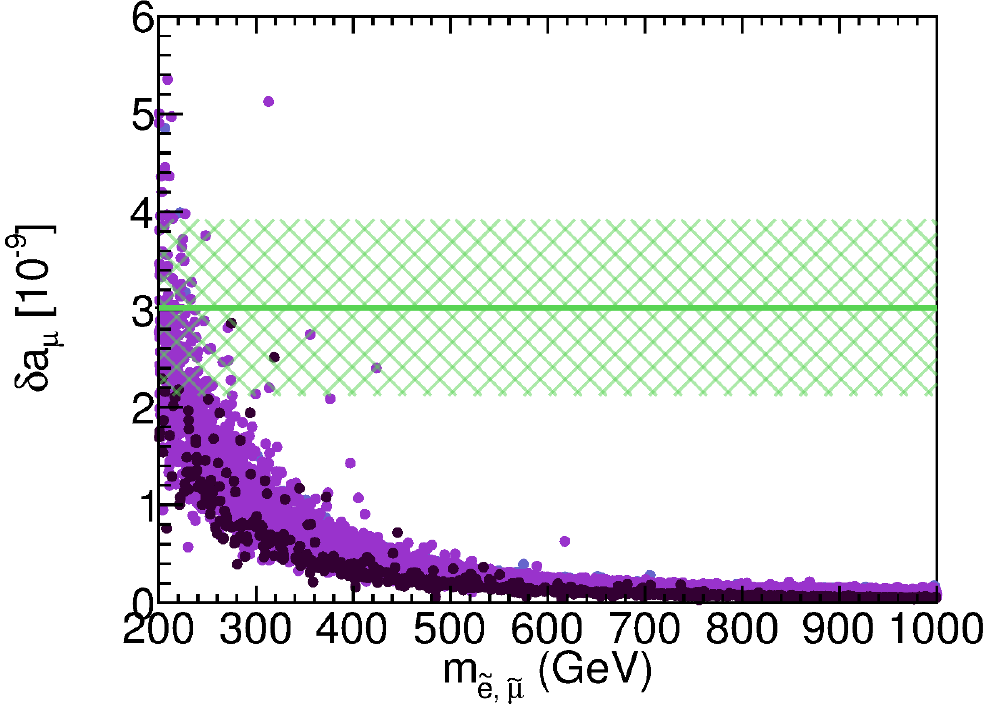}
\caption{The SUSY contribution to the anomalous magnetic moment of the
muon, $\delta a_\mu$, as a function of the (1st and 2nd generation) slepton
mass, $m_{\tilde{e},\tilde{\mu}}$, for the favored points in the fit
without taking into account the LEOs,
for all scan points (\emph{left}) and for the low-$M_A$
selection, $M_A < 350 \gev$ (\emph{right}). The green line 
indicates the desired new physics contribution needed to achieve
agreement with the observed deviation from the SM, 
and the hatched region
corresponds to the $1\sigma$ experimental uncertainty. The color coding of
the displayed points is the same as in \reffi{fig:h_Atmu_tanb}.}
\label{fig:h_LEO_2}
\vskip -0.05in
\end{figure} 

Finally, \reffi{fig:h_LEO_2} shows the prediction for the SUSY contribution to the anomalous magnetic moment of the muon, $\delta a_\mu$, against the first and second generation slepton mass, $m_{\tilde{e},\tilde{\mu}}$ for all scan points (left) and in the low-$M_A$ selection (right).\footnote{Note that in order to
be able to separately analyze the corresponding effects in the fit to the Higgs rates and the fit to $a_\mu$, we have chosen to treat $M_{\tilde{\ell}_{1,2}}$ and $M_{\tilde{\ell}_{3}}$ as independent fit parameters in this work. Indeed, light staus can significantly influence the Higgs rates, in particular the $\gamma\gamma$ rate~\cite{Carena:2011aa,Bechtle:2012jw,Liebler:2015ddv}.}
The MSSM contributions to $a_\mu$ consist of smuon-neutralino ($\tilde{\mu}^\pm$-$\tilde{\chi}^0$) and sneutrino-chargino ($\tilde{\nu}$-$\tilde{\chi}^\pm$) loops. 
Clearly $\delta a_\mu$ is strongly correlated with the slepton mass, whereas the correlation with the chargino and neutralino masses (not shown) is less pronounced.
In the full scan the favored region extends to large $\tb$ values, implying that sizable contributions to $\delta a_\mu$ are possible also for moderately large slepton masses. 
Parameter points within the $1\sigma$ region around the desired $\delta a_\mu$ value are found all the way up to 
$m_{\tilde{e},\tilde{\mu}} \sim 1000$ GeV for large $\tb$.
In the low-$M_A$ selection, $\tb$ is small ($\lesssim 10$) and therefore the 
sleptons need to be very light, $\lesssim$ 300 GeV, in order to give a large enough contribution to $\delta a_\mu$.
This has interesting consequences for the direct SUSY searches, discussed below, where we 
find that a larger 
fraction of the preferred points in the alignment region is excluded compared to the decoupling region.
This is because SUSY searches with multilepton final states targeting direct stop, gaugino or slepton production yield stronger exclusion if the 1st/2nd generation sleptons are light. 

\subsubsection{Impact of direct LHC SUSY searches}
\label{Sect:ImpactDirectSUSYSearchedLighth}

As discussed in \refse{Sect:Constraints}, we have included all relevant 8 TeV SUSY searches in
our analysis.%
\footnote{{\tt CheckMATE-1.2.2} does not  yet include the first 13 TeV
analyses of ATLAS and CMS. However so far the 13 TeV exclusion limits are
comparable (or still weaker) than the 8 TeV limits.} 
Overall, we find that the
LHC SUSY searches impact the validity of the scan points in the light
Higgs interpretation considerably. In total $38\%$ of the $2.4 \times 10^5$ tested \pMSSM\ parameter
points are excluded by the limits from LHC SUSY searches.\footnote{Recall that only points with a $\Delta\chi^2\leq 10$ with respect to the BF point are fed into the \texttt{CheckMATE} analysis.} 
Restricting ourselves to the points within the (approximate)
\CL{95\%} favored region, LHC SUSY searches exclude $43\%$ of these points.
The impact of the direct LHC SUSY searches is mostly ``orthogonal'' to the 
impact of the Higgs constraints in the sense that the direct SUSY searches have no impact on the $\chi^2$ profile.  This observation can already be inferred from the light gray points of
Fig.~\ref{fig:totchi2h_rates}.

\begin{figure}[H]
\centering
\includegraphics[width=0.44\columnwidth]{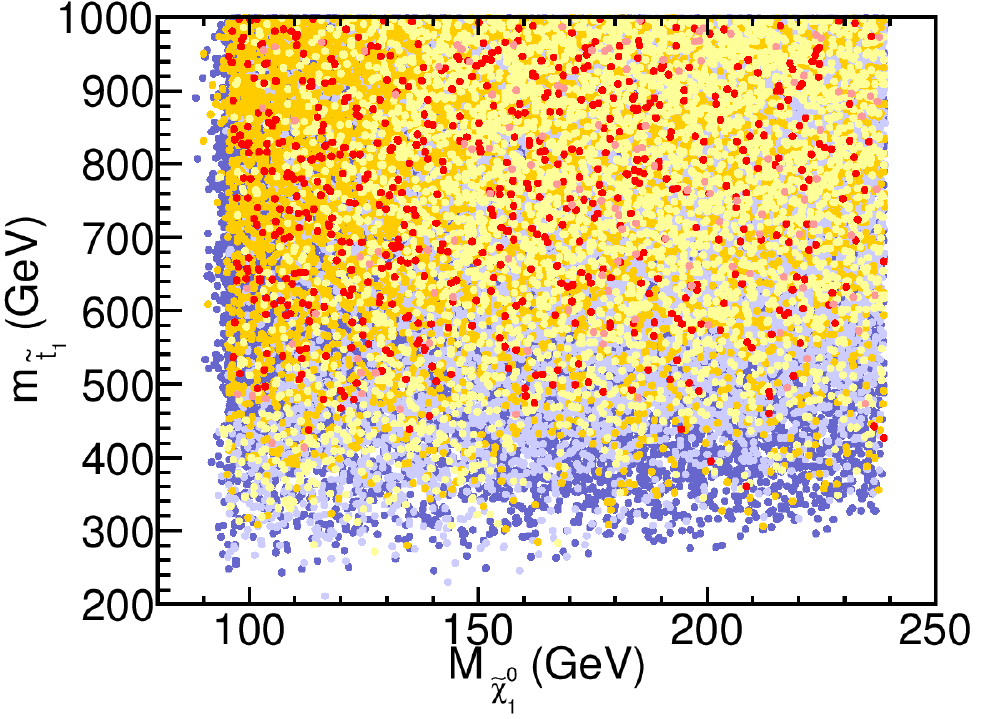}\hfill
\includegraphics[width=0.44\columnwidth]{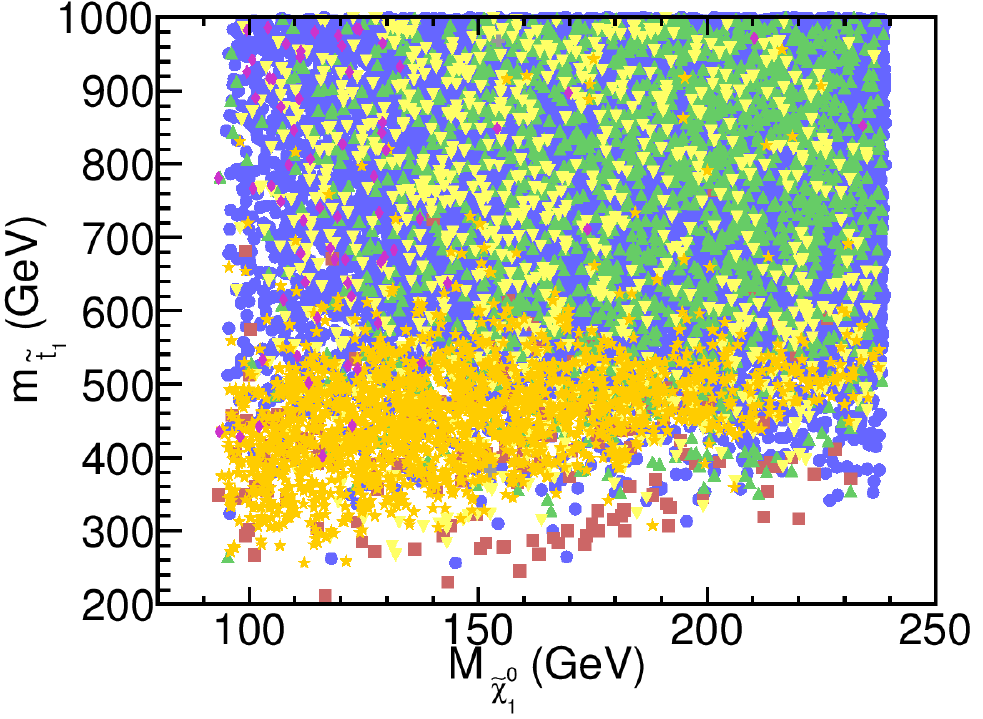}
\includegraphics[width=0.44\columnwidth]{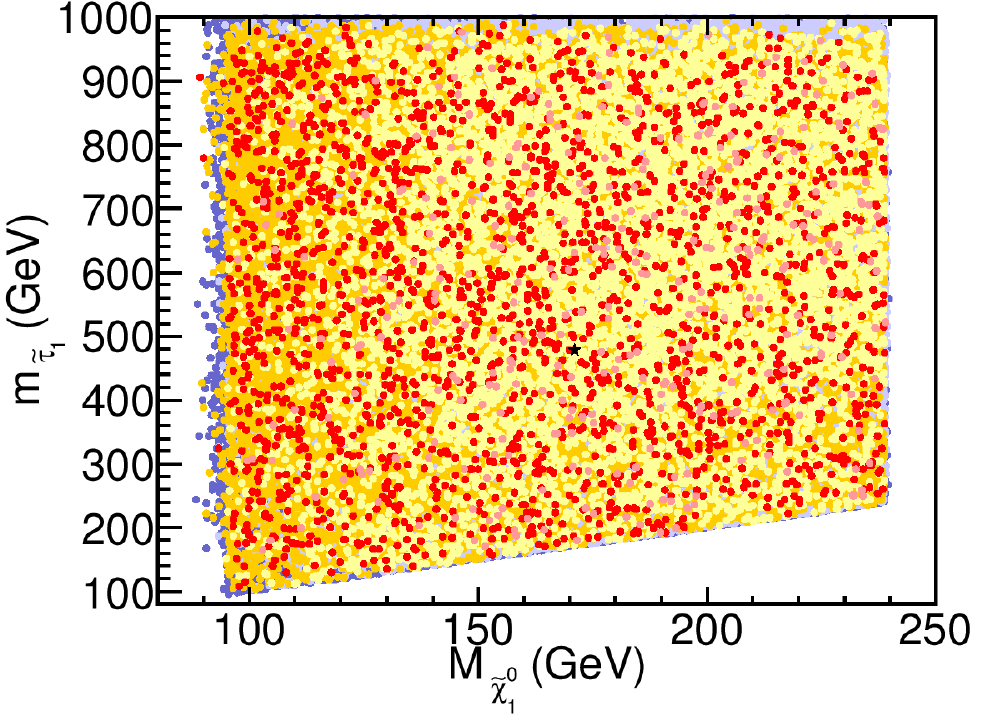}\hfill
\includegraphics[width=0.44\columnwidth]{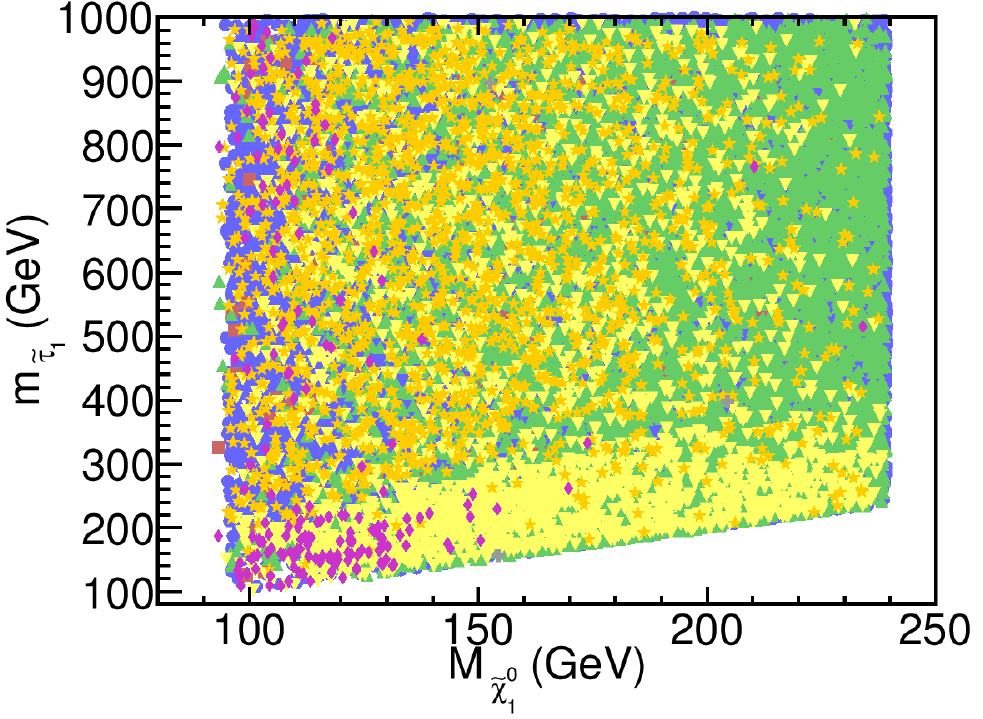}
\includegraphics[width=0.44\columnwidth]{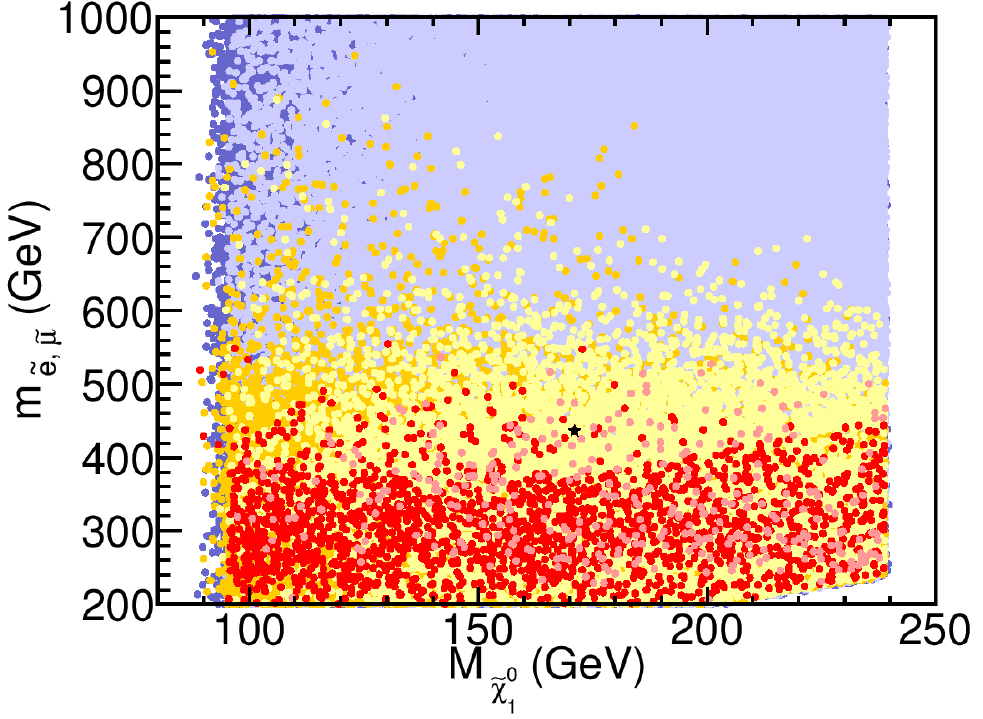}\hfill
\includegraphics[width=0.44\columnwidth]{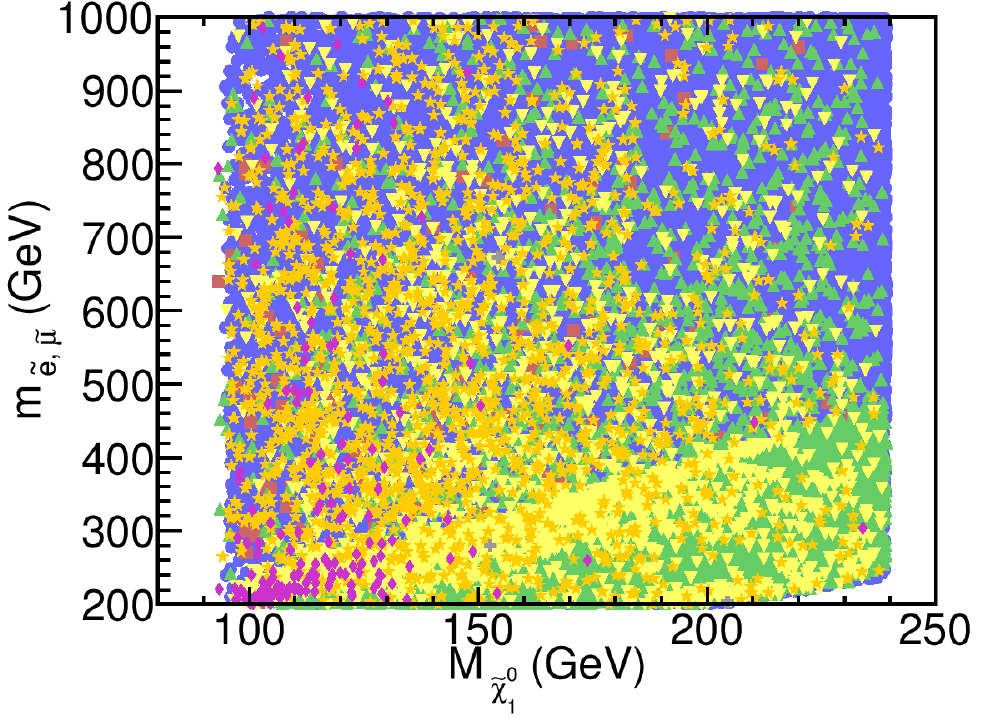}
\caption{
Impact of $8\tev$ LHC SUSY searches on the fit, in dependence of the
lightest stop (\emph{top row}), stau (\emph{middle row}) and selectron/smuon
(\emph{bottom row}) mass and the lightest neutralino mass. In the three left panels, we compare the 
 \CM\ allowed (\emph{bright colored}) and excluded (\emph{pale
colored}) points for all scan points (\emph{blue}), the (approximate) \CL{95\%} (\emph{yellow}) and \CL{68\%} (\emph{red}) preferred points (plotted in this order, with excluded points on top of the allowed points in each step). In the three right panels, we exhibit LHC analyses that yield the exclusion (in the order of plotting): 
  \emph{(i)}  $N~\text{jets}+\ETmiss$ searches (\emph{blue circles})~\cite{Aad:2013wta,Aad:2014wea}, \emph{(ii)} hadronic
  $\tilde{t}\tilde{t}$ searches with $b$-jets (\emph{red squares})~\cite{Aad:2013ija,Aad:2014bva,Aad:2014lra,Chatrchyan:2013mys},
  \emph{(iii)} $\chi^\pm_1\chi^0_2\to 3\ell+\ETmiss$ searches (\emph{green
  up-triangles})~\cite{Aad:2014nua}, \emph{(iv)} $\tilde{t}\tilde{t}\to 2\ell+\ETmiss$
  searches (\emph{yellow down-triangle})~\cite{Aad:2014qaa}, \emph{(v)} $\tilde{t}\tilde{t} \to
  1\ell + N~\text{jets}+\ETmiss$ searches (\emph{orange stars})~\cite{Aad:2014kra}, \emph{(vi)}
  $\chi^\pm_1\chi^0_2, \tilde{\ell}\tilde{\ell} \to \ell^+\ell^-
  +\ETmiss$ searches (\emph{magenta diamonds})~\cite{Aad:2014vma}, \emph{(vii)}~$2\ell +
  N~\text{jets} + \ETmiss$ (\emph{gray plus})~\cite{Aad:2015mia,Khachatryan:2015lwa}.}
\label{fig:CMresults_h}
\end{figure}

This point is further exemplified in the left plots of
Fig.~\ref{fig:CMresults_h} in two-dimensional mass planes of the 
lightest stop, stau, selectron/smuon and neutralino, which are the most
important SUSY particles in this context. The impact of the
LHC searches for SUSY particles has 
been tested with \texttt{CheckMATE} for all displayed points.
The points excluded by direct SUSY search limits are plotted in pale
colors above the points allowed by the direct SUSY searches, which are shown
in bright colors. The excluded points are found with a similar
distribution as the allowed points in all projections of these
  mass parameters.
The reduced density of excluded points for low values of the lightest neutralino mass,
  $m_{\tilde{\chi}^0_1}$, is explained by a reduction of the signal acceptance in most of
  the searches applied (the $\tilde{t}\tilde{t}\to 1\ell +
  N~\text{jets}+\ETmiss$ search~\cite{Aad:2014kra} being a notable
  exception). This is caused by the fact that often
  $m_{\tilde{\chi}^0_1}\approx
  \frac{1}{2}m_{\tilde{\chi}^{\pm}_1}\approx\frac{1}{2}m_{\tilde{\chi}^0_2}$ due to the assumption of \refeq{def:Mone}. For small neutralino masses the signal acceptance of $\tilde{\chi}^\pm/\tilde{\chi}^0\to (W^\pm/Z)\chi^0_1$
final states reduces as the phase space of the decay decreases, $m_{\tilde{\chi}^0/\tilde{\chi}^\pm}-m_{\tilde{\chi}^0_1} \rightarrow \MW/\MZ$
(see e.g.~Fig.~11 of Ref.~\cite{Aad:2015baa}), yielding reduced final state activity. Furthermore, the missing energy is smaller for small neutralino masses, affecting in particular the signal acceptance of searches for hadronic final states. Nevertheless, we do not find any specific effect of this feature on the allowed Higgs sector phenomenology.

Although the impact of the Higgs data seems orthogonal to the impact of
the LHC SUSY searches, the impact of some of the included LEOs, in
particular $a_\mu$, is not. The latter favors light first/second generation
sleptons and light gauginos, which can be probed by searches at the LHC. As
a result, we find a significant amount of favored points being excluded by multilepton searches for direct slepton, electroweak gaugino or stop pair production. As already mentioned above, this effect is particularly prominent in the low-$\MA$ region, where the requirement of light 1st/2nd generation sleptons to accommodate $\delta a_\mu$ is much stricter than in the decoupling region (cf.~\figref{fig:h_LEO_2}).


For the SUSY exclusion using \texttt{CheckMATE}, only $8\tev$ searches
and LO cross sections are used with a global $k$-factor of $1.5$ (see \refse{Sect:Constraints}), due to
their availability in the applied computer codes 
(as mentioned above, the version of \texttt{CheckMATE} used for our
analysis does not yet include the first limits from ATLAS and CMS searches
at 13 TeV).
The results shown in the left plots of Fig.~\ref{fig:CMresults_h} imply that 
possible limits from upcoming searches at $13\tev$ are not expected to
significantly alter
the Higgs signal interpretation in the \pMSSM:
Due to the described ``orthogonality'' of the SUSY searches to the
Higgs rate constraints, a refinement of the $k$-factor
calculation or a further strengthening of the limits, both in terms of
rate constraints and in terms of mass reach, are not expected to change
the parameter ranges preferred by the Higgs rate observables
significantly. Instead, it can be expected that strengthened SUSY
limits will only decrease the point density, thus
indicating that parameters need an increasingly refined tuning in order to still
find a good fit for the Higgs mass and the Higgs rates.

The sensitivity of the various SUSY searches in the different kinematical regions of the parameter space is shown in the right plots of \reffi{fig:CMresults_h}, where we give for every excluded parameter point the relevant LHC SUSY search (see figure caption for a detailed list and references).
 In the upper right plot in
Fig.~\ref{fig:CMresults_h}, hadronic and one-lepton stop searches (red squares and orange stars) can
be easily identified to be the most sensitive searches for $m_{\tilde{t}}\lesssim
500\gev$, while for larger masses a more mixed distribution of all
searches is observed. Likewise, in the middle and bottom right plots in
Fig.~\ref{fig:CMresults_h} it can be seen that the sensitivity of LHC searches for 
direct electroweak gaugino and slepton production with dilepton final states (magenta diamonds)
is centered at low mass parameters for sleptons, staus and neutralinos. Dilepton searches for stop pairs (yellow down-triangles) and multilepton searches for electroweak gauginos (green up-triangles) also provide important constraints at low stau and slepton masses, and in particular remain sensitive at larger neutralino masses.


\subsection{The heavy Higgs interpretation}
\label{sec:HeavyHiggsResults}

We now consider the more exotic MSSM
interpretation where the heavy $\cp$-even Higgs boson is identified as
the observed Higgs boson at $\MHexp\gev$.
We recall that this interpretation has a similarly good fit quality
as the light Higgs interpretation (see Table~\ref{tab:totchi2}). As in the
light Higgs case we first show the
predictions for the Higgs signal rates and their correlations and then
discuss the preferred parameter region. After commenting on the impact
of the low energy observables and LHC SUSY searches on the fit we
conclude this section by discussing the discovery prospects for the
other neutral and charged Higgs states in the heavy Higgs
interpretation.

\subsubsection{Higgs signal rates}

\begin{figure}[t]
\centering
\includegraphics[width=0.44\columnwidth]{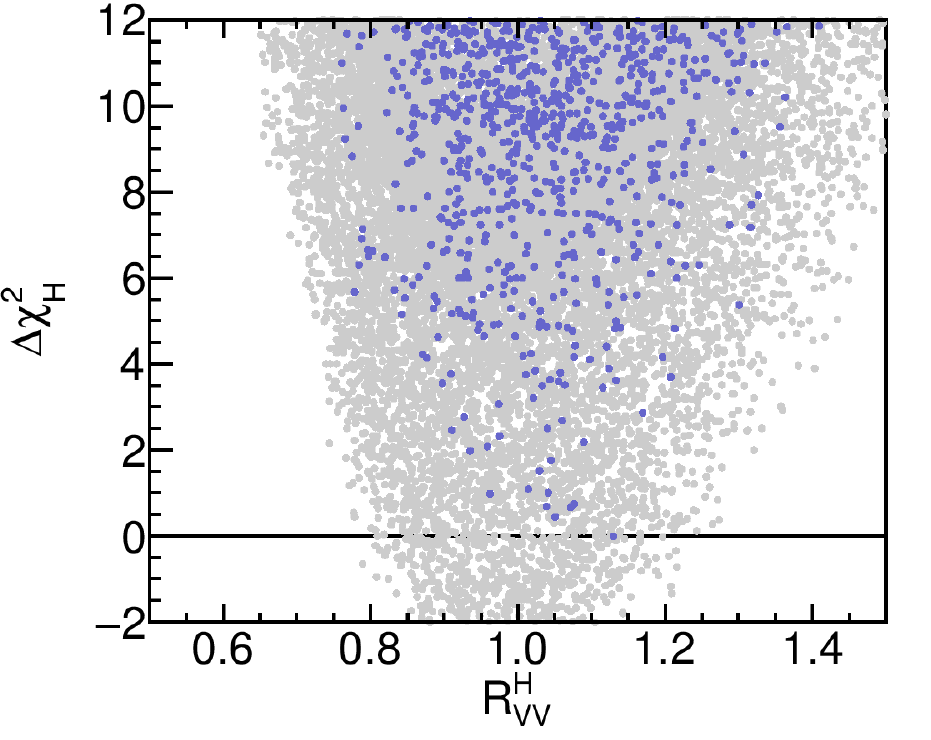}\hspace{0.5cm}
\includegraphics[width=0.44\columnwidth]{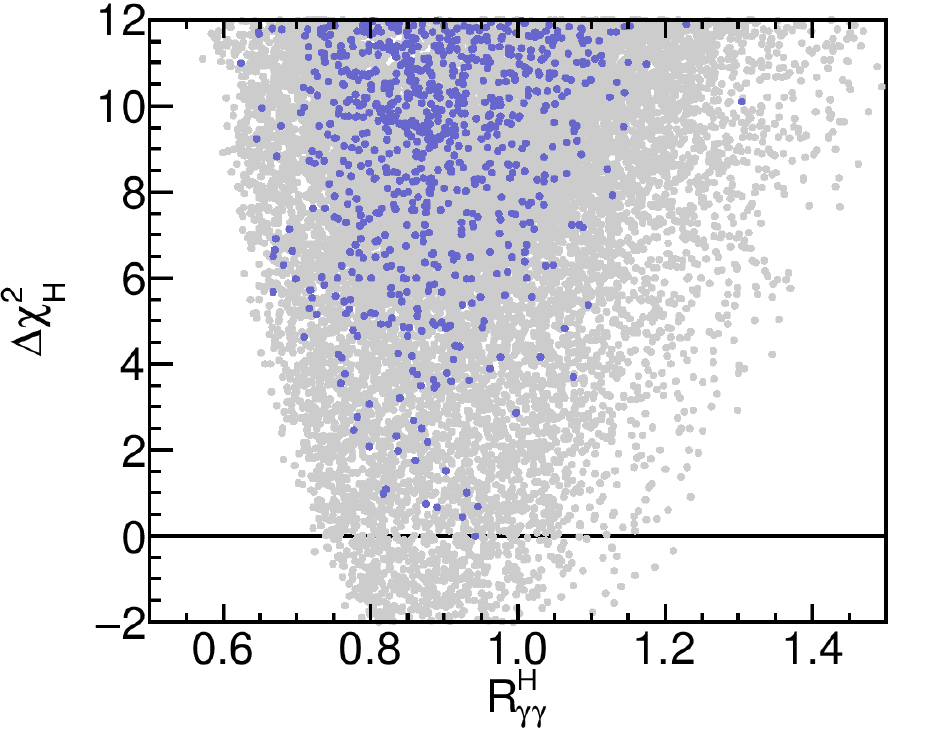}\\
\includegraphics[width=0.44\columnwidth]{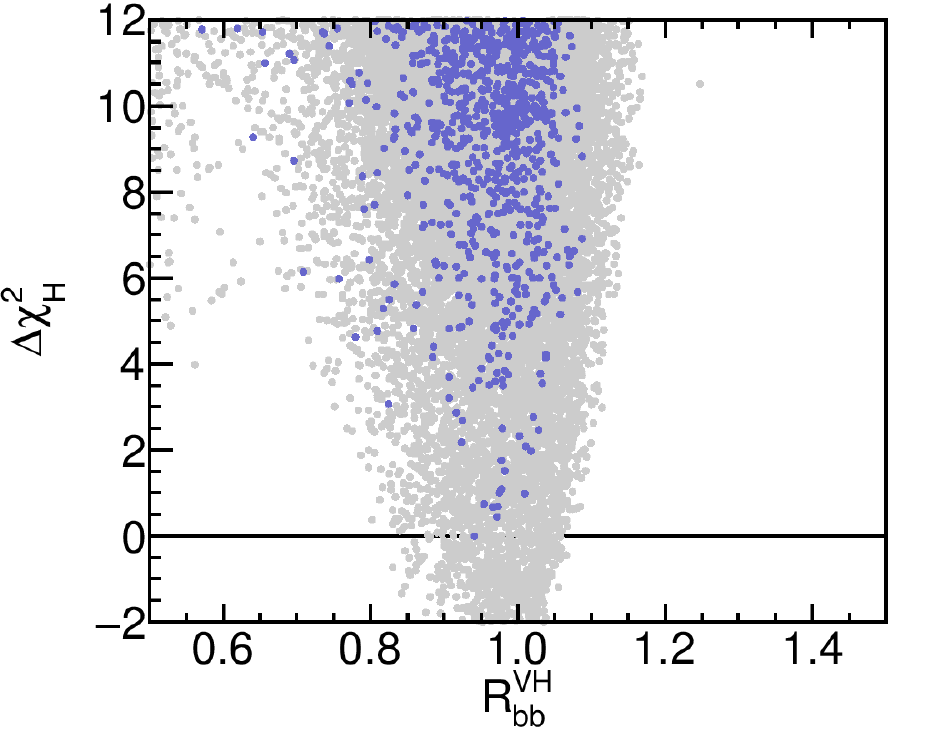}\hspace{0.5cm}
\includegraphics[width=0.44\columnwidth]{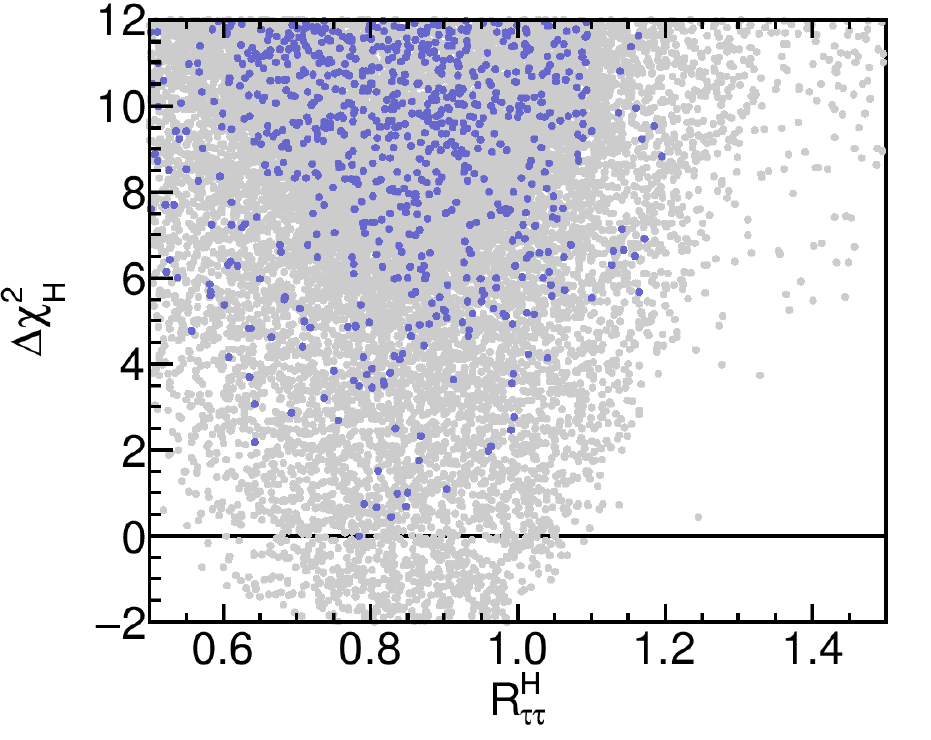}
\caption{$\Delta\chi^2$ distributions for the most important Higgs signal
  rates (defined in the text) from the complete scan for the heavy Higgs interpretation. The colors show
  all points in the scan (\emph{gray}), and points that pass the direct constraints from Higgs
  searches from  \HB\ (v.~4.2.1) and   from LHC sparticle searches from \CM\ (v.1.2.2) (\emph{blue}).}  
\label{fig:totchi2hh_rates}
\end{figure}

We show the $\Delta\chi^2$ profiles in the most important Higgs signal
rates (cf.~\refeq{Eq:Rvalues}) in
Fig.~\ref{fig:totchi2hh_rates}. From the sparseness of the allowed
points (\emph{blue}) in these distributions it is evident that the heavy
Higgs interpretation is quite constrained. Moreover, the actual $\chi^2$
minimum of all scan points lies significantly deeper than the determined
best-fit point, however, all points with a lower total $\chi^2$ than the
best-fit point are excluded by either direct Higgs search limits, SUSY
search limits, or other more technical requirements such as the
$\mathbf{Z}$-matrix criterion (see Sec.~\ref{sec:sampling}). 

We find the preferred ranges for the Higgs signal rates to be
\begin{align}
R_{VV}^H \in [0.95, 1.13],\qquad  R_{\gamma\gamma}^H \in [0.81, 0.94],
\qquad R_{bb}^{VH} \in [0.94, 1.03],\qquad R_{\tau\tau}^{H} \in [0.78,
  0.90]. 
\end{align}
The Higgs rates $R_{XX}^{P(H)}$ are defined as in the light Higgs case, see \refeq{Eq:Rvalues}, but with $h \leftrightarrow H$. Due to the sparseness of points in the vicinity of the $\chi^2$ minimum,
these ranges should be taken only as indicative results for the actual
$68\%~\mathrm{C.L.}$ range 
(and therefore we also refrain from giving the central values here).
Nevertheless, these distributions indicate on the one hand 
rather good agreement
with the SM prediction ($R_{XX}^H=1$) for the $H\to VV$ and $VH\to
Vb\bar{b}$ channels, and on the other hand a modest suppression of the
$H\to \gamma\gamma$ and $H\to \tau\tau$ signal rates with respect to the
SM prediction. These tendencies appear not only for the most favored
points (with $\Delta\chi^2 \le 1$) but also for the bulk of allowed
points with larger $\Delta\chi^2$ values.
They may thus be tested in the current and upcoming LHC runs.

In \citere{Bechtle:2012jw} we analyzed two mechanisms
for modifying $R_{\gamma\gamma}$. It can be enhanced (suppressed) either via a 
suppression (enhancement) of the total width, which is dominated by $H \to b \bar{b}$,
or via a directly enhancement (suppression) of the $H\to \gamma \gamma$ decay width through light charged particles, e.g.~the lightest stau.
Since we observe $R_{bb}^{VH} \sim 1$ here, the suppression 
of $R_{\gamma\gamma}^H$ is due to a direct suppression of the $H\to\gamma\gamma$ width induced by SUSY loops in the $H \to \gamma \gamma$ decay amplitude.
Indeed we find for the most favored points $\Gamma (H \to \gamma \gamma) /
\Gamma (H \to \gamma \gamma)_{\text{SM}} \sim 0.80 - 0.95$, from loops of
a light charged Higgs boson and a moderately light stop,
$m_{\tilde{t}_1} \sim (350-650)\gev$, where the stop mixing parameter
is close to $X_t \sim -1.5 M_S$ (see \refse{Sect:HH_paramspace})~\cite{Liebler:2015ddv}.

\begin{figure}
\centering
\includegraphics[width=0.44\columnwidth]{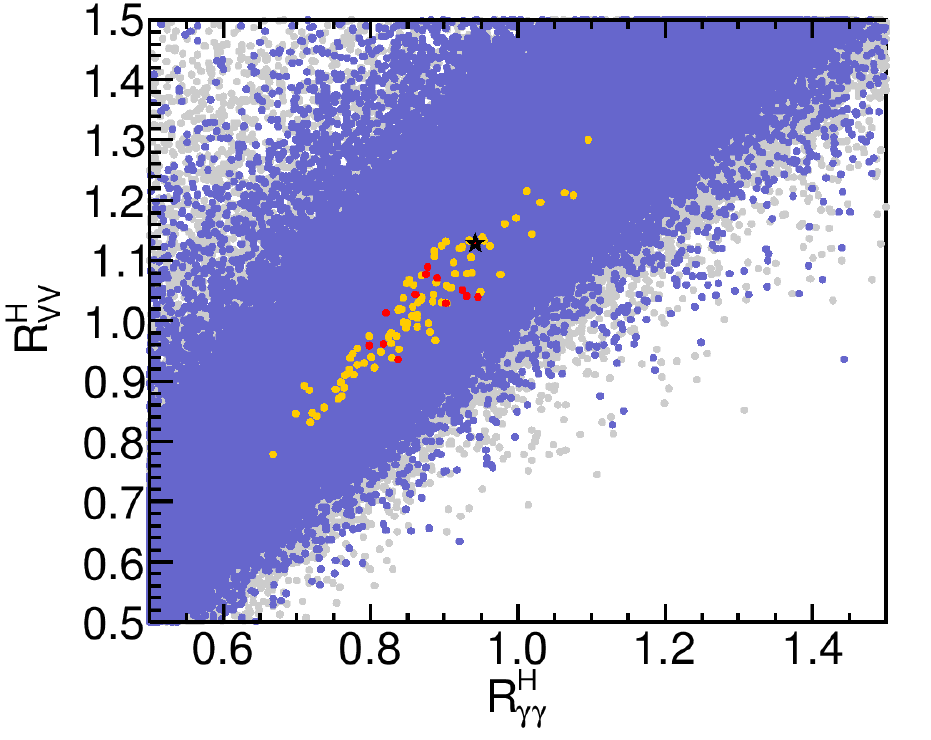}\hspace{0.5cm}
\includegraphics[width=0.44\columnwidth]{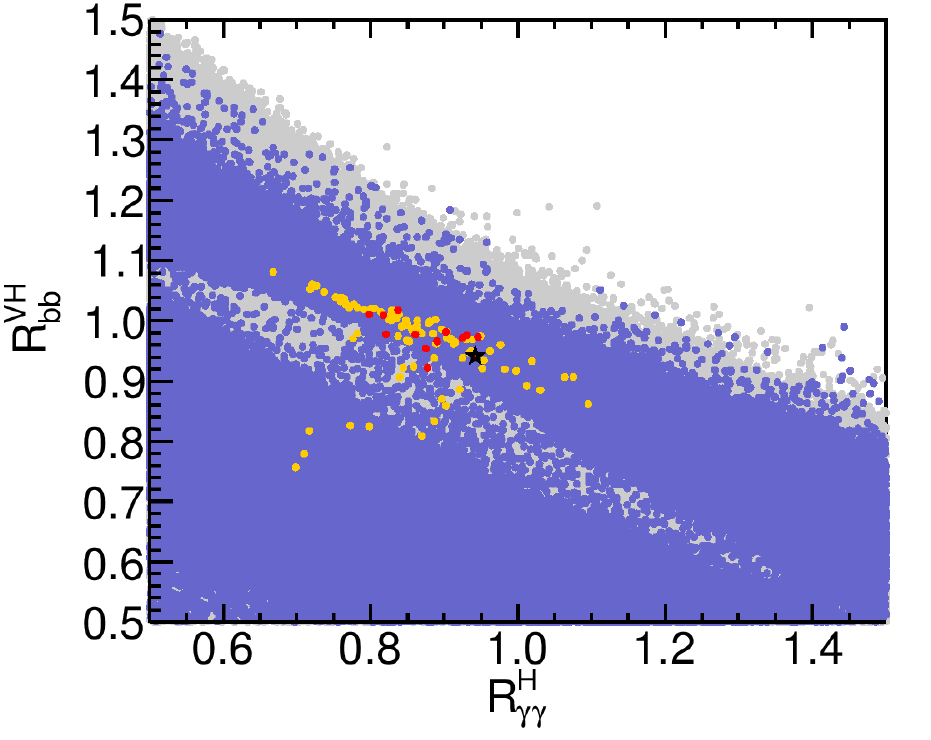}\\
\includegraphics[width=0.44\columnwidth]{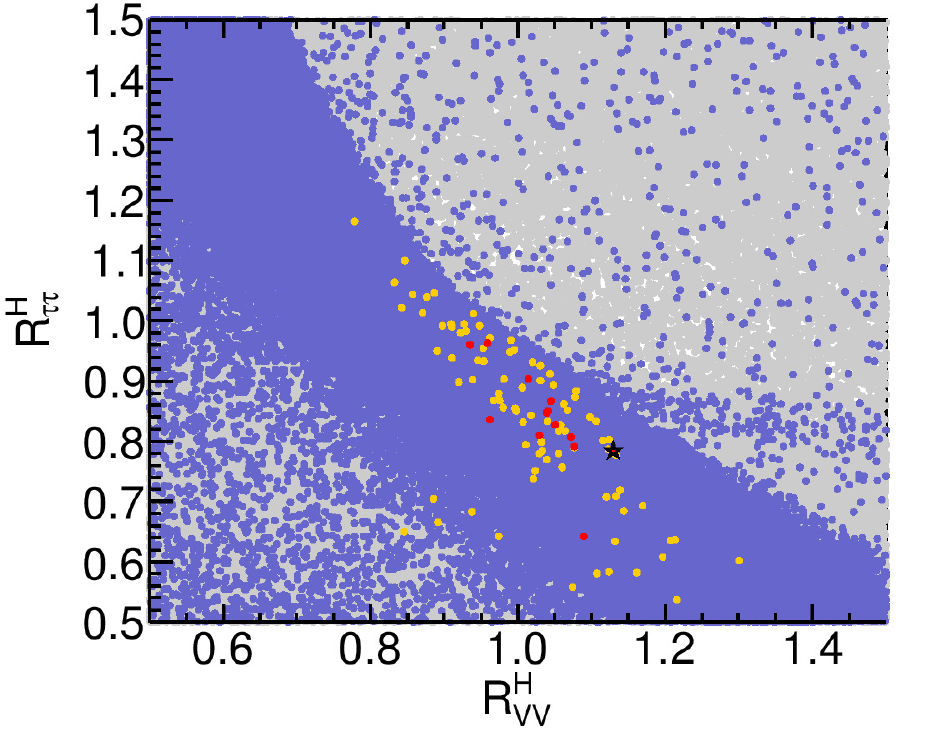}\hspace{0.5cm}
\includegraphics[width=0.44\columnwidth]{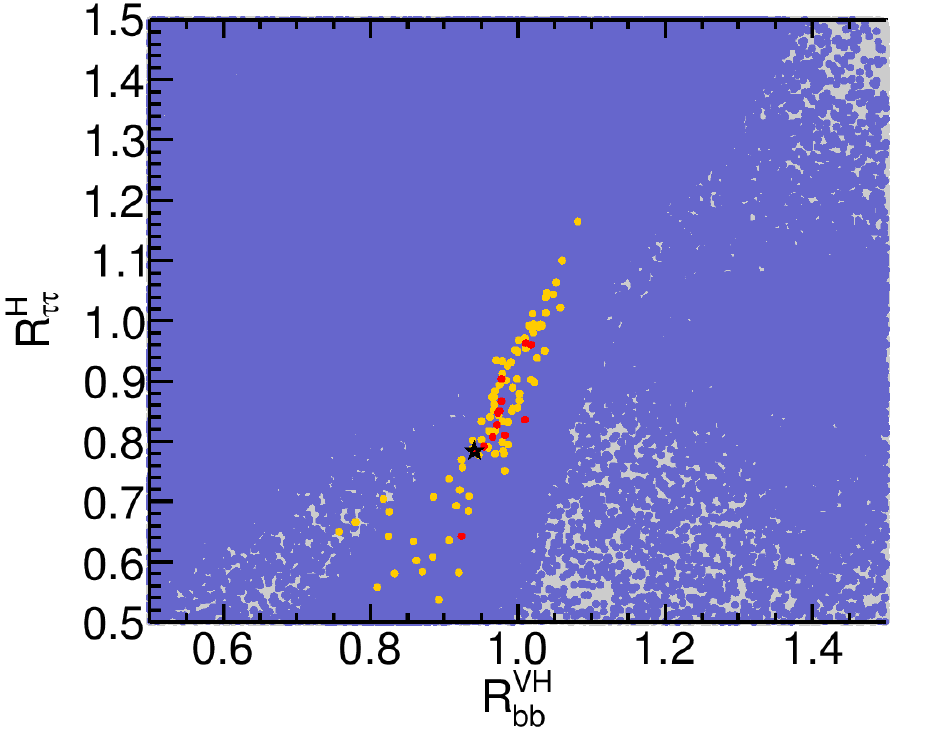}
\caption{Correlations between signal rates for the
  heavy Higgs case. The color coding follows that of
  \figref{fig:totchi2hh_rates}, with the addition of the favored regions
  with $\Delta \chi_H^2 < 2.3$ (\emph{red}) and $\Delta \chi_H^2 <
5.99$ (\emph{yellow}). The best fit point is indicated by a \emph{black star}.} 
\label{fig:Hrates_corr}
\end{figure}

We show some of the correlations of the four Higgs signal rates in
\figref{fig:Hrates_corr}. Interestingly, all preferred parameter points
have $R_{VV}^H > R_{\gamma\gamma}^H$, with almost all of them following
a strong linear correlation approximately given by $R_{VV}^H \approx 0.05 +
1.23 \cdot R_{\gamma\gamma}^H$. Note that in the light Higgs case we
observed a similar linear correlation, however, $R_{VV}^h <
R_{\gamma\gamma}^h$ for most of the scan points in that case.
Precision determination of the $H\to VV$ and $H\to \gamma\gamma$ rates might therefore help to distinguish between the two
interpretations.
The other
rate correlations, namely $(R_{\gamma\gamma}^H, R_{bb}^{VH})$,
$(R_{VV}^H, R_{\tau\tau}^{H})$ and $(R_{bb}^{VH}, R_{\tau\tau}^{H})$,
are very similar to those found for the light Higgs interpretation
(cf.~\figref{fig:hrates_corr}). We note that the (approximate)
$95\%~\mathrm{C.L.}$
region extends over smaller values for the $H\to \tau^+\tau^-$ rate than
in the light Higgs interpretation,  down to values of $R_{\tau\tau}^H
\gtrsim 0.5$.


\subsubsection{Parameter space}
\label{Sect:HH_paramspace}
\begin{figure}
\centering
\includegraphics[width=0.44\columnwidth]{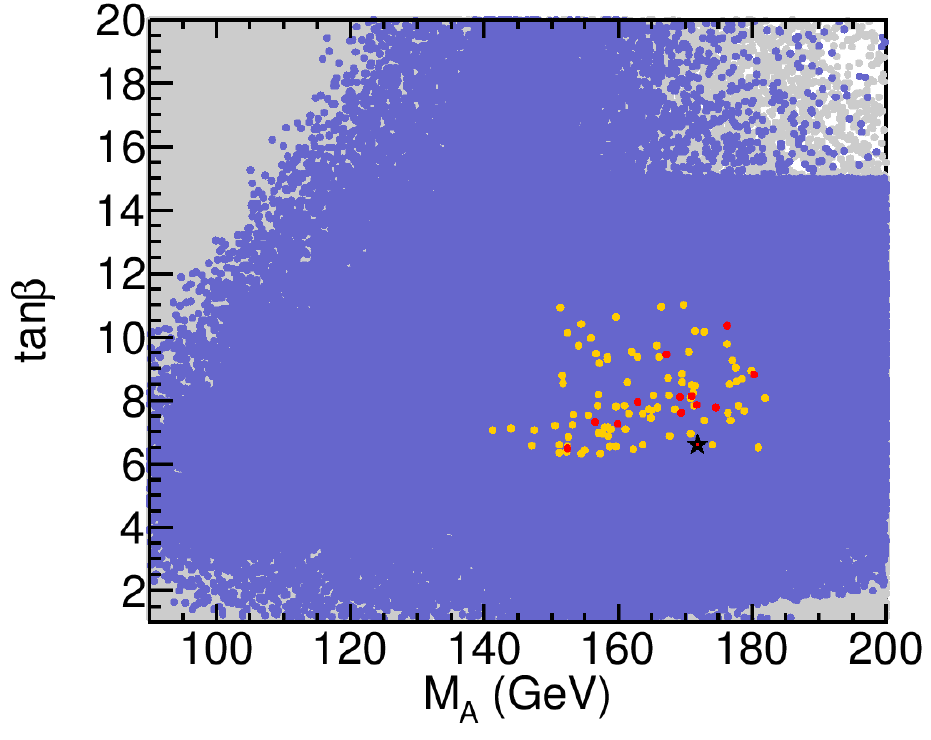}\hspace{0.5cm}
\includegraphics[width=0.44\columnwidth]{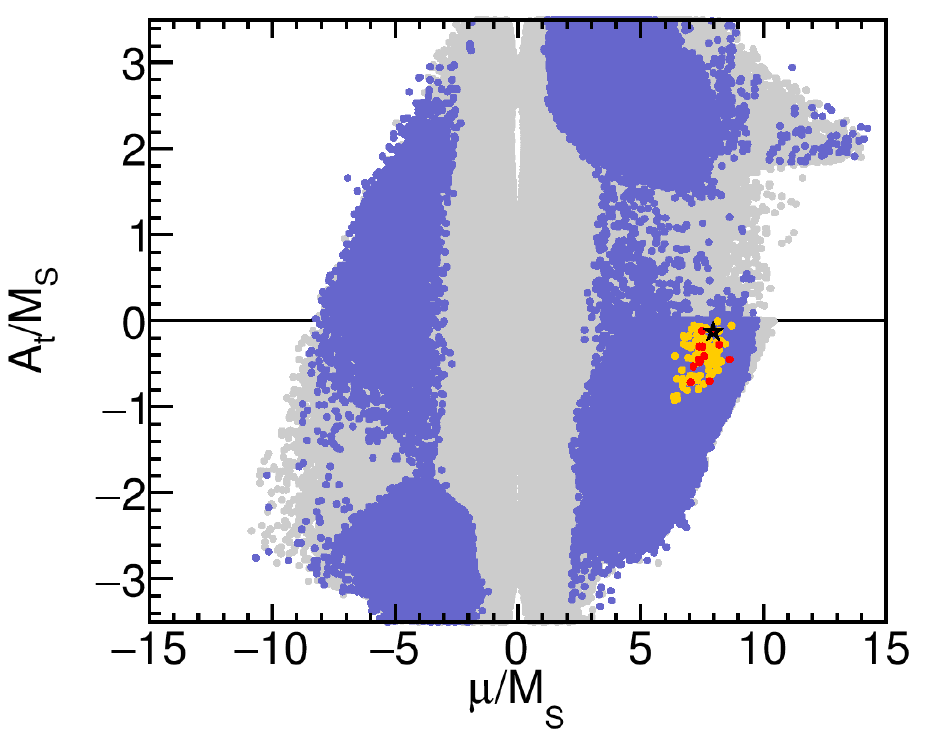}\\
\caption{Preferred parameter regions in the ($\MA$, $\tb$) plane (\emph{left})
and the ($\mu/M_S$, $A_t/M_S$) plane (\emph{right}) 
in the heavy Higgs case. The color coding is the same as in \reffi{fig:Hrates_corr}.
  } 
\label{fig:hh_pars}
\end{figure} 

We show the fit results for the heavy Higgs interpretation in
\figref{fig:hh_pars} in the ($M_A, \tan\beta$) plane (left) and the
($\mu/M_S, A_t/M_S$) plane (right). The preferred parameter points
expand over only a narrow range in the parameters
determining the Higgs sector at lowest order,
$M_A \sim (140, 185)\gev$ and $\tan\beta\sim 6-11$. Compared to our previous
results~\cite{Bechtle:2012jw}, where we found smaller values $M_A \sim
(110-140)\gev$ being  
preferred, the favored parameter region has shifted towards larger $M_A$
values, caused by several reasons. Firstly, at small values
$M_A \lesssim 150\gev$ the $\cp$-odd Higgs boson $A$ potentially
contributes to the predicted signal rate at $125\gev$ in the
$\tau^+\tau^-$ channel.\footnote{\HS\ automatically adds the signal
  rates of Higgs bosons that overlap within the combined experimental
  and theoretical mass uncertainties. For most Higgs channels with
  $\tau^+\tau^-$ final states, the experimental mass resolution is
  assumed to be $20\% \cdot m_H \approx 25\gev$, thus the signals of a
  $125\gev$ heavy Higgs $H$ and a $150\gev$ $\cp$-odd Higgs $A$ would
  be added.} In that case, the predicted signal rate would tend to be
higher than the total observed
$\tau^+\tau^-$ rate,
resulting in a larger $\chi^2$
from \HS. In Ref.~\cite{Bechtle:2012jw} we also took a possible signal
overlap of $H$ and $A$ in the $\tau^+\tau^-$ channel into account;
the measurements at that time, however, were not accurate enough to
notably affect the fit outcome. Secondly, parameter points with charged
Higgs masses $M_{H^+}$ below $160\gev$ are strongly constrained by
exclusion limits from LHC searches for a charged Higgs boson in top
quark decays, $t\to H^+ b$, with successive decay to $\tau$ leptons,
$H^+\to \tau^+\nu_\tau$~\cite{Aad:2014kga,Khachatryan:2015qxa}.
At tree-level, the $\cp$-odd and charged
Higgs masses are related as $M_{H^{\pm},\text{tree}}^2 = \MA^2
+ M_W^2$, thus, these constraints apply in particular at low values
$M_A \lesssim 140\gev$.  
In Ref.~\cite{Bechtle:2012jw} we found good discovery prospects for the
heavy Higgs case in $t\to H^+ b \to (\tau\nu_\tau)b$ searches. 
Based on the most recent limits from 
such searches performed by ATLAS and CMS~\cite{Aad:2014kga,Khachatryan:2015qxa}
the favored parameter regions of Ref.~\cite{Bechtle:2012jw} are now
excluded and the new preferred parameter space has moved
towards larger 
$M_A$ values in the light of the updated limits.  
Thirdly, another reason for disfavoring $M_A$ values below $\sim
150\gev$ is the prediction of somewhat too large values of \bsg, as
  will be discussed below.

The distribution of preferred parameter points in the ($\mu/M_S,
A_t/M_S$) plane, \figref{fig:hh_pars} (right), singles out values of
$\mu \sim (6-9) M_S$ and $A_t$ from $ -M_S$ to zero.
This is due to an
interplay of various observables and constraints, as we will outline in the following.
Agreement of the heavy
Higgs boson $H$ with the LHC Higgs rate measurements requires firstly that $M_H \sim \MHexp \gev$ and $m_h < M_H$ (corresponding to $M_A < M_{A,c}$, see \refse{Sec:alignment}) and secondly that the
alignment condition, $Z_6 v^2 = 0$,
is approximately
fulfilled. Furthermore, the LHC $H/A\to \tau^+\tau^-$ constraints impose
that this alignment must occur at not too large $\tan\beta$ values.
As we analyzed in detail in \refse{Sec:alignment}, these requirements single out small regions of the ($\mu$, $A_t$) plane, 
namely either $\mu/M_S \sim A_t/M_S \approx \pm (3-4) $ or 
$|\mu| / M_S \approx 4-9$ (or even larger) and negative $A_t$ with $|A_t|/M_S \lesssim 1$.
Quite generically, in the heavy Higgs interpretation we have a
slightly too high prediction of $\bsg$ due to the light charged Higgs
boson. This 
discrepancy, however, decreases for very large positive values of
$\mu/M_S$ (similar to the alignment limit in the light Higgs case, cf.~\reffi{fig:h_LEO_1}).
Combining these arguments we find that large positive $\mu$ values, together with rather small negative $A_t$ are favored.

\begin{figure}
\centering
\includegraphics[width=0.44\columnwidth]{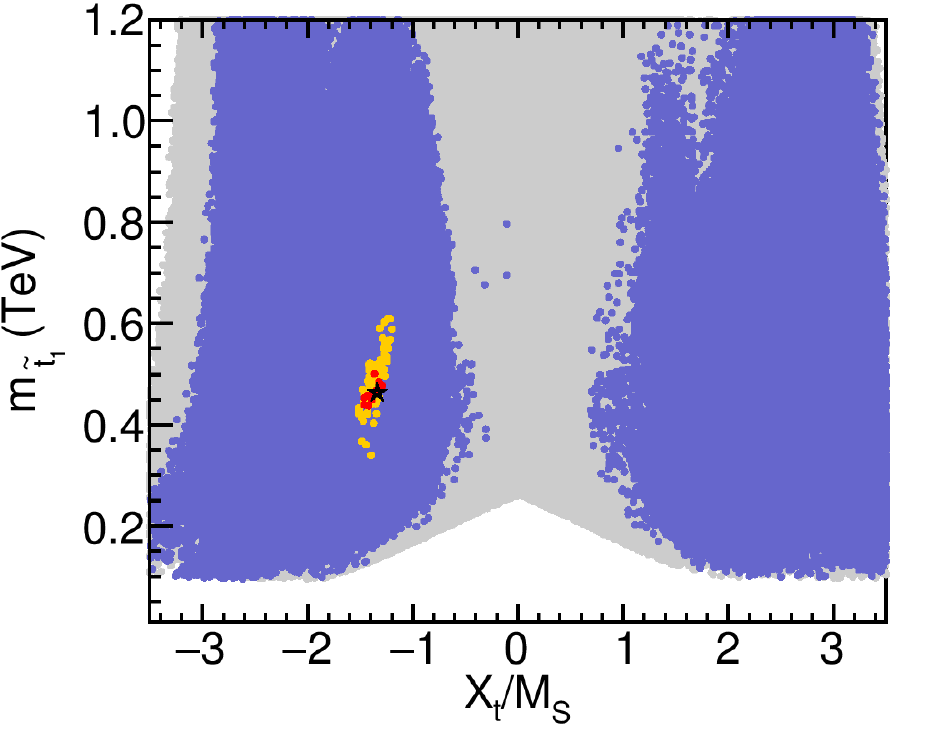}
\caption{
Preferred parameter regions in the ($X_t/M_S$, $m_{\tilde{t}_1}$) plane
in the heavy Higgs case. The color coding is the same as in \reffi{fig:Hrates_corr}.
  } 
\label{fig:Xt-mstop1}
\end{figure}

In order to avoid problems with vacuum instability, we applied a cut of $|\mu|/M_S < 3$ in the light Higgs case.
Obviously, all our favored points in the heavy Higgs case would be cut away by 
such a constraint.\footnote{Taking into account only the parameter points which would survive such a cut ($|\mu|/M_S < 3$), we find a minimum $\chi_H^2$ of 90.7 for the full fit ($\nu=85$), corresponding to 
a $p$-value of 0.32. In this case the best-fit region features values of $M_{\tilde{q}_3} \sim 1$ TeV and $\mu \sim A_t \approx (2.5 - 3) M_S$.}
However we want to stress here that this cut only provides an approximate limit. Testing if any of our 
favored points in the heavy Higgs case are still allowed by vacuum stability would require a 
more thorough analysis, which is beyond the scope of this paper.

The fact that $A_t$ and $\mu$ are predicted to be confined to a narrow
range in the heavy Higgs interpretation
also results in a definite prediction for $X_t$ and a quite small range for
the light stop mass,
as shown in \reffi{fig:Xt-mstop1}. The favored region has
$\Xt \sim -1.5\,M_S$, and the light stop mass is found between
$350 \gev$ and $650 \gev$ --- a prediction that can be tested at the
upcoming LHC stop searches. However, note that the upper limit on the light stop mass is a consequence of our restricted scan range, $\mu\le 5\tev$, and the fact that large $\mu/M_S$ values are favored.


\subsubsection{Impact of low energy observables}\label{sec:HH_leo}

The discussion of the low energy observables in the heavy Higgs case closely follows the corresponding discussion in the 
alignment limit of the light Higgs case, see \refse{sec:h_leo}, as the parameters most relevant for the low energy observables are similar, i.e.~we again have small $M_A$, $\MHp$ and small $\tb$.

\begin{figure}[t!]
\centering
\includegraphics[width=0.46\columnwidth]{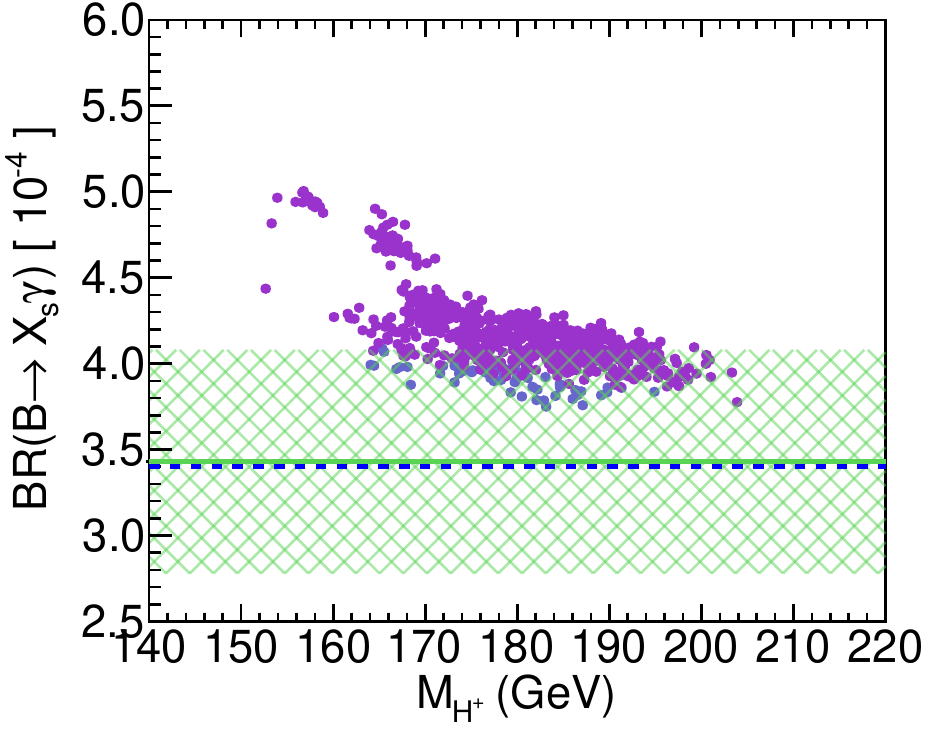}\hfill
\includegraphics[width=0.46\columnwidth]{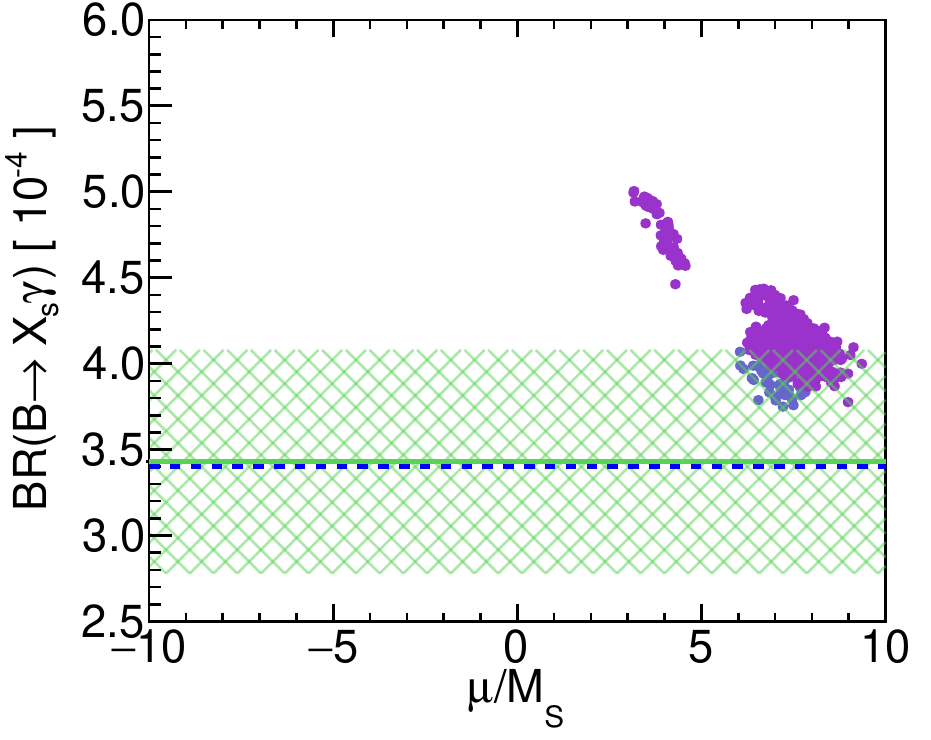}\hfill
\caption{$\bsg$ vs.\ $\MHp$ (\emph{left}) and
$\mu/M_S$ (\emph{right})
for the favored points 
in the heavy Higgs case in the fit without taking into account the LEOs.
The green line and hatched region show the experimental measurement and the total $1\sigma$ uncertainty region, while the SM prediction is indicated by the blue dashed line.
The color coding of the displayed points is the same as in \reffi{fig:h_Atmu_tanb}.}
\label{fig:HH_LEO_1}
\end{figure} 

\begin{figure}[h!]
\centering
\includegraphics[width=0.46\columnwidth]{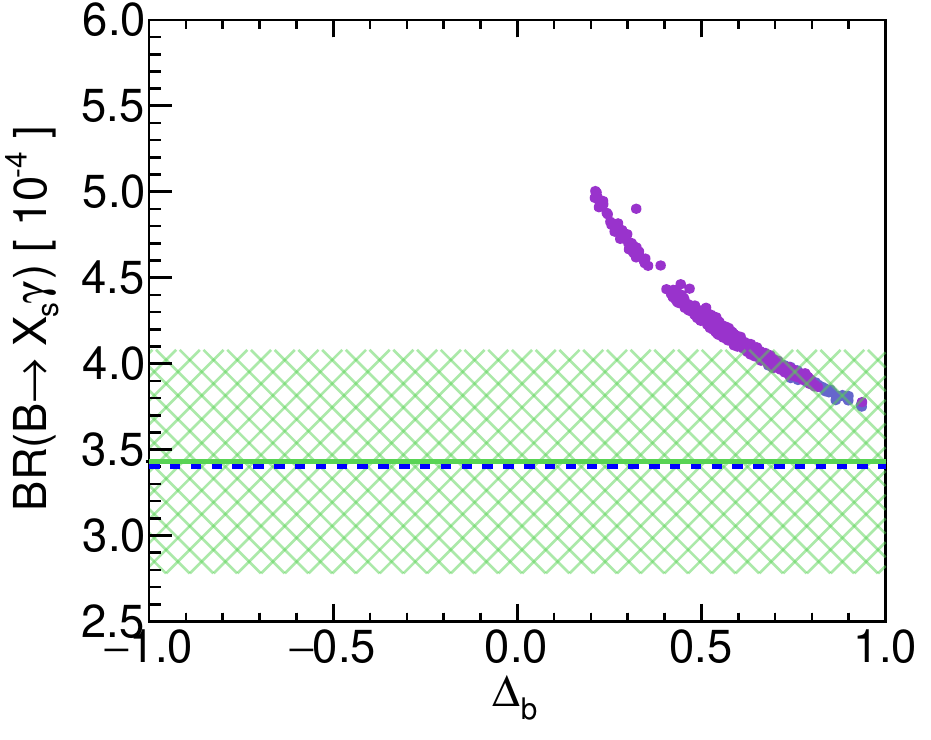}
\caption{$\bsg$ vs.~$\Delta_b$ for the favored points in the heavy Higgs case in the fit 
without taking into account the LEOs. The green and blue line and the hatched region are the same as in Fig.~\ref{fig:HH_LEO_1}. The color coding of the displayed points is the same as in \reffi{fig:h_Atmu_tanb}.}
\label{fig:HH_bsg_Deltab}
\end{figure} 

\reffi{fig:HH_LEO_1} shows $\bsg$ as a function of $\MHp$ (left plot) and $\mu/M_S$ (right plot) for the favored points of the fit to the Higgs signal rates and Higgs mass (including the exclusion likelihood from LEP and LHC searches, but without LEOs).
As discussed above the parameter points with $\mu / M_S \approx 3-4$ feature $A_t \sim \mu$, whereas the points with $\mu / M_S \gtrsim 6$ correspond to small and negative $A_t$ values. By comparing the two plots in \reffi{fig:HH_LEO_1} we can also see that the parameter points with $\mu / M_S \sim A_t / M_S \approx 3-4$ have relatively light charged Higgs masses, $\MHp < \mt$.
Again we find the expected increase of $\bsg$ when going to small charged Higgs masses.
Overall, we observe surprisingly good agreement with the experimental measurement. Even parameter points with a charged Higgs mass below the top quark mass
 lie mostly within the 2$\sigma$ region of the experimental measurement.

\begin{figure}[t!]
\centering
\includegraphics[width=0.46\columnwidth]{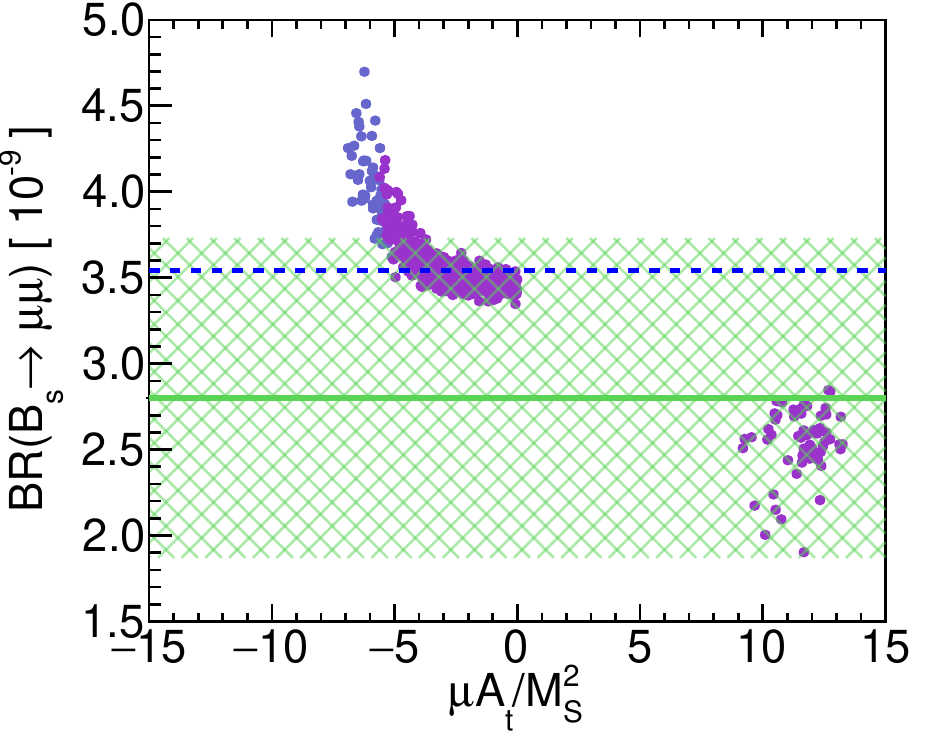}\hfill
\includegraphics[width=0.46\columnwidth]{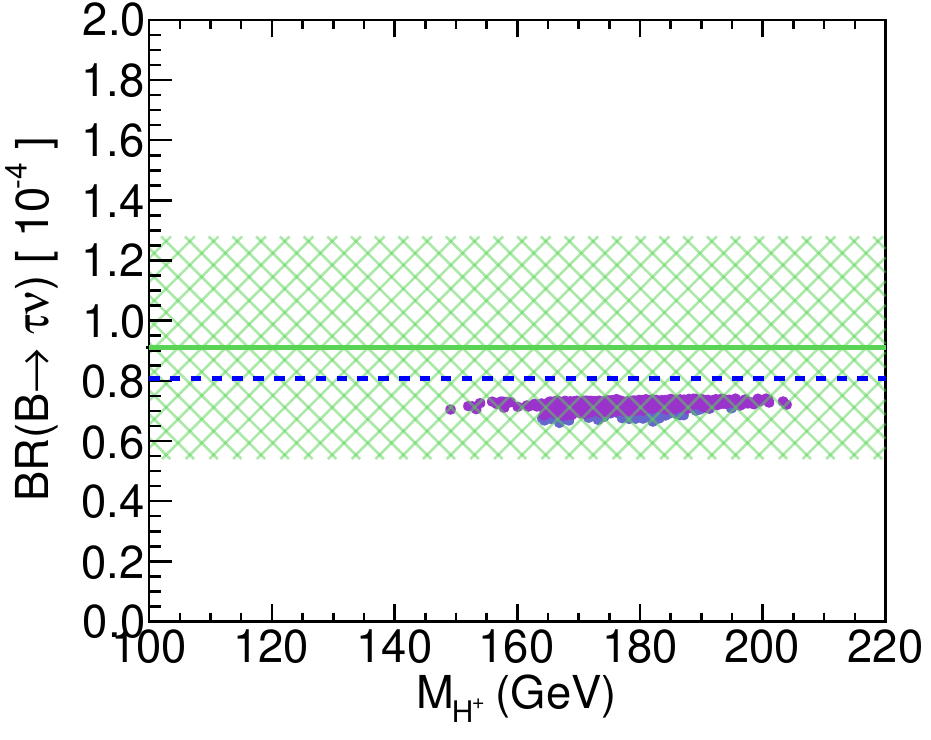}\hfill
\caption{
$B_s\to \mu^+\mu^-$ vs. $\mu A_t / M_S^2$ (\emph{left}) and 
$\mathrm{BR}(B^+\to \tau^+\nu)$ vs.\ $\MHp$ (\emph{right}) 
for the favored points 
in the heavy Higgs case in the fit without taking into account the LEOs.
The green line and hatched region show the experimental measurements and the total $1\sigma$ uncertainty region, the SM prediction is indicated by the blue dashed line.
The color coding of the displayed points is the same as in \reffi{fig:h_Atmu_tanb}.}
\label{fig:HH_LEO_1b}
\end{figure} 
\begin{figure}[t]
\centering
\includegraphics[width=0.46\columnwidth]{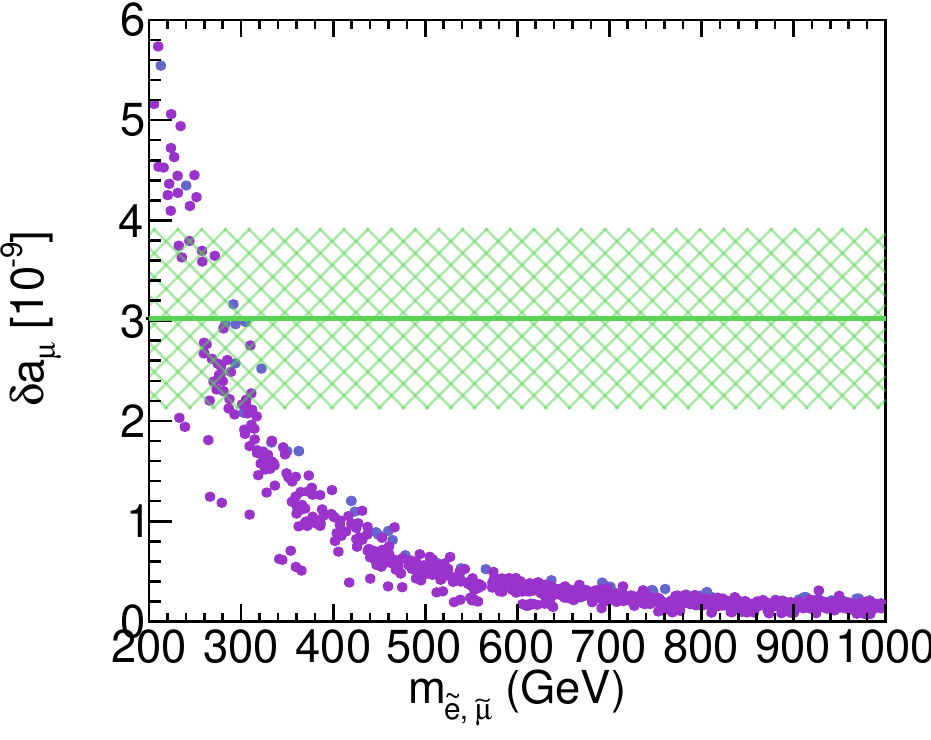}\hfill
\caption{The SUSY contribution to the anomalous magnetic moment of the
muon, $\delta a_\mu$, as a function of the (1st and 2nd generation) slepton
mass, $m_{\tilde{e},\tilde{\mu}}$, for the favored points 
in the heavy Higgs case in the fit without taking into account the LEOs.
The green line and hatched region
show the desired new physics contribution needed to achieve
agreement with the observed deviation from the SM and the $1\sigma$ experimental uncertainty.
The colors of the displayed points
is the same as in \reffi{fig:h_Atmu_tanb}.}
\label{fig:HH_LEO_2}
\end{figure} 

As in the light Higgs case [cf.~\reffi{fig:h_LEO_1}] we again find that large positive $\mu$ values give the best agreement with the measurement (within the $1\sigma$ region or just above). However, whereas it takes $\mu/M_S\gsim 3$ in order that the $\bsg$ prediction be within the $1\sigma$ region of the measurement in the light Higgs case, here we need much larger values $\mu/M_S \gtrsim 6$ to achieve this. Again, this can be understood from the sizable $\Delta_b$ corrections to the charged Higgs coupling to top and bottom quarks, as discussed in \refse{Sec:wrong}, and shown in Fig.~\ref{fig:HH_bsg_Deltab}, which impressively resembles the relation in \eq{BRratio}.

The \bmm\ prediction is shown in the left plot of \reffi{fig:HH_LEO_1b} in dependence of $\mu A_t/M_S^2$. We find that the few points with positive $\mu A_t$ are in excellent agreement with \bmm,
whereas for negative $\mu A_t$ the predictions are slightly too high, in particular for $\mu A_t / M_S^2 \lesssim -5$.
Again this can be compared to \reffi{fig:h_LEO_1b} where we found a similar result (even though there the typical $\mu A_t$ values of the favored points were quite different).
Again we find only very small SUSY corrections to \btn\ (despite the light charged Higgs) due to the small $\tan\beta$ values, as one can see in the right plot of \reffi{fig:HH_LEO_1b}.

The prediction for the SUSY contribution to the anomalous magnetic moment of the muon, $\delta a_\mu$, as a function of the slepton mass are shown in \reffi{fig:HH_LEO_2}. Again we see a very strong correlation, i.e.~the value for $\delta a_\mu$ is almost entirely determined by the slepton mass. Light slepton masses $\lesssim 300\gev$ are favored in the heavy Higgs case.


\vskip 2in

\subsubsection{Impact of direct LHC SUSY searches}
\label{Sect:ImpactDirectSUSYSearchedHeavyHH}

\begin{figure}[p]
\centering
\includegraphics[width=0.44\columnwidth, height=0.33\columnwidth]{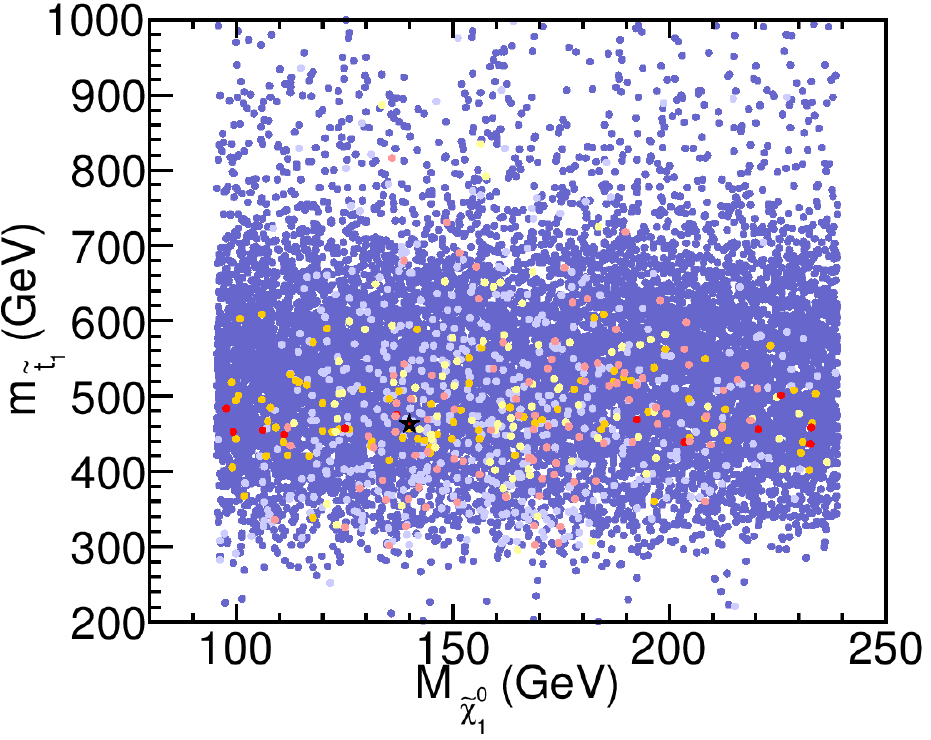}\hfill
\includegraphics[width=0.44\columnwidth, height=0.33\columnwidth]{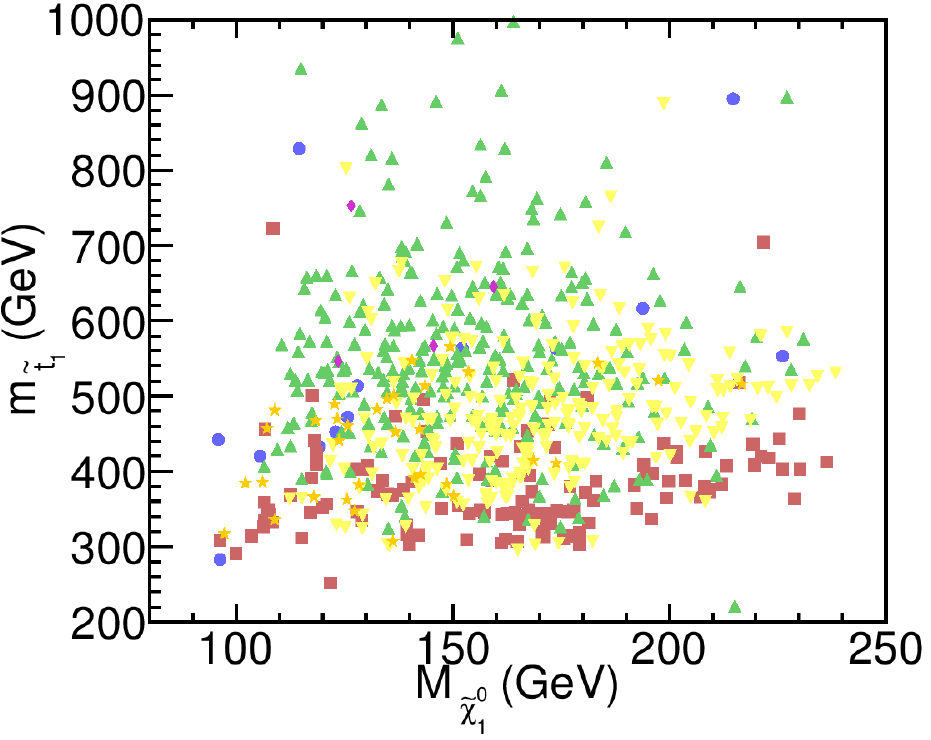}
\includegraphics[width=0.44\columnwidth, height=0.33\columnwidth]{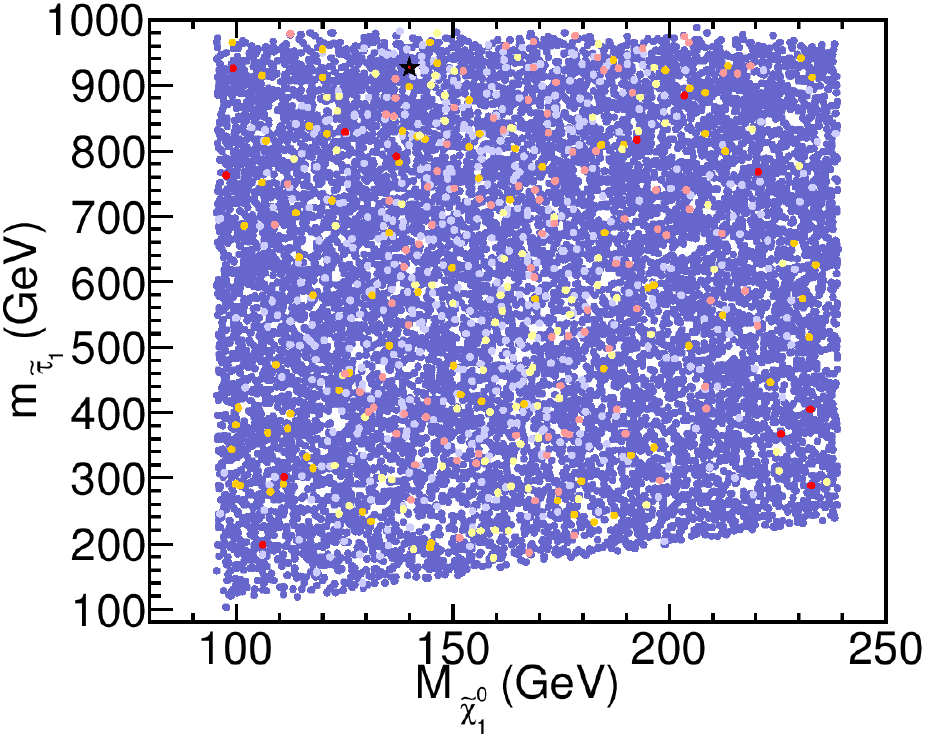}\hfill
\includegraphics[width=0.44\columnwidth, height=0.33\columnwidth]{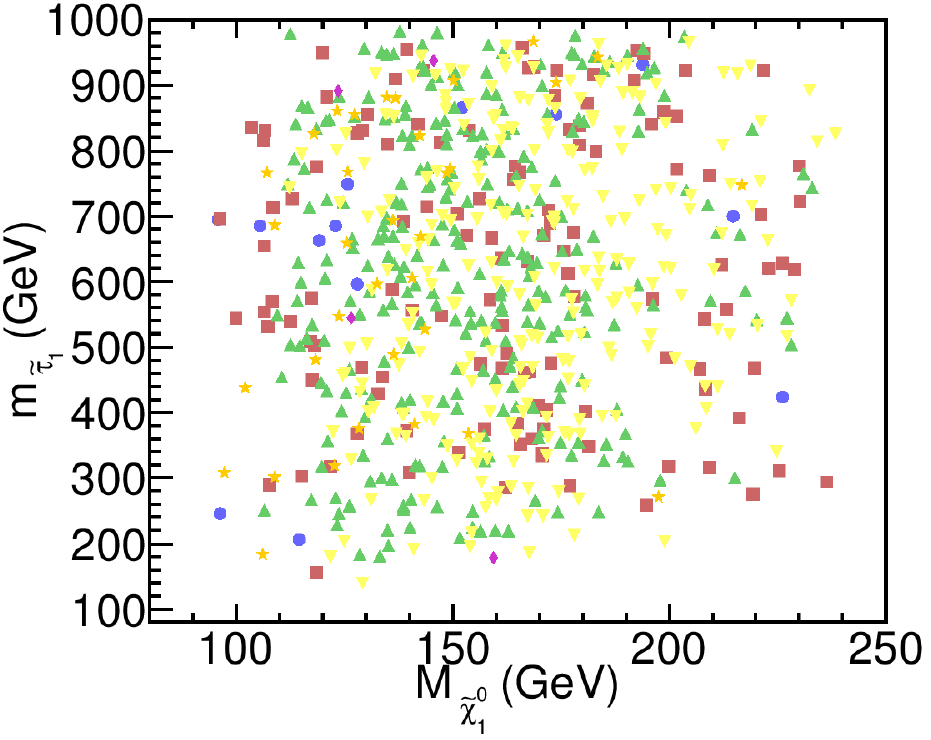}
\includegraphics[width=0.44\columnwidth, height=0.33\columnwidth]{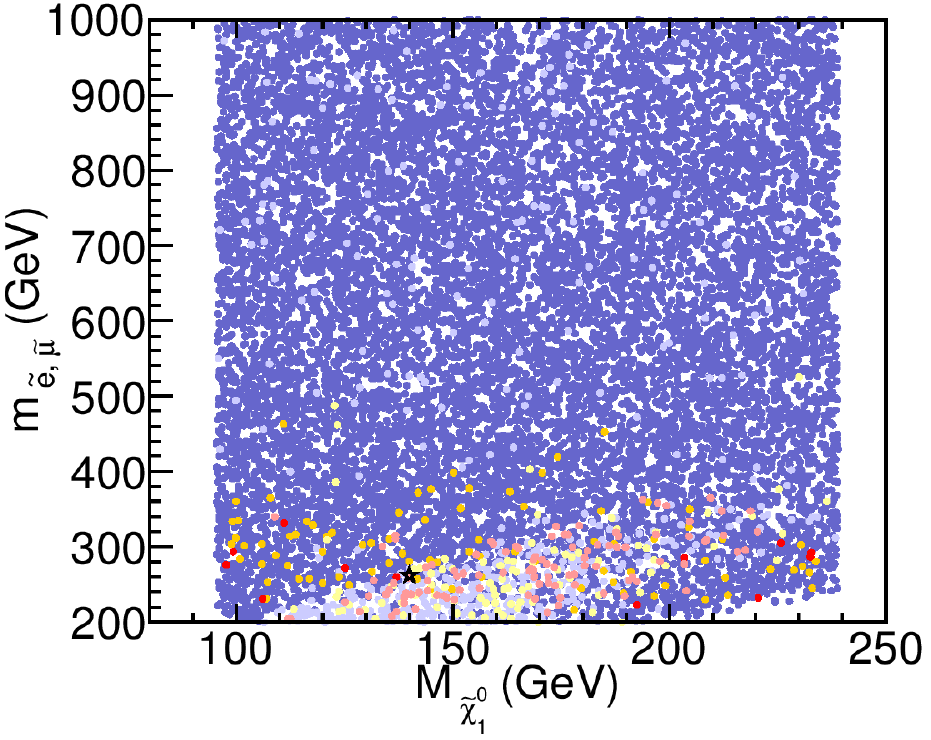}\hfill
\includegraphics[width=0.44\columnwidth, height=0.33\columnwidth]{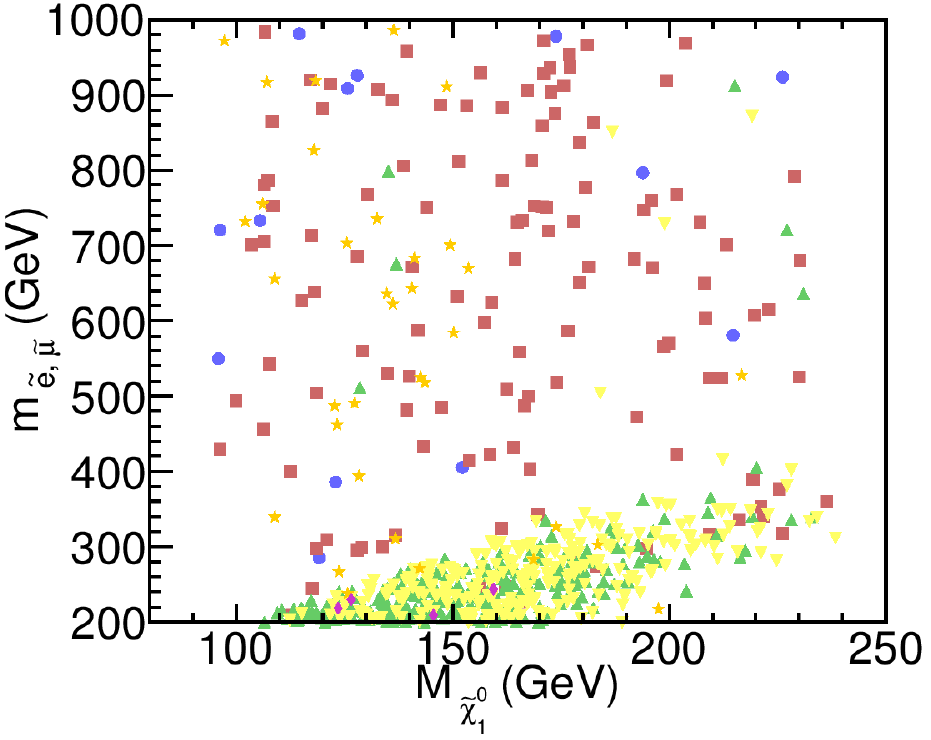}
\caption{
Impact of $8\tev$ LHC SUSY searches on the fit, in dependence of the lightest stop (\emph{top row}), stau (\emph{middle row}) and selectron/smuon (\emph{bottom row}) mass and the lightest neutralino mass. 
In the three left panels, we compare the
 \CM\ allowed (\emph{bright colored}) and excluded (\emph{pale colored}) points for all scan points (\emph{blue}), the \CL{95\%} (\emph{yellow}) and \CL{68\%} (\emph{red}) preferred points (plotted in this order, with excluded points on top of the allowed points in each step). In the three right panels, we exhibit LHC analyses that yield the exclusion (in the order of plotting): 
  \emph{(i)}
  $N~\text{jets}+\ETmiss$ searches (\emph{blue circles})~\cite{Aad:2013wta,Aad:2014wea}, \emph{(ii)} hadronic
  $\tilde{t}\tilde{t}$ searches with $b$-jets (\emph{red squares})~\cite{Aad:2013ija,Aad:2014bva,Aad:2014lra,Chatrchyan:2013mys},
  \emph{(iii)} $\chi^\pm_1\chi^0_2\to 3\ell+\ETmiss$ searches (\emph{green
  up-triangles})~\cite{Aad:2014nua}, \emph{(iv)} $\tilde{t}\tilde{t}\to 2\ell+\ETmiss$
  searches (\emph{yellow down-triangle})~\cite{Aad:2014qaa}, \emph{(v)} $\tilde{t}\tilde{t} \to
  1\ell + N~\text{jets}+\ETmiss$ searches (\emph{orange stars})~\cite{Aad:2014kra}, \emph{(vi)}
  $\chi^\pm_1\chi^0_2, \tilde{\ell}\tilde{\ell} \to \ell^+\ell^-
  +\ETmiss$ searches (\emph{magenta diamonds})~\cite{Aad:2014vma}, \emph{(vii)} $2\ell +
  N~\text{jets} + \ETmiss$ (\emph{gray plus})~\cite{Aad:2015mia,Khachatryan:2015lwa}. }
\label{fig:CMresults_HH}
\end{figure}

The comparison of the impact of the direct SUSY searches with the
impact of the other constraints qualitatively
agrees between the light Higgs
interpretation (see
Section~\ref{Sect:ImpactDirectSUSYSearchedHeavyHH}) and the heavy
Higgs interpretation. Also for the heavy Higgs interpretation, the SUSY search limits
are thinning the parameter space in an ``orthogonal'' way to the Higgs
observables and limits from the Higgs searches. 
We again show the distribution of
\CM\ allowed and excluded points in terms of
the most relevant mass parameters in
Fig.~\ref{fig:CMresults_HH} (left), and indicate the corresponding
experimental search that yields the exclusion in
Fig.~\ref{fig:CMresults_HH} (right). Similar to the light Higgs
case, the
\texttt{CheckMATE} exclusion is correlated with low-energy constraints, particularly
the constraints from the anomalous magnetic moment of the muon.
As discussed in the previous section, $a_\mu$ strongly favors low slepton masses. Hence we observe a large fraction of favored parameter points excluded by multilepton searches for gaugino pair
production (green up-triangles) as long as $M_2 -
m_{\tilde{e},\tilde{\mu}} \gtrsim 20\gev$ is fulfilled. As we have
relatively light stops in the preferred region of parameter space,
constraints from dilepton searches (yellow down-triangles) as well as
hadronic searches with $b$-jets (red squares) targeting stop pair production
are also relevant. Overall, LHC SUSY searches exclude only $\sim 5\%$
of all tested parameter points (i.e.~points with $\Delta\chi^2 \le 10$), but
$\sim 65\%$ of the favored ($\Delta\chi^2<5.99$) parameter points. This
illustrates again the complementarity between the impact of $a_\mu$ and the
constraints from SUSY searches sensitive to light sleptons.


\clearpage

\subsubsection{Phenomenology of the other Higgs states}

\begin{figure}[b!] 
\centering
\includegraphics[width=0.45\columnwidth]{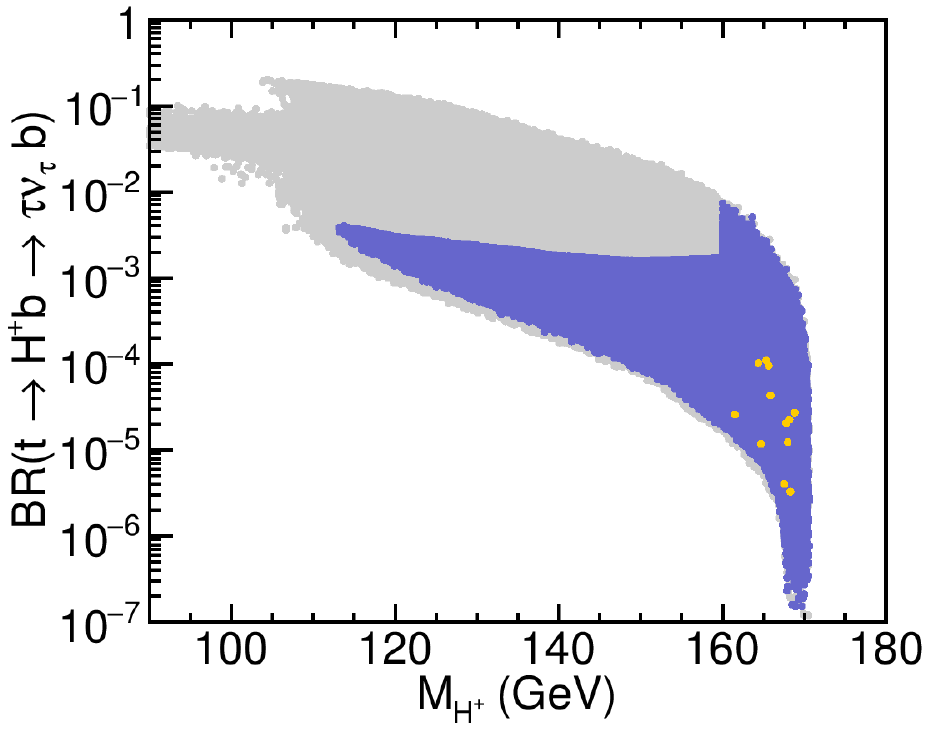}\hfill
\caption{
  Branching ratio of the top quark decay into a charged Higgs boson and
  a bottom quark, with the successive decay of the charged Higgs boson
  into a tau lepton and neutrino, in the heavy Higgs case.
  The color coding is the same as in \reffi{fig:Hrates_corr}.}
\label{fig:chargedHiggs}
\end{figure} 

Here we discuss the prospects for the discovery of the other Higgs
bosons at the LHC in the heavy Higgs interpretation. We start with the
phenomenology of the charged Higgs boson, which is a crucial test of
this scenario. In fact, the previous benchmark scenario for the 
heavy Higgs interpretation~\cite{Bechtle:2012jw} 
has been excluded with limits that have meanwhile been obtained from searches for charged Higgs 
bosons~\cite{Aad:2014kga,Khachatryan:2015qxa}.

In \reffi{fig:chargedHiggs} we show the rate for the main production and decay channel of a light charged Higgs boson with $\MHp < \mt$, $\br(t \to H^\pm b) \times \br(H^\pm \to \tau \nu_\tau)$, as a function
of $\MHp$. The impact of the current limits from charged Higgs searches in
this channel~\cite{Aad:2014kga,Khachatryan:2015qxa} can be seen by the gray
area in \reffi{fig:chargedHiggs}, cutting out the region with $\MHp < 160
\gev$ and $\br(t \to H^\pm b \to \tau \nu_\tau b) \gtrsim (2-4) \times
10^{-3}$. Only a few favored points (and none of the most favored points) have
$\MHp < \mt$ and are therefore displayed in \reffi{fig:chargedHiggs}.
As one can see, these points have charged Higgs masses close to the top
quark mass and thus $\br(t \to H^\pm b) \times \br(H^\pm \to \tau \nu_\tau)$
is strongly suppressed due to the limited phase space of the top quark
decay. Also the decay $\br(H^\pm \to \tau \nu_\tau)$ is suppressed by the
competing decay $H^\pm \to h W^\pm$, which is open for most of these favored
points (we will discuss the $m_h$ range of the favored points below).
Consequently it is very difficult 
to detect charged Higgs bosons in this mass range in the $t\to H^+b\to
(\tau^+\nu_\tau)b$ channel at the LHC.

Many of our favored and most favored points have $\MHp > \mt$ (and are thus not visible in \reffi{fig:chargedHiggs}).
Charged Higgs bosons with masses above the top quark mass are 
searched for in the $pp \to t H^\pm$ production channel with $H^\pm \to \tau \nu_\tau$ \cite{Aad:2014kga,Khachatryan:2015qxa,Aaboud:2016dig} or $H^\pm \to t b$ \cite{Aad:2015typ,Khachatryan:2015qxa}. 
These searches, although concentrating on the charged Higgs mass region that is relevant for the
heavy Higgs interpretation, are not yet sensitive to constrain the favored parameter space.
However, they will become more sensitive with increasing integrated luminosity. Furthermore, we emphasize again that the decay $H^\pm \to hW^\pm$ is possible and unsuppressed in large parts of the parameter space,
but currently not directly searched for at the LHC.
In \refse{Section:HHbenchmark} we will present specific benchmark scenarios, inspired by our best-fit point in the heavy Higgs case, that can be employed to study the sensitivity of these searches.

\begin{figure}[t!] 
\centering
\includegraphics[width=0.44\columnwidth,height=0.33\columnwidth]{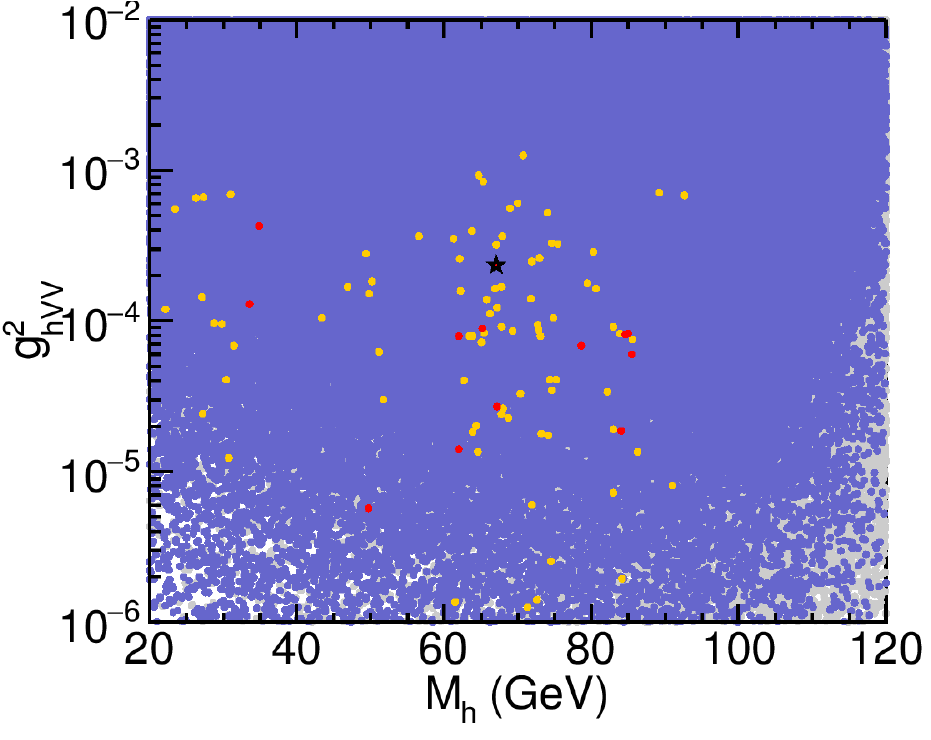}\hfill
\includegraphics[width=0.44\columnwidth,height=0.33\columnwidth]{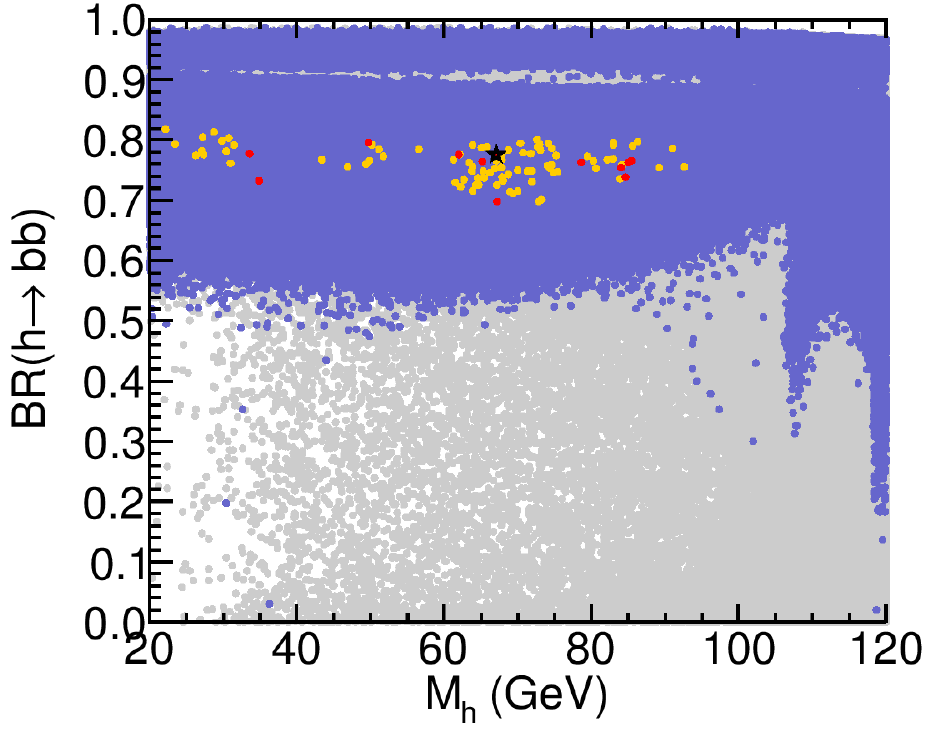}
\includegraphics[width=0.44\columnwidth,height=0.33\columnwidth]{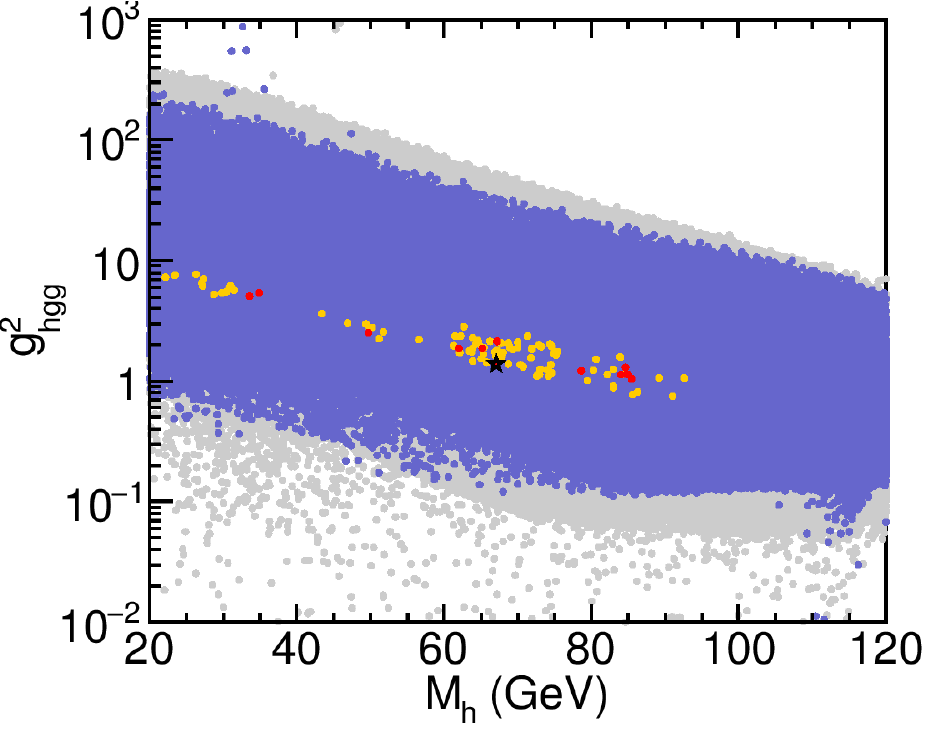}\hfill
\includegraphics[width=0.44\columnwidth,height=0.33\columnwidth]{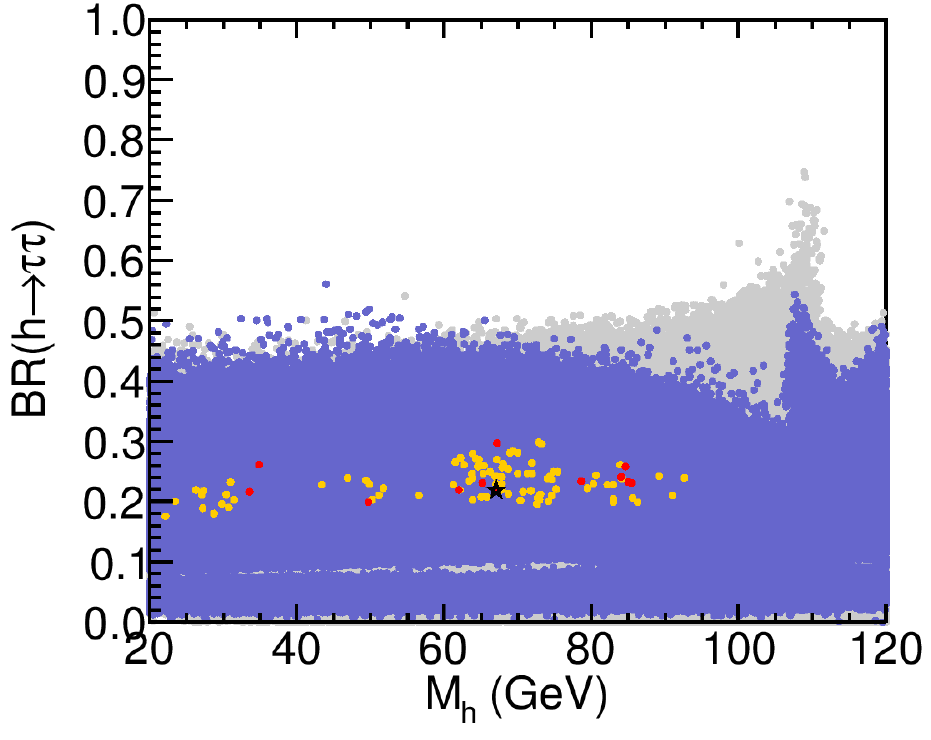}
\includegraphics[width=0.44\columnwidth,height=0.33\columnwidth]{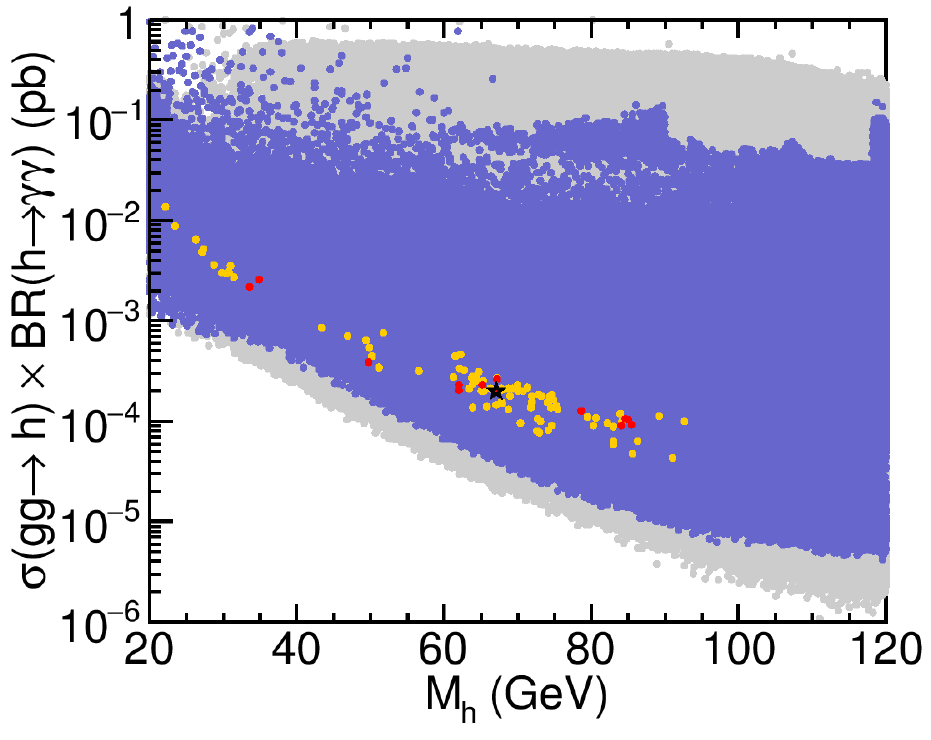}\hfill
\includegraphics[width=0.44\columnwidth,height=0.33\columnwidth]{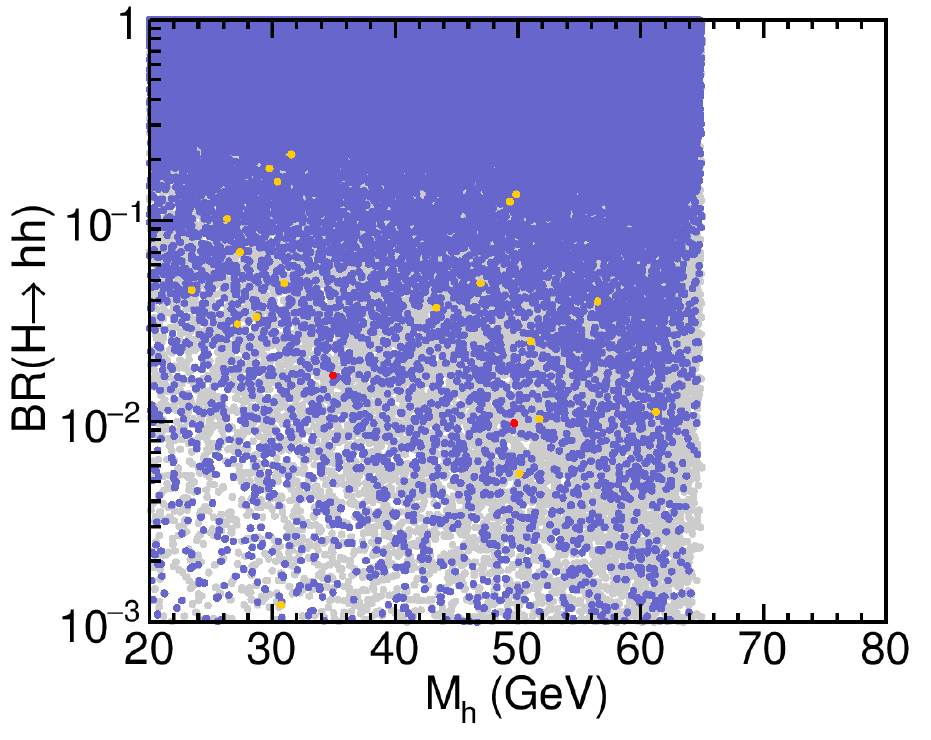}
\caption{
Light Higgs boson ($h$) phenomenology, in dependence of the light Higgs mass $M_h$, in the heavy Higgs interpretation: (SM normalized) squared $hVV$ coupling, $g_{hVV}^2$, (\emph{top left}) and $hgg$ coupling, $g_{hgg}^2$, (\emph{middle left}), LHC $8\tev$ signal rate for the process $gg\to h\to \gamma\gamma$ (\emph{bottom left}), branching fractions for the decays $h\to b\bar{b}$ (\emph{top right}), $h\to \tau^+\tau^-$ (\emph{middle right}) and the Higgs-to-Higgs decay $H\to hh$ (\emph{bottom right}). The color coding is the same as in \reffi{fig:Hrates_corr}.
 }
\label{fig:lightHiggs1}
\end{figure}

We will now turn to the discussion of the phenomenology of the light
$\cp$-even Higgs boson, $h$, in the preferred parameter region in the heavy
Higgs case. The light $\cp$-even Higgs boson has a mass in the range
$(20-90)\gev$ and a strongly reduced coupling to vector bosons. This is
shown in the top left plot of \reffi{fig:lightHiggs1}, where the
squared coupling $g_{hVV}^2$ is displayed, normalized to the corresponding
coupling in the SM with the same value of the Higgs boson mass. One can see
that the squared coupling is
reduced by a factor of $10^3$ or more with respect to the SM, as the 
heavy $\cp$ even
Higgs boson $H$ in this scenario acquires the coupling to vector bosons
with approximately SM Higgs strength.
This results in a strongly reduced cross section for the LEP
Higgs-Strahlung process, $e^+e^-\to Zh$. Consequently, the light Higgs boson
in this case would have
escaped detection in corresponding LEP Higgs searches.
The limits from the Higgs
searches at LEP occur for higher values of the
relative squared coupling $g_{hVV}^2$ and are not visible in this plot.

The reduced light Higgs coupling to vector bosons furthermore leads to a
reduced rate of the 
$h\to\gamma\gamma$ decay, which happens through a $W$-boson loop (amongst other contributing diagrams).
In contrast, the light Higgs coupling to gluons is up to ten times stronger than the SM Higgs boson coupling at very low light Higgs masses, $M_h$, as shown in the center left plot of \reffi{fig:lightHiggs1}, where the (SM normalized) squared light Higgs-gluon coupling, $g_{hgg}^2$, is shown in dependence of $M_h$. This results in an
abundant production of the light $\cp$-even Higgs boson via gluon fusion.
The resulting LHC cross section for $gg \to h$ (at $8\tev$) with subsequent decay $h \to \gamma \gamma$ is shown in the bottom left plot of \reffi{fig:lightHiggs1}. Limits from LHC searches in this channel~\cite{Aad:2014ioa} 
have been taken into account in our analysis (using {\tt HiggsBounds}).
Their effect can be seen in this plot as the gray excluded region above
$\sim 0.1$ pb. Clearly these searches are currently very
far from being sensitive to detect the light Higgs boson in this scenario.

The light Higgs boson predominantly decays to
bottom quarks ($\sim (70-80)\%$ of the time) or tau leptons
($\sim(15-30)\%$ of the time), as shown in the top and middle right
plots in \reffi{fig:lightHiggs1}, respectively. Given the mass range
$20\leq M_h \leq 90\,\mathrm{GeV}$ we expect
 direct LHC searches in both channels to be rather challenging, given the huge SM background and the difficulty to trigger the events. 
The majority of the favored points have $M_h > M_H/2$, however we also
observe some favored points with lower $M_h$ (down to 20 GeV) for which the
Higgs-to-Higgs decay channel $H \to hh$ is kinematically open. If this
decay rate is sizable the rates of heavy Higgs $H$ decays to SM particle
final states can be significantly affected. We find that for our preferred
points the branching ratio of this decay is at most $\sim 20\%$, but
for most points $\lesssim 10\%$, as shown in the bottom right plot of
\reffi{fig:lightHiggs1}. Branching fractions of this size lead to
only moderate changes of the Higgs decay rates to SM particles.
Consequently we still find a good fit in this scenario even for the
case $M_h < M_H/2$ (the allowed points for $M_h > M_H/2$ with vanishing
branching ratio of $H \to hh$ are not visible in the plot).

The $\cp$-odd Higgs boson has a mass between $140 \gev$ and $185\gev$,
as shown already in \reffi{fig:hh_pars}, for $\tb$ in the range
between $6$ and $11$. Consequently, in this scenario
it should signify itself with higher luminosity at
the LHC in $A\to\tau\tau$ searches. At the lower end of its mass range
it might be visible possibly as an enhanced
decay rate of the SM-like Higgs boson due to the limited mass
resolution. At the higher end of its mass range it would appear as a new resonance
decaying to $\tau^+\tau^-$. 

\subsection{Updated \boldmath{\lowMH} benchmark scenarios}
\label{Section:HHbenchmark}

In the previous section we demonstrated that the heavy Higgs
interpretation in which the heavy $\cp$-even Higgs boson is identified with the
observed   
signal~\cite{Heinemeyer:2011aa,Hagiwara:2012mga,Benbrik:2012rm,Drees:2012fb,Han:2013mga}
is still a viable scenario.
While the original \emph{low-$\MH$} benchmark
scenario~\cite{Carena:2013ytb} has meanwhile been ruled out by ATLAS and
CMS via the search for a light charged Higgs boson in top quark
decays~\cite{Aad:2014kga,Khachatryan:2015qxa}, we 
showed that viable realizations of the heavy Higgs interpretation exist outside the parameter region where these searches are sensitive. In this section we 
define new versions of the \lowMH\ benchmark scenario that are
valid after taking into account all current 
experimental constraints. They are inspired by the best-fit point found in our global analysis of the
heavy Higgs case, see Table~\ref{tab:BFparameters}, however, we slightly increased the stop mass scale (while roughly retaining the preferred $\mu$ to $M_S$ ratio) in order to evade potential exclusion limits on the stop mass from the upcoming $13\tev$ LHC results. These new scenarios could provide a useful benchmark for the ongoing light charged Higgs boson searches in the MSSM. In particular, they exhibit the not-yet-sought-for MSSM decay signature $H^+\to W^+ h$ and, in some parameter regions, even the decay $H^+\to W^+H$.


We define three different scenarios, which we call \emph{\lowMHup} in
order to distinguish them from the previous \emph{low-$\MH$} benchmark
scenario~\cite{Carena:2013ytb}.
The first two \lowMHup\
scenarios follow the original idea in \citere{Carena:2013ytb} and fix a
``heavy'' Higgs boson mass, in this case the charged Higgs boson mass,
$\MHp$, to a certain value, whereas $\mu$ and $\tb$ are taken as free
parameters. We suggest two variants given by different
choices of $\MHp$ below (\lowMHuplow) and above
(\lowMHuphigh) the top quark mass, $\mt$. The second
scenario fixes $\mu$ and explores the ($\MHp$, $\tb$) plane. This scenario
may be utilized for charged Higgs boson searches in the relevant mass
range. The parameters of these benchmark scenarios are given
in~Table~\ref{Tab:benchmarks}.

\begin{table}[b]
\centering
\begin{tabularx}{0.95\textwidth}{l>{\centering\arraybackslash}X>{\centering\arraybackslash}X>{\centering\arraybackslash}X}
\toprule
Benchmark scenario		& 	$\MHp~[\mathrm{GeV}]$ & $\mu~[\mathrm{GeV}]$ & $\tb$ \\
\midrule
\lowMHuplow &	$155$		&	$3800$ -- $6500$	&		$4$ -- $9$ \\
\lowMHuphigh &	$185$		&	$4800$ -- $7000$	&		$4$ -- $9$ \\
\lowMHupvar  &		$140$ -- $220$ &     $6000$			&		$4$ -- $9$ \\
\midrule
fixed parameters: &\multicolumn{3}{l}{$\mt = 173.2\gev$, \quad $\At = \Atau = \Ab = -70\gev$,\quad $M_2 = 300 \gev$,}\\ 
&\multicolumn{3}{l}{$\MsqL = \MsqR =  1500 \gev$~($q = c, s, u, d$),\quad $\mgl = 1500 \gev$,} \\
&\multicolumn{3}{l}{$M_{\tilde{q}_3} = 750\gev$,\quad$\mslez = 250 \gev$,\quad $\msld = 500 \gev$} \\
\bottomrule
\end{tabularx}
\caption{Parameters of the updated low-$M_H$ benchmark scenarios. All
parameters are given in the on-shell (OS) definition. The lower row gives the fixed parameters that are common to all three benchmark scenarios. $M_1$ is fixed via \refeq{def:Mone}.}
\label{Tab:benchmarks}
\end{table}

\begin{figure}[th!]
\centering
\includegraphics[width=0.45\columnwidth]{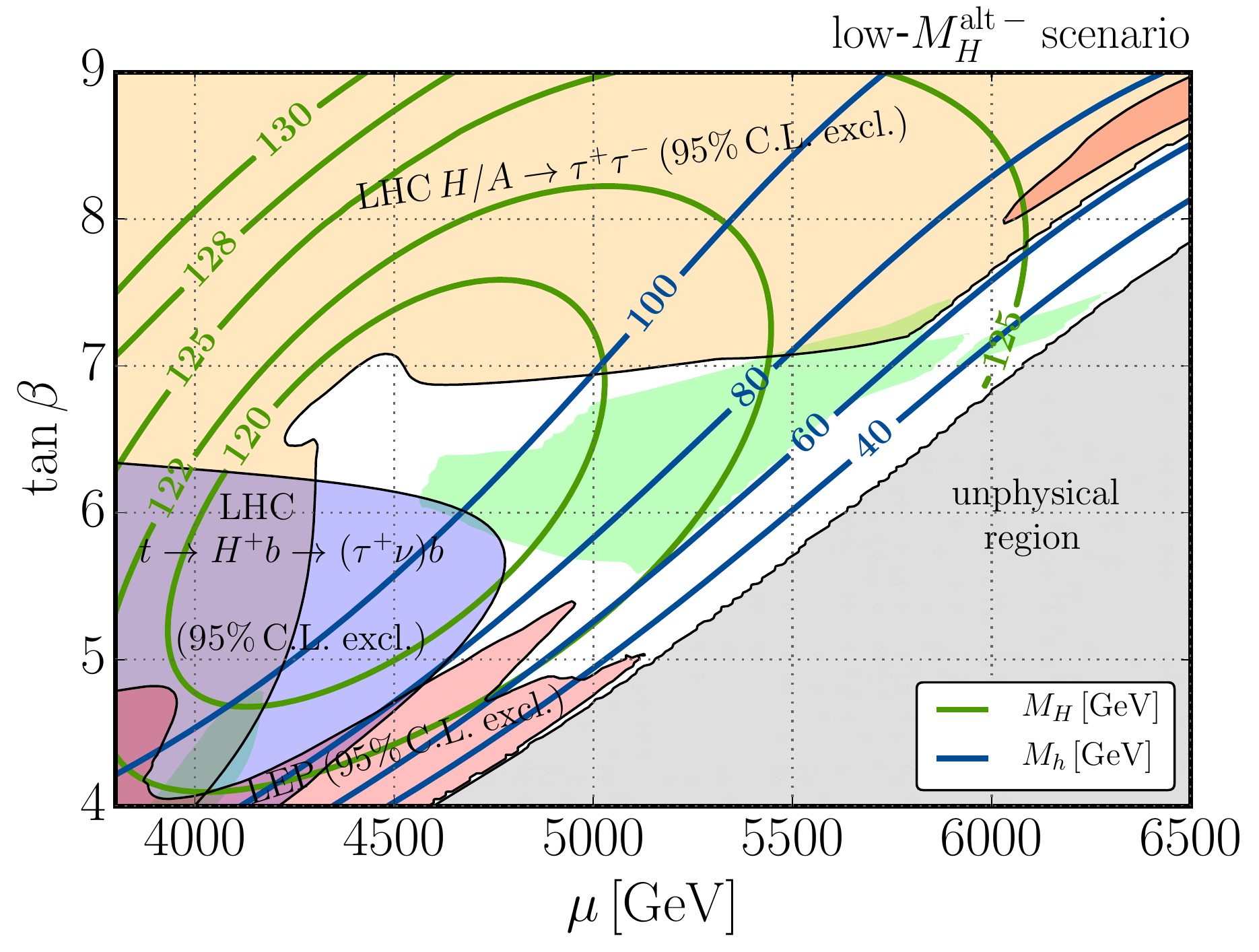}\hfill
\includegraphics[width=0.45\columnwidth]{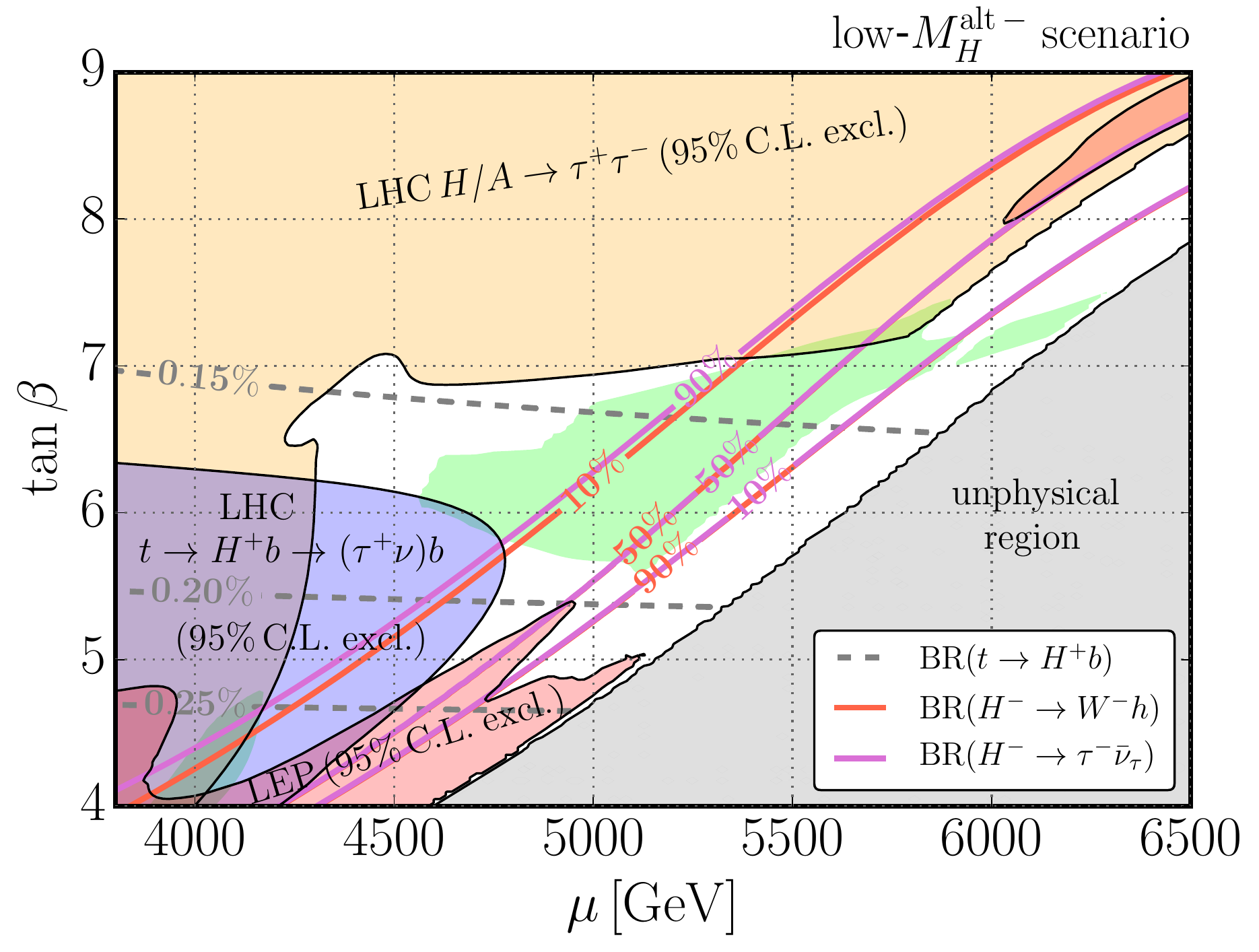}
\includegraphics[width=0.45\columnwidth]{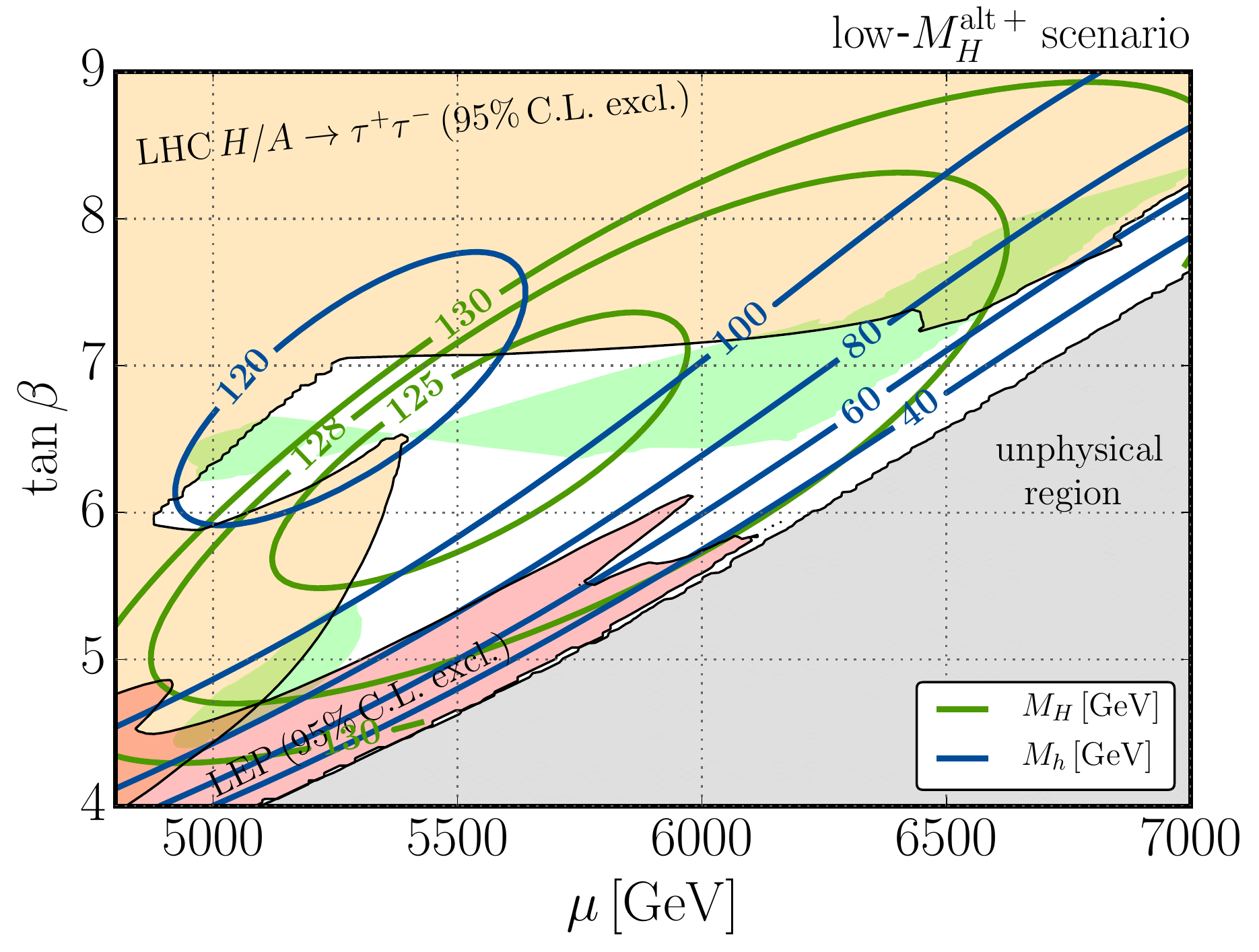}\hfill
\includegraphics[width=0.45\columnwidth]{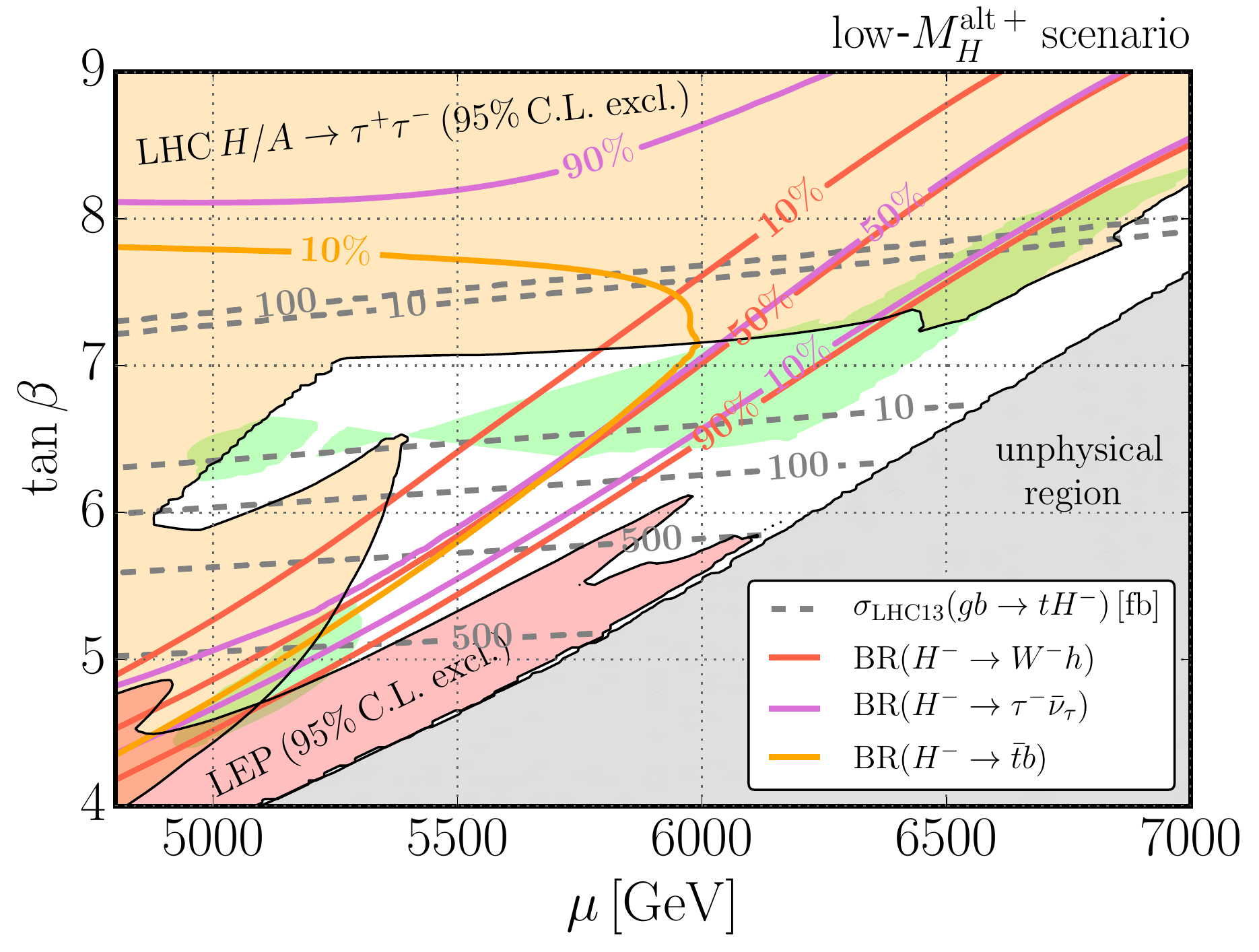}
\caption{The \lowMHuplow\ and \lowMHuphigh\ benchmark scenarios in
  the ($\mu$, $\tb$) plane with $\MHp = 155\gev$ (\emph{upper row}),
  and with $\MHp = 185\gev$ (\emph{lower row}), respectively. The red,
  orange and blue regions are disfavoured at the \CL{95\%} by LEP light
  Higgs $h$ searches~\cite{Schael:2006cr}, LHC $H/A\to \tau^+\tau^-$
  searches~\cite{Khachatryan:2014wca,CMS:2015mca} and LHC $t\to
  H^+b\to(\tau\nu)b$ searches~\cite{Aad:2014kga,Khachatryan:2015qxa},
  respectively. The green area indicates parameter regions that are
  compatible with the Higgs signal (at $\sim$ \CL{95\%}, see text for
  details), unphysical regions are displayed in gray (see text). 
  In the two left panels, contour lines indicate the Higgs masses $M_h$ and
  $M_H$ (in GeV). In the two right panels, contours lines indicate the charged Higgs
  branching ratios, as well as the branching ratio for the top quark decay $t\to
  H^+b$ (\emph{upper row}) or the $13\tev$ LHC cross section for
  charged Higgs production in association with a top quark,
  $\sigma_\text{LHC13}(gb\to tH^+)$ (in
  fb)~\cite{Berger:2003sm,Flechl:2014wfa} (\emph{lower row}). }  
\label{fig:lowMH-mutb}
\end{figure} 

In the following we discuss the compatibility of these benchmark
planes with the current experimental constraints. In
\reffi{fig:lowMH-mutb} we present the results in the ($\mu$,
  $\tb$) plane in the \lowMHuplow\ scenario with $\MHp = 155\gev$
  (\emph{upper row}) and in the \lowMHuphigh\ scenario with
$\MHp = 185\gev$ (\emph{lower row}). The blue, red and orange areas
indicate the \CL{95\%} excluded regions by LHC
$t\to H^+b\to(\tau\nu)b$
searches~\cite{Aad:2014kga,Khachatryan:2015qxa}, LEP light Higgs
searches~\cite{Schael:2006cr} (mostly from the
$e^+e^- \to Zh \to Z(b\bar{b})$ channel) and LHC $H/A\to \tau^+\tau^-$
searches~\cite{Khachatryan:2014wca,CMS:2015mca}, respectively. For the
latter two we again employ the $\chi^2$ implementation of these
results in \HB, see~\refse{Sect:Constraints}, and define the \CL{95\%}
excluded region by $\Delta\chi^2 \ge 6.0$ (given the two free
parameters of the model). Moreover, we indicate the regions compatible
with the Higgs signal based on the total $\chi^2$ constructed from
Higgs signal rates, Higgs mass and the two exclusion likelihoods from
LEP light Higgs searches and LHC $H/A\to \tau^+\tau^-$
searches. The green area indicates where the $p$-value --- estimated from this total $\chi^2$ value under the assumption of independent and Gaussian observables --- is above $5\%$.

In most of the parameter region of the 
\lowMHuplow\  and \lowMHuphigh\ scenarios,
the heavy Higgs mass, $M_H$, ranges between
$120\gev$ and $130\gev$, whereas the light Higgs mass, $M_h$, varies
between $\sim0\gev$ (at the edge of the unphysical, gray
region) and $120\gev$.\footnote{In the evaluation of these benchmark scenarios, the two-loop corrections 
to the relation between $\MA$ and $\MHp$ has been omitted. Taking them into
account will lead to a slight shift in the $\Mh$ prediction.} 
In the \lowMHuplow\ (\lowMHuphigh) scenario the $\cp$-odd Higgs mass is $\MA \sim 137~(169)\gev$
(within a few $\mathrm{GeV}$).

In the \lowMHuplow\ scenario the branching fraction for the
top quark decay into a charged Higgs boson, $t\to H^+b$, ranges between
$\sim 0.1\%$ and $0.25\%$. The charged Higgs $H^+$ successively decays
either to $\tau^+\nu_{\tau}$ or $W^+ h$, where the branching ratios are
highly dependent on the kinematical phase space of the $H^+\to W^+h$
decay, and thus on the light Higgs mass, $\Mh$. Either decay can be
completely dominating while the other is suppressed, depending on the
parameter space. The charged Higgs boson decay $H^+\to c\bar{s}$ is
negligible.

In the \lowMHuphigh\ scenario, the charged Higgs boson is predominantly produced in
association with a top quark via
$gb \to t H^\pm$.
The charged Higgs boson branching fractions in the \lowMHuphigh\
scenario are very similar to the \lowMHuplow\ scenario, with the exception that the decay
$H^+\to t\bar{b}$ is present. However, its decay rate amounts to at most
$\sim 20\%$ (at the low end of the $\mu$ range). An interesting situation in this benchmark scenario
occurs in the region
around $\mu\sim (5.0 - 5.5)\tev$ and $\tb\sim 6-7$, where both the light
and heavy Higgs boson have masses between $120$ and $130\gev$. In our
analysis, part of this region is even compatible with the Higgs signal
and at the same time not directly excluded by limits from Higgs
searches, i.e.~this part of the parameter space gives rise to 
{\em two} $\cp$-even Higgs bosons close to $125\gev$. 
A dedicated experimental analysis of such a scenario, taking into account
also interference effects between the two nearly mass-degenerate Higgs
bosons, see \citere{Fuchs:2014ola}, would be
desirable.

\begin{figure}[t!]
\centering
\includegraphics[width=0.45\columnwidth]{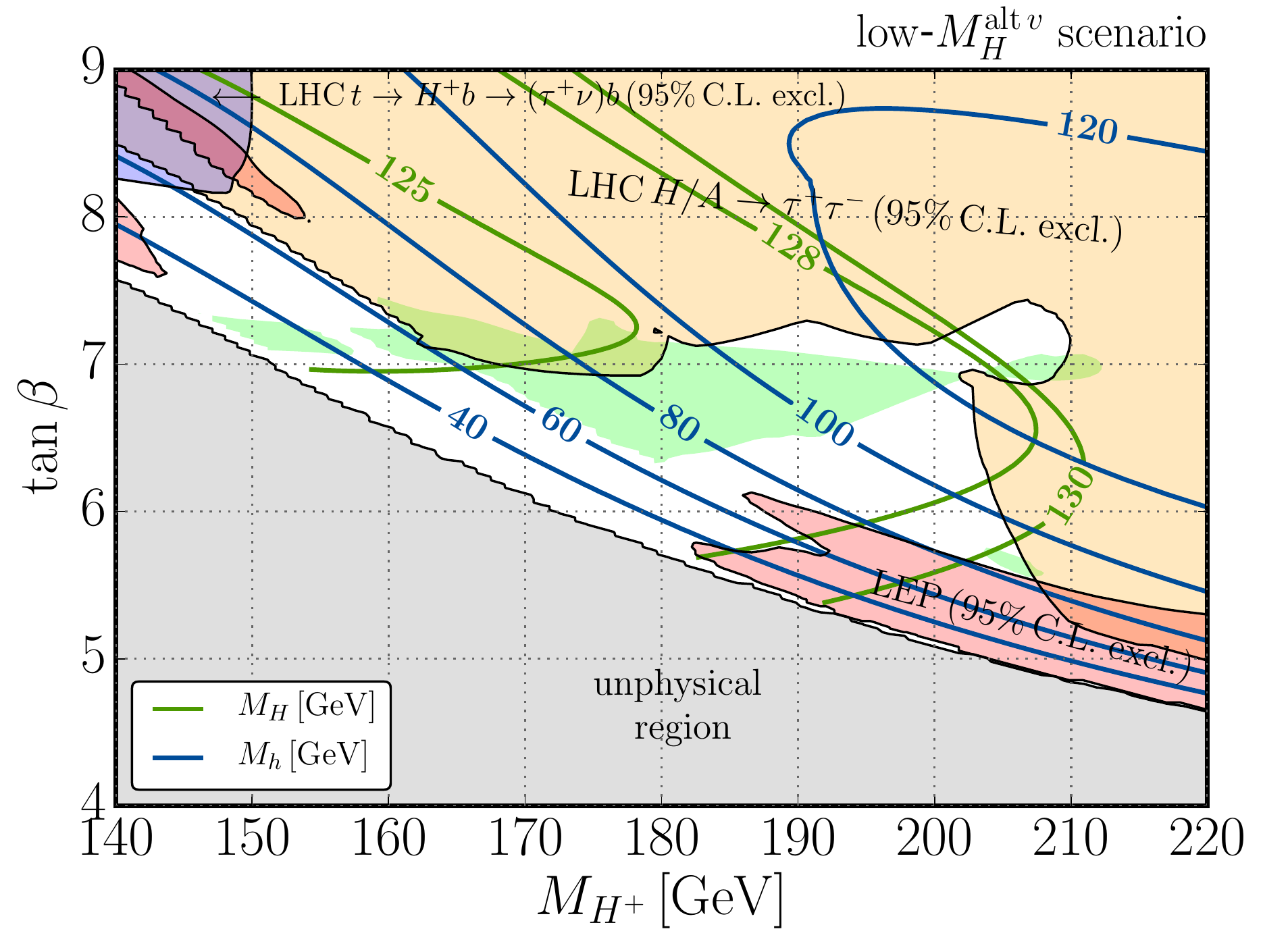}\hfill
\includegraphics[width=0.45\columnwidth]{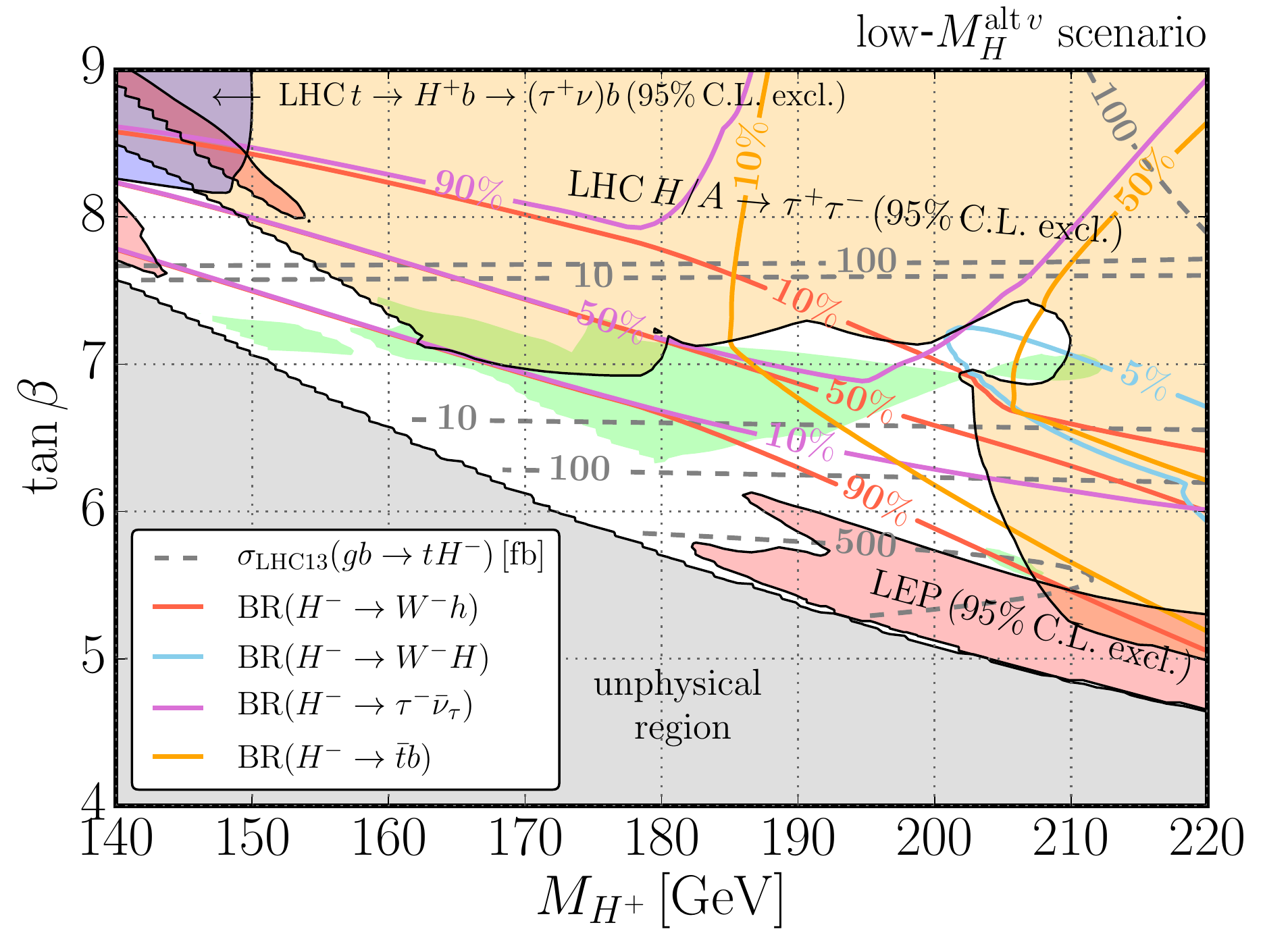}
\caption{The \lowMHupvar\  benchmark scenario in the ($\MHp$, $\tb$) plane
  (with $\mu = 6000 \gev$). 
The colored regions follow the definitions in Fig.~\ref{fig:lowMH-mutb}. 
In the left panel, contour lines indicate the Higgs masses $M_h$ and
$M_H$ (in GeV). In the right panel, contours lines indicate the charged 
Higgs branching ratios and the $13\tev$ LHC cross section for charged Higgs
production in association with a top quark, $\sigma_\text{LHC13}(gb\to
tH^+)$ (in fb).}   
\label{fig:lowMH-MHptb}
\end{figure} 

The \lowMHupvar\ benchmark scenario is illustrated in
\reffi{fig:lowMH-MHptb}. Here it can nicely be seen that the limit of
alignment without decoupling occurs roughly at $\tb\sim 7$,
as the green area is centered around this value. The charged Higgs
phenomenology is quite rich: At lower charged Higgs masses, $\MHp
\lesssim 180\gev$, the decay modes $H^+\to \tau^+\nu_\tau$ and $H^+\to
W^+h$ completely dominate. At larger $\MHp$, the decay mode $H^+\to
t\bar{b}$ and even $H^+\to W^+H$ become non-negligible, albeit they
remain small for most of the unexcluded parameter space. 

The best prospects for exploring these scenarios may still be via LHC
searches for $H/A\to \tau^+\tau^-$. However, these benchmark scenarios
will hopefully also provide useful guidance for upcoming charged Higgs
searches and in particular motivate searches for new charged Higgs
signatures such as $H^\pm \to W^\pm h$, with the light Higgs $h$
decaying into bottom quark or tau lepton pairs, $h\to b\bar{b},
\tau^+\tau^-$. 
Excluding these scenarios, which appear to be cornered from all
sides, with the upcoming searches at the LHC would strongly restrict the heavy Higgs
interpretation in the MSSM, approaching an exclusion of this interesting 
possibility, whose phenomenology drastically differs from the most commonly
considered light Higgs scenario.


\section{Conclusions and Outlook}\label{Sect:Summary}

We have analyzed the compatibility of the phenomenological Minimal
Supersymmetric Standard Model (pMSSM) with the SM-like Higgs boson with a mass $\sim
125 \gev$ as measured by ATLAS and CMS.
We performed a parameter scan of the pMSSM with the eight most relevant
parameters varied freely (\pMSSM): the $\cp$-odd Higgs boson mass, $\MA$, the
ratio of the two neutral Higgs vacuum expectation values, $\tb$, a common soft SUSY-breaking
parameter for the scalar top- and bottom quarks, $\msqd$, a soft SUSY-breaking
parameter for the scalar tau and neutrino sector, $\msld$, and similarly for
the first and second generation of sleptons, $\mslez$, a common trilinear
coupling for the third generation, $\Af$, the 
Higgsino mass parameter, $\mu$, as well as the SU(2) gaugino mass
parameter, $M_2$. The U(1) gaugino mass parameter $M_1$ was fixed from
the value of $M_2$ using the GUT relation.  
The other parameters have been set to fixed values that are 
generically in agreement with recent SUSY searches at the LHC. 

A random parameter scan with
\order{10^7} scan points has been performed. For each scan point
a $\chi^2$ function was evaluated, taking into account the combined Higgs boson mass measurement of the LHC experiments and the measured rates in $85$ individual Higgs search channels from ATLAS, CMS, and the
Tevatron experiments (via the code {\tt HiggsSignals}), the exclusion bounds from the search for
additional Higgs bosons (via the code {\tt HiggsBounds}, both relying on the
evaluations done by {\tt FeynHiggs}), exclusion bounds from the direct search for SUSY
particles (via the code {\tt CheckMate}), as well as the following low-energy observables \bsg, \bmm, \btn, \gmt\ (as obtained from {\tt SuperIso}) and $\MW$ (via the prediction of \citeres{Heinemeyer:2013dia,Stal:2015zca}).

Taking into account only the Higgs measurements and direct searches,
we find that the SM, the MSSM scenario where the light $\cp$-even Higgs boson
corresponds to the observed signal (``light Higgs case''), as 
as well as the MSSM scenario where the heavy $\cp$-even Higgs boson
corresponds to the observed signal (``heavy Higgs case'')
provide similarly good fits to the data
with a $\chi^2/$d.o.f.\ of $70.2/86$, $67.9/79$ and $70.0/80$, respectively. 
In a naive evaluation of $p$~values that neglects the correlations
between different Higgs observables, this translates into $p$~values of $89\%$, $81\%$ and $78\%$, respectively. 
Including also the low-energy observables we find (via the same
evaluation) $p$~values of $69\%$, $89\%$ and $80\%$, where the SM suffers in particular from the inclusion of
\gmt. Thus, the ``light Higgs case'' of the MSSM and even the rather
exotic ``heavy Higgs case'' of the MSSM provide a slightly better
descriptions of the data in a global fit than the SM.

Within the MSSM, a SM-like Higgs boson at $\sim 125 \gev$ can be realized in
three ways and with similarly good fit qualities. For $\MA \gg \MZ$ the
light $\cp$-even Higgs boson is SM-like in the \emph{decoupling limit}. For
$\MA \lsim 350 \gev$ (the ``low-$\MA$ case'') the light $\cp$-even Higgs boson
can be SM-like in the limit of 
\emph{alignment without decoupling}. This limit also offers additionally the unique
possibility that the heavy $\cp$-even Higgs boson can have a mass of $\sim 125
\gev$ with SM-like couplings. We have analyzed analytically in which parts of
the MSSM parameter space the limit of alignment without decoupling can be
realized.   In this latter scenario all MSSM Higgs bosons are relatively light, 
offering good prospects for the searches for additional Higgs bosons at
the LHC and future colliders.  Our analytic expressions contain the leading two-loop contributions of
$\mathcal{O}(\als h_t^2)$~\cite{preparation}, which somewhat modify the leading one-loop contributions
of $\mathcal{O}(h_t^2)$ that had been considered previously~\cite{Carena:2014nza}.

For the light Higgs case in the decoupling limit and in the limit of alignment without
decoupling, as well as in the heavy Higgs case we have analyzed the
predictions for the various Higgs boson production and decay rates. In the
light Higgs case the various rates are predicted to be close to the SM Higgs
boson rates, where the largest allowed deviation to smaller values is found in
the $h \to \tau^+\tau^-$ channel. The light scalar top is found to have masses
down to $\sim 300 \gev$, while the preferred region in the fit provides
no upper limit on the scalar top quarks. For the low-$\MA$ case
we find lighter stop masses down to $\sim 500 \gev$ and 
$\Xt/\MS \sim +2$. While
in the decoupling limit the parameters $\mu$ and $\At$ can vary from very
small to very large values, the low-$\MA$ case (and thus the limit of alignment without
decoupling in the light Higgs interpretation) can be realized only for 
$\mu/\MS \sim 1.4$ to $3$ and $\At/\MS \sim 2.4$ to $3$, where the upper values correspond to the upper scan ranges imposed in our study.
For larger $\MA$ values also relatively large values of $\tb$ are still
allowed, leading to a sizable contribution to \gmt\ even for
relatively large chargino/neutralino and slepton masses. 
Concerning the impact of the low-energy observable, it is interesting
to note that a clear preference for positive $\mu$ (and also positive $\At$)
already emerges when combining the Higgs- and $B$-physics observables in the
fit, i.e.\ already \emph{before} including \gmt.
We have furthermore found that the preferred region in the fit \emph{without}
taking into account the low-energy observables includes predictions for 
${\rm BR}(B_s \to \mu^+\mu^-)$ that are close to the experimental central
value (i.e., below the SM prediction) both for the decoupling and the
alignment without decoupling region of the light Higgs case.

We included the limits from direct LHC searches for SUSY particles from the $8\tev$ run via \texttt{CheckMATE}. These searches constrain the parameter space of the \pMSSM\ in an orthogonal way to the Higgs mass and signal rate constraints and therefore do not directly alter the Higgs phenomenology, neither in the light nor the heavy Higgs case. Furthermore, due to this orthogonality we also expect that future stronger SUSY limits from the $13\tev$ run would not substantially alter the conclusions found in this paper.

In the heavy Higgs case the preferred rates are also SM-like, however with a
possibly larger suppression of $H \to \ga\ga$ and/or $H \to \tau^+\tau^-$. We
find that in the heavy Higgs case the $\cp$-odd Higgs boson $A$ is restricted to have a mass of 
$140 \gev \lsim \MA \lsim 185 \gev$, with the charged Higgs boson being the
heaviest Higgs boson with $\MHp \lsim 210 \gev$, and $6 \lsim \tb \lsim 11$. We
furthermore find 
$\mu/\MS \sim 6$ to $9$ and $-1 \lsim \At/\MS \lsim 0$.
The light scalar top
is predicted to have a mass of $350 \gev \lsim \mste \lsim 650 \gev$. Due to
the relatively small $\tb$ values the scalar leptons have to be relatively
light with masses below $\sim 450 \gev$ to bring the prediction into
agreement with the observed discrepancy of the experimental measurement of \gmt\ with the SM prediction.
In particular, we have checked that the charged Higgs corrections to the
$B$-physics observables, $B\to X_s\gamma$, $B_u\to\tau\nu_\tau$ and $B_s\to\mu^+\mu^-$, are consistent
with the experimental results. The preferred region in the
fit has mostly light Higgs boson masses above $\MH/2$, in which case $H \to hh$ decays are kinematically closed. However, also smaller $\Mh$ values are possible, where $\br(H \to hh)$ does not exceed 20\%
(for the most favored parameter region we find $\br(H \to hh) < 2\%$.) The
coupling of the light Higgs boson to $W^\pm$ and $Z$~bosons is strongly suppressed, 
much below the existing bounds from LEP Higgs searches. Note that such a light Higgs boson re-opens the possibility of light neutralino dark matter in the sub-GeV to $65\gev$ range by acting as $s$-channel (near-)resonance in dark matter pair-annihilation~\cite{Profumo:2016zxo}.

As a guidance for the Higgs boson searches in the heavy Higgs interpretation we provide
a new set of benchmark scenarios that can be employed to maximize the
sensitivity of the experimental analysis to this interpretation. In the 
($\mu$, $\tb$) plane we define the \lowMHuplow\ and the \lowMHuphigh\ scenario
with $\MHp = 155 \gev < \mt$ and $\MHp = 185 \gev > \mt$,
respectively. In the ($\MHp$, $\tb$) plane we define the \lowMHupvar\ scenario
with $\mu = 6 \tev$. We have shown that in all three scenarios a parameter
regime exists, where the mass and rates of the heavy $\cp$-even Higgs boson 
are in agreement with all available measurements, and which is also not
excluded by searches for additional Higgs bosons. In the \lowMHuphigh\ scenario we find a
very restricted part of the parameter space in which 
$\Mh \sim \MH \sim 125 \gev$, with both $\cp$-even Higgs bosons contributing
to the Higgs boson rates. The proposed benchmark scenarios
will be of interest for upcoming charged Higgs boson
searches and will provide motivation for searches for new charged Higgs
signatures such as $H^\pm \to W^\pm h$, with the light Higgs $h$
decaying into bottom quark or tau lepton pairs, $h\to b\bar{b},
\tau^+\tau^-$.

New data from the ATLAS and CMS Higgs measurements and the search for
new Higgs boson states are now rapidly emerging in the current run of the LHC. 
It is critical to improve the precision of the measurements of the properties of the SM-like Higgs boson,
while improving the sensitivity of the searches for new Higgs bosons
with masses either above or below the observed Higgs boson mass.  In particular, these searches
will yield new constraints on the parameter space of the MSSM.   The
observation of the SM-like Higgs boson already implies that the MSSM Higgs sector lies
close to the alignment limit.  Indeed, 
the regions of the MSSM parameter space in which the approximate
alignment limit is realized provide a description of the data that is as good (or in some cases
slightly better) than the SM.  In addition to the possibility that the observed
Higgs boson corresponds to the lighter $\cp$-even Higgs boson of the MSSM, the more exotic
possibility in which the heavier $\cp$-even Higgs boson is identified as the observed Higgs
boson cannot yet be ruled out.   Higgs studies at Run 2 of the LHC may prove decisive in determining
whether the cracks in the Standard Model facade finally shatter, and whether a supersymmetric
interpretation of Higgs phenomena is ultimately viable.





\section*{Acknowledgments}
We thank
Thomas Hahn, Ben O'Leary, Stefan Liebler, Sebastian Passehr and Florian Staub
for helpful discussions and Daniel Schmeier for help with \CM. We are grateful to Oliver Ricken and Peter Wienemann for their technical support and invaluable help with the computer cluster in Bonn.
H.E.H.\ and T.S.\ are supported in part by the U.S. Department of Energy grant number DE-SC0010107. T.S.~greatly acknowledges additional support from the Alexander von Humboldt foundation through a Feodor-Lynen research fellowship.
The work of S.H.\ is supported in part by CICYT (Grant FPA 2013-40715-P) 
and by the Spanish MICINN's Consolider-Ingenio 2010 Program under Grant 
MultiDark CSD2009-00064.
L.Z.\ is supported by the Netherlands Organization for
Scientific Research (NWO) through a VENI grant.
G.W.\ acknowledges
support by  the DFG through  the SFB~676 ``Particles, Strings  and the
Early Universe''.  
This research was supported in part by the European
Commission  through   the  ``HiggsTools''  Initial   Training  Network
PITN-GA-2012-316704.


\appendix


\section*{Appendix A: How tuned is approximate Higgs alignment without decoupling in the MSSM?}
\label{App:tuning}

The precision Higgs data implies that the properties of one of the Higgs bosons in the MSSM Higgs sector (which is to be identified with the observed Higgs boson of mass 125 GeV) approximate the predicted properties of the SM Higgs boson.  
This corresponds to the approximate alignment limit described in \refse{Sec:alignment}.   Approximate alignment can be easily achieved in the parameter regime in which all other (non-SM-like) Higgs boson states are significantly heavier than 125 GeV.   However, it is also possible that a parameter region of the MSSM exists in which approximate alignment without decoupling  
is satisfied. In this parameter regime, the other non-SM-like Higgs states are not significantly separated in mass from the observed Higgs boson, which provides additional opportunities for the discovery of new scalar states at the LHC.  As shown in \refse{Sec:alignment}, alignment without decoupling is achieved when the Higgs basis parameter $|Z_6|\ll1$ and $\MA\lsim 350$~GeV.   
However, a skeptical reader might wonder how difficult it is to achieve regions of the MSSM parameter space with very small values of $Z_6$.  That is, how tuned is approximate Higgs alignment without decoupling in the MSSM?

Consider the case of exact alignment in the two Higgs doublet model (2HDM), which corresponds to $Z_6=0$.  If exact alignment is a consequence of a global (discrete or continuous) symmetry, then it is natural to consider 2HDM Higgs sectors that exhibit approximate alignment without decoupling.   The possibility of achieving exact alignment by a symmetry of the scalar potential was considered in Refs.~\cite{Dev:2014yca, Pilaftsis:2016erj}.  However, it is not clear whether these symmetries survive once a realistic Higgs-fermion Yukawa sector is considered~\cite{Ferreira:2010bm}.  The one known example of a realistic 2HDM with exact alignment is the inert doublet model~\cite{Barbieri:2006dq}, where $Z_6=0$ is a consequence of an exact $\mathbb{Z}_2$ discrete symmetry, under which the inert doublet field is odd and all other fields 
are even.

In the MSSM, there is no symmetry associated with the parameter regime corresponding to exact alignment.  Indeed, as explained in  
\refse{Sec:alignment}, exact alignment in the MSSM is a result of an accidental cancellation between tree-level and radiatively corrected loop effects that contribute to the effective $Z_6$ parameter.    In an approximate one-loop expression, this cancellation can be explicitly seen in \eq{z6zero}, where the expression inside the braces vanishes for a particular choice of $\tan\beta$.
Of course, given the limited statistics of the present day Higgs data, a region of the MSSM space that exhibits approximate alignment without decoupling can still be consistent with all known experimental constraints.  Thus, in any comprehensive scan of the MSSM parameter space, one must necessarily encounter regions of approximate alignment without decoupling.  In the absence of a fundamental underlying theory of supersymmetry breaking, the correct measure that governs the MSSM parameter space is unknown.  Thus, in order to get a sense of the extent of the tuning associated with the parameter regime of alignment without decoupling, the best one can do is to examine the frequency that such parameter points occur in a comprehensive parameter scan (with uniform priors).

\begin{figure}[t!]
\centering
\includegraphics[width=0.45\columnwidth,height=6cm]{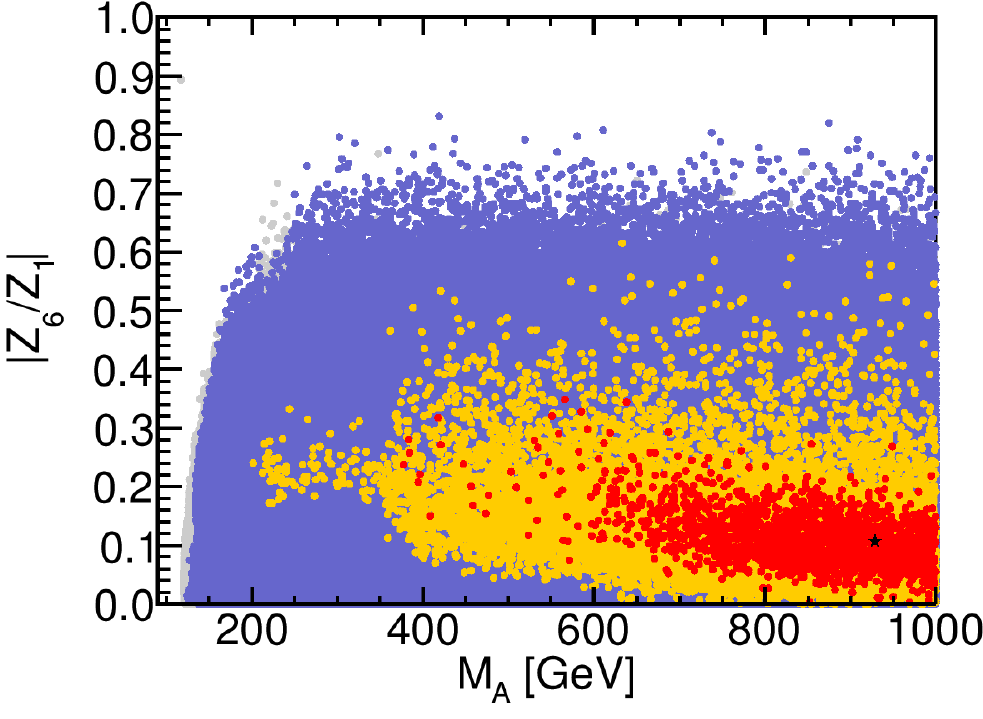}\hfill
\includegraphics[width=0.45\columnwidth,height=6cm]{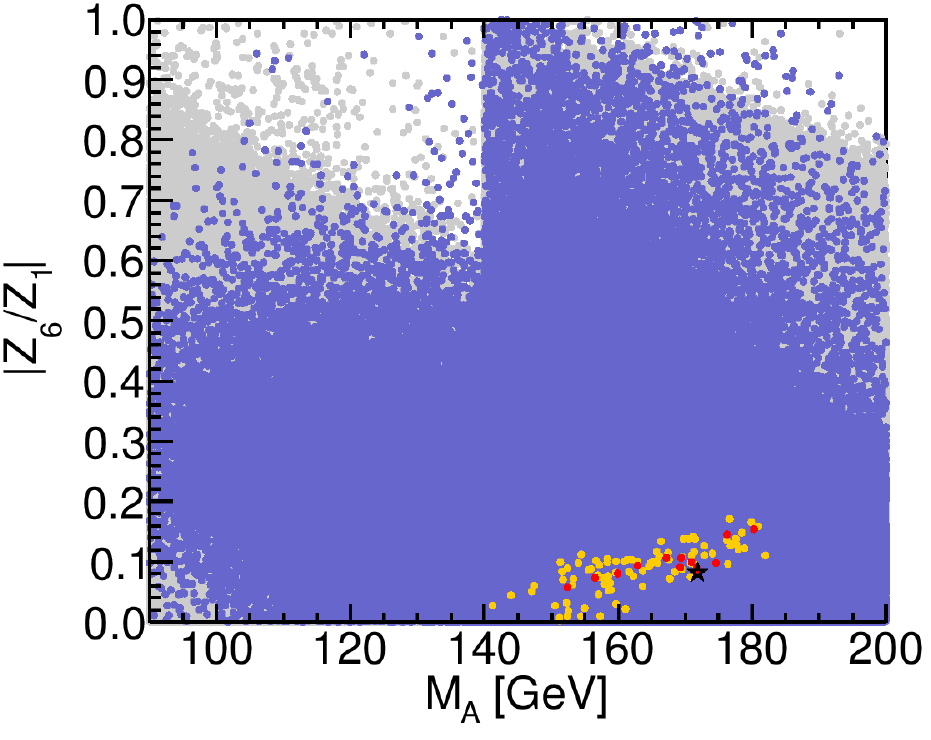}
\caption{$|Z_6/Z_1|$ in dependence of the pseudoscalar Higgs mass,
  $M_A$, for the light Higgs (\emph{left}) and heavy Higgs
  (\emph{right}) interpretation. The parameters $Z_1$ and $Z_6$ are
  calculated using the approximate two-loop formulas as described in
  Section~\ref{Sec:alignment}.  
The value of $Z_1$ is fixed by $M_{h/H} \sim 125 \gev$ (see text).  The color coding is the same as in Fig.~\ref{fig:hrates_corr}.}
\label{fig:tuning_MA_Z6Z1}
\end{figure}

In this Appendix, we address this last point from a practical point of view.  We examine the values of $Z_6$ as a function of $\MA$ that arise in our \pMSSM\ parameter scans.
Note that in the approximate alignment limit, 
$m_h^2\simeq Z_1 v^2=(125 \gev)^2$, i.e., $Z_1 \simeq 0.26$, which
sets the ``natural'' size for the other Higgs basis quartic self-couplings (including $Z_6$). 
In \reffi{fig:tuning_MA_Z6Z1} we show $|Z_6/Z_1|$ as a function of $\MA$
for the light Higgs (left) and heavy Higgs (right)
interpretation. 
The parameters $Z_1$ and $Z_6$ are calculated using the
approximate two-loop formulas as described in \refse{Sec:alignment}.
In the case where $h$ is SM-like,  one can see that $|Z_6/Z_1| \sim 0.2$
in the region of approximate alignment without decoupling.
That is, $Z_6$ is suppressed by less than one order of magnitude
relative to $Z_1$.\footnote{Indeed, in the case where $h$ is SM-like, regions of exact alignment 
are ruled out, as these regions correspond to values of $\tan\beta$ that are excluded by the 
LHC searches for $H, A\to\tau^+\tau^-$~\cite{Khachatryan:2014wca,Carena:2014nza,Aaboud:2016cre}.}
In the case where $H$ is SM-like,
values between $0$ and $0.2$ are found for $|Z_6/Z_1|$, with the
best-fit value of $|Z_6/Z_1| \simeq 0.1$. 
In both the light and heavy Higgs case, the values of $|Z_6|$ in the regions of approximate alignment without decoupling
are not unnaturally small.  Indeed, even in the decoupling regime of large $\MA$, the typical
values of $|Z_6|$ in the preferred MSSM parameter regime are not significantly different in magnitude.
We conclude that given the present precision of the Higgs data, approximate alignment without
decoupling can be achieved without resorting to an excessive fine tuning of the MSSM parameters. 



\section*{Appendix B: Higgs measurements from Tevatron and LHC}
\label{App:Rates}
Tables~\ref{Tab:HSobs1} and~\ref{Tab:HSobs2} list the 85 signal strength
measurements from ATLAS, CMS and the Tevatron (D\O\ and CDF), which we
include in our analysis via \HSv{1.4.0}. For each analysis, we give the
measured signal strength value, $\hat{\mu}$, its $1\sigma$ uncertainty,
$\Delta\hat{\mu}$, as well as the signal composition for the production
of a SM Higgs with mass $\sim 125\gev$. 

\begin{table}[h!]
\scalebox{0.86}{
\renewcommand{\arraystretch}{1.0}
\begin{threeparttable}[b]
\footnotesize
 \begin{tabular}{lccrrrrr}
\midrule
 Analysis & \qquad energy $\sqrt{s}$\qquad & \qquad$\hat{\mu} \pm \Delta \hat{\mu}$\qquad & \multicolumn{5}{c}{\quad SM signal contamination [in \%]\quad} \\
 & & & \quad ggH \quad &  \quad VBF \quad & \quad WH \quad &   \quad ZH \quad & \quad $t\bar{t}H$\quad \\
\midrule
ATLAS $h\to WW\to \ell\nu\ell\nu~\mathrm{(VBF)}$~\cite{ATLAS:2014aga} & $7/8\tev$ & $  1.27\substack{+  0.53\\ -  0.45}$ & $  24.1$ & $  75.9$ & $   0.0$ & $   0.0$ & $   0.0$\\ 
ATLAS $h\to WW\to \ell\nu\ell\nu~\mathrm{(ggH)}$~\cite{ATLAS:2014aga} & $7/8\tev$ & $  1.01\substack{+  0.27\\ -  0.25}$ & $  97.8$ & $   1.2$ & $   0.6$ & $   0.3$ & $   0.1$\\ 
ATLAS $h\to ZZ\to 4\ell~\mathrm{(VBF/VH)}$~\cite{Aad:2014eva} & $7/8\tev$ & $  0.26\substack{+  1.64\\ -  0.94}$ & $  37.8$ & $  35.7$ & $  16.8$ & $   9.7$ & $   0.0$\\ 
ATLAS $h\to ZZ\to 4\ell~\mathrm{(ggH)}$~\cite{Aad:2014eva} & $7/8\tev$ & $  1.66\substack{+  0.51\\ -  0.44}$ & $  91.6$ & $   4.6$ & $   2.2$ & $   1.3$ & $   0.4$\\ 
ATLAS $h\to \gamma\gamma~\mathrm{(VBF, loose)}$~\cite{Aad:2014eha} & $7/8\tev$ & $  1.33\substack{+  0.92\\ -  0.77}$ & $  39.0$ & $  60.0$ & $   0.6$ & $   0.3$ & $   0.1$\\ 
ATLAS $h\to \gamma\gamma~\mathrm{(VBF, tight)}$~\cite{Aad:2014eha} & $7/8\tev$ & $  0.68\substack{+  0.67\\ -  0.51}$ & $  18.2$ & $  81.5$ & $   0.1$ & $   0.1$ & $   0.1$\\ 
ATLAS $h\to \gamma\gamma~(Vh,E_T^\text{miss})$~\cite{Aad:2014eha} & $7/8\tev$ & $  3.51\substack{+  3.30\\ -  2.42}$ & $   8.7$ & $   3.7$ & $  35.8$ & $  44.8$ & $   7.1$\\ 
ATLAS $h\to \gamma\gamma~(Vh,2j)$~\cite{Aad:2014eha} & $7/8\tev$ & $  0.23\substack{+  1.67\\ -  1.39}$ & $  45.0$ & $   3.3$ & $  31.9$ & $  19.8$ & $   0.1$\\ 
ATLAS $h\to \gamma\gamma~(Vh,1\ell)$~\cite{Aad:2014eha} & $7/8\tev$ & $  0.41\substack{+  1.43\\ -  1.06}$ & $   0.7$ & $   0.2$ & $  91.4$ & $   5.9$ & $   1.8$\\ 
ATLAS  $h\to \gamma\gamma~\mathrm{(central,high}~p_{Tt})$~\cite{Aad:2014eha} & $7/8\tev$ & $  1.62\substack{+  1.00\\ -  0.83}$ & $  72.6$ & $  16.4$ & $   6.1$ & $   3.7$ & $   1.2$\\ 
ATLAS $h\to \gamma\gamma~\mathrm{(central,low}~p_{Tt})$~\cite{Aad:2014eha} & $7/8\tev$ & $  0.62\substack{+  0.42\\ -  0.40}$ & $  93.2$ & $   4.1$ & $   1.6$ & $   1.0$ & $   0.1$\\ 
ATLAS $h\to \gamma\gamma~\mathrm{(forward,high}~p_{Tt})$~\cite{Aad:2014eha} & $7/8\tev$ & $  1.73\substack{+  1.34\\ -  1.18}$ & $  71.4$ & $  16.7$ & $   6.9$ & $   4.1$ & $   0.9$\\ 
ATLAS $h\to \gamma\gamma~\mathrm{(forward,low}~p_{Tt})$~\cite{Aad:2014eha} & $7/8\tev$ & $  2.03\substack{+  0.57\\ -  0.53}$ & $  92.5$ & $   4.2$ & $   2.0$ & $   1.2$ & $   0.1$\\ 
ATLAS $h\to \gamma\gamma~(tth,\mathrm{hadr.)}$~\cite{Aad:2014eha} & $7/8\tev$ & $ -0.84\substack{+  3.23\\ -  1.25}$ & $  15.0$ & $   1.3$ & $   1.3$ & $   1.4$ & $  81.0$\\ 
ATLAS $h\to \gamma\gamma~(tth,\mathrm{lep.)}$~\cite{Aad:2014eha} & $7/8\tev$ & $  2.42\substack{+  3.21\\ -  2.07}$ & $   8.4$ & $   0.1$ & $  14.9$ & $   4.0$ & $  72.6$\\ 
ATLAS $h\to \tau\tau~\mathrm{(VBF,hadr.hadr.)}$~\cite{Aad:2015vsa} & $7/8\tev$ & $  1.40\substack{+  0.90\\ -  0.70}$ & $  30.1$ & $  69.9$ & $   0.0$ & $   0.0$ & $   0.0$\\ 
ATLAS $h\to \tau\tau~\mathrm{(boosted,hadr.hadr.)}$~\cite{Aad:2015vsa} & $7/8\tev$ & $  3.60\substack{+  2.00\\ -  1.60}$ & $  69.5$ & $  13.3$ & $  11.3$ & $   5.8$ & $   0.0$\\ 
ATLAS $h\to \tau\tau~\mathrm{(VBF,lep.hadr.)}$~\cite{Aad:2015vsa} & $7/8\tev$ & $  1.00\substack{+  0.60\\ -  0.50}$ & $  17.2$ & $  82.8$ & $   0.0$ & $   0.0$ & $   0.0$\\ 
ATLAS $h\to \tau\tau~\mathrm{(boosted,lep.hadr.)}$~\cite{Aad:2015vsa} & $7/8\tev$ & $  0.90\substack{+  1.00\\ -  0.90}$ & $  73.0$ & $  13.3$ & $   9.1$ & $   4.6$ & $   0.0$\\ 
ATLAS $h\to \tau\tau~\mathrm{(VBF,lep.lep.)}$~\cite{Aad:2015vsa} & $7/8\tev$ & $  1.80\substack{+  1.10\\ -  0.90}$ & $  15.4$ & $  84.6$ & $   0.0$ & $   0.0$ & $   0.0$\\ 
ATLAS $h\to \tau\tau~\mathrm{(boosted,lep.lep.)}$~\cite{Aad:2015vsa} & $7/8\tev$ & $  3.00\substack{+  1.90\\ -  1.70}$ & $  70.9$ & $  21.4$ & $   5.7$ & $   2.1$ & $   0.0$\\ 
ATLAS $Vh\to V(bb)~(0\ell)$~\cite{Aad:2014xzb} &  $7/8\tev$ &$ -0.35\substack{+  0.55\\ -  0.52}$ & $   0.0$ & $   0.0$ & $  20.8$ & $  79.2$ & $   0.0$\\ 
ATLAS $Vh\to V(bb)~(1\ell)$~\cite{Aad:2014xzb} & $7/8\tev$ & $  1.17\substack{+  0.66\\ -  0.60}$ & $   0.0$ & $   0.0$ & $  96.7$ & $   3.3$ & $   0.0$\\ 
ATLAS $Vh\to V(bb)~(2\ell)$~\cite{Aad:2014xzb} & $7/8\tev$ & $  0.94\substack{+  0.88\\ -  0.79}$ & $   0.0$ & $   0.0$ & $   0.0$ & $ 100.0$ & $   0.0$\\ 
ATLAS $Vh\to V(WW)~(2\ell)$~\cite{Aad:2015ona} & $7/8\tev$ & $  3.70\substack{+  1.90\\ -  1.80}$ & $   0.0$ & $   0.0$ & $  83.3$ & $  16.7$ & $   0.0$\\ 
ATLAS $Vh\to V(WW)~(3\ell)$~\cite{Aad:2015ona} & $7/8\tev$ & $  0.72\substack{+  1.30\\ -  1.10}$ & $   0.0$ & $   0.0$ & $  86.5$ & $  13.5$ & $   0.0$\\ 
ATLAS $Vh\to V(WW)~(4\ell)$~\cite{Aad:2015ona} & $7/8\tev$ & $  4.90\substack{+  4.60\\ -  3.10}$ & $   0.0$ & $   0.0$ & $   0.0$ & $ 100.0$ & $   0.0$\\ 
ATLAS $tth\to \mathrm{multilepton}~(1\ell, 2\tau_h)$~\cite{Aad:2015iha} & $7/8\tev$ & $ -9.60\substack{+  9.60\\ -  9.70}$ & $   0.0$ & $   0.0$ & $   0.0$ & $   0.0$ & $   100.0$\tnote{1}\\ 
ATLAS $tth\to \mathrm{multilepton}~(2\ell, 0\tau_h)$~\cite{Aad:2015iha} & $7/8\tev$ & $  2.80\substack{+  2.10\\ -  1.90}$ & $   0.0$ & $   0.0$ & $   0.0$ & $   0.0$ & $   100.0$\tnote{2}\\ 
ATLAS $tth\to \mathrm{multilepton}~(2\ell, 1\tau_h)$~\cite{Aad:2015iha} & $7/8\tev$ & $ -0.90\substack{+  3.10\\ -  2.00}$ & $   0.0$ & $   0.0$ & $   0.0$ & $   0.0$ & $   100.0$\tnote{3}\\ 
ATLAS $tth\to \mathrm{multilepton}~(3\ell)$~\cite{Aad:2015iha} &  $7/8\tev$ &$  2.80\substack{+  2.20\\ -  1.80}$ & $   0.0$ & $   0.0$ & $   0.0$ & $   0.0$ & $   100.0$\tnote{4}\\ 
ATLAS $tth\to \mathrm{multilepton}~(4\ell)$~\cite{Aad:2015iha} & $7/8\tev$ & $  1.80\substack{+  6.90\\ -  6.90}$ & $   0.0$ & $   0.0$ & $   0.0$ & $   0.0$ & $   100.0$\tnote{5}\\ 
ATLAS $tth\to tt(bb)$~\cite{Aad:2015gra} & $7/8\tev$ & $  1.50\substack{+  1.10\\ -  1.10}$ & $   0.0$ & $   0.0$ & $   0.0$ & $   0.0$ & $ 100.0$\\ 
\midrule
CDF $h\to WW$~\cite{Aaltonen:2013ipa} & $1.96\tev$& $  0.00\substack{+  1.78\\ -  1.78}$ & $  77.5$ & $   5.4$ & $  10.6$ & $   6.5$ & $   0.0$\\ 
CDF $h\to \gamma\gamma$~\cite{Aaltonen:2013ipa} & $1.96\tev$& $  7.81\substack{+  4.61\\ -  4.42}$ & $  77.5$ & $   5.4$ & $  10.6$ & $   6.5$ & $   0.0$\\ 
CDF $h\to \tau\tau$~\cite{Aaltonen:2013ipa} & $1.96\tev$& $  0.00\substack{+  8.44\\ -  8.44}$ & $  77.5$ & $   5.4$ & $  10.6$ & $   6.5$ & $   0.0$\\ 
CDF $Vh\to V(bb)$~\cite{Aaltonen:2013ipa} & $1.96\tev$& $  1.72\substack{+  0.92\\ -  0.87}$ & $   0.0$ & $   0.0$ & $  62.0$ & $  38.0$ & $   0.0$\\ 
CDF $tth\to tt(bb)$~\cite{Aaltonen:2013ipa} & $1.96\tev$& $  9.49\substack{+  6.60\\ -  6.28}$ & $   0.0$ & $   0.0$ & $   0.0$ & $   0.0$ & $ 100.0$\\ 
\bottomrule
 \end{tabular}
   \begin{tablenotes}
 \footnotesize
 \item[1] The SM Higgs signal composition is $h\to \tau\tau$ ($93.0\%$), $h\to WW$ ($4.0\%$), $h\to bb$ ($3.0\%$).
 \item[2] The SM Higgs signal composition is $h\to WW$ ($80.1\%$), $h\to \tau\tau$ ($14.9\%$), $h\to ZZ$ ($3.0\%$), $h\to bb$ ($2.0\%$).
 \item[3] The SM Higgs signal composition is $h\to \tau\tau$ ($61.8\%$), $h\to WW$ ($35.2\%$), $h\to ZZ$ ($2.0\%$), $h\to bb$ ($1.0\%$).
 \item[4] The SM Higgs signal composition is $h\to WW$ ($74.1\%$), $h\to \tau\tau$ ($14.9\%$), $h\to ZZ$ ($7.0\%$), $h\to bb$ ($3.9\%$).
 \item[5] The SM Higgs signal composition is $h\to WW$ ($68.1\%$), $h\to \tau\tau$ ($13.9\%$), $h\to ZZ$ ($14.0\%$), $h\to bb$ ($4.0\%$).
 \end{tablenotes}
 \end{threeparttable}
}
 \caption{Higgs signal strengths measurements from the LHC ATLAS and Tevatron CDF collaboration.}
 \label{Tab:HSobs1}
\end{table}

\begin{table}
\scalebox{0.86}{
\renewcommand{\arraystretch}{1.0}
\begin{threeparttable}[b]
\footnotesize
 \begin{tabular}{lccrrrrr}
\midrule
 Analysis & \qquad energy $\sqrt{s}$\qquad & \qquad$\hat{\mu} \pm \Delta \hat{\mu}$\qquad & \multicolumn{5}{c}{\quad SM signal contamination [in \%]\quad} \\
 & & & \quad ggH \quad &  \quad VBF \quad & \quad WH \quad &   \quad ZH \quad & \quad $t\bar{t}H$\quad \\
\midrule
CMS $h\to WW\to 2\ell2\nu~(0/1j)$~\cite{Chatrchyan:2013iaa} &  $7/8\tev$ & $  0.74\substack{+  0.22\\ -  0.20}$ & $  85.8$ & $   8.9$ & $   3.3$ & $   1.9$ & $   0.0$\\ 
CMS $h\to WW\to 2\ell2\nu~\mathrm{(VBF)}$~\cite{Chatrchyan:2013iaa} &  $7/8\tev$ & $  0.60\substack{+  0.57\\ -  0.46}$ & $  24.1$ & $  75.9$ & $   0.0$ & $   0.0$ & $   0.0$\\ 
CMS $h\to ZZ\to 4\ell~(0/1j)$~\cite{Chatrchyan:2013mxa}  &  $7/8\tev$& $  0.88\substack{+  0.34\\ -  0.27}$ & $  91.9$ & $   8.1$ & $   0.0$ & $   0.0$ & $   0.0$\\ 
CMS $h\to ZZ\to 4\ell~(2j)$~\cite{Chatrchyan:2013mxa} &  $7/8\tev$ & $  1.55\substack{+  0.95\\ -  0.66}$ & $  76.1$ & $  23.9$ & $   0.0$ & $   0.0$ & $   0.0$\\ 
CMS $h\to \gamma\gamma~(\text{untagged}~0)$~\cite{Khachatryan:2014ira}&  $7\tev$  & $  1.97\substack{+  1.51\\ -  1.25}$ & $  80.8$ & $   9.7$ & $   5.8$ & $   3.2$ & $   0.6$\\ 
CMS $h\to \gamma\gamma~(\text{untagged}~1)$~\cite{Khachatryan:2014ira} &  $7\tev$ & $  1.23\substack{+  0.98\\ -  0.88}$ & $  92.3$ & $   4.1$ & $   2.3$ & $   1.2$ & $   0.1$\\ 
CMS $h\to \gamma\gamma~(\text{untagged}~2)$~\cite{Khachatryan:2014ira} &  $7\tev$ & $  1.60\substack{+  1.25\\ -  1.17}$ & $  92.3$ & $   4.0$ & $   2.3$ & $   1.3$ & $   0.1$\\ 
CMS $h\to \gamma\gamma~(\text{untagged}~3)$~\cite{Khachatryan:2014ira} &  $7\tev$ & $  2.61\substack{+  1.74\\ -  1.65}$ & $  92.5$ & $   3.9$ & $   2.3$ & $   1.2$ & $   0.1$\\ 
CMS $h\to \gamma\gamma~(\mathrm{VBF,dijet}~0)$~\cite{Khachatryan:2014ira} &  $7\tev$ & $  4.85\substack{+  2.17\\ -  1.76}$ & $  19.9$ & $  79.6$ & $   0.3$ & $   0.2$ & $   0.1$\\ 
CMS $h\to \gamma\gamma~(\mathrm{VBF,dijet}~1)$~\cite{Khachatryan:2014ira} &  $7\tev$ & $  2.60\substack{+  2.16\\ -  1.76}$ & $  39.0$ & $  58.9$ & $   1.2$ & $   0.7$ & $   0.3$\\ 
CMS $h\to \gamma\gamma~(Vh,E_T^\text{miss})$~\cite{Khachatryan:2014ira} &  $7\tev$ & $  4.32\substack{+  6.72\\ -  4.15}$ & $   4.9$ & $   1.2$ & $  43.2$ & $  44.4$ & $   6.3$\\ 
CMS $h\to \gamma\gamma~(Vh,\text{dijet})$~\cite{Khachatryan:2014ira} &  $7\tev$ & $  7.86\substack{+  8.86\\ -  6.40}$ & $  28.6$ & $   2.9$ & $  43.8$ & $  23.3$ & $   1.5$\\ 
CMS $h\to \gamma\gamma~(Vh,\text{loose})$~\cite{Khachatryan:2014ira} &  $7\tev$ & $  3.10\substack{+  8.29\\ -  5.34}$ & $   3.8$ & $   1.1$ & $  79.7$ & $  14.6$ & $   0.7$\\ 
CMS $h\to \gamma\gamma~(tth, \text{tags})$~\cite{Khachatryan:2014ira} &  $7\tev$ & $  0.71\substack{+  6.20\\ -  3.56}$ & $   4.3$ & $   1.5$ & $   2.9$ & $   1.6$ & $  89.7$\\ 
CMS $h\to \gamma\gamma~(\text{untagged}~0)$~\cite{Khachatryan:2014ira} &  $8\tev$& $  0.13\substack{+  1.09\\ -  0.74}$ & $  75.7$ & $  11.9$ & $   6.9$ & $   3.6$ & $   1.9$\\ 
CMS $h\to \gamma\gamma~(\text{untagged}~1)$~\cite{Khachatryan:2014ira} &  $8\tev$& $  0.92\substack{+  0.57\\ -  0.49}$ & $  85.1$ & $   7.9$ & $   4.0$ & $   2.4$ & $   0.6$\\ 
CMS $h\to \gamma\gamma~(\text{untagged}~2)$~\cite{Khachatryan:2014ira}&  $8\tev$ & $  1.10\substack{+  0.48\\ -  0.44}$ & $  91.1$ & $   4.7$ & $   2.5$ & $   1.4$ & $   0.3$\\ 
CMS $h\to \gamma\gamma~(\text{untagged}~3)$~\cite{Khachatryan:2014ira}&  $8\tev$ & $  0.65\substack{+  0.65\\ -  0.89}$ & $  91.5$ & $   4.4$ & $   2.4$ & $   1.4$ & $   0.3$\\ 
CMS $h\to \gamma\gamma~(\text{untagged}~4)$~\cite{Khachatryan:2014ira}&  $8\tev$ & $  1.46\substack{+  1.29\\ -  1.24}$ & $  93.1$ & $   3.6$ & $   2.0$ & $   1.1$ & $   0.2$\\ 
CMS $h\to \gamma\gamma~(\mathrm{VBF,dijet}~0)$~\cite{Khachatryan:2014ira} &  $8\tev$ & $  0.82\substack{+  0.75\\ -  0.58}$ & $  17.8$ & $  81.8$ & $   0.2$ & $   0.1$ & $   0.1$\\ 
CMS $h\to \gamma\gamma~(\mathrm{VBF,dijet}~1)$~\cite{Khachatryan:2014ira} &  $8\tev$& $ -0.21\substack{+  0.75\\ -  0.69}$ & $  28.4$ & $  70.6$ & $   0.6$ & $   0.2$ & $   0.2$\\ 
CMS $h\to \gamma\gamma~(\mathrm{VBF,dijet}~2)$~\cite{Khachatryan:2014ira} &  $8\tev$& $  2.60\substack{+  1.33\\ -  0.99}$ & $  43.7$ & $  53.3$ & $   1.4$ & $   0.8$ & $   0.8$\\ 
CMS $h\to \gamma\gamma~(Vh,E_T^\text{miss})$~\cite{Khachatryan:2014ira} &  $8\tev$& $  0.08\substack{+  1.86\\ -  1.28}$ & $  16.5$ & $   2.7$ & $  34.4$ & $  35.3$ & $  11.1$\\ 
CMS $h\to \gamma\gamma~(Vh,\text{dijet})$~\cite{Khachatryan:2014ira} &  $8\tev$& $  0.39\substack{+  2.16\\ -  1.48}$ & $  30.4$ & $   3.1$ & $  40.5$ & $  23.3$ & $   2.6$\\ 
CMS $h\to \gamma\gamma~(Vh,\text{loose})$~\cite{Khachatryan:2014ira} &  $8\tev$& $  1.24\substack{+  3.69\\ -  2.62}$ & $   2.7$ & $   1.1$ & $  77.9$ & $  16.8$ & $   1.5$\\ 
CMS $h\to \gamma\gamma~(Vh,\text{tight})$~\cite{Khachatryan:2014ira} &  $8\tev$& $ -0.34\substack{+  1.30\\ -  0.63}$ & $   0.2$ & $   0.2$ & $  76.9$ & $  19.0$ & $   3.7$\\ 
CMS $h\to \gamma\gamma~(tth, \text{multijet})$~\cite{Khachatryan:2014ira} &  $8\tev$& $  1.24\substack{+  4.23\\ -  2.70}$ & $   4.1$ & $   0.9$ & $   0.8$ & $   0.9$ & $  93.3$\\ 
CMS $h\to \gamma\gamma~(tth, \text{lepton})$~\cite{Khachatryan:2014ira} &  $8\tev$& $  3.52\substack{+  3.89\\ -  2.45}$ & $   0.0$ & $   0.0$ & $   1.9$ & $   1.9$ & $  96.1$\\ 
CMS $h\to \mu\mu$~\cite{Khachatryan:2014aep} &  $7/8\tev$& $  2.90\substack{+  2.80\\ -  2.70}$ & $  94.1$ & $   5.9$ & $   0.0$ & $   0.0$ & $   0.0$\\ 
CMS $h\to \tau\tau~(0j)$~\cite{Chatrchyan:2014nva}&  $7/8\tev$ & $  0.40\substack{+  0.73\\ -  1.13}$ & $  98.5$ & $   0.8$ & $   0.4$ & $   0.3$ & $   0.0$\\ 
CMS $h\to \tau\tau~(1j)$~\cite{Chatrchyan:2014nva} &  $7/8\tev$& $  1.06\substack{+  0.47\\ -  0.47}$ & $  79.7$ & $  12.1$ & $   5.2$ & $   3.0$ & $   0.0$\\ 
CMS $h\to \tau\tau~\mathrm{(VBF)}$~\cite{Chatrchyan:2014nva} &  $7/8\tev$& $  0.93\substack{+  0.41\\ -  0.41}$ & $  20.9$ & $  79.1$ & $   0.0$ & $   0.0$ & $   0.0$\\ 
CMS $Vh\to V(\tau\tau)$~\cite{Chatrchyan:2014nva}&  $7/8\tev$ & $  0.98\substack{+  1.68\\ -  1.50}$ & $   0.0$ & $   0.0$ & $  47.1$\tnote{1} & $  27.3$\tnote{1} & $   0.0$\\ 
CMS $Vh\to V(bb)$~\cite{Chatrchyan:2013zna}&  $7/8\tev$ & $  1.00\substack{+  0.51\\ -  0.49}$ & $   0.0$ & $   0.0$ & $  63.3$ & $  36.7$ & $   0.0$\\ 
CMS $Vh\to V(WW)\to 2\ell2\nu$~\cite{Chatrchyan:2013iaa} &  $7/8\tev$& $  0.39\substack{+  1.97\\ -  1.87}$ & $  60.2$ & $   3.8$ & $  22.8$ & $  13.2$ & $   0.0$\\ 
CMS $Vh\to V(WW)~(\text{hadr.})$~\cite{Chatrchyan:2013iaa} &  $7/8\tev$& $  1.00\substack{+  2.00\\ -  2.00}$ & $  63.7$ & $   3.3$ & $  21.9$ & $  11.1$ & $   0.0$\\ 
CMS $Wh\to W(WW)\to 3\ell3\nu$~\cite{Chatrchyan:2013iaa} &  $7/8\tev$& $  0.56\substack{+  1.27\\ -  0.95}$ & $   0.0$ & $   0.0$ & $ 100.0$ & $   0.0$ & $   0.0$\\ 
CMS $tth\to 2\ell~\text{(same-sign)}$~\cite{Khachatryan:2014qaa} &  $7/8\tev$& $  5.30\substack{+  2.10\\ -  1.80}$ & $   0.0$ & $   0.0$ & $   0.0$ & $   0.0$ & $  100.0$\tnote{2}\\ 
CMS $tth\to 3\ell$~\cite{Khachatryan:2014qaa}&  $7/8\tev$ & $  3.10\substack{+  2.40\\ -  2.00}$ & $   0.0$ & $   0.0$ & $   0.0$ & $   0.0$ & $  100.0$\tnote{3}\\ 
CMS $tth\to 4\ell$~\cite{Khachatryan:2014qaa} &  $7/8\tev$& $ -4.70\substack{+  5.00\\ -  1.30}$ & $   0.0$ & $   0.0$ & $   0.0$ & $   0.0$ & $  100.0$\tnote{4}\\ 
CMS $tth\to tt(bb)$~\cite{Khachatryan:2014qaa}&  $7/8\tev$ & $  0.70\substack{+  1.90\\ -  1.90}$ & $   0.0$ & $   0.0$ & $   0.0$ & $   0.0$ & $ 100.0$\\ 
CMS $tth\to tt(\gamma\gamma)$~\cite{Khachatryan:2014qaa}&  $7/8\tev$ & $  2.70\substack{+  2.60\\ -  1.80}$ & $   0.0$ & $   0.0$ & $   0.0$ & $   0.0$ & $ 100.0$\\ 
CMS $tth\to tt(\tau\tau)$~\cite{Khachatryan:2014qaa}&  $7/8\tev$ & $ -1.30\substack{+  6.30\\ -  5.50}$ & $   0.0$ & $   0.0$ & $   0.0$ & $   0.0$ & $ 100.0$\\ 
\midrule
D\O\ $h\to WW$~\cite{Abazov:2013gmz} & $1.96\tev$& $  1.90\substack{+  1.63\\ -  1.52}$ & $  77.5$ & $   5.4$ & $  10.6$ & $   6.5$ & $   0.0$\\ 
D\O\  $h\to bb$~\cite{Abazov:2013gmz} & $1.96\tev$& $  1.23\substack{+  1.24\\ -  1.17}$ & $   0.0$ & $   0.0$ & $  62.0$ & $  38.0$ & $   0.0$\\ 
D\O\  $h\to \gamma\gamma$~\cite{Abazov:2013gmz} & $1.96\tev$& $  4.20\substack{+  4.60\\ -  4.20}$ & $  77.5$ & $   5.4$ & $  10.6$ & $   6.5$ & $   0.0$\\ 
D\O\  $h\to \tau\tau$~\cite{Abazov:2013gmz} & $1.96\tev$& $  3.96\substack{+  4.11\\ -  3.38}$ & $  77.5$ & $   5.4$ & $  10.6$ & $   6.5$ & $   0.0$\\ 
\bottomrule
 \end{tabular}
   \begin{tablenotes}
 \footnotesize
 \item[1] The signal is contaminated to $16.2\%$ [$9.4\%$] by $WH\to WWW$ [$ZH\to ZWW$] in the SM.
 \item[2] The SM Higgs signal composition is $h\to WW$ ($73.3\%$), $h\to \tau\tau$ ($23.1\%$), $h\to ZZ$ ($3.6\%$)
 \item[3] The SM Higgs signal composition is $h\to WW$ ($71.8\%$), $h\to \tau\tau$ ($23.8\%$), $h\to ZZ$ ($4.4\%$).
 \item[4] The SM Higgs signal composition is $h\to WW$ ($53.0\%$), $h\to \tau\tau$ ($30.1\%$), $h\to ZZ$ ($16.9\%$).
 \end{tablenotes}
 \end{threeparttable}
 }
 \caption{Higgs signal strengths measurements from LHC CMS and Tevatron D\O\ collaboration.} 
  \label{Tab:HSobs2}
\end{table}


\clearpage

\bibliography{hifi}
\bibliographystyle{JHEP}

\end{document}